\documentclass[english,12pt]{article}
\usepackage{graphicx}
\usepackage{epsfig} 
\usepackage{times}
\usepackage{cite}
\usepackage{afterpage}
\usepackage{hyperref}
\usepackage{color}
\usepackage[T1]{fontenc}
\usepackage[latin9]{inputenc}
\usepackage{amsmath}
\usepackage{graphicx}
\usepackage{amssymb}
\usepackage{babel}
\usepackage{lipsum}

\usepackage{ulem} 
\usepackage{color}

\newcommand{\gx}{\textsc{GlueX}}

\def\lambdabar{\lambda\kern-1ex\raise0.65ex\hbox{-}}
\newcommand{\be}{\begin{eqnarray}}
\newcommand{\ee}{\end{eqnarray}}
\newcommand{\bc}{\begin{center}}
\newcommand{\ec}{\end{center}}
\newcommand{\beq}{\begin{eqnarray}}
\newcommand{\eea}{\end{eqnarray}}

\usepackage[margin=1in]{geometry}
\begin{document}
\hfill{\bf Date:} {\today}
\begin{center}
{\large Proposal for JLab PAC45}
\end{center}
\begin{center}
{\large PR12--17--001}
\end{center}
\hspace{0.05in}
\begin{center}
\Large{Strange Hadron Spectroscopy with a Secondary $K_L$ Beam at GlueX}
\end{center}
\hspace{0.05in}
\begin{center}
S.~Adhikari$^{12}$,
H.~Al~Ghoul$^{13}$,
A.~Ali$^{17}$,
M.~J.~Amaryan$^{40,\ast,\dagger}$,
E.~G.~Anassontzis$^2$,
A.~V.~Anisovich$^{20,14}$,
A.~Austregesilo$^{32}$,
M.~Baalouch$^{40}$,
F.~Barbosa$^{32}$,
A.~Barnes$^9$
% ...
\end{center}
\vspace{0.15in}
\noindent 
% ...
$^2$    National and Kapodistrian University of Athens, Athens 15771, 
	Greece \\
% ...
$^9$    University of Connecticut, Storrs, CO 06269, USA \\
% ...
$^{12}$ Florida International University, Miami, FL 33199, USA \\
$^{13}$ Florida State University, Tallahassee, FL 32306, USA \\
$^{14}$ National Research Centre "Kurchatov Institute", Petersburg Nuclear
        Physics Institute, Gatchina 188300, Russia \\
% ...
$^{17}$ GSI Helmholtzzentrum f\"ur Schwerionenforschung GmbH, Darmstadt 
	64291, Germany \\
% ...
$^{20}$ Helmholtz-Institut f\"ur Strahlen- und Kernphysik, Universit\"at
        Bonn, Bonn 53115, Germany \\
% ...
$^{32}$ Thomas Jefferson National Accelerator Facility, Newport News,
        VA 23606, USA \\
% ...
$^{40}$ Old Dominion University, Norfolk, VA 23529, USA \\
% ...
$^\ast$ Contact Person: mamaryan@odu.edu \\
$^\dagger$ Spokesperson \\
\noindent

\hspace{0.1in}
\begin{center}
\large{(The \textsc{GlueX} Collaboration)}
\end{center}

\newpage
\begin{center}{\bf Abstract}\end{center}

We propose to create a secondary beam of neutral
kaons in Hall~D at Jefferson Lab to be used with the GlueX experimental
setup for strange hadron spectroscopy.  A flux on the order of 
$3\times 10^4~K_L$/s will allow a broad range of measurements to be made 
by improving the statistics of previous data obtained on hydrogen targets 
by three orders of magnitude. Use of a deuteron target will provide first 
measurements on the neutron which is {\it terra incognita}.

The experiment will measure both differential cross sections
and self-analyzed polarizations of the produced $\Lambda$,
$\Sigma$, $\Xi$, and $\Omega$ hyperons using the GlueX detector 
at the Jefferson Lab Hall~D. The measurements will span c.m.
$\cos\theta$ from $-0.95$ to 0.95 in the c.m. range above $W = 
1490$~MeV and up to 3500~MeV. These new GlueX data will greatly 
constrain partial-wave analyses and reduce model-dependent 
uncertainties in the extraction of strange resonance properties 
(including pole positions), and provide a new benchmark for 
comparisons with QCD-inspired models and lattice QCD 
calculations.

The proposed facility will also have an impact in the strange
meson sector by providing measurements of the final-state $K\pi$
system from threshold up to 2~GeV invariant mass to establish 
and improve on the pole positions and widths of all $K^{\ast}(K\pi)$ 
P-wave states as well as for the S-wave scalar meson $\kappa(800)$.

\newpage
\tableofcontents

\newpage
\pagenumbering{arabic}
\setcounter{page}{1}
\section{Executive Summary}
\label{sec:Exec}

We propose to establish a secondary $K_L$ beam line at JLab Hall~D for 
scattering experiments on both proton and neutron (for the first time) 
targets in order to determine the differential cross sections and the 
self-polarization of strange hyperons with the GlueX detector to enable 
precise partial wave analysis in order to determine all the resonances 
up to 2400~MeV in the spectra of the $\Lambda, \Sigma, \Xi$, and 
$\Omega$ hyperons. 

In addition, we intend to do strange meson spectroscopy by studies of 
the $\pi -K$ interaction to locate the pole positions in the $I=1/2$ 
and $3/2$ channels.

The $K_L$ beam will be generated by directing a high energy, high 
intensity photon beam onto a Be target in front of the GlueX detector. 
The flux of the $K_L$ beam will be of the order $3\times 10^4~K_L$/s 
on a liquid hydrogen/deuterium cryotarget within the GlueX detector, 
which has a large acceptance with coverage of both charged and neutral 
particles. This flux will allow statistics in the case of hydrogen 
targets to exceed that of earlier experiments by almost three orders 
of magnitude. The main components of the experimental setup are the 
compact photon source, the Be target with a beam plug, sweeping magnet 
and a pair spectrometer.

The physics case for the experiments is aligned with the \textit{2015 
Long Range Plan for Nuclear Science}~\cite{LRP}: ``\textit{...a better 
understanding of the role of strange quarks became an important 
priority}". Knowledge of the hyperon spectra is an important component 
for this. The empirical knowledge of the low lying spectra of 
the $\Lambda$ and $\Sigma$ hyperons remains very poor in comparison 
with that of the nucleon, and in the case of the $\Xi$ hyperons 
extremely poor. The structure of these hyperon resonances cannot be 
understood without empirical determination of their pole positions 
and decays, which is the goal of the proposed experiments. The 
determination of the strange hyperon spectra in combination with 
the current measurements of the spectra of the charm and beauty 
hyperons at the LHCb experiment at CERN should allow a clear 
understanding of soft QCD matter and the approach to heavy quark 
symmetry.

As the first stage of the GlueX program the focus will be on two-body 
and quasi-two-body: elastic $K_Lp\to K_Sp$ and charge-exchange 
$K_Lp\to K^+n$ reactions, then on two-body reactions producing $S = 
-1 (S = -2)$ hyperons as $K_Lp \to \pi^+\Lambda$, $K_Lp \to \pi^+
\Sigma^0$, and $K_Lp \to \pi^0\Sigma^+ (K_Lp \to K^+\Xi^0)$, as well 
as three body $K_Lp \to K^+K^+\Omega-$.

For analyzing the data a coupled channel partial wave analysis will 
be done of the GlueX data in parallel with an analysis of the data 
from the J-PARC $K^-$ measurements, when available. The best fit will 
determine the partial wave amplitudes and the resonance pole positions, 
residues and Breit-Wigner parameters. These will provide a benchmark 
for results of forthcoming QCD lattice calculations and lead to the 
desired understanding of the structure of the strange hyperons.

Our timeline is to begin $K_L$ beam experiments at the completion of 
the current GlueX physics program.

\newpage
\section{Scope of the Proposal} 
\label{sec:Scope}

The nature of QCD confinement continues to provide a
challenge to our understanding of soft QCD. Studies of the 
baryon spectrum provide one obvious avenue to understand 
this region since the location and properties of excited 
states reflect the dynamics and relevant degrees-of-freedom 
of hadrons.

Through analyses of decades worth of data, from both hadronic
and electromagnetic (EM) scattering experiments,
numerous baryon resonances have been observed, with their masses, 
widths, and quantum numbers fully determined. There are 109 baryons 
in the PDG2016 listings but only 58 of them are $4^\ast$ or
$3^\ast$~\cite{PDG2016}. Many more states are predicted 
by quark models (QMs). For example, in the case of $SU(6)\times 
O(3)$, 434 resonances would be required, if all partly revealed 
multiplets were completed (three 70 and four 56).

Three light quarks can be arranged in six baryonic families, 
$N^\ast$, $\Delta^\ast$, $\Lambda^\ast$, $\Sigma^\ast$, $\Xi^\ast$, 
and $\Omega^\ast$. The possible number of members in a family 
is not arbitrary~\cite{Nefkens97}. If the $SU(3)_F$ symmetry of 
QCD is controlling, then for the octet: $N^\ast$, $\Lambda^\ast$, 
and $\Sigma^\ast$, and for the decuplet: $\Delta^\ast$, $\Sigma^\ast$, 
$\Xi^\ast$, and $\Omega^\ast$. The number of experimentally 
identified resonances in each baryon family in PDG2016 summary 
tables is 17 $N^\ast$, 24 $\Delta^\ast$, 14 $\Lambda^\ast$, 12 
$\Sigma^\ast$, 7 $\Xi^\ast$, and 2 $\Omega^\ast$. Constituent QMs, 
for instance, predict the existence of no fewer than 64 $N^\ast$ 
and 22 $\Delta^\ast$ states with mass less than 3~GeV. The 
seriousness of the ``missing-states" problem~\cite{KI80} is obvious 
from these numbers. To complete $SU(3)_F$ multiplets, one needs no 
fewer than 17 $\Lambda^\ast$s, 41 $\Sigma^\ast$s, 41 $\Xi^\ast$s, 
and 24 $\Omega^\ast$s.

If these ``missing resonances" exist, they have either eluded 
detection or have produced only weak signals in the existing data 
sets. The search for such resonances provides a natural motivation 
for future measurements at Jefferson Lab.  As stated in the 
\textit{2015 Long Range Plan for Nuclear Science}~\cite{LRP}: 
\textit{For many years, there were both theoretical and experimental 
reasons to believe that the strange sea-quarks might play a 
significant role in the nucleon's structure; a better understanding 
of the role of strange quarks became an important priority.}

Low-lying baryon resonances, both hyperons and non-strange states, 
are usually considered to be three-quark systems; however, those 
quarks are constituent, not current ones. This prevents their 
description by the well-understood perturbative QCD. It seems, 
however, that some qualitative consequences of perturbative QCD 
still apply even for the non-perturbative constituent quarks. One 
of them is the suppression of the effective strong interaction for 
the heavier strange quark in comparison with the lighter up and 
down flavored quarks (due to the asymptotic freedom). This is 
revealed, e.g., in smaller widths of hyperon resonances as compared 
with similar non-strange baryon resonances. The same phenomenon is 
seen also for meson resonances (compare the widths of $K^\ast$ and 
$\rho$ meson resonances). Further investigation of this and other 
similar properties may help to improve our understanding of the 
nature of the constituent quarks and other non-perturbative effects.

The JLab 12~GeV energy upgrade, with the new Hall~D, is an ideal 
tool for extensive studies of non-strange and, specifically, strange
baryon resonances~\cite{Ghoul16,AlekSejevs13}. Our plan is to take 
advantage of the existing high-quality photon beam line and 
experimental area in the Hall~D complex at Jefferson Lab to deliver 
a beam of $K_L$ particles onto a liquid hydrogen/deuterium 
cryotarget (LH$_2$/LD$_2$) within the GlueX detector. 

The recently constructed GlueX detector in Hall~D is a large 
acceptance spectrometer with good coverage for both charged and 
neutral particles that can be adapted to this purpose. Obviously, 
a $K_L$ beam facility with good momentum resolution is crucial 
to provide the data needed to identify and characterize the 
properties of hyperon resonances. The masses and widths of the 
lowest $\Lambda$ and $\Sigma$ baryons were determined mainly with 
kaon beam experiments in the 1970s~\cite{PDG2016}.  First 
determinations of the pole position in the complex-energy plane 
for a hyperon, for instance for the $\Lambda(1520)3/2^-$, has 
been made only recently~\cite{Qiang2010}. An intense $K_L$ beam 
would open a new window of opportunity, not only to locate 
``missing resonances", but also to establish their properties by 
studying  different decay channels systematically.

The recent white paper, dedicated to the physics with meson beams
and endorsed by a broad physics community, \underline{summarized}
unresolved issues in hadron physics, and outlined the vast
opportunities and advances that only become possible with a
``secondary beam facility"~\cite{Briscoe2015}.  The Hall~D GlueX 
K-long Facility (KLF) measurements will allow studies of very 
poorly known multiplets of $\Lambda^\ast$, $\Sigma^\ast$, 
$\Xi^\ast$, and even $\Omega^\ast$ hyperons with unprecedented
statistical precision.  These measurements also have the potential 
to observe dozens of predicted (but heretofore unobserved) states 
and to establish the quantum numbers of already observed hyperons 
listed in PDG2016~\cite{PDG2016}. Interesting puzzles exist for 
PDG-listed excited hyperons that do not fit into any of the 
low-lying excited multiplets, and these need to be further 
revisited and investigated. Excited $\Xi$s, for instance, are very 
poorly known. Establishing and discovering new states is important, 
in particular, for determination of the multiplet structure of
excited baryons.

We have organized three Workshops: \textit{Physics with Neutral Kaon
Beam at JLab} (KL2016) (February 2016)~\cite{KL2016}, \textit{Excited
Hyperons in QCD Thermodynamics at Freeze-Out} (YSTAR2016) (November
2016)~\cite{YSTAR2016}, and \textit{New Opportunities with
High-Intensity Photon Sources} (HIPS2017) (February
2017)~\cite{HIPS2017}.  They were dedicated to the physics of hyperons
produced by the neutral kaon beam.  The KL2016 Workshop~\cite{ProcKL}
followed our LoI--12--15--001~\cite{LoI} to help address the
comments made by PAC43 and to prepare the full proposal for PAC45.
The proposed GlueX KLF program is complementary, for instance, to the 
CLAS12 baryon spectroscopy experiments~\cite{VSP,KY} and would operate 
in Hall~D for several years.  The YSTAR2016 Workshop~\cite{ProcYS} was 
a successor to the recent KL2016 Workshop and considered the influence 
of possible ``missing" hyperon resonances on QCD thermodynamics, on 
freeze-out in heavy ion collisions and in the early universe, and in 
spectroscopy. Finally, the HIPS2017 Workshop~\cite{ProcHI} aimed at 
producing an optimized photon source concept with potential increase 
of scientific output at Jefferson Lab, and at refining the science for 
hadron physics experiments benefitting from such a high-intensity 
photon source.

Additionally, the proposed facility will also have a great impact in
the strange meson sector by measurements of the final-state $K\pi$ 
system from threshold up to 2~GeV in invariant mass to establish
and improve on pole positions and widths of all $K^{\ast}(K\pi)$
$P$-wave states and the $S$-wave scalar meson $\kappa(800)$. In 
particular, the $\kappa(800)$ meson has been under discussion for 
decades and still remains to be unequivocally confirmed with 
corresponding quantum numbers by doing detailed phase-shift analysis 
with high statistics data~\cite{kappa}. A detailed study of the 
$K\pi$ system is very important to extract the so-called $K\pi$ 
vector and scalar form factors
to be compared with $\tau\to K\pi\nu_{\tau}$ decay and can be used to
constrain the $V_{us}$ Cabibbo-Kobayashi-Maskawa (CKM) matrix
element as well as to be used in testing CP violation in decays of
heavy $B$ and $D$ mesons into $K\pi\pi$ final states.

The proposal is organized in the following manner. We give an Executive 
Summary in Sec.~\ref{sec:Exec} and the Scope of the proposal in 
Sec~\ref{sec:Scope}.  Then the Brief 
Case of Hyperon Spectroscopy is given in Sec.~\ref{sec:Brief} while 
Hyperons in Lattice studies are presented in Sec.~\ref{sec:Lattice}.
An overview of the Interest of the RHICH/LHC community in Hyperon 
measurements is summarized in Sec.~\ref{sec:friez}. The short 
overview of previous bubble chamber measurements is given in 
Sec.~\ref{sec:data}. Partial-wave phenomenology is considered in 
Sec.~\ref{sec:PWA} and Theory for the ``Neutron" Target in 
Sec.~\ref{sec:Maxim}. A short overview for Strange Meson 
Spectroscopy is given in Sec.~\ref{sec:Moskov}.  Our Proposed 
measurements are reported in Sec.~\ref{sec:run}.  It describes
a Compact Photon Source, $K_L$ 
production and $K_L$ beam properties, Start Counter Resolution, 
measurements of $K_L$ flux, and cryotarget description.  Running 
conditions are described in Sec.~\ref{sec:RC}. 
Sec.~\ref{sec:Cover} contains a Cover Letter for the KLF proposal 
submission. The Appendixes contain many technical details for 
our proposal: Analysis of Three-Body Final States in 
Appendix~A1~\ref{sec:A1}, 
Determination of Pole Positions in Appendix~A2~\ref{sec:A2}, 
Statistics Tools for Spectroscopy of Strange Resonances in 
	Appendix~A3~\ref{sec:A3}, 
Neutron Background in Appendix~A4~\ref{sec:A4}, 
Details of Monte Carlo Study in Appendix~A5~\ref{sec:A5}, 
Current Hadronic Projects in Appendix~A6~\ref{sec:A6}, 
Additional Physics Potential with a $K_L$ Beam~Appendix in 
	A7~\ref{sec:A7}, 
and List of New Equipment and of Changes in Existing Setup Required 
	in Appendix~A8~\ref{sec:A8}.

\section{The Brief Case for Hyperon Spectroscopy} 
\label{sec:Brief}

Our present experimental knowledge of the strange hyperon
spectrum is deplorably incomplete, despite the fact that
the ground states of the strange hyperons have been known
since the 1960s. In the case of the $\Lambda$ hyperon
resonance spectrum, only the lowest negative-parity doublet
is well established even though the structure of these
resonances remains under discussion. In the case of the
$\Sigma$ and $\Xi$ hyperons, only the lowest decuplet
resonance states $\Sigma(1385)$ and $\Xi(1530)$ are well
understood.

The masses of the lowest positive-parity resonances in the 
spectrum of the $\Lambda$ and $\Sigma$ hyperons, 
the $\Lambda(1600)$ and $\Sigma(1660)$ are 
experimentally known, but their structure is not. In the case 
of the $\Xi$ hyperon, the lowest positive-parity resonance 
remains unobserved.

To settle the nature of the hyperon resonances, their main
decay modes have to be determined by experiment. A clear
example of how the decay modes can settle the structure of
the resonances is provided by the $\pi$-decay widths of
the decuplets $\Delta(1232)$, $\Sigma(1385)$, and $\Xi(1530)$.
The ratio of these decay widths is 13:4:1, whereas if they
were simple three-quark states, with 3, 2, and 1 light quarks 
each, the ratio should be 9:4:1. A comparison of these ratios
indicates that the $\Sigma(1385)$ and $\Xi(1530)$ appear to
be three-quark states, while the $\Delta(1232)$ is more complex
and formed by a three-quark core with a surrounding meson (or
multiquark) cloud. This conclusion is well supported by
extensive theoretical calculations~\cite{Julia07,Sato09}.

\subsection{The $\Lambda(1405) - \Lambda(1520)$ $1/2^--3/2^-$ 
	Doublet}

In the simplest constituent quark model, the most natural $-$ 
and the oldest $-$ interpretation, is that the $\Lambda(1405) - 
\Lambda(1520)$ $1/2^--3/2^-$ doublet is a low-lying flavor
singlet multiplet of three quarks (\textit{uds}). Dynamical
versions of this model, with two-body interactions between the
quarks can describe the low mean energy of this multiplet, but
not the 115~MeV splitting between them. This has led to
suggestions that there may even be two different 1/2$^-$ states
$-$ one dynamical low $\overline{K}N$ resonance at 1405~MeV,
and an unresolved higher state close to 1520~MeV~\cite{Liu17}.
If so, it is high time that the ``missing" 1/2$^-$ higher-energy
state be empirically identified. This problem indicates that the
$\Lambda(1405)$ has a more complex multiquark structure. 
This structure is tested in modern theoretical 
approached, including contraints from unitarity and chiral symmerty.
Confirmed by multiple calculations later on, a two pole structure 
of $\Lambda(1405)$ was found in Ref.~\cite{Oller2001}. The narrow 
pole lies slightly below $\bar KN$ threshold fixed by the scattering 
data rather well, see Ref.~\cite{Maxim16} for the comparison of 
different modern coupled-channel approaches.  However, the position 
of the second pole is determined less precisely, lying much further 
below $\bar KN$ threshold and deeper in the complex plane. Recent 
photoproduction data on $\pi\Sigma$ by CLAS~\cite{Moriya:2013eb}
can be used to reduce the theoretical ambiguity on this (second) 
pole of $\Lambda(1405)$ as demonstrated in Ref.~\cite{Mai:2014xna}.
Modern lattice QCD (LQCD) calculations also support the view that 
its structure is a $\overline{K}N$ state~\cite{Kamleh16}.  In 
Skyrme's topological soliton model for the baryons, the 
low-lying $\Lambda(1405)$ state also appears naturally as a 
mainly 5-quark state~\cite{Scoccola88,Callan88}.  That model is 
consistent with QCD in the large color number ($N_C$) limit.  

There are similar low-lying 
flavor-singlet parity doublets in both the charm and bottom 
hyperon spectra: $\Lambda_c(2595)$--$\Lambda_c(2625)$ $1/2^-$ -- 
$3/2^-$ and $\Lambda_b(5912)$--$\Lambda_b(5920)$ $1/2^-$ -- $3/2^-$
doublets~\cite{PDG2016}. The ratio between the $1/2^-$ -- $3/2^-$
splittings in these three doublets are 8.2:2.1:1, which is not
far from the corresponding inverse ratios of the $K$, $D$, and $B$
meson masses: 10.7:2.8:1. The latter is what one should expect from
the gradual approach to heavy-quark symmetry with increasing
meson (or constituent quark) mass if the quark structure of
these three multiplets is similar. This pattern is also consistent 
with the large N$_C$ limit of QCD.

\subsection{The Low-Lying Positive-Parity Resonances}

In the spectra of the nucleon and the $\Lambda$ and $\Sigma$
hyperons, the lowest positive-parity resonances all lie below 
the lowest negative-parity multiplets except for the flavor
singlet doublet $\Lambda(1405)-\Lambda(1520)$ $1/2^--3/2^-$.
This reversal of normal ordering cannot be achieved in the 
constituent quark model with purely color-spin-dependent quark 
interactions. These low-lying positive-parity resonances are 
the $N(1440)$, $\Lambda(1600)$, and the $\Sigma(1660)$ 1/2$^+$ 
states. Their low masses do however
appear naturally, if the interactions between the quarks are
flavor dependent~\cite{Glozman96}.

Present day LQCD calculations have not yet converged
on whether these low-lying states can be described as having
a mainly three-quark structure~\cite{Liu14}. This may reflect 
that there is a collective nature in the quark content of all
these resonances, which have a low soft vibrational mode
Skyrme's topological soliton model for the baryons, which
represents one version of the large $N_C$ limit of QCD,
describes these low-lying states as such vibrational states.

In the spectrum of the $\Xi$, the $\Xi(1690)$ may be such a
1/2$^+$ state as well, although the quantum numbers of that
state are yet to be determined.

In the corresponding decuplet spectra, a similar low-lying
positive-parity state has so far only been definitely
identified in the $\Delta(1232)$ spectrum: namely, the
$\Delta(1600)3/2^+$. The $\Sigma(1840)3/2^+$ resonance very 
likely represents the corresponding positive-parity 
$\Sigma^\ast$ state. It should be important to identify the 
corresponding $3/2^+$ state in the spectrum of the $\Xi^\ast$.

It is of course very probable that corresponding low-lying
positive-parity states will be found in the spectra of the
$\Lambda_c$ and $\Lambda_b$ hyperons, given the fact that
they have low-lying negative-parity states akin to those of
the $\Lambda$ hyperon as described above. The experimental
identification of those is an important task. Even if the
still tentative resonance $\Lambda_c(2765)$ turns out to be
a 1/2$^+$ state, its energy appears to be too high for being
the equivalent of the $\Lambda(1600)$ in the charm hyperon
spectrum.

In the spectrum of the $\Sigma_c$, the decuplet state
$\Sigma_c(2520)$ is well established. The tentative resonance
$\Sigma_c(2800)$ may, should it turn out to be a 1/2$^+$
state, correspond to the $\Sigma(1660)$ in the strange
hyperon spectrum.

\subsection{The Negative-Parity Hyperon Resonances}

In the spectrum of the nucleon, two well-separated groups of
negative-parity resonances appear above the 1/2$^+$ state
$N(1440)$. In the three-quark model, the symmetry of the lowest
energy group is [21]$_{FS}$[21]$_F$[21]$_S$; i.e., it has
mixed flavor (F) and spin (S) symmetry as well as mixed
flavor-spin (FS) symmetry~\cite{Glozman96,Coester98}. This
group consists of the $N(1535)1/2^-$ and the $N(1520)3/2^-$
resonances.  There is a direct correspondence in
the $\Lambda(1670)1/2^-$ and the $\Lambda(1690)3/2^-$
resonances. There is also a repeat of this group in the
spectrum of the $\Sigma$ hyperon in the two resonances
$\Sigma(1620)1/2^-$ (tentative) and $\Sigma(1670)3/2^-$.

These spin $1/2^-$ and $3/2^-$ states in the spectum of the nucleon
have intriguing decay patterns. The $N(1535)$ resonance has a
large (32-52\%) decay branch to $\eta N$, even though its energy
lies very close to the $\eta N$ threshold. This pattern
repeats in the case of the the $\Lambda(1670)$, which
also has a substantial (10-25\%) decay branch to the
corresponding $\eta\Lambda$ state, even though it lies even
closer to the threshold for that decay. As the still uncertain
$\Sigma(1620)1/2^-$ resonance is located almost exactly at the
threshold for $\eta\Sigma$, there is naturally no signal for an
$\eta\Sigma$ decay from it. The ratio of the $\eta$
decay widths of the $N(1535)$ and the $\Lambda(1670)$ is about
6:1, which suggests that the $\eta$ decay might involve a pair of 
quarks rather than a single constituent quark as in the $\pi$ 
decay of the decuplet resonances.

In the spectrum of the $\Xi$ hyperon, none of the negative-parity
multiplets is complete. The state $\Xi(1820)3/2^-$ may be the
analog in the $\Xi$ spectrum of the states $N(1520)$,
$\Lambda(1670)$, and $\Sigma(1670)$. It should be important to
identify the lowest $1/2^-$ resonance in the $\Xi$ spectrum. If
that resonance lacks an $\eta$ decay branch, it would demonstrate
that the $\eta$ decay of the $1/2^-$ resonances in the spectra of
the nucleon, $\Lambda$ and $\Sigma$ involves two quarks.

It should also be important to determine whether the uncertain
``bumps" referred to in the Particle Data Tables labelled
$\Sigma(1480)$, $\Sigma(1560)$, and $\Xi(1620)$ represent true
resonances.

About 120~MeV above the $1/2^--3/2^-$ pair of nucleon resonances
$N(1535)$ and $N(1520)$, the nucleon spectrum has three negative-parity
resonances close in energy to one another. This multiplet
is formed of the $N(1650)1/2^-$, $N(1700)3/2^-$, and
$N(1675)5/2^-$ resonances. In the three-quark model the symmetry
configuration of these states are [21]$_{FS}$[21]$_F$[21]$_S$;
i.e., their spin configuration is completely symmetric.

The analogs in the spectrum of the $\Lambda$ of the first and
last of these nucleon resonances are the $\Lambda(1800)1/2^-$
and the $\Lambda(1830)5/2^-$ resonances. This correspondence
remains uncertain, however, because the missing 3/2$^-$ state
in this $\Lambda$ resonance multiplet has not yet been identified.

A common feature of all the 1/2$^-$ resonances in these
multiplets is their substantial $\eta$ decay branches.

Our present knowledge of the spectrum of the $\Xi$ hyperons
remains too incomplete to identify any member of the negative-parity
multiplet with the symmetry structure
[21]$_{FS}$[21]$_F$[21]$_S$.

\subsection{Summary for the Brief Case}

This overview shows that the present empirical knowledge of the
spectrum of the strange hyperons remains remarkably incomplete.
As a consequence, the quark structure of even the lowest-energy
resonances remains uncertain. Only an experimental determination
of the lowest-energy positive- and negative-parity hyperon
resonances and their decay branches would settle the main open
issues.

In the spectrum of the $\Lambda$ hyperon, there remains a
question of the existence of a 1/2$^-$ partner to the
$\Lambda(1520)3/2^-$ resonance. In addition, it should be
important to search for the missing 3/2$^-$ $\Lambda$ resonance
near 1700~MeV. Equally important would be the search for the
apparently ``missing" 3/2$^-$ state near 1750~MeV in the
spectrum of the $\Sigma$ hyperon.

Our present knowledge of the spectrum of the $\Sigma$ hyperons
remains too incomplete to identify any member of the
corresponding negative-parity multiplet formed of 1/2$^-$, 3/2$^-$,
and 5/2$^-$ resonances.

It should also be important to determine, whether the uncertain
``bumps" referred to in the Particle Data Tables labelled
$\Sigma(1480)$, $\Sigma(1560)$, and $\Sigma(1620)$ represent true
resonances~\cite{PDG2016}.

\section{Strange Hadrons from the Lattice} 
\label{sec:Lattice}

\begin{figure}[htpb]
\centering
{
    \includegraphics[width=0.7\textwidth,keepaspectratio]{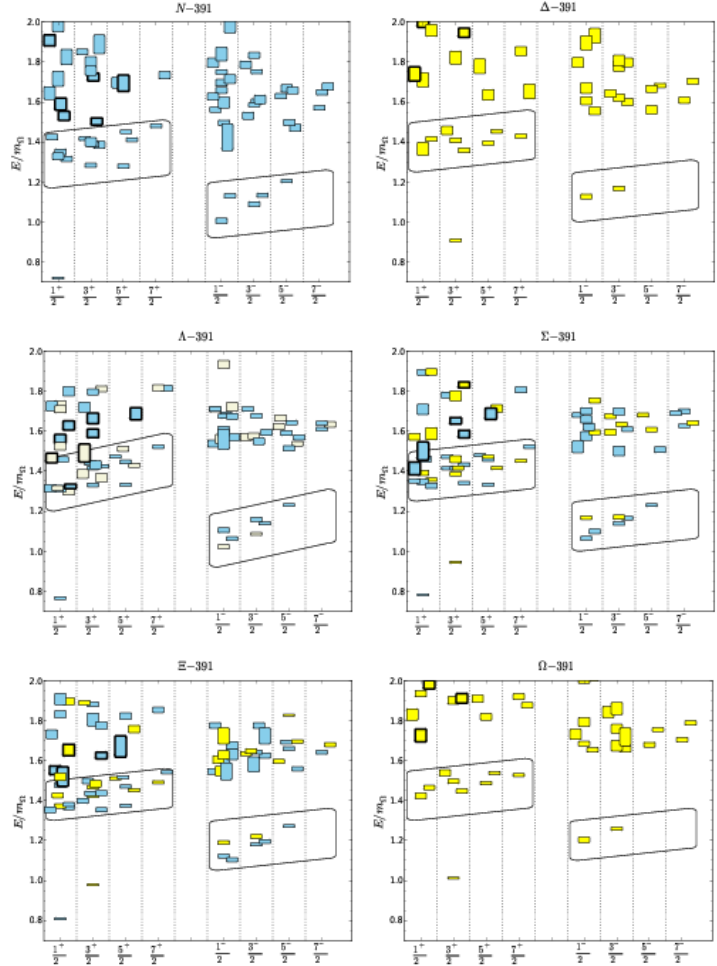} }

\centerline{\parbox{0.80\textwidth}{
    \caption[] {\protect\small Results for baryon excited states
       using an ensemble with $m_\pi = 391$~MeV are shown versus
       $J^P$~\protect\cite{Edwards2013}. Colors are used to display
       the flavor symmetry of dominant operators as follows: blue
       for \textbf{$8_F$} in $N$, $\Lambda$, $\Sigma$, and $\Xi$;
       beige for \textbf{$1_F$} for $\Lambda$; yellow for
       \textbf{$10_F$} in $\Delta$, $\Sigma$, $\Xi$, and $\Omega$.
       The lowest bands of positive- and negative-parity states are
       highlighted within slanted boxes.  Hybrid states, in
       which the gluons play a substantive role, are shown for
       positive parity by symbols with thick
       borders.} \label{fig:lqcd3}}}
\end{figure}

Our knowledge of the excited-state spectrum of QCD through the
solution of the theory on a Euclidean space-time lattice has undergone
tremendous advances over the past several years. What we characterize
as excited states are resonances that are unstable under the strong
interaction, and their properties are encapsulated in
momentum-dependent scattering amplitudes.  The means of calculating
such momentum-dependent phase shifts for elastic scattering on a
Euclidean lattice at finite volume was provided many years 
ago~\cite{Luscher:1990ux} and extended to systems in 
motion~\cite{Rummukainen:1995vs}, but its implementation for
QCD remained computationally elusive until recently. A combination of
theoretical, algorithmic, and computational advances has changed this
situation dramatically, notably in the case of mesons.  There have
been several lattice calculations of the momentum-dependent phase
shift of the $\rho$ mesons~\cite{Aoki:2007rd,Feng:2010es,Dudek:2012xn,
Guo:2016zos,Alexandrou:2017mpi,Bulava:2016mks,Lang:2011mn}.  
More recently, the formulation to extract amplitude information has 
been extended to the coupled-channel case\cite{Guo:2012hv,Briceno:2012yi,
Meissner:2014dea,Liu:2005kr,Lage:2009zv}, and applied to the case of the 
coupled $K\overline{K} - 
\pi\pi$~\cite{Wilson:2015dqa} system describing the $\rho$ resonance, 
and to the $\eta K - \eta\pi$\cite{Wilson:2014cna,Dudek:2014qha}.

The application to baryons is far more limited but, nonetheless,
important insights have been gained.  In an approach in which the
excited-state hadrons are treated as stable particles, a spectrum of
baryons at least as rich as that of the quark model is
revealed~\cite{Edwards:2011jj,Engel:2013ig}, and evidence has been 
presented for
``hybrid" baryon states, beyond those of the quark model, in which gluon 
degrees of freedom are essential~\cite{Dudek:2012ag}.  Notably, this 
picture extends to the spectrum of $\Lambda, \Sigma, \Xi$, and $\Omega$ 
states where the counting of states relects $SU(6) \times O(3)$ symmetry, 
and the presence of hybrids is common across the spectrum.  As indicated
above, these calculations are incomplete in that the momentum-dependent 
scattering amplitudes remain to be extracted. In Fig.~\ref{fig:lqcd3},
baryon spectra from~\cite{Edwards2013} are presented in units of
$\Omega$ mass from LQCD calculations with ensemble $m_\pi = 391$~MeV
(not yet at physical $m_\pi$).  However, in comparison with the case of 
mesons cited above, the challenges are more computational than theoretical 
or conceptual, and the progress made in the meson sector will be reflected 
for the case of baryons in the coming years.

\section{The Interest of the RHIC/LHC Community in Excited Hyperon 
	Measurements} 
\label{sec:friez}

The relativistic heavy-ion community at RHIC and the LHC has recently
embarked on specific analyses to address the issue of
strangeness hadronization. LQCD calculations in the QCD
crossover transition region between a deconfined phase of quark and
gluons and a hadronic resonance gas have revealed a potentially
interesting sub-structure related to the hadronization process.
Studies of flavor-dependent susceptibilities, which can be equated
to experimental measurements of conserved quantum-number fluctuations,
seem to indicate a slight flavor hierarchy in the three-quark sector
(u,d,s) in thermalized systems. Specifically, the ratios of higher-order 
susceptibilities in the strange sector show a higher transition
temperature than in the light sector~\cite{Bellwied:2013cta}. Both
pseudo-critical temperatures are still within the error bars of the
quoted transition temperature based on all LQCD order
parameters~\cite{Borsanyi:2010bp,Bazavov:2014pvz}, which is
154$\pm$9~MeV, but the difference of the specific susceptibilities is
around 18~MeV and well outside their individual uncertainties.

This difference seems to be confirmed by statistical thermal-model
calculations that try to describe the yields of emitted hadrons from
a QGP based on a common chemical freeze-out temperature. Although
the yields measured by ALICE at the LHC in 2.76~TeV PbPb collisions
can be described by a common temperature of 156$\pm$2~MeV, with a
reasonable $\chi$$^{2}$, the fit improves markedly if one allows the
light quark baryons to have a lower temperature than the strange
quark baryons~\cite{Floris:2014pta}. A similar result has been found
when the thermal fluctuations of particle yields as measured by
STAR Collaboration~\cite{Adamczyk:2013dal,Adamczyk:2014fia}, which can 
be related to the light quark dominated susceptibilities of the electric 
charge and the baryon number on the lattice, have been compared to 
statistical model calculations~\cite{Alba:2014eba}.

If one assumes that strange and light quarks indeed prefer different
freeze-out temperatures, then the question arises how this could impact
the hadronization mechanism and abundance of specific hadronic species.
In other words, is the production of strange particles, in particular
excited resonant states, enhanced in a particular temperature range in
the crossover region? Strange ground-state particle production shows
evidence of enhancement, but the most likely scenario is that the
increased strange quark abundance will populate excited states;
therefore, the emphasis of any future experimental program trying to
understand hadron production is shifting towards strange baryonic
resonance production. Furthermore, recent LHC measurements in small
systems, down to elementary proton-proton collisions, have revealed
that even in these small systems there is evidence for deconfinement,
if the achieved energy density, documented by the measured charged
particle multiplicity is large enough~\cite{Adam:2016emw}. Therefore,
future measurements of elementary collisions in the K-Long Facility 
experiment
at JLab might well provide the necessary link to future analysis of
strange resonance enhancements in heavy-ion collisions at RHIC and
the LHC and a deeper understanding of the hadronization process.

\begin{figure}[h!]
\centering
{
    \includegraphics[width=0.6\textwidth,keepaspectratio]{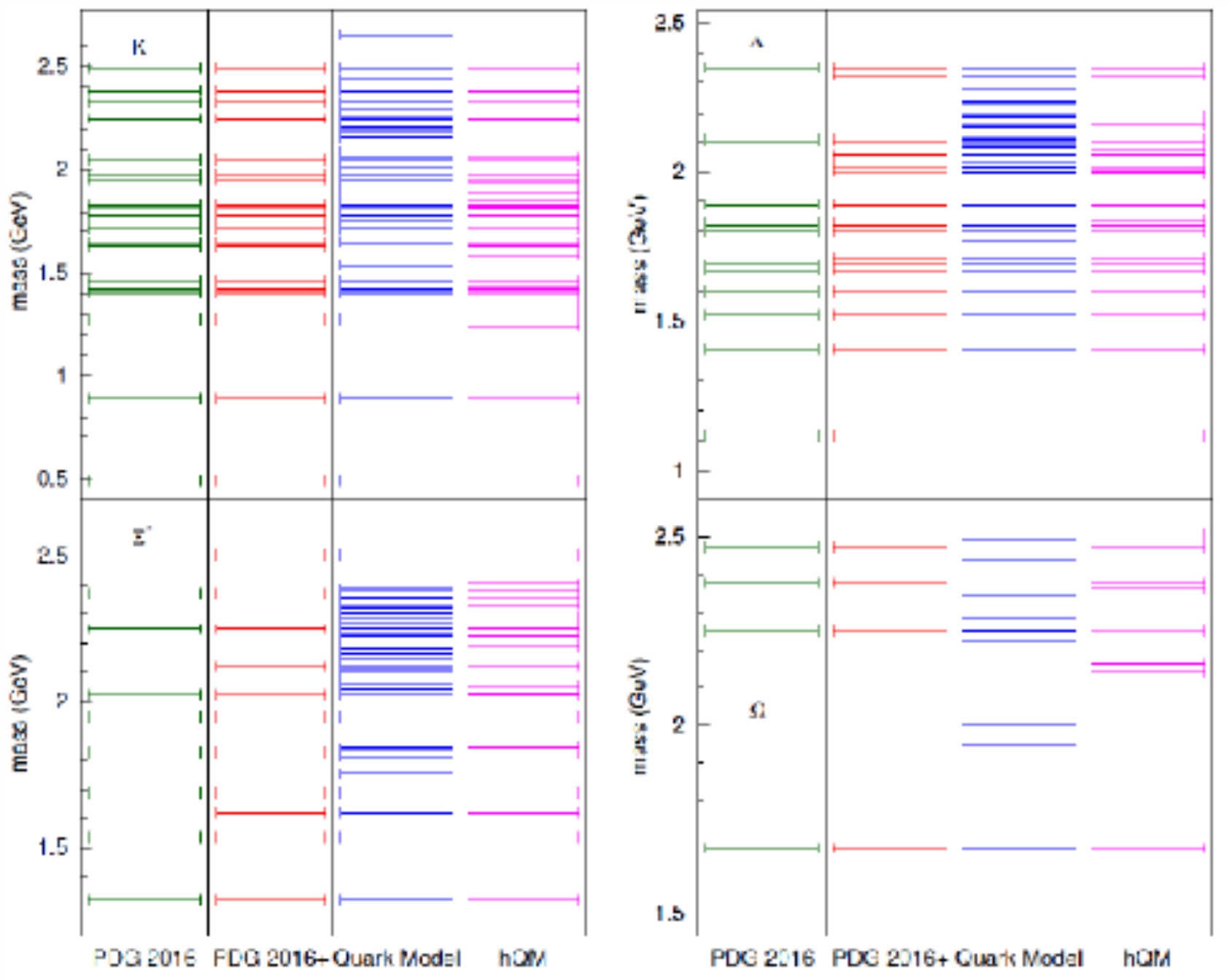} }
{
    \includegraphics[width=0.6\textwidth,keepaspectratio]{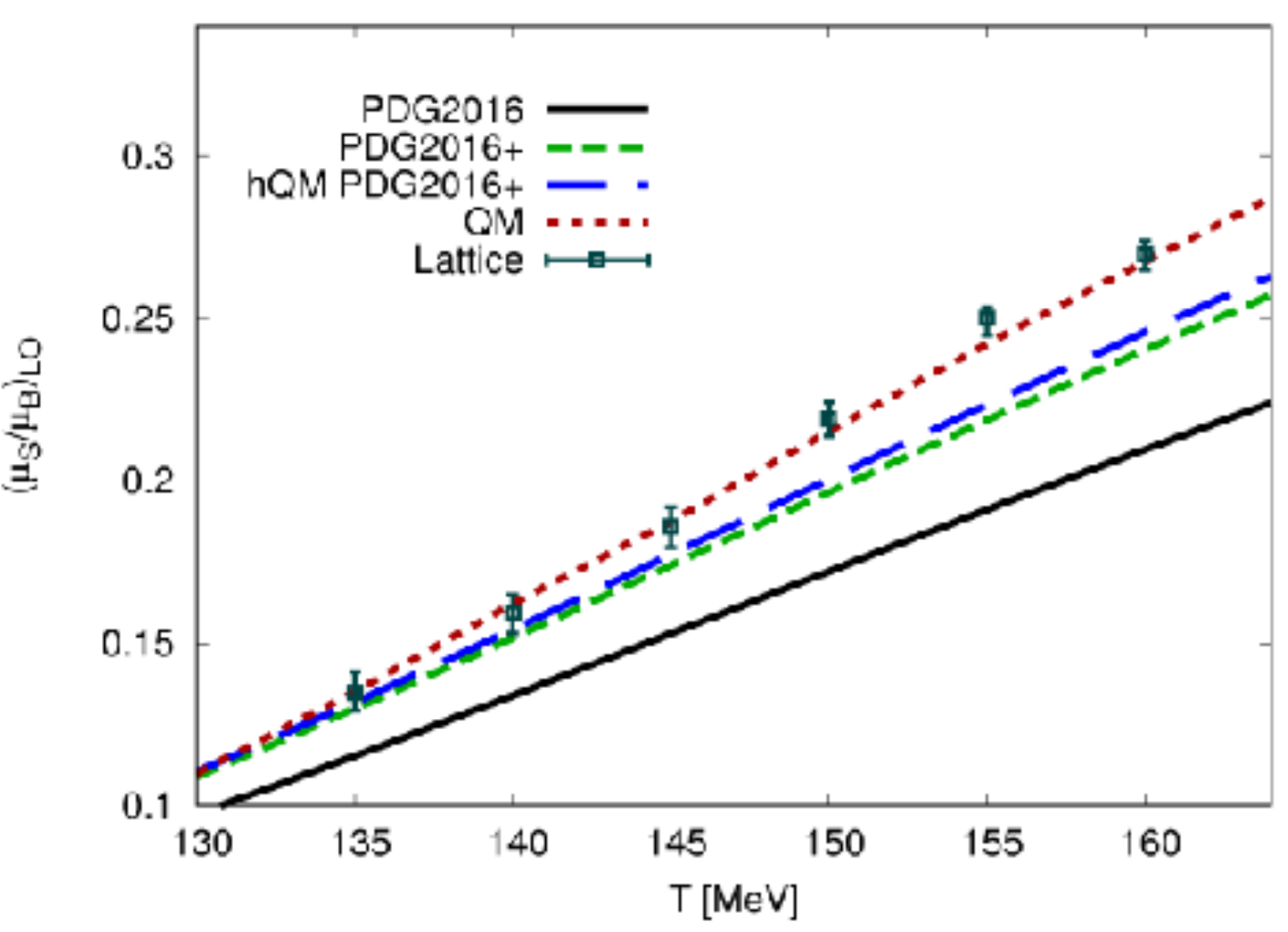} }

    \caption{Upper plot: Comparison of predicted and measured
    excited strange hadronic states in PDG-2016, PDG-2016+ (including
    one star states), QM, and hQM. 
    Lower plot: Lattice QCD calculation of the temperature dependence of 
    the leading order susceptibility ratio ($\mu_s$/$\mu_B$) compared to 
    results from HRG model calculations with varying number of hadronic 
    states.} \label{fig:figxx}
\end{figure}

This statement is also supported by comparisons between the
aforementioned LQCD calculations and model predictions based
on a non-interacting hadronic resonance gas.  The Hadron Resonance Gas
(HRG) model~\cite{Dashen:1969ep,Venugopalan:1992hy,Karsch:2003zq,
Tawfik:2004sw} yields a good description of most thermodynamic
quantities in the hadronic phase up to the pseudo-critical temperature.
The idea that strongly interacting matter in the ground state can be
described in terms of a non-interacting gas of hadrons and resonances,
which effectively mimics the interactions of hadrons by simply
increasing the number of possible resonant states exponentially as a
function of temperature, was proposed early on by
Hagedorn~\cite{Hagedorn:1976ef}. The only input to the model is the
hadronic spectrum: usually it includes all well-known hadrons in the
{\it Review of Particle Physics} (RPP), namely the ones rated with at
least two stars.
Recently, it has been noticed that some more differential observables
present a discrepancy between lattice and HRG model results. The
inclusion of not-yet-detected states, such as the ones predicted
by the original Quark Model (QM)~\cite{Capstick:1986bm,Ebert:2009ub}
has been proposed to improve the agreement~\cite{Majumder:2010ik,
Bazavov:2014xya}. A systematic study based on a breakdown of contributions
to the thermodynamic pressure given by particles grouped according to
their quantum numbers (in particular baryon number and strangeness)
enables us to infer in which hadron sector more states are needed
compared to the well-known ones from the RPP~\cite{Alba:2017mqu}.
In case of a flavor hierarchy in the transition region one would expect
the number of strange resonances to increase, due to a higher freeze-out
temperature, compared to the number of light-quark resonances.
Figure~\ref{fig:figxx} shows the effect of different strange hadron input
spectra to the HRG model in comparison to LQCD.
Figure~\ref{fig:figxx}(upper plot) shows the number of states in
PDG-2016~\cite{PDG2016}, PDG-2016+ (including one star states), the
standard QM, and a Quark Model with enhanced quark interactions in the
hadron (hyper-central model hQM ~\cite{Giannini:2015zia}).
Fig.~\ref{fig:figxx}(lower plot) shows a comparison
of the HRG results to a leading-order LQCD calculation of
$\mu_{s}$/$\mu_{B}$; i.e., the ratio to strange to baryon number
susceptibility\cite{Alba:2017mqu}.

An interesting conclusion that arises from these studies is that the
improvement in the listing of strange resonances between
PDG-2008~\cite{Amsler:2008zzb} and PDG-2016 definitely brought the
HRG calculations closer to the LQCD data. By looking at details
in the remaining discrepancy, which is in part remedied by including
one-star rated resonances in PDG-2016, it seems that the effect is more
carried by singly strange resonances rather than multi-strange resonances,
also in light of comparisons to quark models that include di-quark
structures~\cite{Santopinto:2014opa} or enhanced quark interactions in
the baryon (hypercentral models~\cite{Giannini:2015zia}). This is good
news for the experiments since the $\Lambda$ and $\Sigma$ resonances
below 2~GeV/$c^{2}$ are well within reach of the KLF experiment and,
to a lesser significance, the RHIC/LHC experiments. In this context it
is also important to point out that the use of both hydrogen and
deuterium targets in KLF is crucial since it will enable the measurement
of charged and neutral hyperons. A complete spectrum of singly strange
hyperon states is necessary to make a solid comparison to first-principle
calculations.

\underline{In summary:} Any comparisons between experimentally verified
strange quark-model states from YSTAR and LQCD will shed light
on a multitude of interesting questions relating to hadronization in
the non-perturbative regime, exotic particle production, the interaction
between quarks in baryons and a possible flavor hierarchy in the
creation of confined matter.

\section{Previous Measurements} 
\label{sec:data}

\begin{figure}[h!]
\centering
{
    \includegraphics[width=0.3\textwidth,keepaspectratio,angle=90]{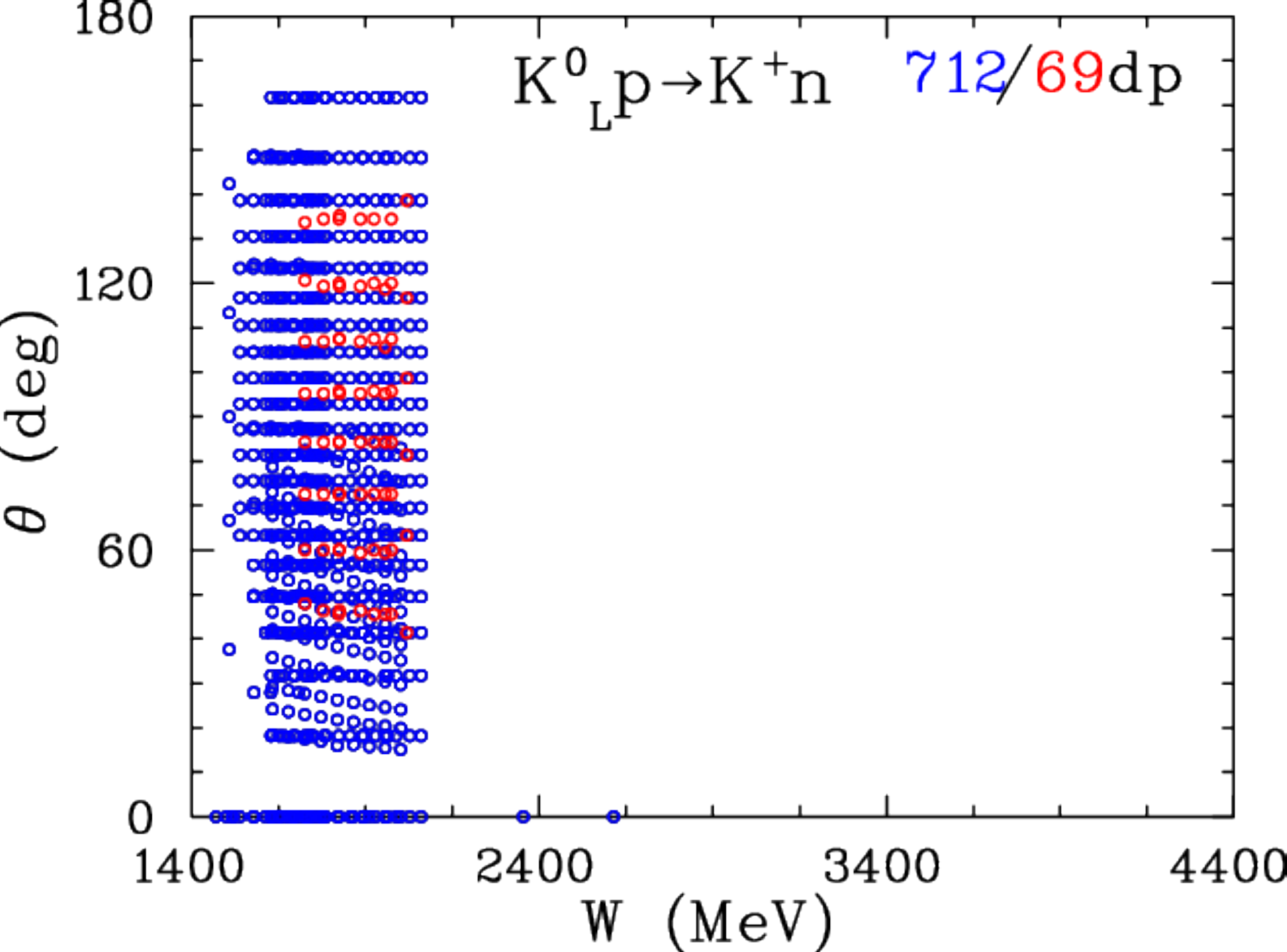} }
{
    \includegraphics[width=0.3\textwidth,keepaspectratio,angle=90]{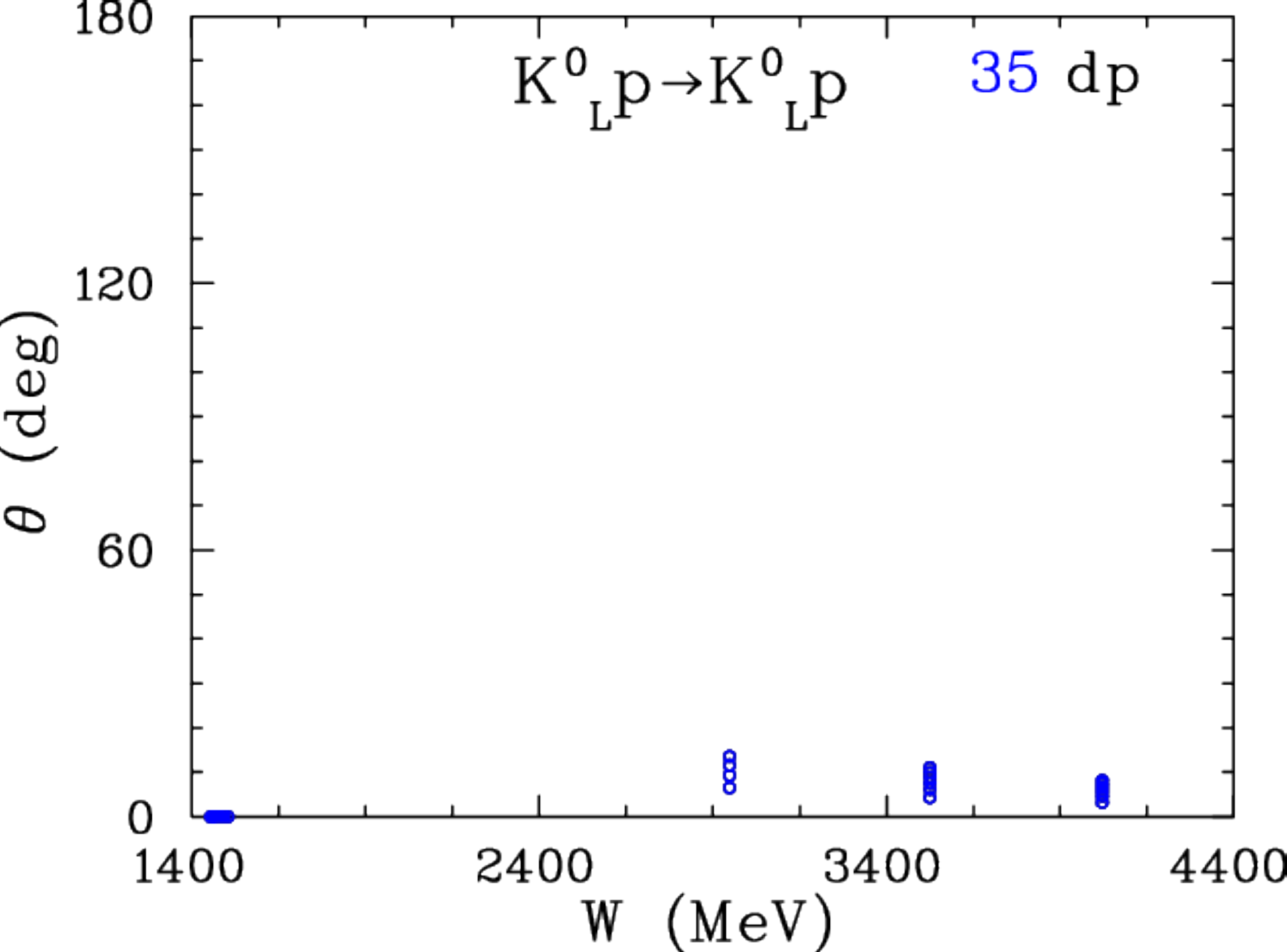} }
{
    \includegraphics[width=0.3\textwidth,keepaspectratio,angle=90]{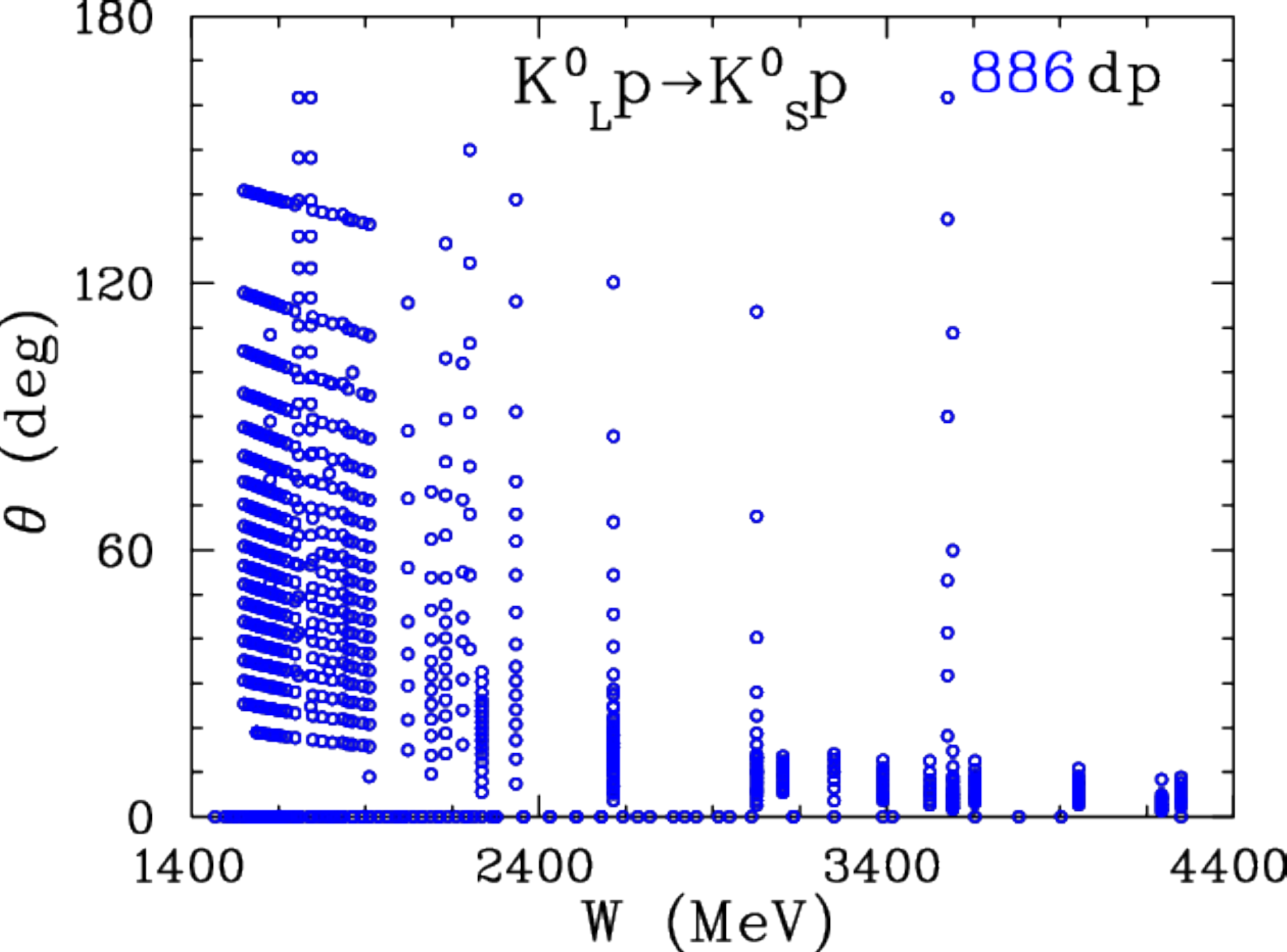} }
{
    \includegraphics[width=0.3\textwidth,keepaspectratio,angle=90]{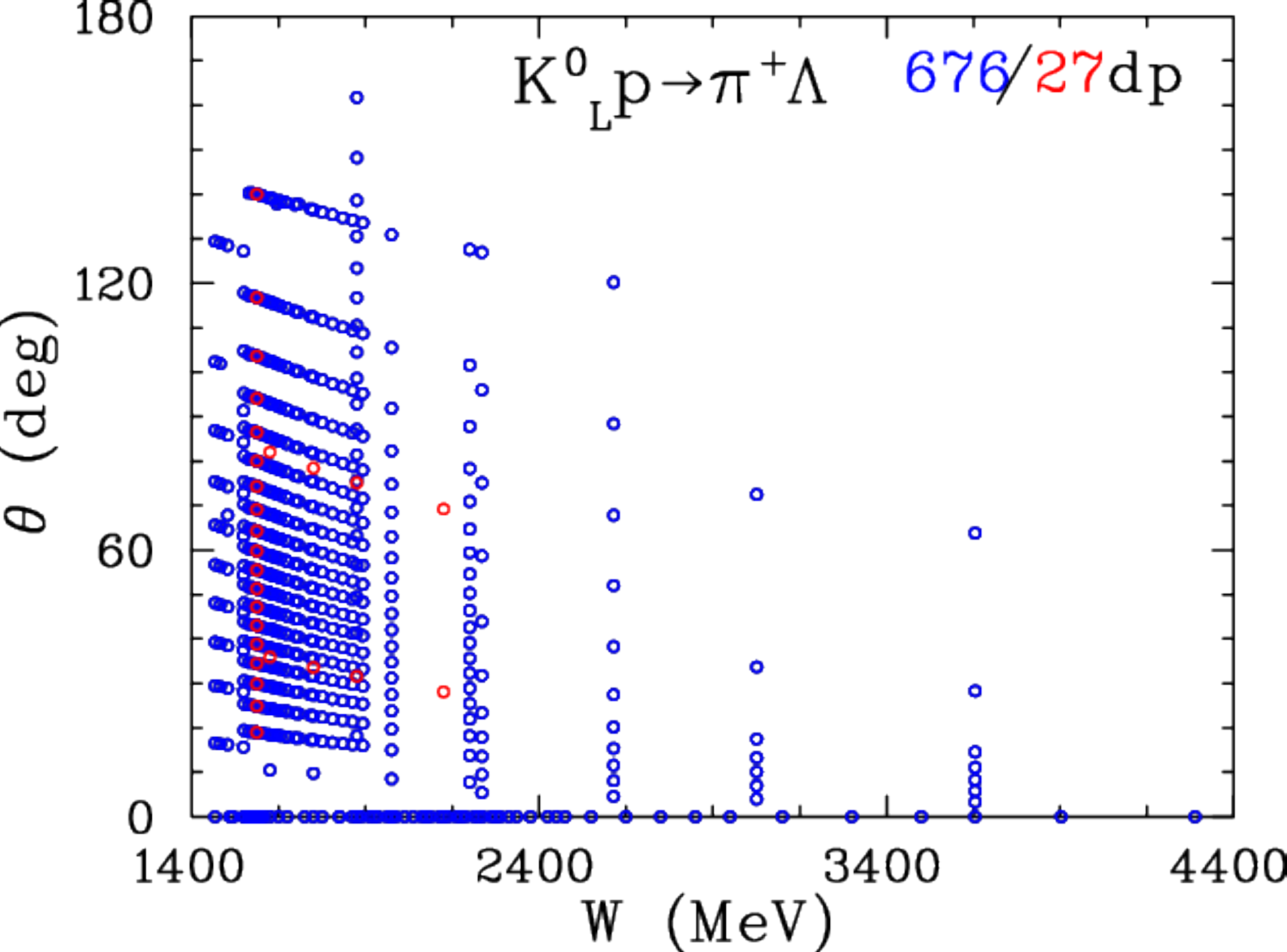} }
{
    \includegraphics[width=0.3\textwidth,keepaspectratio,angle=90]{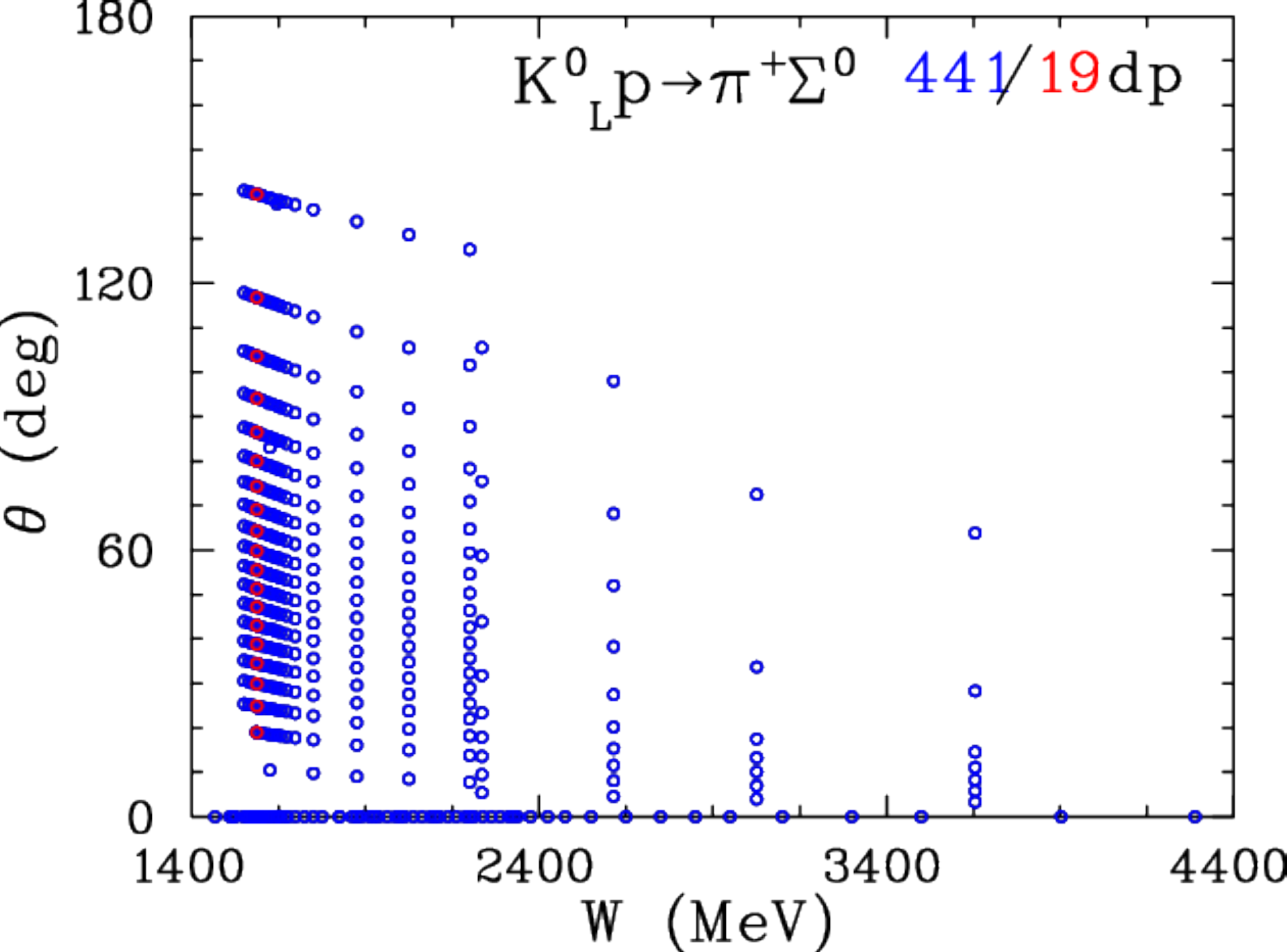} }
{
    \includegraphics[width=0.3\textwidth,keepaspectratio,angle=90]{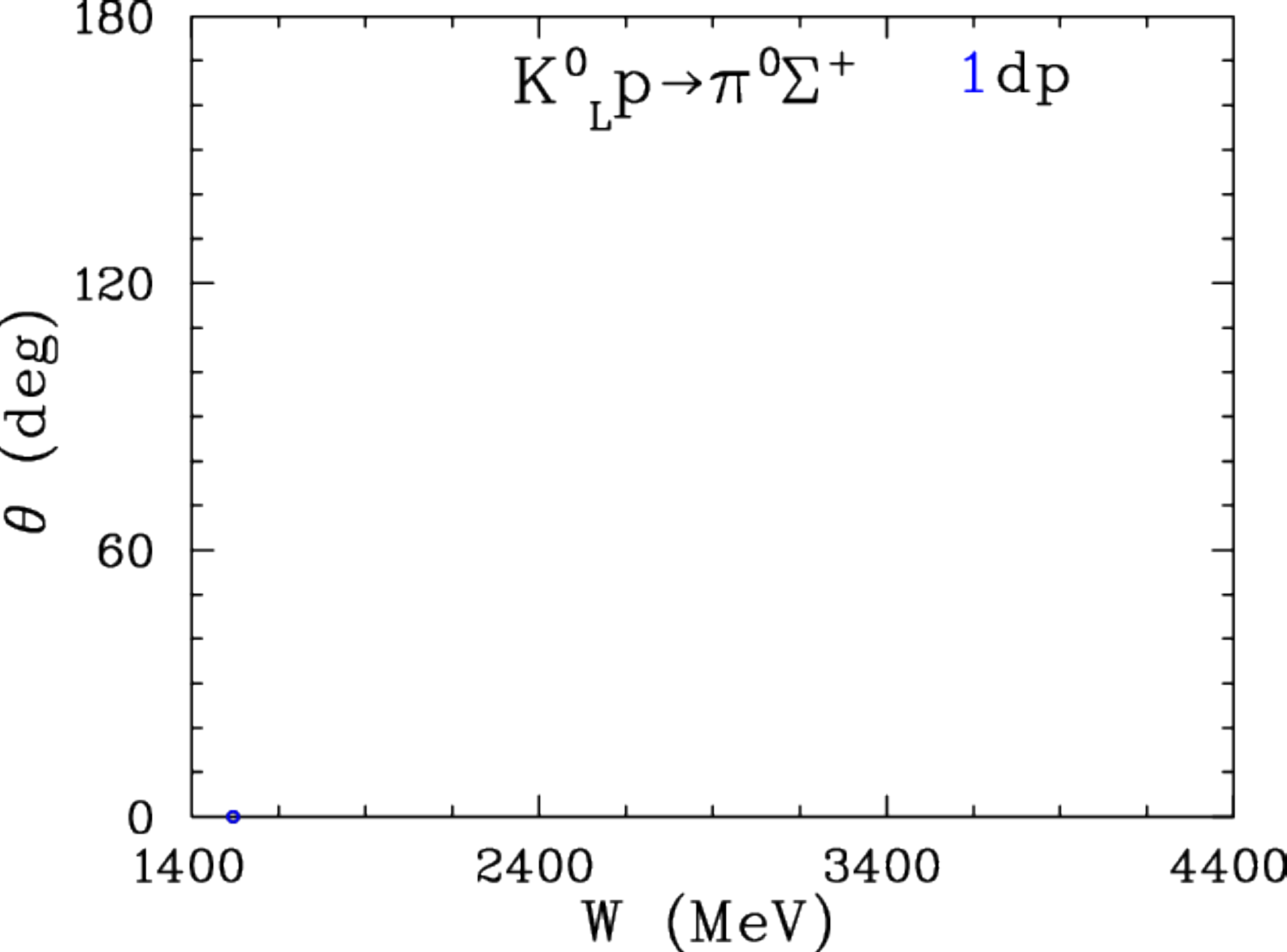} }

    \caption{Experimental data available for $K_Lp\to K^+n$, $K_Lp\to 
    K_Lp$, $K_Lp\to K_Sp$, $K_Lp\to\pi^+\Lambda$, $K_Lp\to\pi^+\Sigma^0$, 
    and $K_Lp\to\pi^0\Sigma^+$ as a function of c.m.\ energy 
    $W$~\protect\cite{Durham}. The number of data points (dp) is given in 
    the upper righthand side of each subplot [blue (red) shows amount 
    of unpolarized (polarized) observables]. Total cross sections are 
    plotted at zero degrees.} \label{fig:data}
\end{figure}

While a formally complete experiment requires the measurement, at
each energy, $W$, and angle, $\theta$, of at least three independent 
observables,
the current database for $K_Lp\to\pi Y$ and $KY$ is populated
mainly by unpolarized cross sections. Figure~\ref{fig:data}
illustrates this quite clearly.

The initial studies of the KLF program at GlueX will likely focus on
two-body and quasi-two-body processes: elastic $K_Lp\to K_Sp$ and 
charge-exchange
$K_Lp\to K^+n$ reactions, then two-body reactions producing $S=-1$
($S=-2$) hyperons as $K_Lp\to\pi^+\Lambda$, $K_Lp\to\pi^+\Sigma^0$, and
$K_Lp\to\pi^0\Sigma^+$ ($K_Lp\to K^+\Xi^0$). Most of the previous
measurements induced by a $K_L$ beam, were collected for $W = 1454$~MeV
and with some data up to $W = 5054$~MeV. Experiments were performed 
between 1961 and 1982 with
mostly hydrogen bubble chambers at ANL, BNL, CERN, DESY, KEK, LRL,
NIMROD, NINA, PPA, and SLAC. Note that some of data were taken at EM
facilities at NINA~\cite{NINA} (a short overview about NINA experiments
is given by Albrow recently~\cite{Albrow}) and SLAC~\cite{Brody}. The
goal of the Manchester University group that worked at the Daresbury
5-GeV electron synchrotron NINA was CP-violation, which was a hot topic
back to the mid 1960s. The main physics topics that the SLAC group 
addressed were
studies of the systematics for particle/anti-particle processes through 
the intrinsic properties of the K-longs.

The first paper that discussed the possibility of creating a practical
neutral kaon beam at an electron synchrotron through
photoproduction was an optimistic prediction for SLAC
by Drell and Jacob in 1965~\cite{DJ}.  With significant developments 
in technology, high-quality EM facilities, such as JLab~\cite{LoI}, 
are now able to realize a complete hyperon spectroscopy program.

The overall systematics of previous $K_L$p experiments varies
between 15\% and 35\%, and the energy binning is much broader than
hyperon widths. The previous number of $K_L$-induced measurements
(2426 $d\sigma/d\Omega$, 348 $\sigma^{tot}$, and 115 $P$
observables)~\cite{Durham} was very limited. 
Additionally, we are not aware of any measurements on a ``neutron" 
target.

Our knowledge about the non-strange sector is more advanced vs.\ the 
strange one~\cite{PDG2016}.  For the non-strange case, for instance,
phenomenology has access to 51k data of $\pi N\to\pi N$ and 39k data
of $\gamma N\to\pi N$ below $W = 2.5$~GeV~\cite{SAID-website}.

\section{Phenomenology / Partial-Wave Analysis}
\label{sec:PWA}

Here, we \underline{summarize} some of the physics issues
involved with such processes. Following Ref.~\cite{Hoehler84},
the differential cross section and polarization for $K_Lp$
scattering are given by
\begin{equation}
        \frac{d\sigma}{d\Omega} = \lambdabar^2 (|f|^2 + |g|^2),
\end{equation}
\begin{equation}
        P\frac{d\sigma}{d\Omega} = 2\lambdabar^2 {\rm Im}
        (fg^\ast),
\end{equation}
where $\lambdabar = \hbar/k$, with $k$ the magnitude of c.m.
momentum for the incoming meson.  Here $f = f(W,\theta)$ and
$g = g(W,\theta)$ are the usual spin-nonflip and spin-flip
amplitudes at c.m. energy $W$ and meson c.m. scattering angle
$\theta$. In terms of partial waves, $f$ and $g$ can be
expanded as
\begin{equation}
        f(W,\theta) = \sum_{l=0}^\infty [(l+1)T_{l+}
        + lT_{l-}]P_l(\cos\theta),
\end{equation}
\begin{equation}
        g(W,\theta) = \sum_{l=1}^\infty [T_{l+} - T_{l-}]P_l^1
        (\cos\theta),
\end{equation}
where $l$ is the initial orbital angular momentum,
$P_l(\cos\theta)$ is a Legendre polynomial, and
$P_l^1(\cos\theta)$ is an associated Legendre function.
The total angular momentum for the amplitude $T_{l+}$ is
$J=l+\frac{1}{2}$, while that for the amplitude $T_{l-}$
is $J=l-\frac{1}{2}$.  For hadronic scattering reactions,
we may ignore small CP-violating terms and write
\begin{equation}
        K_L = \frac{1}{\sqrt{2}} (K^0 - \overline{K^0}),
\end{equation}
\begin{equation}
        K_S = \frac{1}{\sqrt{2}} (K^0 + \overline{K^0}).
\end{equation}

We may generally have both $I=0$ and $I=1$ amplitudes for
$KN$ and $\overline{K}N$ scattering, so that the amplitudes
$T_{l\pm}$ can be expanded in terms of isospin amplitudes as
\begin{equation}
        T_{l\pm} = C_0 T^0_{l\pm} + C_1 T^1_{l\pm},
\end{equation}
where $T_{l\pm}^I$ are partial-wave amplitudes with isospin
$I$ and total angular momentum $J = l \pm \frac{1}{2}$,
with $C_I$ the appropriate isospin Clebsch-Gordon
coefficients.

We plan to do a coupled-channel partial-wave analysis (PWA) with 
new GlueX data in combination with available new J-PARC $K^-$
measurements when they will come.  Then the best fit will
allow determine model-independent (data-driven)
partial-wave amplitudes and associated resonance parameters 
(pole positions, residues, Breit-Wigner (BW) parameters, etc.)
as the SAID group does, for instance, for the analysis of $\pi
N$-elastic, charge-exchange, and $\pi^-p\to\eta n$
data~\cite{piN}.

\subsection{$KN$ and $\overline{K}N$ Final States}

The amplitudes for reactions leading to $KN$ and $\overline{K}N$
final states are
\begin{eqnarray}
        T(K^-p \to K^-p) &=& \frac{1}{2}T^1({\overline K}N \to
        {\overline K}N) + \frac{1}{2}T^0({\overline K}N \to
        {\overline K}N), \\
	T(K^-p \to \overline{K^0} n) &=& \frac{1}{2}T^1({\overline
	K}N\to {\overline K}N) - \frac{1}{2}T^0({\overline K}N \to
	{\overline K}N), \\
	T(K^+p \to K^+p) &=& T^1(KN \to KN), \\
	T(K^+n \to K^+n) &=& \frac{1}{2}T^1(KN \to KN) +
	\frac{1}{2}T^0(KN \to KN),
\end{eqnarray}
\begin{equation}
        T(K_Lp \to K_Sp) = \frac{1}{2} \left (\frac{1}{2}T^1
        (KN\to KN) + \frac{1}{2}T^0(KN \to KN) \right ) -
        \frac{1}{2}T^1({\overline K}N \to {\overline K}N),
\end{equation}
\begin{equation}
        T(K_Lp\to K_Lp) = \frac{1}{2} \left (\frac{1}{2}T^1
        (KN\to KN) + \frac{1}{2}T^0(KN \to KN) \right ) +
        \frac{1}{2}T^1({\overline K}N\to {\overline K}N),
\end{equation}
\begin{equation}
	T(K_Lp\to K^+n) = \frac{1}{\sqrt{2}} \left (\frac{1}{2}
	T^1(KN\to KN) - \frac{1}{2}T^0(KN\to KN) \right ).
\end{equation}
\begin{figure}
\begin{center}
{
    \includegraphics[width=0.3\textwidth,keepaspectratio,angle=90]{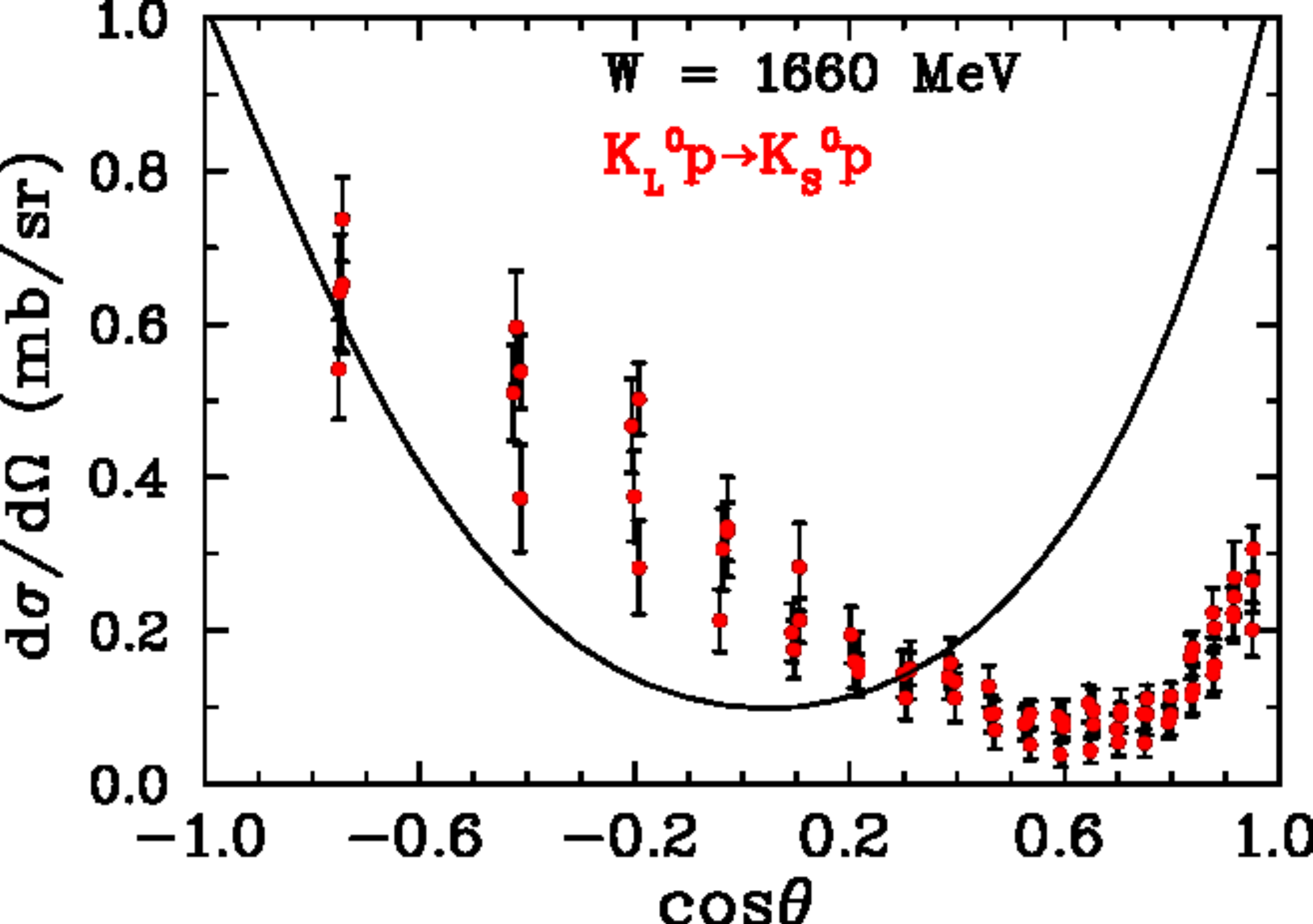} }
{
    \includegraphics[width=0.3\textwidth,keepaspectratio,angle=90]{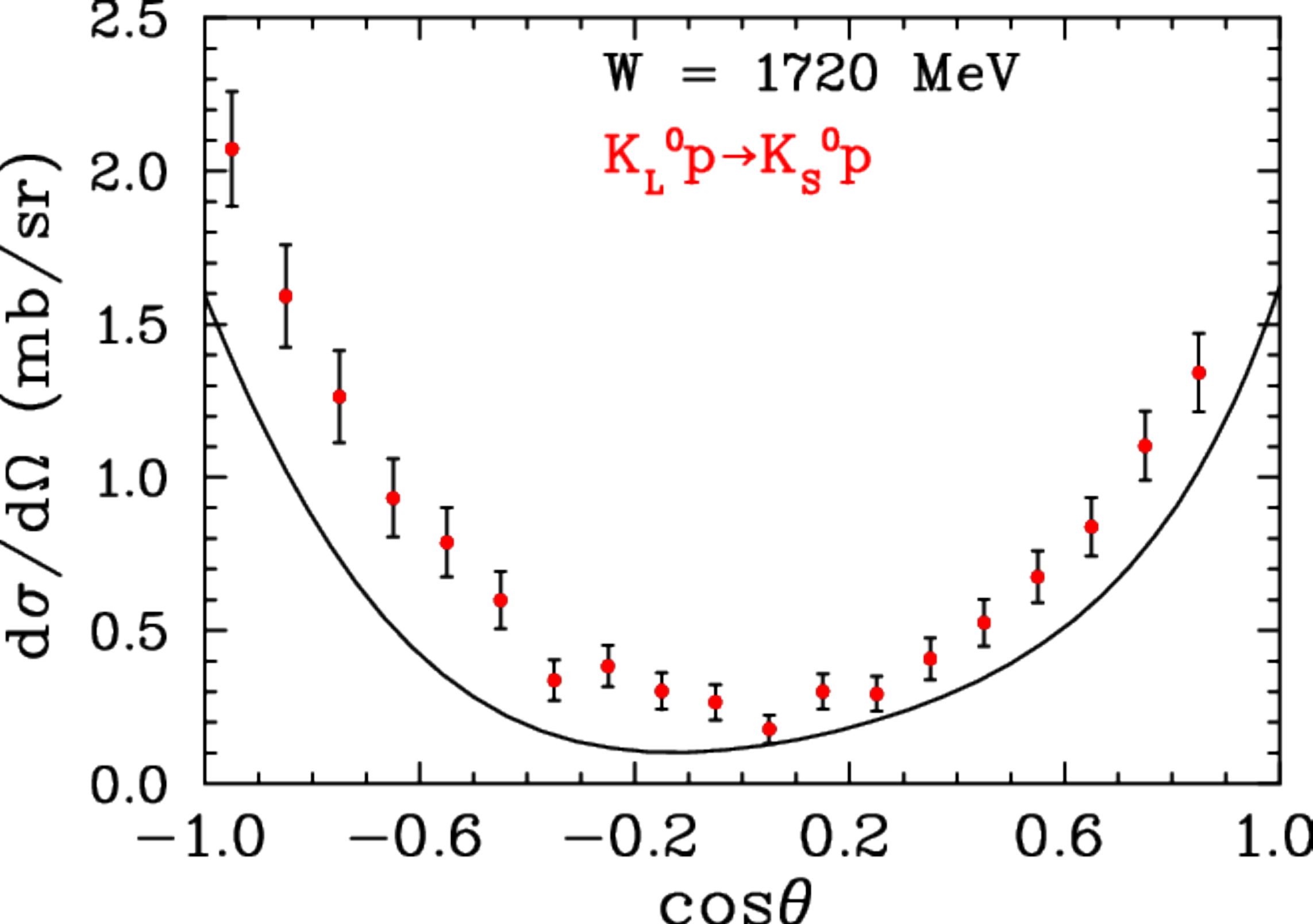} }
{
    \includegraphics[width=0.3\textwidth,keepaspectratio,angle=90]{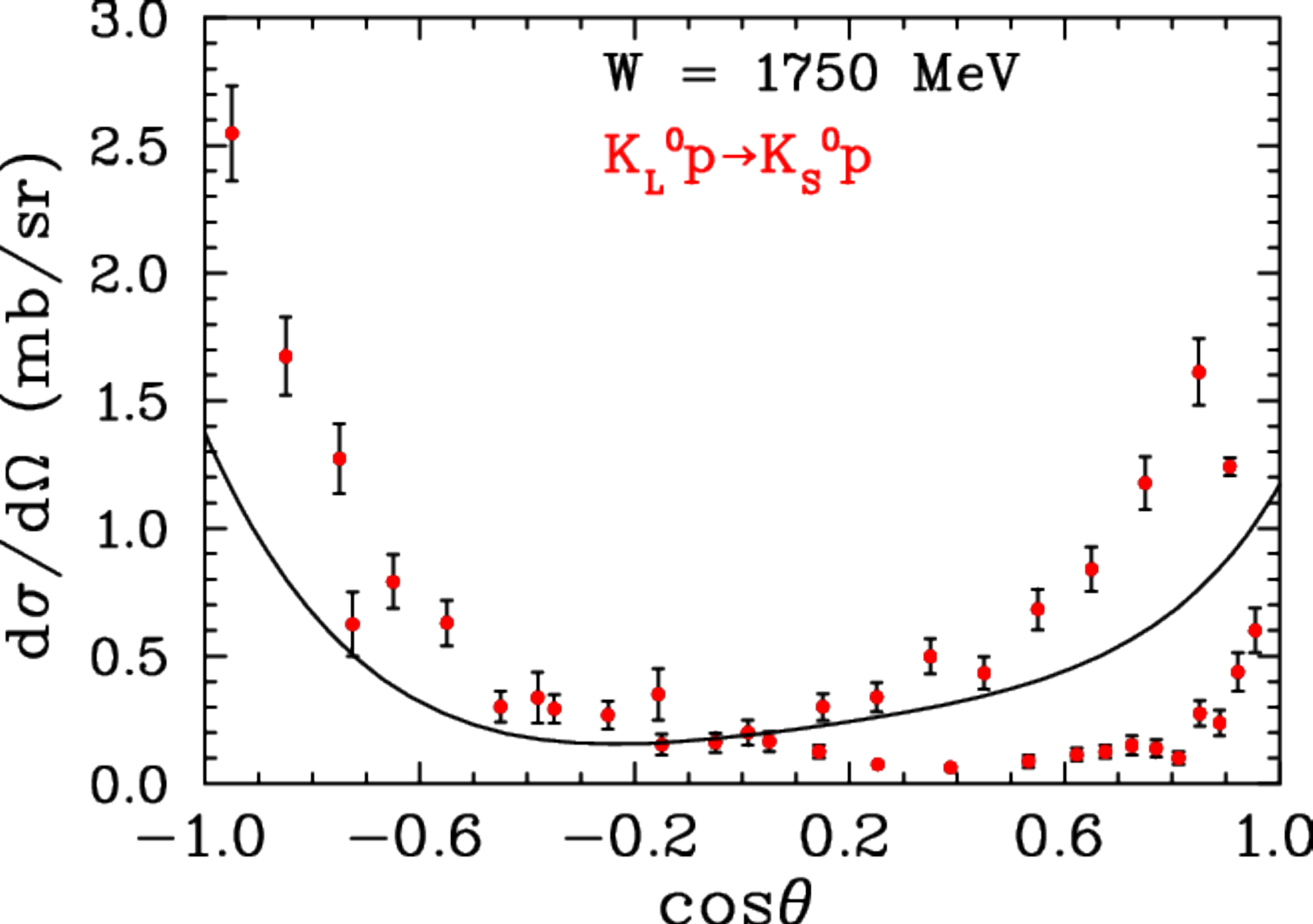} }
{
    \includegraphics[width=0.3\textwidth,keepaspectratio,angle=90]{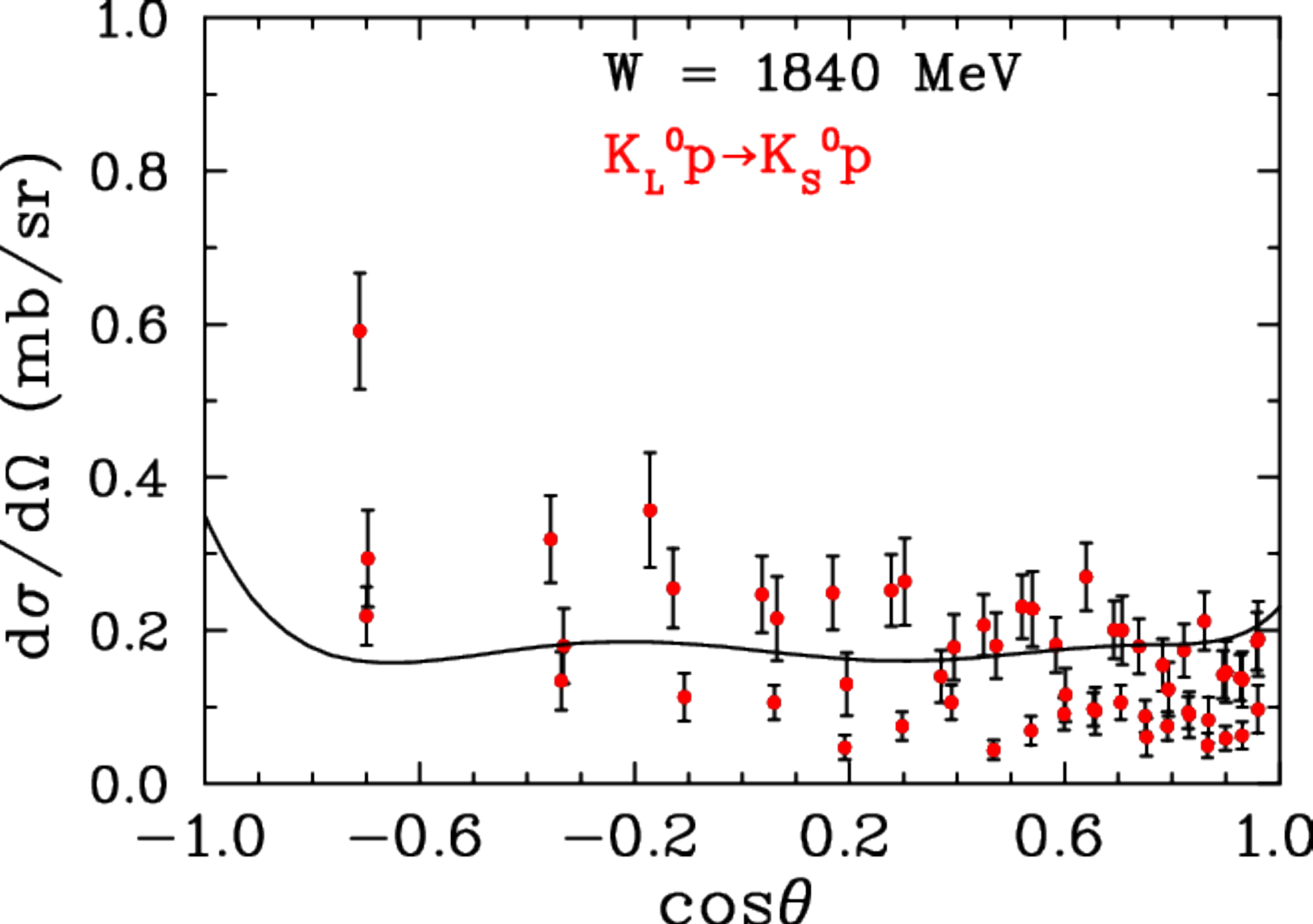} }
\end{center}

\centerline{\parbox{0.80\textwidth}{
 \caption[] {\protect\small Selected differential cross section 
	data for $K_Lp\to K_Sp$ at $W = 1660$~MeV, 1720~MeV, 
	1750~MeV, and 1840~MeV, from Ref.~\protect\cite{Mark2016}. 
	The plotted points from previously published experimental 
	data are those data points within 20~MeV of the kaon
	c.m. energy indicated on each 
	panel~\protect\cite{SAID-website}.
	Plotted uncertainties are statistical only.
	The curves are predictions using amplitudes from a
	recent PWA of $\overline{K}N\to\overline{K}N$ 
	data~\protect\cite{Zhang2013a,Zhang2013b}, combined with 
	$KN\to KN$ amplitudes from the SAID 
	database~\protect\cite{SAID-website}. } 
	\label{fig:KLp_KSp} } }
\end{figure}

No differential cross-section data are available for $K_Lp
\to K_Lp$ below $W \sim 2948$~MeV.  A fair amount of data
are available for the reaction, $K^+n\to K^0p$, measured on a
deuterium target.  Figure~\ref{fig:KLp_KSp} shows a sample of
available differential cross sections for $K_Lp\to K_Sp$
compared with predictions determined from a recent
PWA of $\overline{K}N\to\overline{K}N$
data~\cite{Zhang2013a,Zhang2013b}, combined with $KN\to KN$
amplitudes from the SAID database~\cite{SAID-website}.  The
predictions at lower and higher energies tend to agree less
well with the data.

\subsection{$\pi\Lambda$ Final States}

\begin{figure}
\begin{center}
{
    \includegraphics[width=0.3\textwidth,keepaspectratio,angle=90]{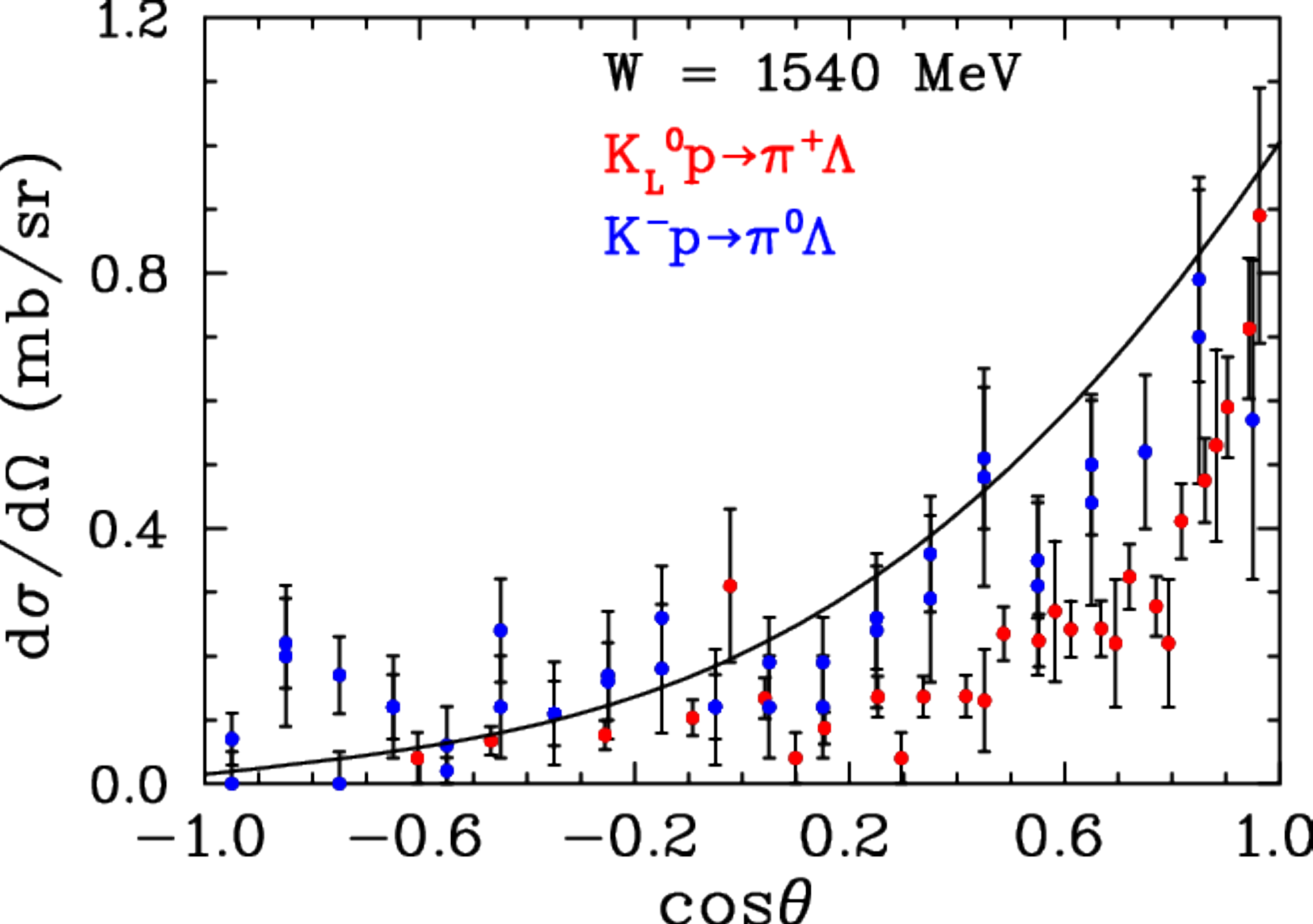} }
{
    \includegraphics[width=0.3\textwidth,keepaspectratio,angle=90]{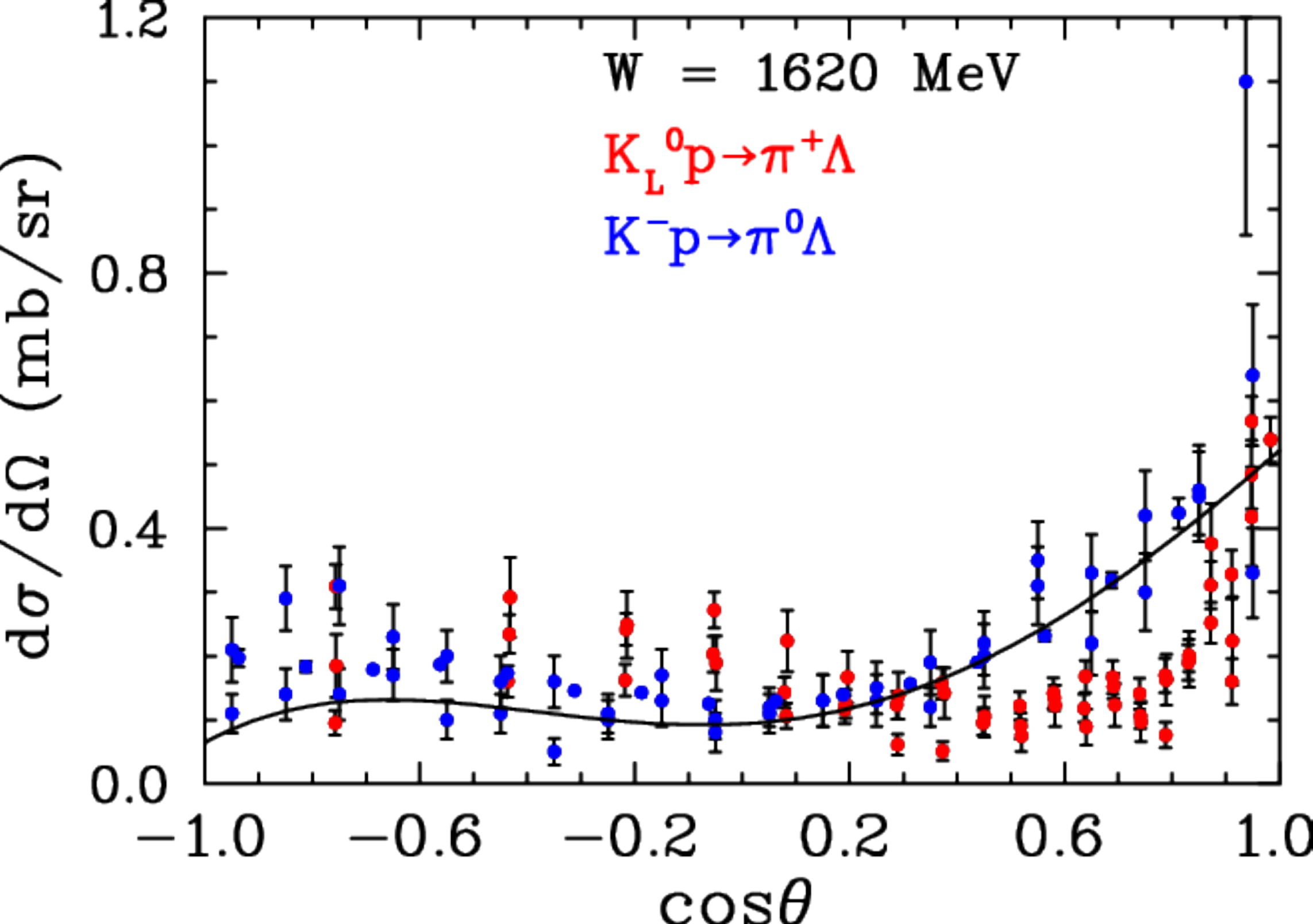} }
{
    \includegraphics[width=0.3\textwidth,keepaspectratio,angle=90]{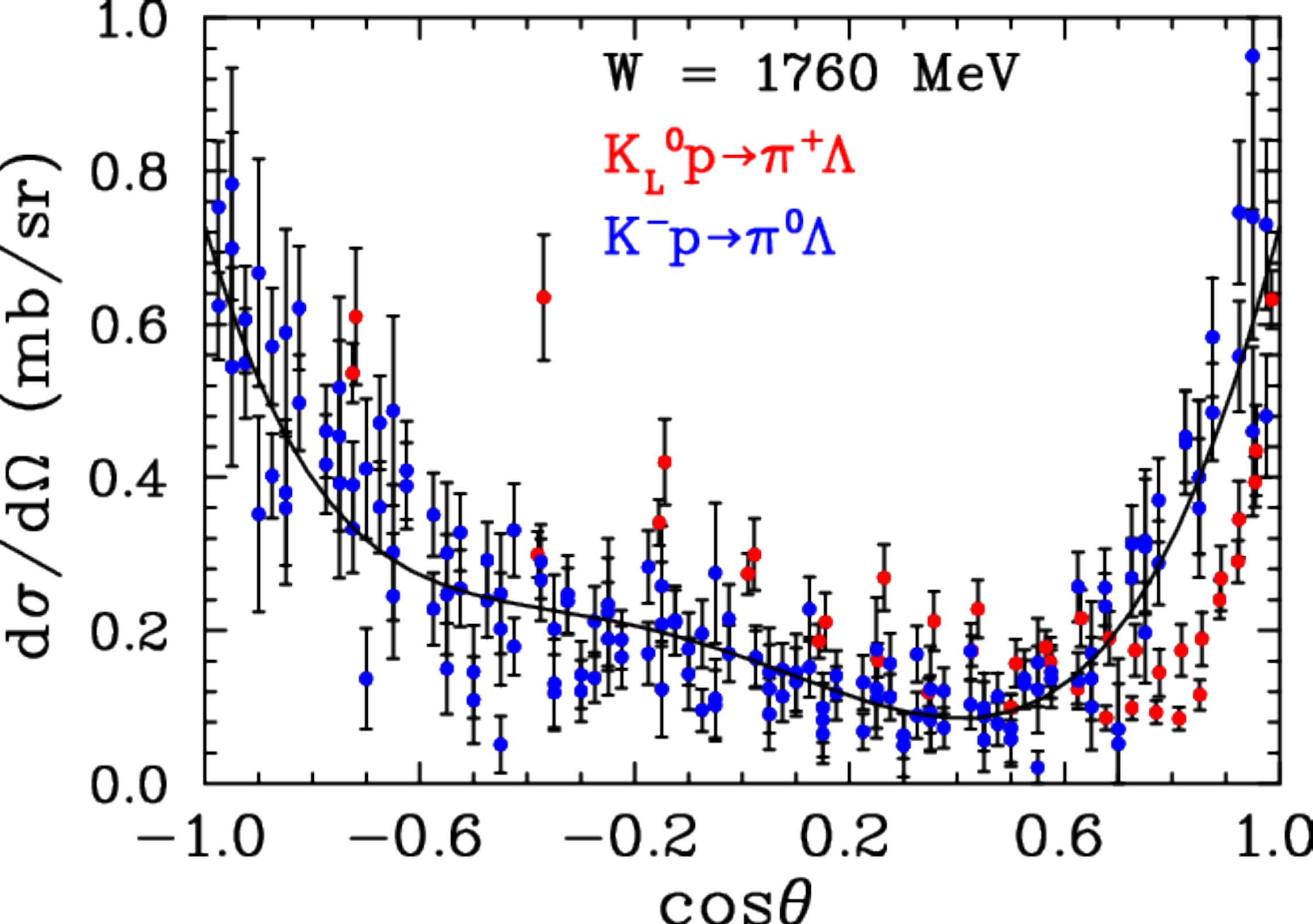} }
{
    \includegraphics[width=0.3\textwidth,keepaspectratio,angle=90]{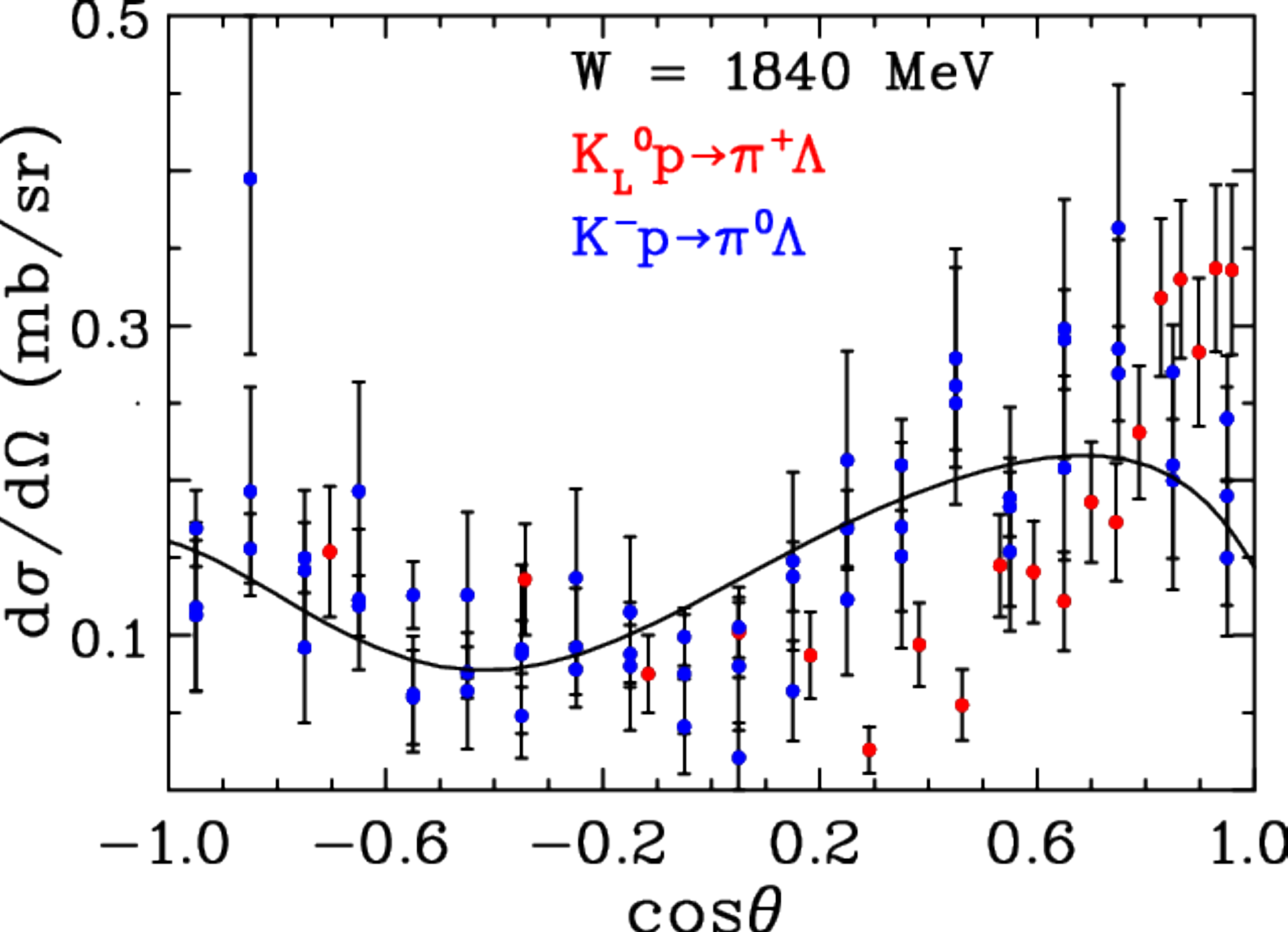} }
\end{center}

\centerline{\parbox{0.80\textwidth}{
 \caption[] {\protect\small Comparison of selected differential
        cross section data for $K^-p\to\pi^0\Lambda$ and
        $K_Lp\to\pi^+\Lambda$ at $W = 1540$~MeV, 1620~MeV, 
	1760~MeV, and 1840~MeV, from Ref.~\protect\cite{Mark2016}.
        The plotted points from previously published experimental
        data are those data points within 20~MeV of the kaon
        c.m. energy indicated on each panel~\protect\cite{SAID-website}.
        Plotted uncertainties are statistical only.
	The curves are from a recent PWA of $K^-p\to\pi^0
	\Lambda$ data~\protect\cite{Zhang2013a,Zhang2013b}. } 
	\label{fig:KLp_piLambda} } }
\end{figure}

\begin{figure}
\begin{center}
{
    \includegraphics[width=0.3\textwidth,keepaspectratio,angle=90]{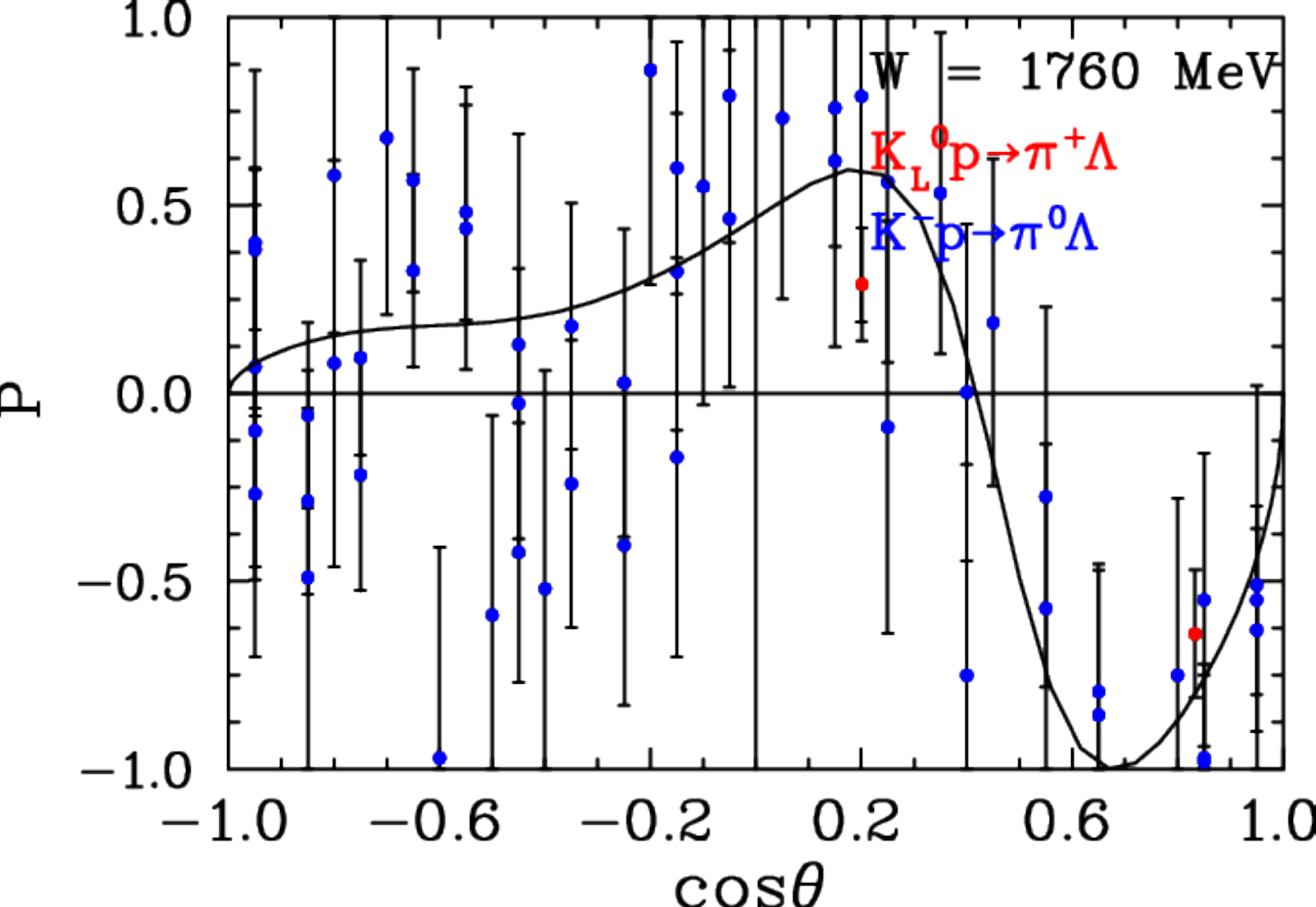} }
{
    \includegraphics[width=0.3\textwidth,keepaspectratio,angle=90]{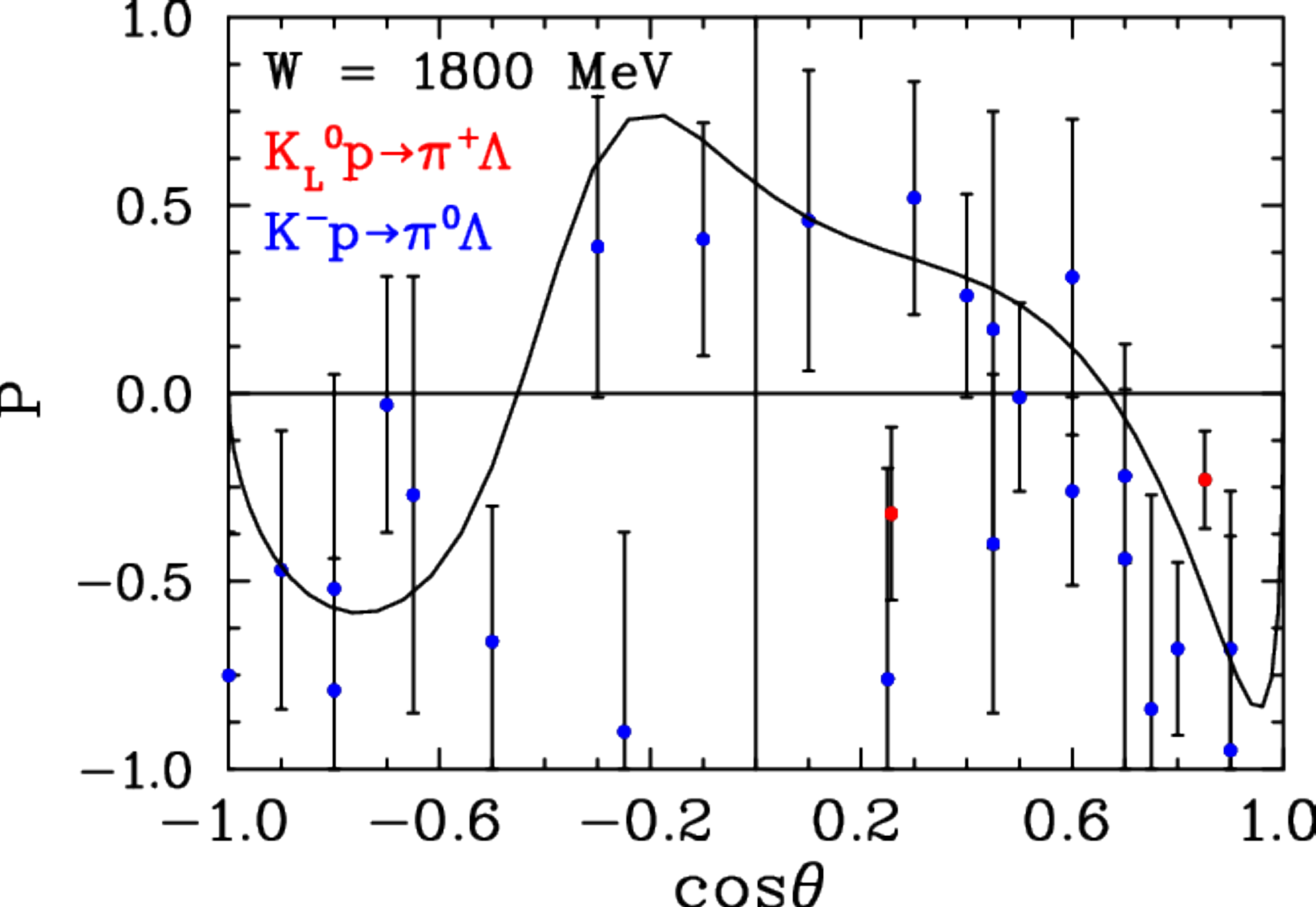} }
\end{center}

\centerline{\parbox{0.80\textwidth}{
 \caption[] {\protect\small Comparison of selected polarization
       data for $K^-p\to\pi^0\Lambda$ and $K_L p\to\pi^+
	\Lambda$ at $W = 1760$~MeV and 1880~MeV, from 
	Ref.~\protect\cite{Mark2016}.  The plotted points 
	from previously published experimental
	data are those data points within 20~MeV of the 
	kaon c.m.\ energy indicated on each 
	panel~\protect\cite{SAID-website}.
	The curves are from a recent PWA of 
	$K^-p\to\pi^0\Lambda$ 
	data~\protect\cite{Zhang2013a,Zhang2013b}. } 
	\label{fig:KLp_piLambda_P} } }
\end{figure}

The amplitudes for reactions leading to $\pi\Lambda$ final
states are
\begin{eqnarray}
	T(K^- p \to \pi^0 \Lambda) &=& \frac{1}{\sqrt{2}}T^1
	({\overline K}N\to\pi\Lambda), \\
	T(K_Lp\to\pi^+\Lambda) &=& -\frac{1}{\sqrt{2}}T^1
	({\overline K}N \to \pi\Lambda).
\end{eqnarray}
The $K^-p\to\pi^0\Lambda$ and $K_Lp\to\pi^+\Lambda$
amplitudes imply that observables for these reactions measured
at the same energy should be the same except for small
differences due to the isospin-violating mass differences in
the hadrons. No differential cross-section data for $K^-p\to
\pi^0\Lambda$ are available at c.m. energies $W < 1540$~MeV,
although data for $K_Lp\to\pi^+\Lambda$ are available at
such energies.  At 1540~MeV and higher energies, differential
cross-section and polarization data for the two reactions are
in fair agreement, as shown in Figs.~\ref{fig:KLp_piLambda}
and \ref{fig:KLp_piLambda_P}.

\subsection{$\pi\Sigma$ Final States}

$SU(3)$ flavor symmetry allows as many $S=-2$ baryon
resonances as there are $N$ and $\Delta$ resonances combined
($\sim27$); however, until now only three states,
$\Xi(1322)1/2^+$, $\Xi(1530)3/2^+$, and $\Xi(1820)3/2^-$, have
their quantum numbers assigned and only a few more states have been
observed~\cite{PDG2016}.

The amplitudes for reactions leading to $\pi\Sigma$ final states
are
\begin{eqnarray}
        T(K^-p\to\pi^-\Sigma^+) &=& -\frac{1}{2}T^1({\overline K}N
        \to\pi\Sigma)-\frac{1}{\sqrt{6}}T^0({\overline K}N\to\pi
        \Sigma),\\
        T(K^-p\to\pi^+\Sigma^-) &=& \frac{1}{2}T^1({\overline K}N
        \to\pi\Sigma)-\frac{1}{\sqrt{6}}T^0({\overline K}N\to\pi
        \Sigma),\\
        T(K^-p\to\pi^0\Sigma^0) &=& \frac{1}{\sqrt{6}}T^0({\overline
        K}N\to\pi\Sigma),\\
        T(K^0_Lp\to\pi^+\Sigma^0) &=& -\frac{1}{2}T^1({\overline K}
	N\to\pi\Sigma),\\
        T(K^0_Lp\to\pi^0\Sigma^+) &=& \frac{1}{2}T^1({\overline K}
	N\to\pi\Sigma).
\end{eqnarray}

\begin{figure}
\begin{center}
{
    \includegraphics[width=0.3\textwidth,keepaspectratio,angle=90]{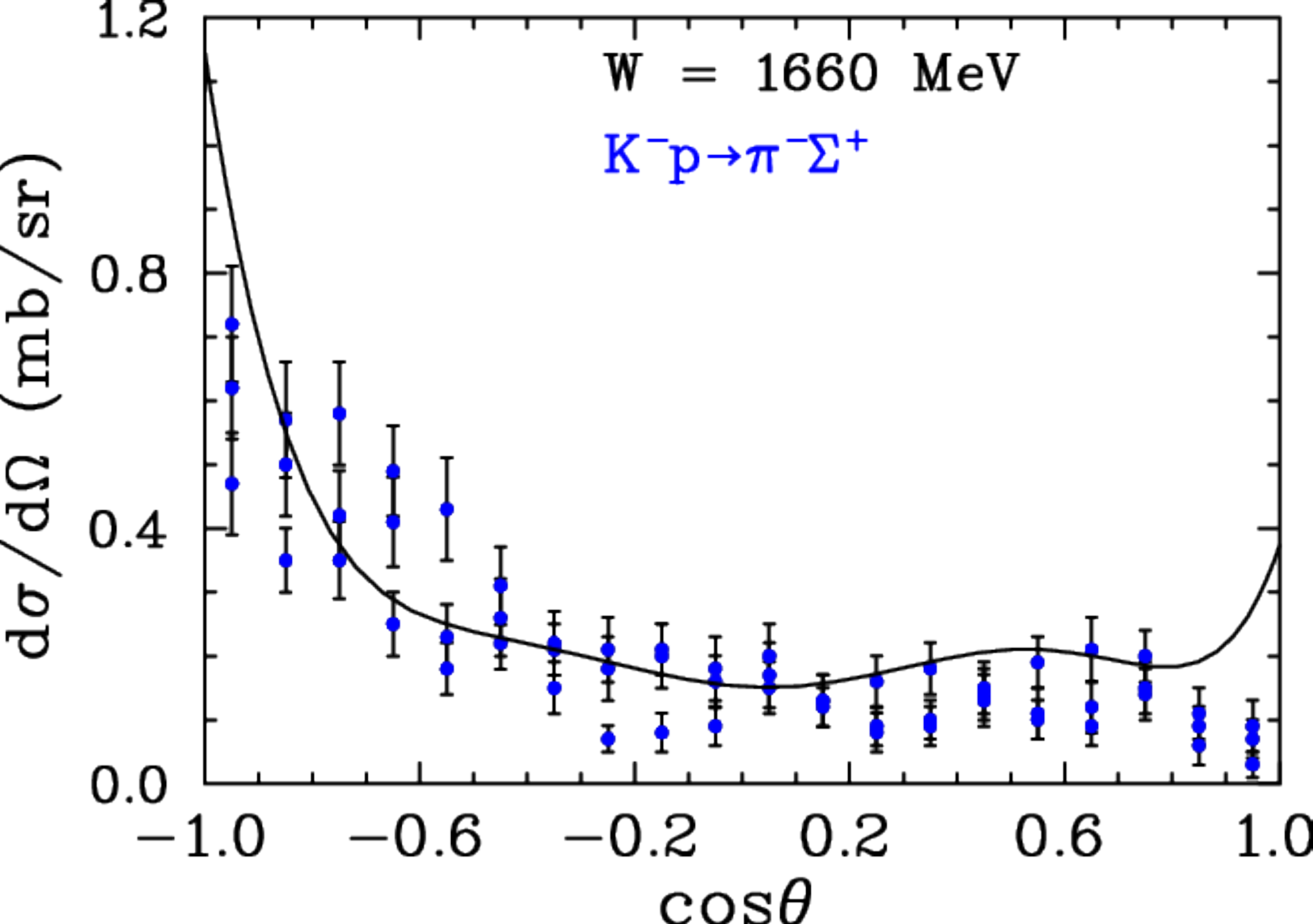} }
{
    \includegraphics[width=0.3\textwidth,keepaspectratio,angle=90]{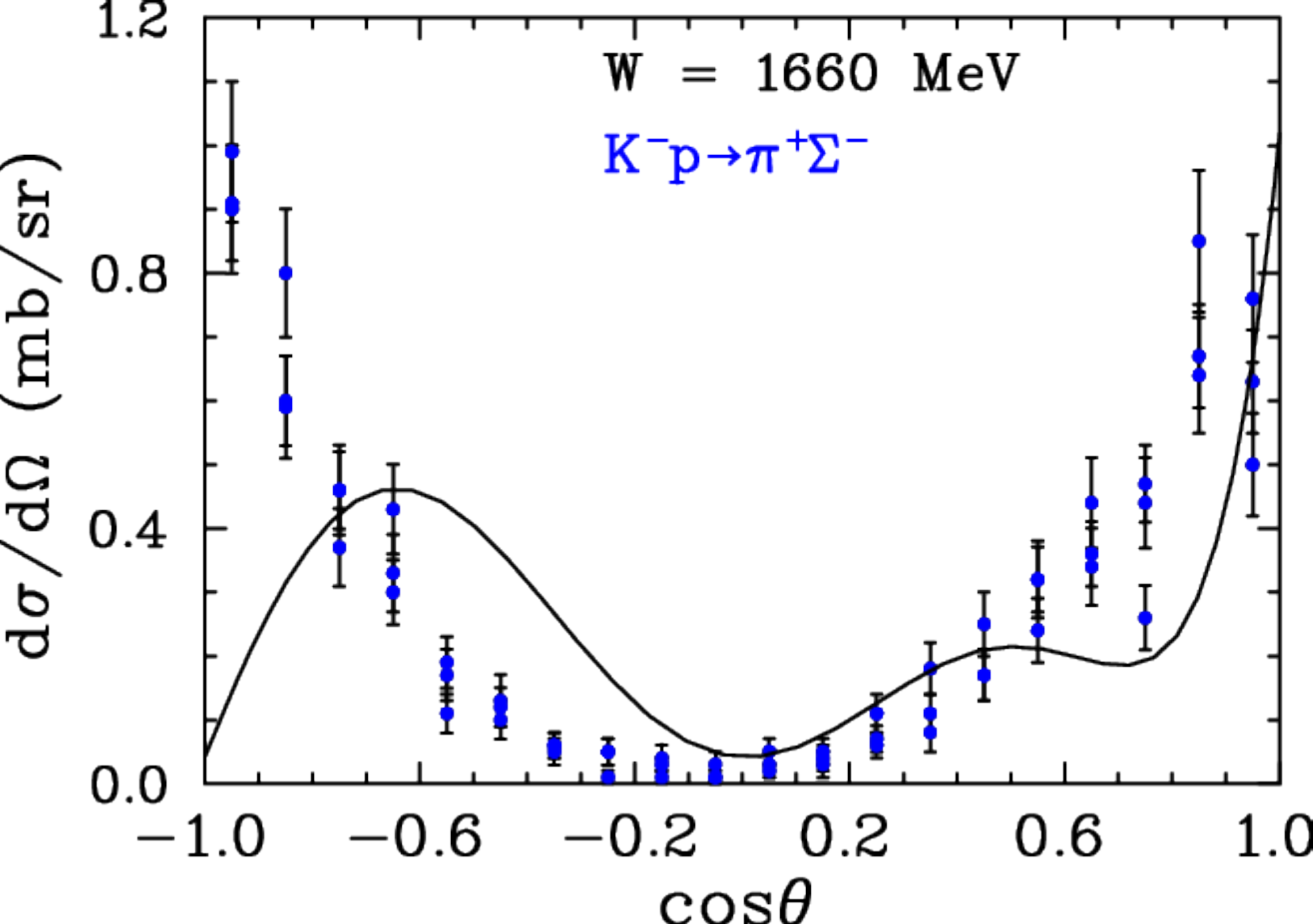} }
{
    \includegraphics[width=0.3\textwidth,keepaspectratio,angle=90]{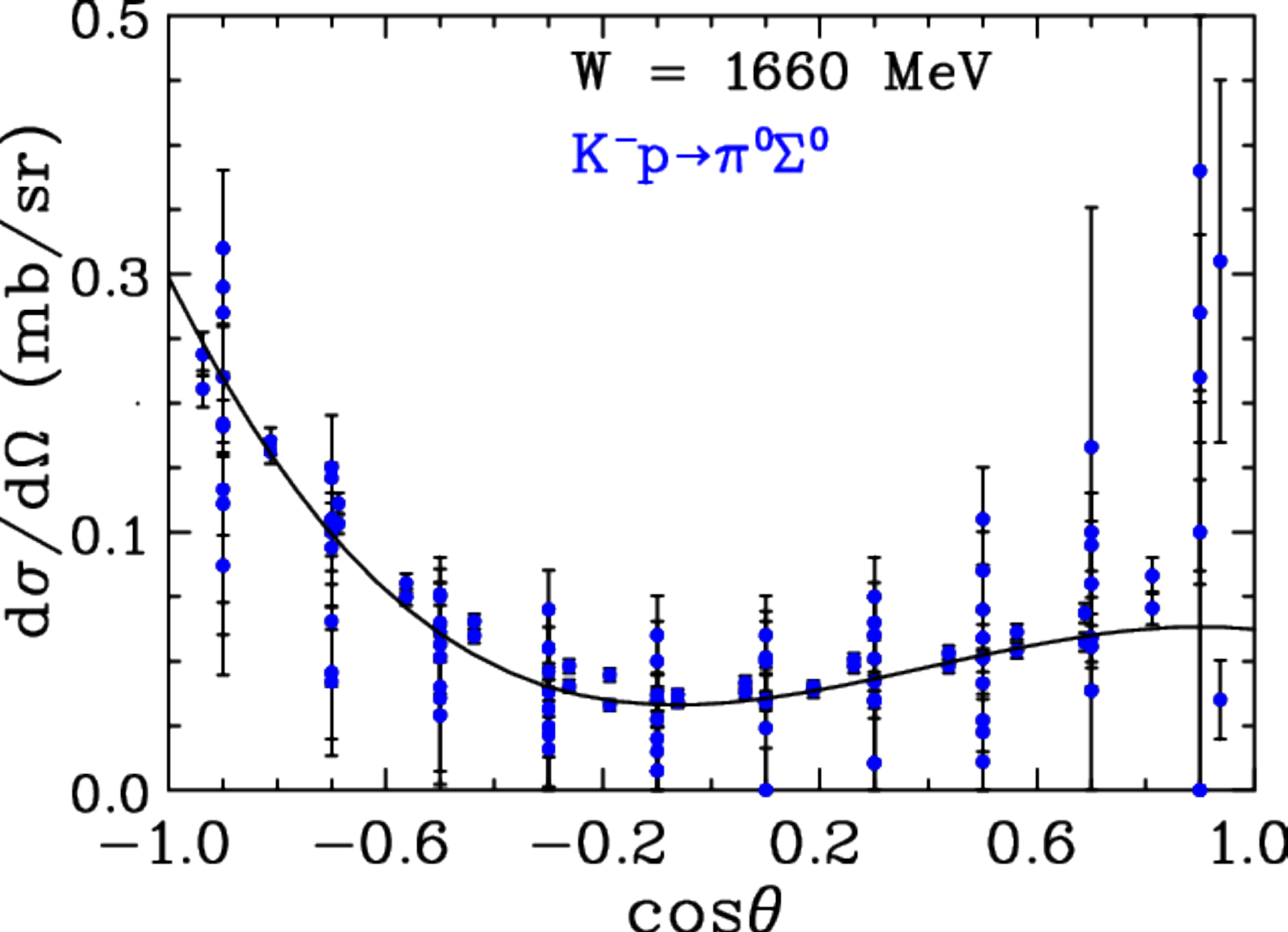} }
{
    \includegraphics[width=0.3\textwidth,keepaspectratio,angle=90]{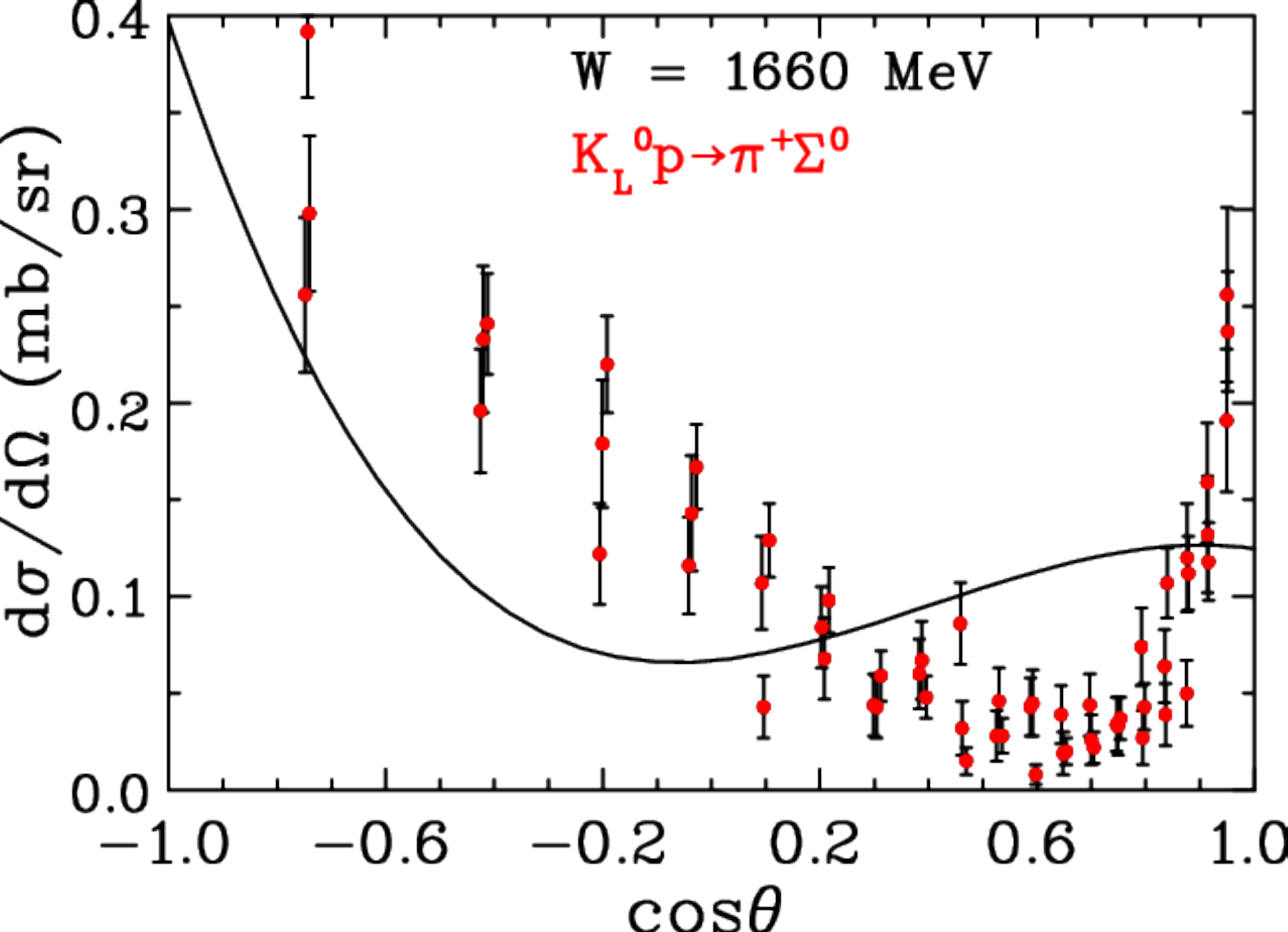} }
\end{center}

\centerline{\parbox{0.80\textwidth}{
 \caption[] {\protect\small Comparison of selected differential
        cross-section data for $K^-p\to\pi^-\Sigma^+$, $K^-p\to
        \pi^+\Sigma^-$, $K^-p\to\pi^0 \Sigma^0$, and $K_Lp
	\to\pi^0\Sigma^+$ at $W = 1660$~MeV, from 
	Ref.~\protect\cite{Mark2016}.  The plotted points
        from previously published experimental
        data are those data points within 20~MeV of the
        kaon c.m.\ energy indicated on each
        panel~\protect\cite{SAID-website}
	The curves are from 
	a recent PWA of $K^-p\to\pi\Sigma$ 
	data~\protect\cite{Zhang2013a,Zhang2013b}. }
        \label{fig:KLp_piSigma} } }
\end{figure}

Figure~\ref{fig:KLp_piSigma} shows a comparison of differential
cross-section data for $K^-p$ and $K_Lp$ reactions leading to
$\pi\Sigma$ final states at $W = 1660$~MeV (or $P_{\rm lab} =
716$~MeV/$c$).  The curves are based on energy-dependent isospin
amplitudes from a recent PWA~\cite{Zhang2013a,Zhang2013b}.
No differential cross-section data are available for $K_Lp\to
\pi^0\Sigma^+$. As this example shows, the quality of the
$K_Lp$ data is comparable to that for the $K^-p$ data.  It
would, therefore, be advantageous to combine the $K_Lp$ data in
a new coupled-channel PWA with available $K^-p$ data. Note that
the reactions $K_Lp\to\pi^+\Sigma^0$ and $K_Lp\to\pi^0
\Sigma^+$ are isospin selective (only $I=1$ amplitudes are
involved) whereas the reactions $K^-p\to\pi^-\Sigma^+$ and
$K^-p\to\pi^+\Sigma^-$ are not.  New measurements with a
$K_L$ beam would lead to a better understanding of
$\Sigma^\ast$ states and would help constrain the amplitudes
for $K^-p$ scattering to $\pi\Sigma$ final states

\subsection{$K\Xi$ Final States}

The amplitudes for reactions leading to $K\Xi$ final states are
\begin{eqnarray}
        T(K^- p \to K^0 \Xi^0) &=& \frac{1}{2}T^1({\overline K}N
        \to K\Xi)+\frac{1}{2}T^0({\overline K}N \to K\Xi),\\
        T(K^- p \to K^+ \Xi^-) &=& \frac{1}{2}T^1({\overline K}N
        \to K\Xi)-\frac{1}{2}T^0({\overline K}N \to K\Xi),\\
        T(K_Lp\to K^+\Xi^0) &=& -\frac{1}{\sqrt{2}}T^1
        ({\overline K}N\to K\Xi).
\end{eqnarray}
The threshold for $K^-p$ and $K_Lp$ reactions leading to
$K\Xi$ final states is fairly high ($W_{\rm thresh} = 1816$~MeV).
In Fig.~\ref{fig:omega}(left), we present the cross section 
for $\Xi$ production using a $K^-$-beam~\cite{Hassall1981}.
There are no differential cross-section data available for
$K_Lp\to K^+\Xi^0$ and very few (none recent) for $K^-p\to
K^0\Xi^0$ or $K^-p\to K^+\Xi^-$.  Measurements for these
reactions would be very helpful, especially for comparing with
predictions from dynamical coupled-channel (DCC)
models~\cite{Kamano2014,Kamano2015} and other effective Lagrangian 
approaches~\cite{Jackson:2015dva}.  The {\it Review of
Particle Physics}~\cite{PDG2016} lists only two states with
branching fractions (BF) to $K\Xi$, namely, $\Lambda(2100)7/2^-$
(BF $<$ 3\%) and $\Sigma(2030)7/2^+$ (BF $<$ 2\%)

\subsection{$KK\Omega$ Final States}

\begin{figure}[h!]
\centering
{
    \includegraphics[width=0.33\textwidth,keepaspectratio]{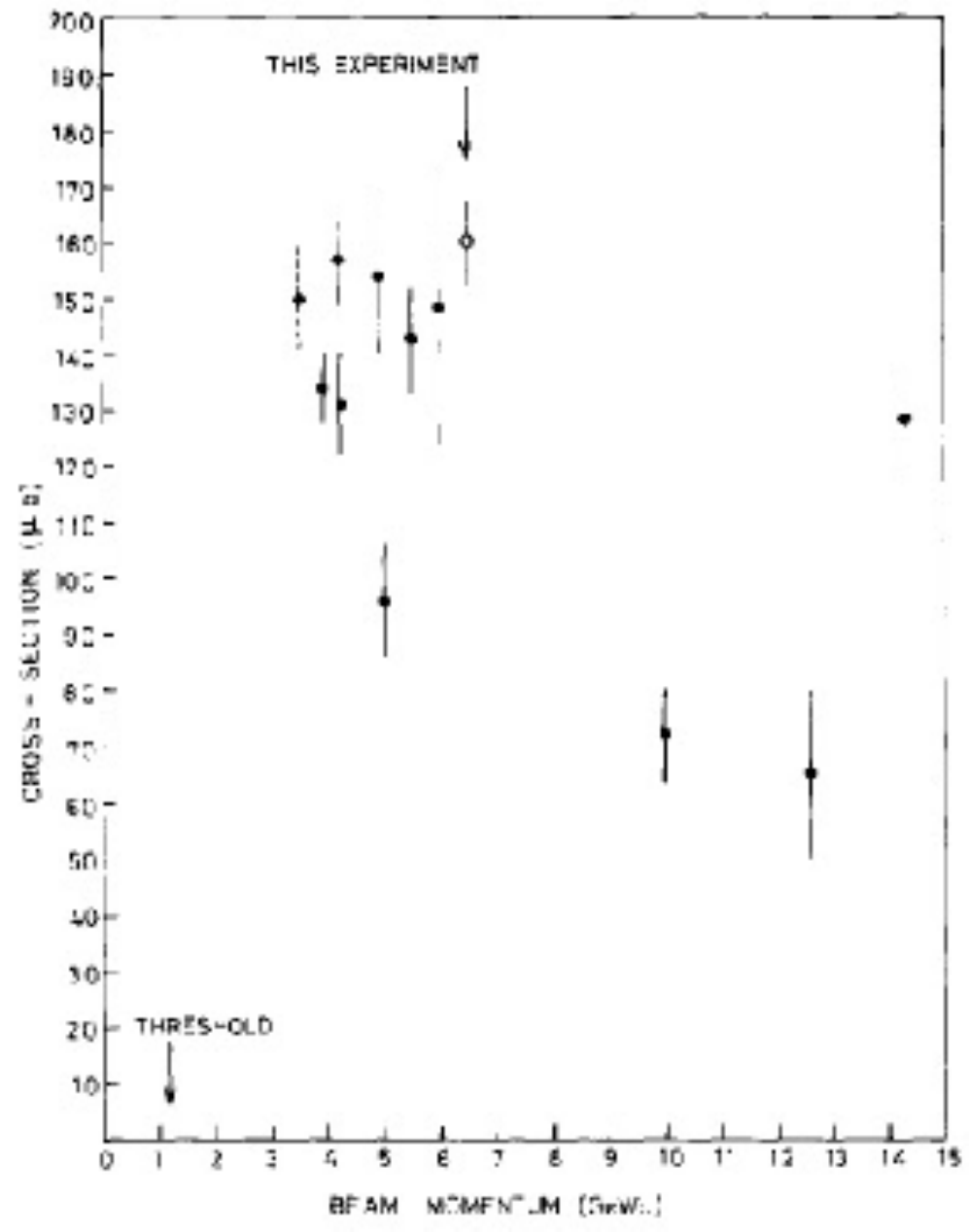} }
{
    \includegraphics[width=0.41\textwidth,keepaspectratio]{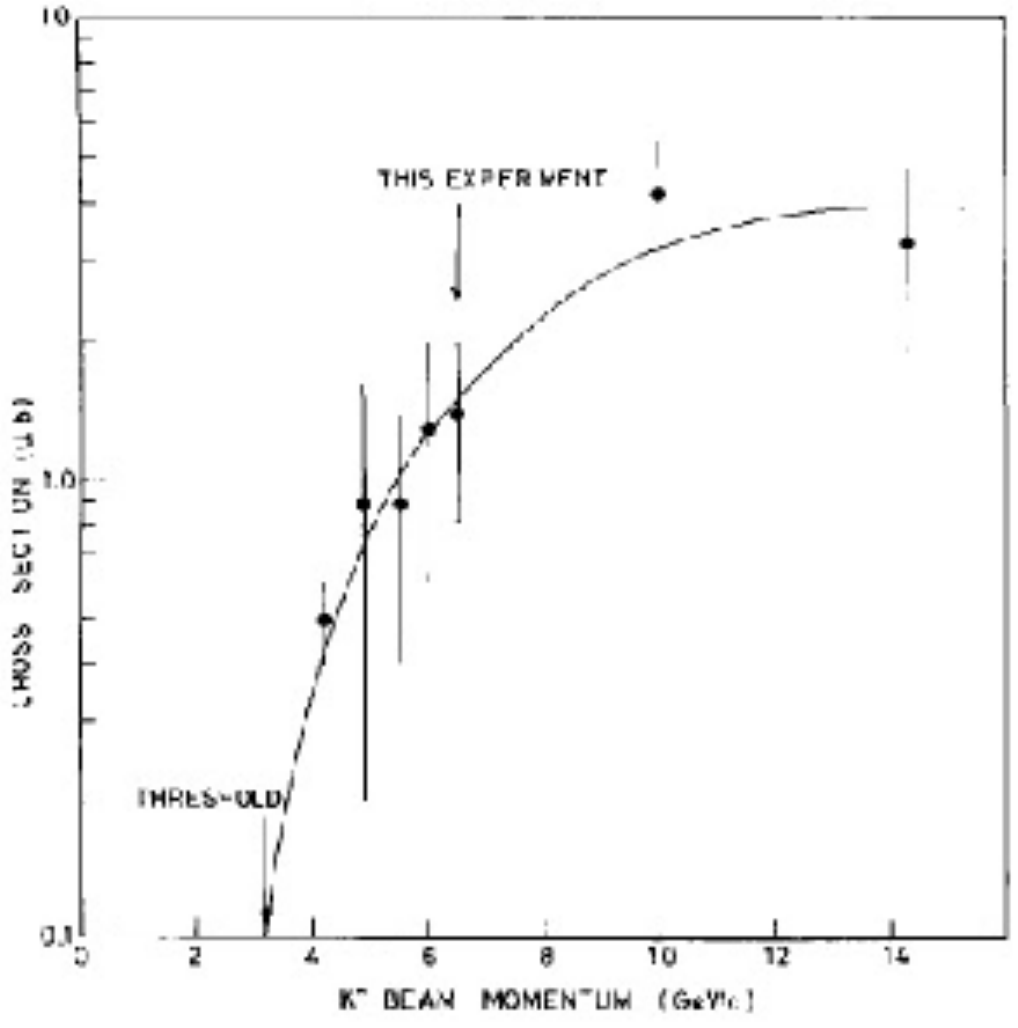} }
    
    \caption{Left panel: Cross section of $\Xi^-$ production,
    $K^-p\to\Xi^-X$, as a function of $K^-$
    momentum~\protect\cite{Hassall1981}.
    Right panel: Cross section of $\Omega^-$ production,
    $K^-p\to\Omega^-K^+K^0$, as a function of $K^-$
    momentum~\protect\cite{Hassall1981}. The curve is a fit by
    eye to the data.}\label{fig:omega}.
\end{figure}

The experimental situation with $\Omega^{-\ast}$s is even worse
than for the $\Xi^\ast$ case -- there are very few data for excited
states. The main reason for such a scarce dataset is the very low 
cross section for their indirect production with pion or photon beams.
In Fig.~\ref{fig:omega}(right), we present the cross section for 
$\Omega$ production using a $K^-$~beam~\cite{Hassall1981}.

A major effort in LQCD calculations involves the determination of
inelastic and multi-hadron scattering amplitudes, and the first 
calculation to study an inelastic channel was recently 
performed~\cite{Wilson2015,Dudek2014}.  For lattice calculations 
involving baryons that contain one or more strange quarks an 
advantage is that the number of open decay channels is generally 
smaller than for baryons comprised only of the light $u$ and $d$ 
quarks.

\subsection{Summary for PWA}

Pole positions have been determined (no uncertainties) for several 
$\Lambda^\ast$s and $\Sigma^\ast$s but information about pole 
positions has not been determined for $\Xi$ or $\Omega$
hyperons~\cite{PDG2016}.  Our plan is to do a coupled-channel PWA 
with new GlueX KLF data in combination with available and new J-PARC 
$K^-p$ measurements when they will be available. Then the best fit 
will allow the determination of data-driven (model independent) 
partial-wave amplitudes and associated resonance parameters (pole 
positions, residues, BW parameters, and so on.  Additionally, PWAs 
with new GlueX data will allow a search for ``missing" hyperons via 
looking for new poles in complex plane positions. It will provide 
a new benchmark for comparisons with QCD-inspired models and LQCD 
calculations.

\section{Theory for ``Neutron" Target Measurements} 
\label{sec:Maxim}

\begin{figure}[h]
\centering
{
    \includegraphics[width=0.6\textwidth,keepaspectratio]{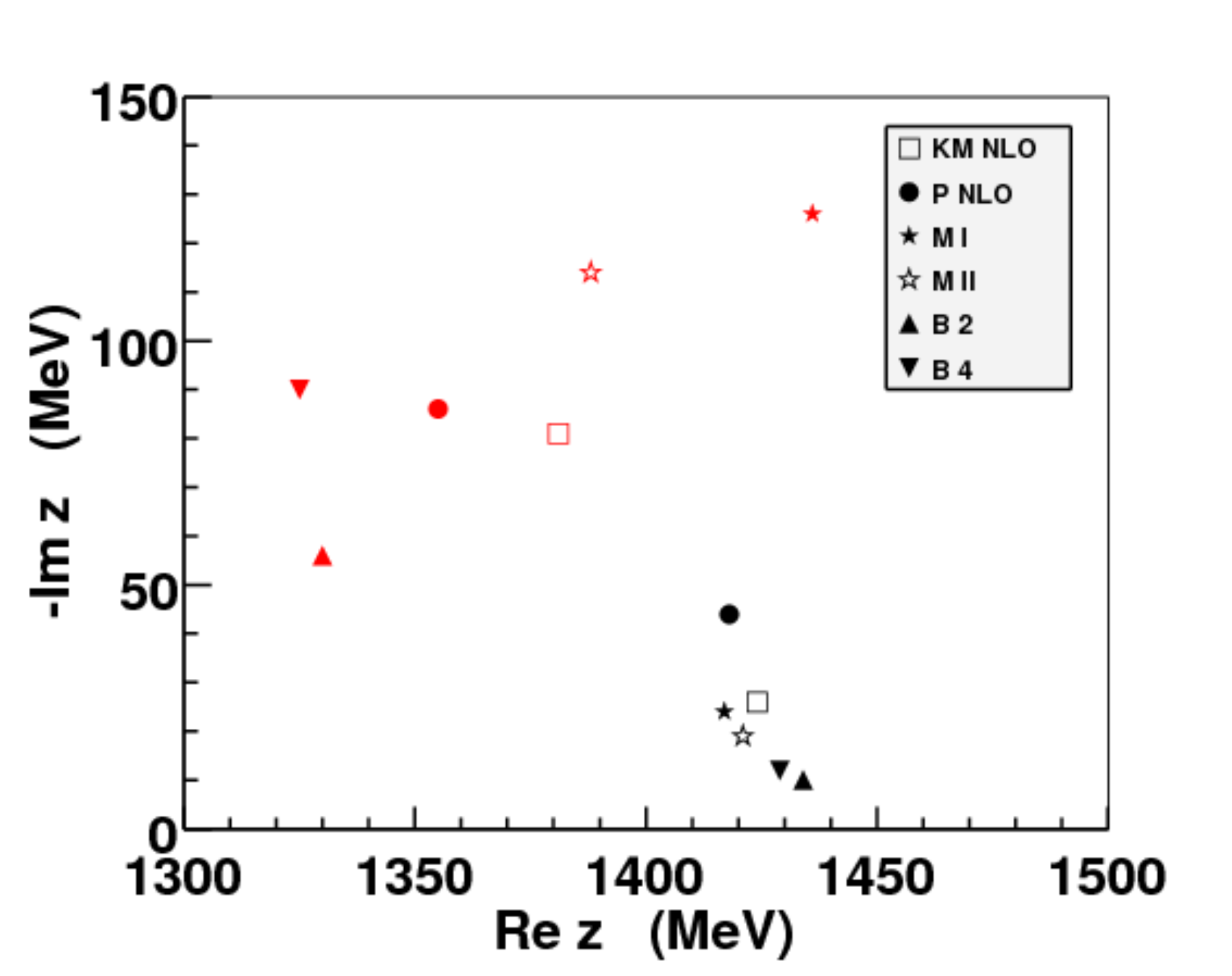} }

    \caption{Pole positions of the $\Lambda(1405)$ for chiral unitary 
    approaches - $KM$ from 
    Ref.~\protect\cite{Ikeda:2012au}, $B$ from Ref.~\protect\cite{Mai:2014xna}, 
    $M$ from Ref.~\protect\cite{Guo:2012vv} and $P$ from 
    Ref.~\protect\cite{Cieply:2011nq} as compared in 
    Ref.~\protect\cite{Maxim16}. Each symbol represents the position of the 
    first (black) and second (red) pole in each model.}\label{fig:maxim3}
\end{figure}

So-called coupled-channel chiral unitary approaches (UChPT) successfully 
describe the properties of the $\overline{K}N$ sub-threshold
resonance $\Lambda(1405)1/2^-$.
Furthermore, such models lead to the prediction that the scattering
amplitude has two poles in the complex-energy plane for the quantum
numbers of this resonance ($I=0, L=0, S=-1$). This coins the so-called
the two-pole structure of the $\Lambda(1405)$; see the current {\it Review 
of Particle Physics}~\cite{PDG2016} for more details.
\begin{figure}[th!]
\centering
{
    \includegraphics[width=\textwidth,keepaspectratio]{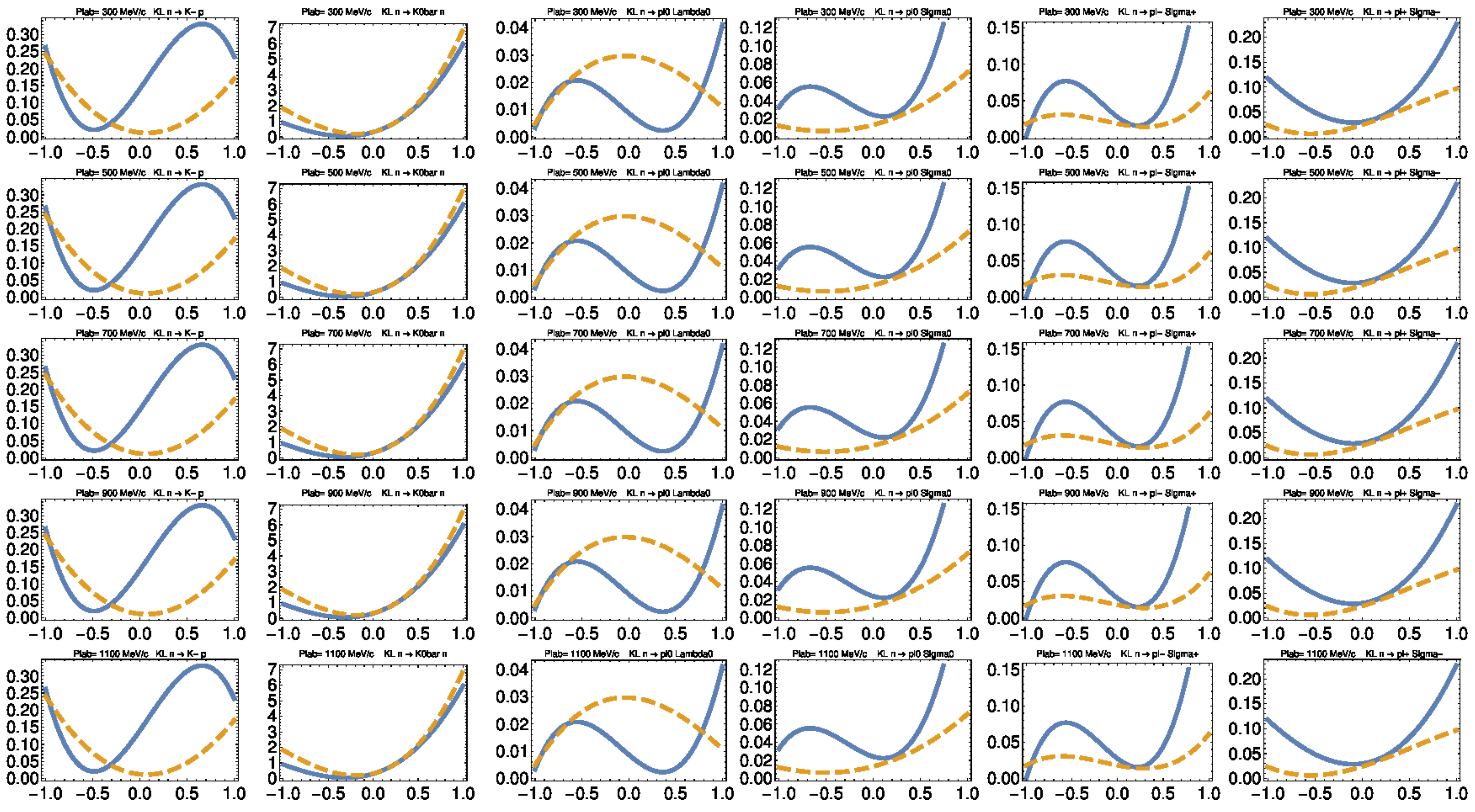} }

    \caption{Theoretical predictions for differential cross sections,
    $d\sigma/d\Omega$, for reactions (columns)
    $K_Ln\to K^-p$,
    $K_Ln\to \overline{K^0}n$,
    $K_Ln\to\pi^0\Lambda$,
    $K_Ln\to\pi^0\Sigma^0$,
    $K_Ln\to\pi^-\Sigma^+$, and
    $K_Ln\to\pi^+\Sigma^-$
    as a function of c.m.\ $\cos\theta$.
    Each row associated with kaon lab momentum  of 300, 400, $\ldots$
    1000~MeV/$c$ of initial neutral kaon beam.
    Orange dashed and blue solid lines show predictions within Model-B2
    and Model-B4, respectively (see text for details). }\label{fig:maxim}
\end{figure}
\begin{figure}[th!]
\centering
{
    \includegraphics[width=\textwidth,keepaspectratio]{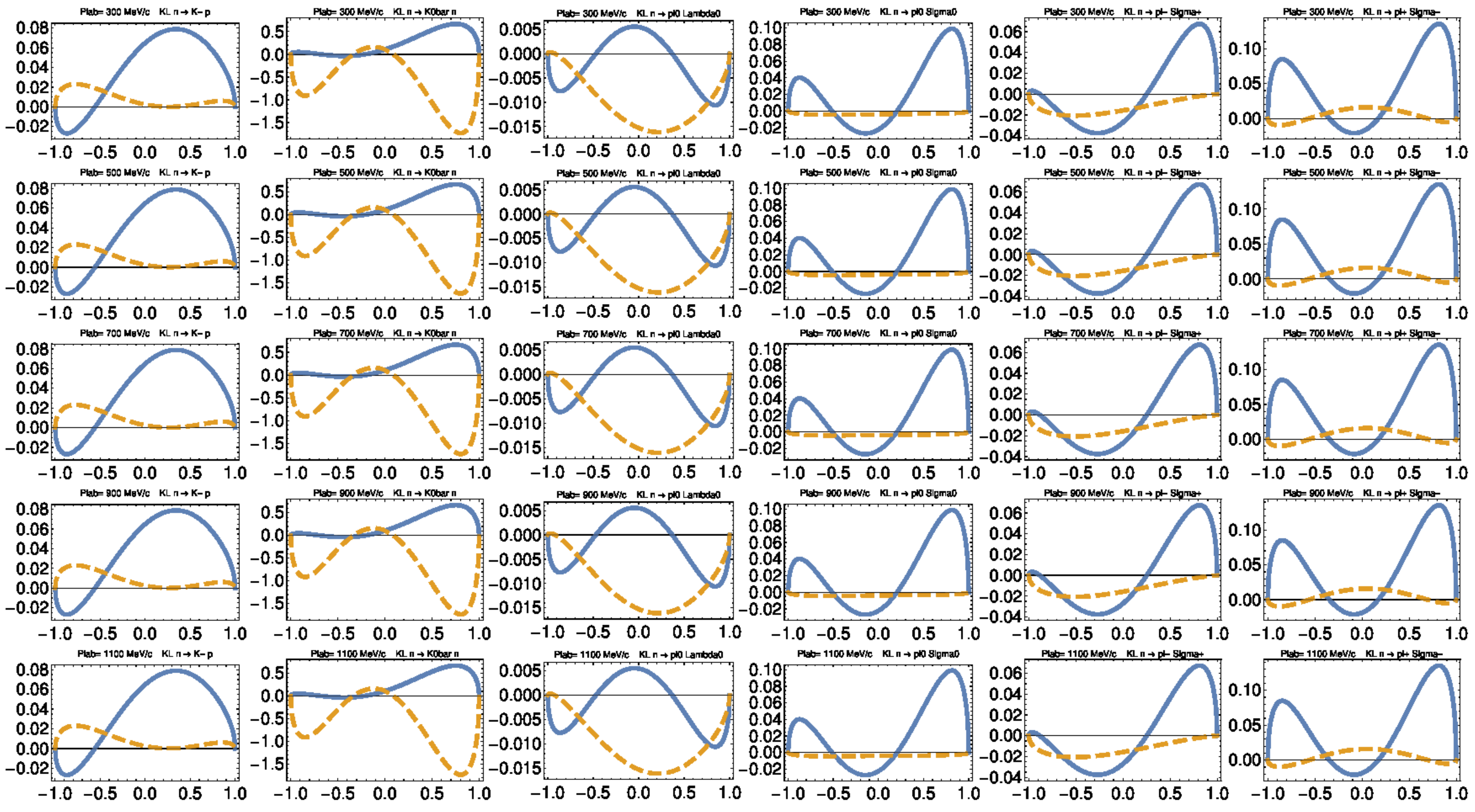} }

    \caption{Theoretical predictions for polarized differential cross
    sections, $P d\sigma/d\Omega$. The notation is the same as in
    Fig.~\protect\ref{fig:maxim}. }\label{fig:maxim1}
\end{figure}

In the most recent formulation, the aforementioned UChPT approaches
rely on a chiral amplitude for the meson-baryon scattering up to
next-to-leading chiral order. Whereas the unitarity constraint is
usually imposed via the Bethe-Salpeter equation either in the full
off-shell formulation~\cite{Mai2012, Bruns:2010sv} or in the so-called
on-shell approximation, e.g, ~\cite{Mai:2014xna,Ikeda:2012au,Guo:2012vv}.
For the analysis of data the former is quite intricate, while as it was
shown in Ref.~\cite{Mai2012} the off-shell effects are rather small.
Therefore,
it is meaningful to use the latter formulation. Recently, a direct
quantitative comparison of the on-shell models~\cite{Mai:2014xna,
Ikeda:2012au,Guo:2012vv,Cieply:2011nq} was performed in
Ref.~\cite{Maxim16}. It was found there that various models, which
typically have many free parameters, adjusted to the same experimental
data, predict very different behavior of the scattering amplitude on
and off the real-energy axis. This systematic uncertainty becomes
evident, when comparing the pole positions of the $\Lambda(1405)$ in
these models (see Fig.~\ref{fig:maxim3}). The position of the narrow
(first) pole seems to be constrained at least in the real part rather
well, while the predictions for the position of the broad (second) pole
cover a very wide region of the complex-energy plane. This uncertainty
is present even within models of the same type. This ambiguity can be
traced back to the fact that the experimental data used to fix the
parameters of the models are rather old and imprecise. It is expected
that the proposed KLF experiment will lead to an improvement
of this situation, as described below.

The $K_L$ beam can be scattered on a ``neutron" target, while measuring
the strangeness $S=-1$ final meson-baryon states (see, e.g.,
Sec.~\ref{sec:PWA}). In such a setup, the proposed experiment can become
a new and very strongly desired new source of experimental data to
pinpoint the properties of the $\overline{K}N$ scattering amplitude.
To make this statement more quantitative, we compare predictions of both
solutions of the model\footnote{The choice of this model for the present
analysis is justified by the fact that it includes the $p$-wave interaction
in the kernel of the Bethe-Salpeter equation explicitly.} from
Ref.~\cite{Mai:2014xna}. These solutions agree with all presently
available scattering, threshold as well as the photoproduction data
for the $\pi\Sigma$ line shapes by the CLAS
Collaboration~\cite{Moriya:2013eb}. The predicted differential cross
sections ($d\sigma/d\Omega$) as well as polarized ones ($P d\sigma/d\Omega$)
for the $K_Ln$ scattering with the final states $K^-p$, $\overline{K^0}n$,
$\pi^0\Lambda$, and $\pi^{0/+/-}\Sigma^{0/-/+}$ are presented in
Figs.~\ref{fig:maxim} and \ref{fig:maxim1}, respectively. There is very
little agreement on the prediction of these observables in the energy
range aimed to study in the proposed $K_L$ experiment. The latter is
very encouraging, meaning that the actual data can sort out one (or
maybe both) solutions as unphysical, which was not possible based on
present experimental data.

\underline{In summary:} The proposed KLF experiment will lead to new
constraints on $\overline{K}N$ models; thus, these data will sharpen our 
understanding of the long-debated nature of strangeness $S=-1$ resonances.

\section{Strange Meson Spectroscopy: $K\pi$ Interaction}
\label{sec:Moskov}

Below we present current status of K-pi scattering summarized in 
Ref.~\cite{Wilson:2014cna}: ``\textit{The bulk of our knowledge of kaon 
scattering amplitudes comes from kaon beam experiments at Stanford 
Linear Accelerator Center (SLAC) in the 1970s and 1980s. $\pi K$
scattering amplitudes were extracted from reactions using a proton 
target by extrapolating to small momentum transfer, $t$, where nearly 
on-shell pion exchange dominates. Phaseshift analysis of the 
flavor-exotic isospin-3/2 amplitudes as extracted from
$K^+p\to K^+\pi^+n$ and $K^-p\to K^-\pi^-\Delta^{++}$ by 
Estabrooks \textit{et al.}~\cite{Estabrooks} indicates a weak 
repulsive interaction in the S-wave and very weak interactions in 
the P-wave and higher.}

\textit{In isospin 1/2, as well as the phase-shift analysis of Estabrooks 
\textit{et al.}, there is a considerable set of $\pi K$ scattering
results provided by the LASS experiment-of particular relevance here 
are the final states $\pi K$~\cite{LASS}, $\eta K$~\cite{LASS1}, and
$\pi\pi K$~\cite{LASS2}. In the partial-wave analysis of $\pi K\to\pi 
K$, a peaking amplitude in the S-wave is interpreted as a broad 
$K_0^\ast(1430)$ resonance which appears to saturate unitarity. The 
narrow elastic vector resonance, $K^\ast(892)$, presents itself as a 
rapid rise in the P-wave phase shift. The D-wave amplitude has a
peak, well below the unitarity limit, that can be interpreted as
an inelastic $K_2^\ast(1430)$ resonance. Further resonances in the
"natural-parity" series ($J^P = 3^-, 4^+, 5^-$) are observed at
higher energies.}

\textit{$\eta K$ is the first inelastic channel to open, but LASS reports
no significant amplitude into $\eta K$ for $E_<cm> < 2$~GeV in S-, 
P-, and D-waves. Indeed the inelasticity in P- and D-waves and
higher appears to come first from the $\pi\pi K$ final state, where
a significant amplitude is seen in $1^-$ above 1.3~GeV and a
peak in $2^+$ at the $K_2^\ast(1430)$, $\pi\pi K$ also couples to the
"unnatural-parity" series, notably to $J^P = 1^+$, where peaking
behavior is observed that is commonly described in terms of two axial 
resonances, $K_1(1270)$, $K_1(1400)$.}"

Recently LQCD studies with $m_{\pi} = 391$~MeV were performed
to search for resonances in coupled $\pi K$ and $\eta K$
scattering~\cite{Wilson2015}.  Scalar $\pi\pi/\overline{K}K$ and 
$K\pi/K\eta$ form factors have been calculated within a variety of 
approaches using (unitarized) chiral perturbation theory~\cite{a46,
a47,a48,a49,a50,a51,a52,a53} and dispersion relations~\cite{a52,a54,
a58}, in many cases using the former to constrain polynomial 
ambiguities of the latter.
\begin{figure}[h!]
\centering
{
    \includegraphics[width=0.5\textwidth,keepaspectratio]{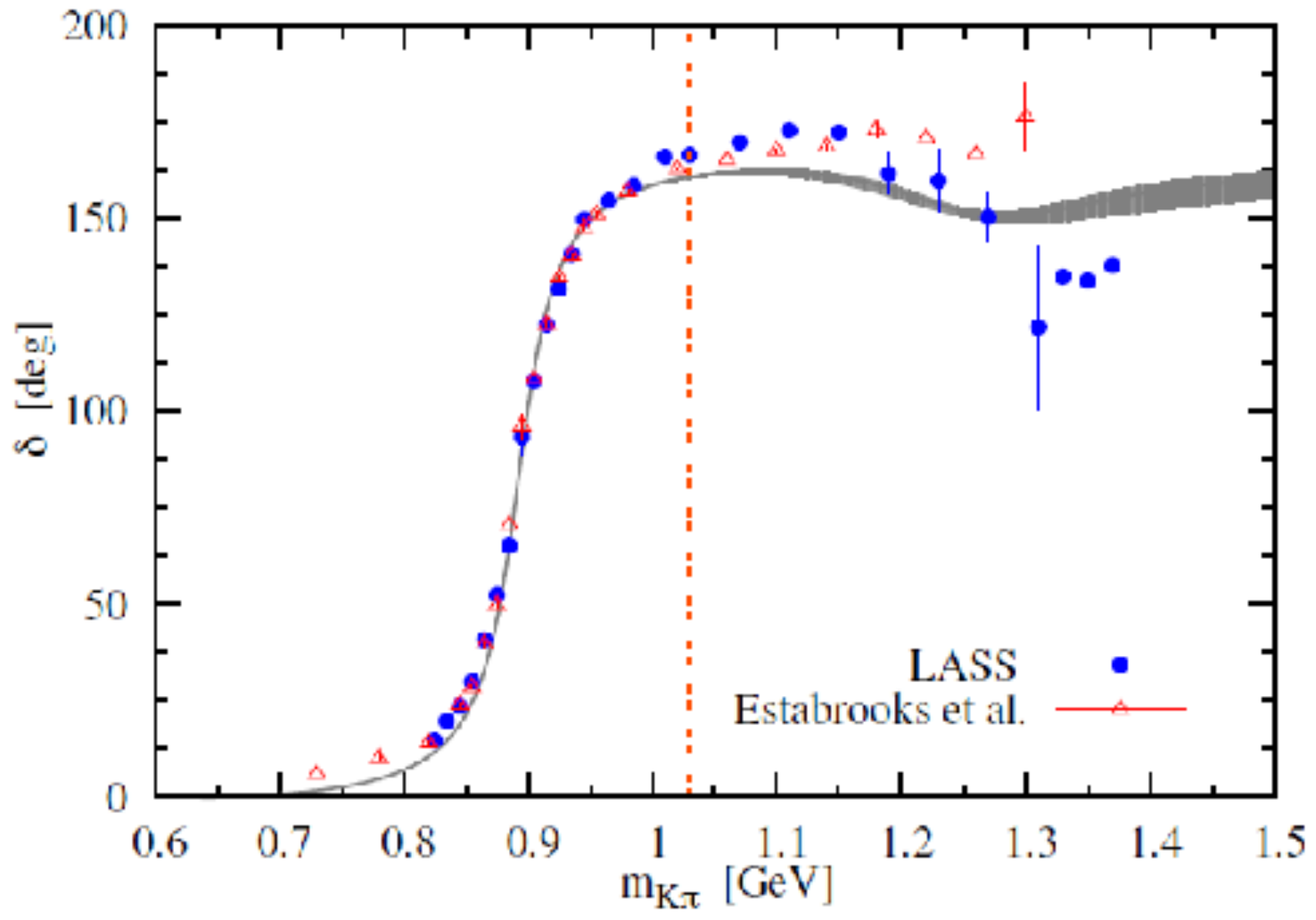} }

        \caption {I=1/2 $K\pi$ scattering P-wave phase shift together
        with experimental results from LASS~\protect\cite{LASS} and
        Estabrooks \textit{et al.}~\protect\cite{Estabrooks}. The opening of
        the first inelastic $\pi K^\ast$ channel is indicated by dashed
        vertical line. The gray band represents the fit results from
        Boito \textit{et al.}~\protect\cite{Boito}.}
       \label{fig:p-wave}
\end{figure}

Measuring $\pi K$ scattering provides a possibility for studying scalar and 
vector $K^\ast$ states, including the S-wave $\kappa(800)$
state (see~\cite{Descotes,Pelaez}), which is not yet well established.  Such
studies are also necessary to get precise vector and scalar $\pi K$ form
factors as an input for the extraction of the Cabibbo-Kobayashi-Maskawa (CKM)
matrix element $V_{us}$ from $\tau\to K\pi\nu$ decay. $\pi K$
scattering amplitudes with high precision are needed to study CP
violation from Dalitz plot analyses of both open charm 
$D$-mesons~\cite{Kubis2015} and the charmless decay of 
$B$-mesons~\cite{Doring:2013wka} into $K\pi\pi$ final state.

In Fig.~\ref{fig:p-wave}, we present the phase of the form factor
$F_{+}(s)$ with experimental results of LASS
Estabrooks~\cite{Estabrooks,LASS} together with the fit of Boito
\textit{et al.} to $\tau$ decay data~\cite{Boito}.

As one can see, all experimental data obtained at SLAC have
very poor statistics above 1.2~GeV; furthermore, the data do not extend 
to higher energies, which are even more important for $B$-meson decays.
Moreover, direct comparison of charged $K^{\pm} \pi^{\mp}$ with $\tau$
assumes isospin invariance as in the $\tau$ decay one has
$K_S\pi^{\pm}$ final state depending on the sign of $\tau$ lepton.

\begin{figure}[h!]
\centering
{
    \includegraphics[width=0.4\textwidth,keepaspectratio]{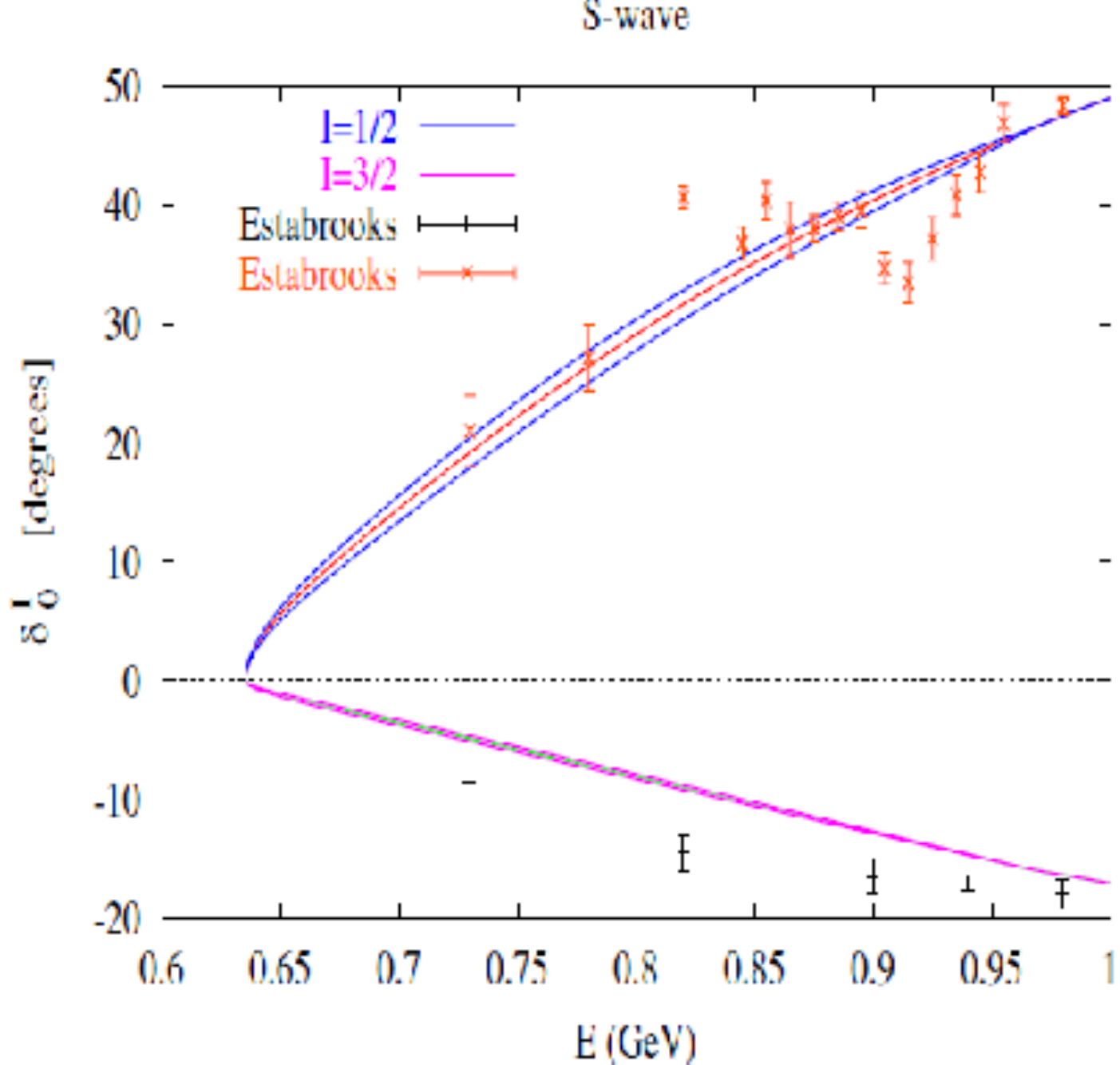} }
{
    \includegraphics[width=0.4\textwidth,keepaspectratio]{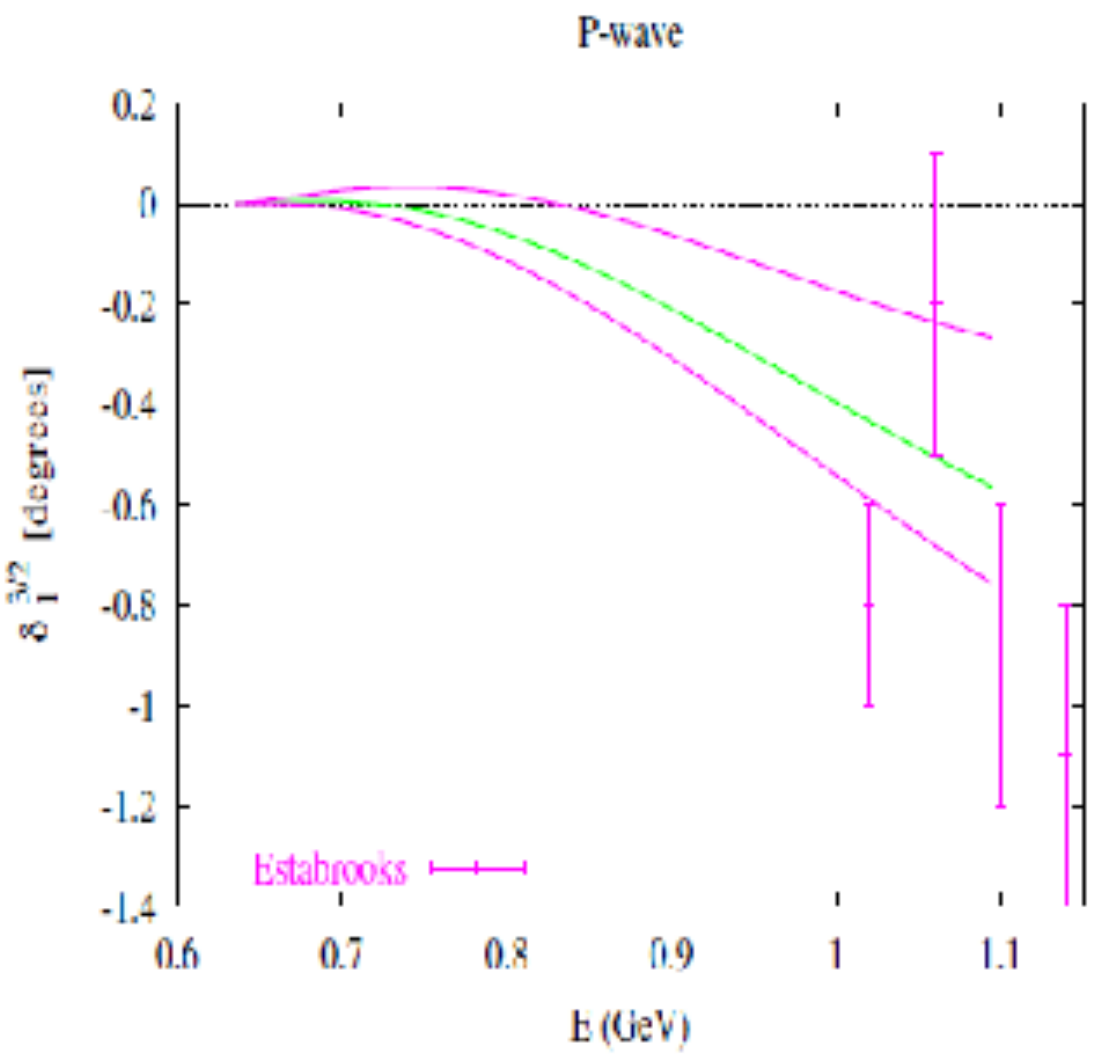} }

    \caption {Left panel: $I=1/2$ S-wave phase shift (curves and data 
    in the upper half of the figure) and the $I=3/2$ S-wave phase shift 
    (curves and data in the lower half).  Experimental data are from 
    SLAC experiments as in previous figure. The curves are obtained 
    from central, upper, and lower values of parameters in the Roy-Steiner
    solutions ellipse~\protect\cite{Bouttiker}.
    Right panel: Same as in previous figure for $I=3/2$. Data points
    are from Estabrooks \textit{et al.}~\protect\cite{Estabrooks}.}
    \label{fig:wave}
\end{figure}

Similarly, as one can see from Fig.~\ref{fig:wave}, the
$I=1/2$ and $I=3/2$ S-wave and $I=3/2$ P-wave phase shifts are very
poorly measured and need more experimental data.

The intensive beam flux of the proposed $K_L$ beam  will
provide high statistics  data on both charged $K\pi$ as well as with 
final-state neutral kaon in the reactions:
\begin{itemize}
\item  $K_Lp\to K^{\pm}\pi^{\mp} p$ (simultaneousely measurable
	with $K_L$ beam).
\item $K_Lp\to K_S\pi^+n$ on a proton target \textit{(for the
	first time)}.
\item $K_Ln\to K_S\pi^-p$ on a deuteron target \textit{(for
	the first time)}.
\end{itemize}

\underline{In summary:} Experimental data obtained in the proposed KLF
experiment at JLab will provide valuable data to 
search for yet not well understood and possibly incomplete scalar, 
vector, and tensor resonances in the strange sector through a phase-shift 
analysis of $\pi K$ and $\eta K$ scattering amplitudes.

\section{Proposed Measurements}
\label{sec:run}

We propose to use the KL Facility with the GlueX spectrometer, in
JLab Hall~D, to perform precision measurements of $K_Lp\to KY^\ast$
from liquid hydrogen and deuterium cryotarget (LH$_2$/LD$_2$) in the
resonance region, $W = 1490$ -- 3500~MeV and c.m.\ $\cos\theta$ from
$-0.95$ to 0.95.  It will operate at a neutral kaon flux of $3\times
10^4~K_L/s$.  The ability of GlueX to measure over wide ranges in
$\theta$ and $\phi$ with good coverage for both charged and neutral
particles, together with the $K_L$ energy information from the KL
Facility, provide an ideal environment for these measurements.

\subsection{$K_L$ Beam in Hall~D}

\begin{figure}[h!]
\centering
{
    \includegraphics[width=1\textwidth]{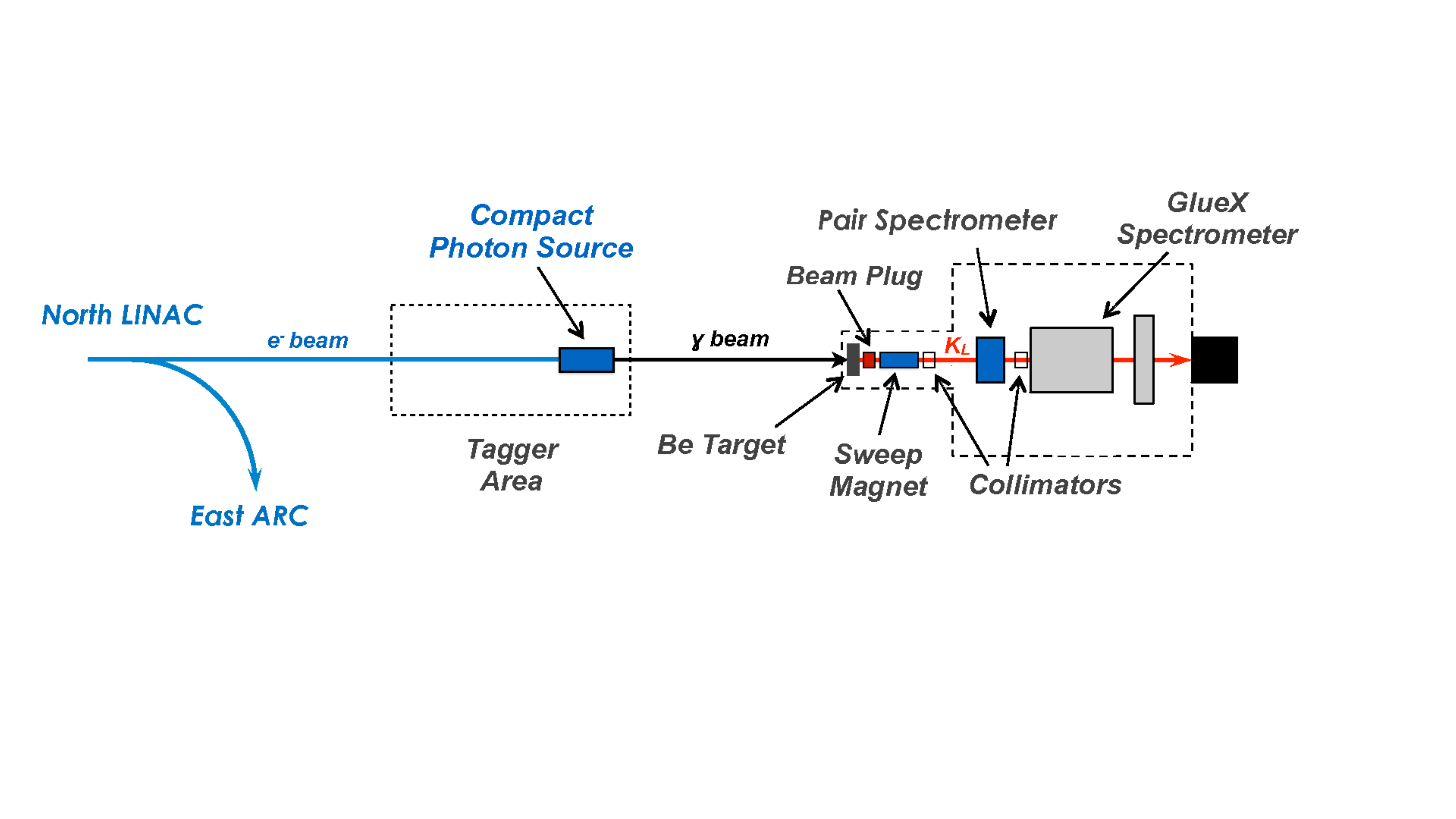} }

    \caption{Schematic view of Hall~D beamline on the way $e\to\gamma\to
    K_L$. Electrons first hit the tungsten radiator, then photons 
    hit the Be~target assembly, and finally, neutral kaons hit the
    LH$_2$/LD$_2$ cryotarget. The main components are CPS, Be~target 
    assembly, beam plug, sweep magnet, and pair spectrometer. See the 
    text for details.} \label{fig:exp}
\end{figure}

A schematic view of the Hall~D beamline for KLF is presented in
Fig.~\ref{fig:exp}. At the first stage, $E = 12$~GeV electrons produced
at CEBAF will scatter in the radiator of the Compact Photon Source 
(CPS), generating an intense beam of untagged bremsstrahlung photons.
The Hall~D tagger magnet and detectors will not be used.
At the second stage bremsstrahlung photons, created by electrons at a 
distance about 75~m upstream, hit the Be~target assembly located in 
the cave, and produce neutral kaons along with neutrons and charged 
particles. Finally, $K_L$ mesons will reach the LH$_2$/LD$_2$ 
cryotarget inside the GlueX spectrometer.

Our calculations have been performed for the JLab Hall~D beamline
geometry. The primary $K_L$ production target has been placed in the Hall~D
collimator cave. For the target material, we selected beryllium because, for
thick targets, the $K_L$ yield is roughly proportional to the radiation length
and density, which gives beryllium as the best candidate. The beam plug
and sweeping magnet are placed right after the target. For our
calculations, we took a simple beam plug: a 15~cm thick piece of lead.
The permanent sweeping magnet cleans up the charged component and has a field
integral of 0.8~Tesla$\cdot$meter, which is enough to remove all charged
background coming out of the beam plug. The vacuum beam pipe has a 7~cm
diameter and prevents neutron rescattering in air. There are two
collimators: one is placed before the wall between collimator cave and
experimental hall, while the other is placed in front of the Hall~D detector. 
The distance between the primary Be~target and the LH$_2$/LD$_2$ target 
(located inside Hall~D detector) was taken as 16~m in our calculations
It can be increased up to 20~m.

\subsubsection{Compact Photon Source: Conceptual Design}
\label{sec:CPS}

\begin{figure}[h!]
\centering
{
    \includegraphics[width=0.6\textwidth,keepaspectratio]{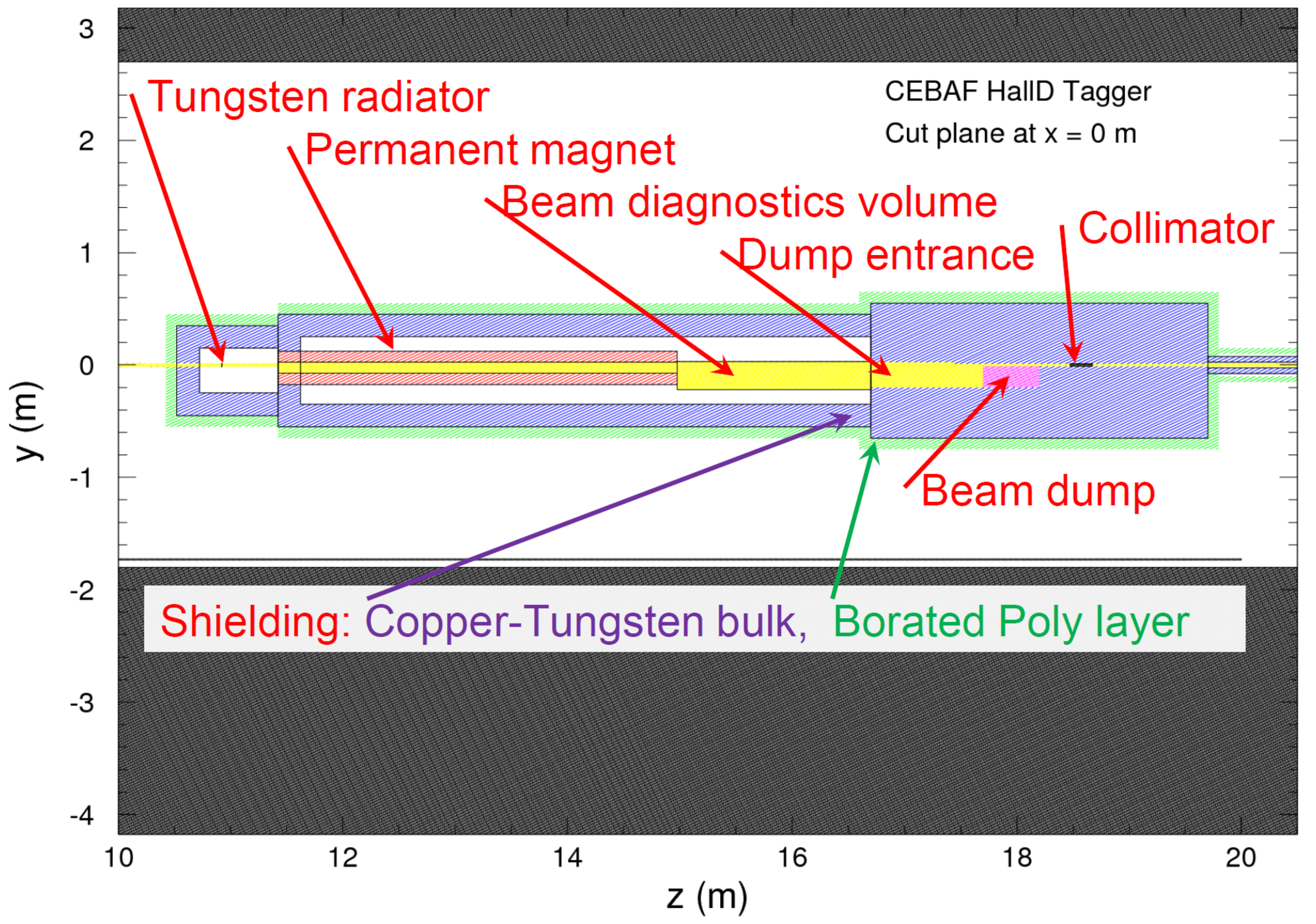} }
{
    \includegraphics[width=0.6\textwidth,keepaspectratio]{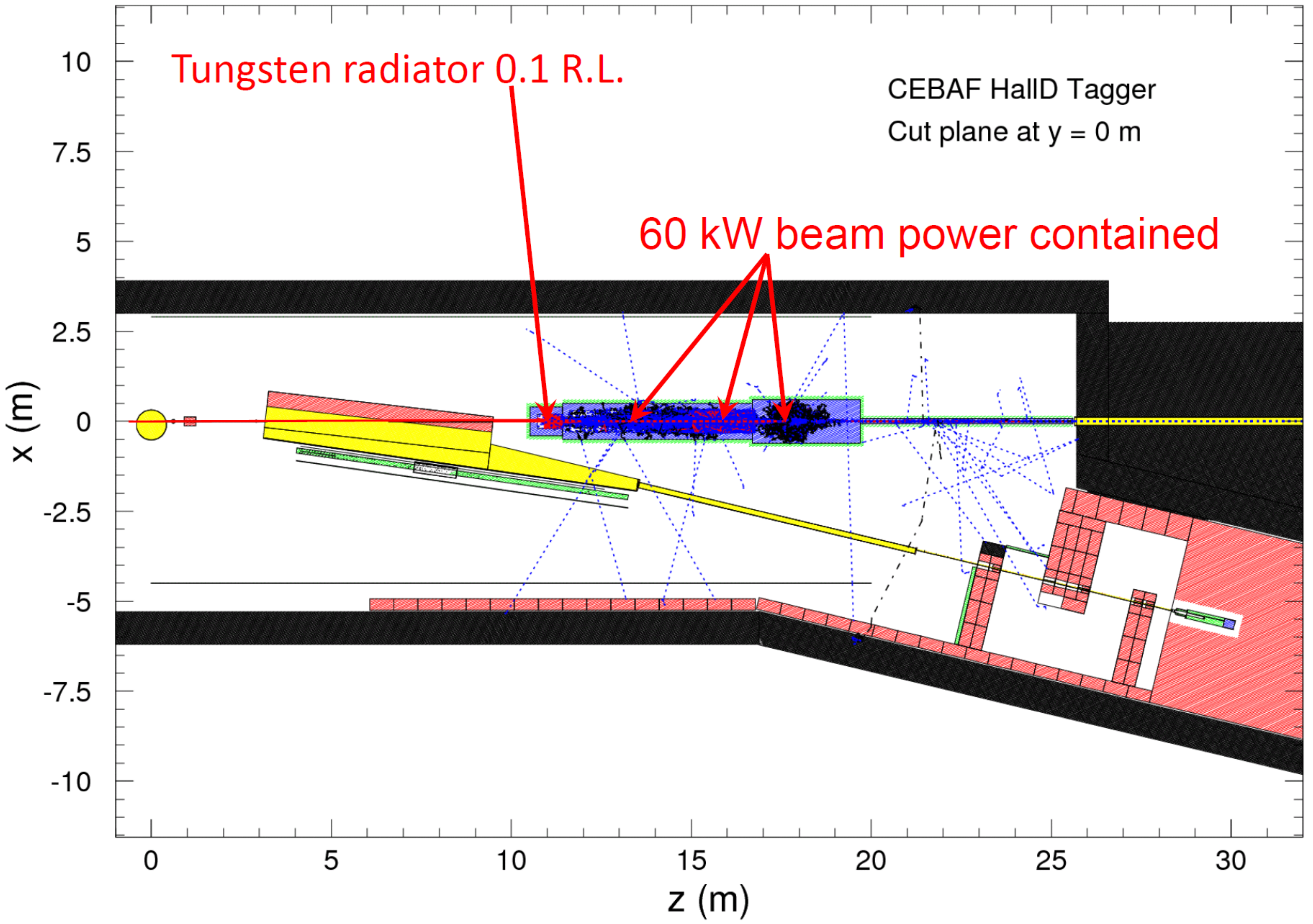} }

    \caption{Elements of the design are indicated in the top panel
    (vertical cut plane of the DINREG/Geant3 model of the CPS).  The 
    bottom panel
    shows the CPS assembly in the Tagger vault and simulations of 2000
    beam electrons at 12~GeV.} \label{fig:CPS}
\end{figure}

An intense high-energy gamma source is a prerequisite for the
production of the $K_L$ beam needed for the new experiments described 
in this proposal. In 2014, Hall~A Collaboration has been discussed
a novel concept of a Compact Photon Source (CPS)~\cite{report1}. It
was developed for a \textit{Wide-Angle Compton Experiment} proposed to 
PAC43~\cite{proposalWACS}.  Based on these ideas, we suggested (see 
Ref.~\cite{PDBW}) to use the new concept in this experiment. A possible 
practical implementation adjusted to the parameters and limitations of 
the available infrastructure is discussed below. The vertical cut of 
the CPS model design, and the horizontal plane view of the present 
Tagger vault area with CPS installed are shown in Fig.~\ref{fig:CPS}.

The CPS design combines in a single properly shielded assembly all
elements necessary for the production of the intense photon beam, such
that the overall dimensions of the setup are limited and the
operational radiation dose rates around it are acceptable.
Compared to the alternative, the proposed CPS solution presents several 
advantages: much lower radiation levels, both prompt and post-operational 
due to the beam line elements' radio-activation at the vault. The new 
design provides much less disturbance of the available infrastructure at 
the Tagger Area, and better flexibility in achieving high-intensity 
photon beam delivery to Hall~D. The new CPS solution will satisfy the 
proposed $K_L$ beam production parameters; we do not envision any 
significant technical or organizational difficulties in the 
implementation of the conceptual design.

The new setup utilizes the Hall~D Tagger vault, properly shielded by
design to accomodate the medium power beam dump capable of accepting up 
to 60~kW of 12~GeV $e^-$ beam, assuming that proper local shielding is 
set around the dump. The presently installed dump is placed behind
the iron labyrinth walls, and is surrounded by a massive iron shielding, 
made of iron blocks available at the time of construction.  
The standard GlueX setup is optimized for operations using 
very thin radiators producing relatively low intensity 
photon beam such that the beam electrons losing energy to 
photon production in the radiator may be detected and counted 
in the tagger hodoscope counters. The present setup is not 
suitable for production of massively more intense photon 
beams needed for the $K_L$ production, due to the expected 
overwhelming radiation and activation levels in the vault.

The new proposed CPS solution solves the problem by incorporating the
new thick radiator and the new beam dump in one assembly installed
along the straight beam line exiting from the tagger magnet (presently the 
line is used as the photon beam line). The new CPS device should be capable 
of taking the same beam power of 60~kW, using optimized shielding made of 
high-Z material, which would make the necessary equivalent shielding 
compact, requiring less total weight of the shielding.  Qualitatively, 
if one needs a sphere of iron (8~g/cm$^3$) of 2~m radius for the shielding, 
it may be roughly replaced by a sphere of 1~m radius made of 
tungsten-copper (16~g/cm$^3$), with its weight actually four times smaller.
The optimized design is able to limit the prompt radiation dose 
rates around the CPS to the present operational levels, while significantly 
limiting the post-operational doses around the heavily shielded assembly. 
Of course, the inner parts of the CPS device will be activated to high
levels, preventing immediate access and disassembly, so the engineering 
requirements to the reliability of all parts inside must be strict. The 
overhead shielding at the CPS location in the tagger vault is about the 
same thickness (13~feet) of concrete and berm as at the present dump 
location. It will keep the radiation doses outside and at the CEBAF 
boundary within the design limits for the site. 

The proposed CPS solution is just conceptual, and a full cycle of 
engineering design is required before the final optimized solution is 
found. The cost and space limitations will determine the choice of 
shielding materials for the CPS.  Details of the dump and magnet design 
will also be included in the overall optimization process, taking into 
account the considerations of cost and reliability of the final device. 
We are considering a possible joint development of the more universal 
CPS solutions in collaboration with other experimental 
projects at JLab interested in implementing similar designs for their 
experiments~\cite{cpsW}.

\subsubsection{Simulations Study of $K_L$ Beam Production}

Neutral kaon production was simulated for a photon bremsstrahlung beam
produced by the 12~GeV electron beam in the Hall~D CPS.  The main mechanism 
of $K_L$ production in our energy range is via $\phi$-meson photoproduction, 
which yields the same number of $K^0$ and $\overline{K^0}$.  Another
mechanism is hyperon photoproduction (yielding only $K^0$), which was
not studied in our simulations separately. Instead, we have taken as
an alternative model the Pythia generator~\cite{pythia}, which includes
hyperon production. Total and differential cross sections for the 
$\phi$-meson photoproduction 
on proton and complex nuclei (coherent and incoherent)
data were taken from Refs.~\cite{Titov2003,McClellan1971}. The angular
distributions that we used for $\phi\to K_LK_S$ decay are from
Ref.~\cite{Titov2003,Titov2007,Mibe2007}. Our calculations show that
the $\phi$ decay in its rest frame is mostly perpendicular to the
axis of $\phi$ momentum. Since $K_L$s need to stay along the original 
photon beam direction to get to the LH$_2$/LD$_2$ cryotarget, this 
condition requires that the $\phi$ production and decay angles in the 
laboratory frame be about the same. That means that we will have only 
$K_L$s from $\phi$-mesons
produced at relatively high momentum transfer $t$ at the Be target.
It suppresses the number of ``useful" $K_L$s by a factor of $\sim 3$ or
more (in comparison with the case if $K_L$ and $K_S$ momenta are
parallel to the $\phi$ momentum).  $K_L$ absorption, used in our
calculations, was studied in Ref.~\cite{Brandenburg1973} very well.
About 80\% of the produced $K_L$s will be absorbed in the
Be~target and following tungsten and water beam plug. The value of
absorbed $K_L$s can be reduced by optimizing the beam plug setup.

\subsubsection{$K_L$ Beam Parameters} 

\begin{figure}[h!]
\centering
{
    \includegraphics[width=0.5\textwidth,keepaspectratio]{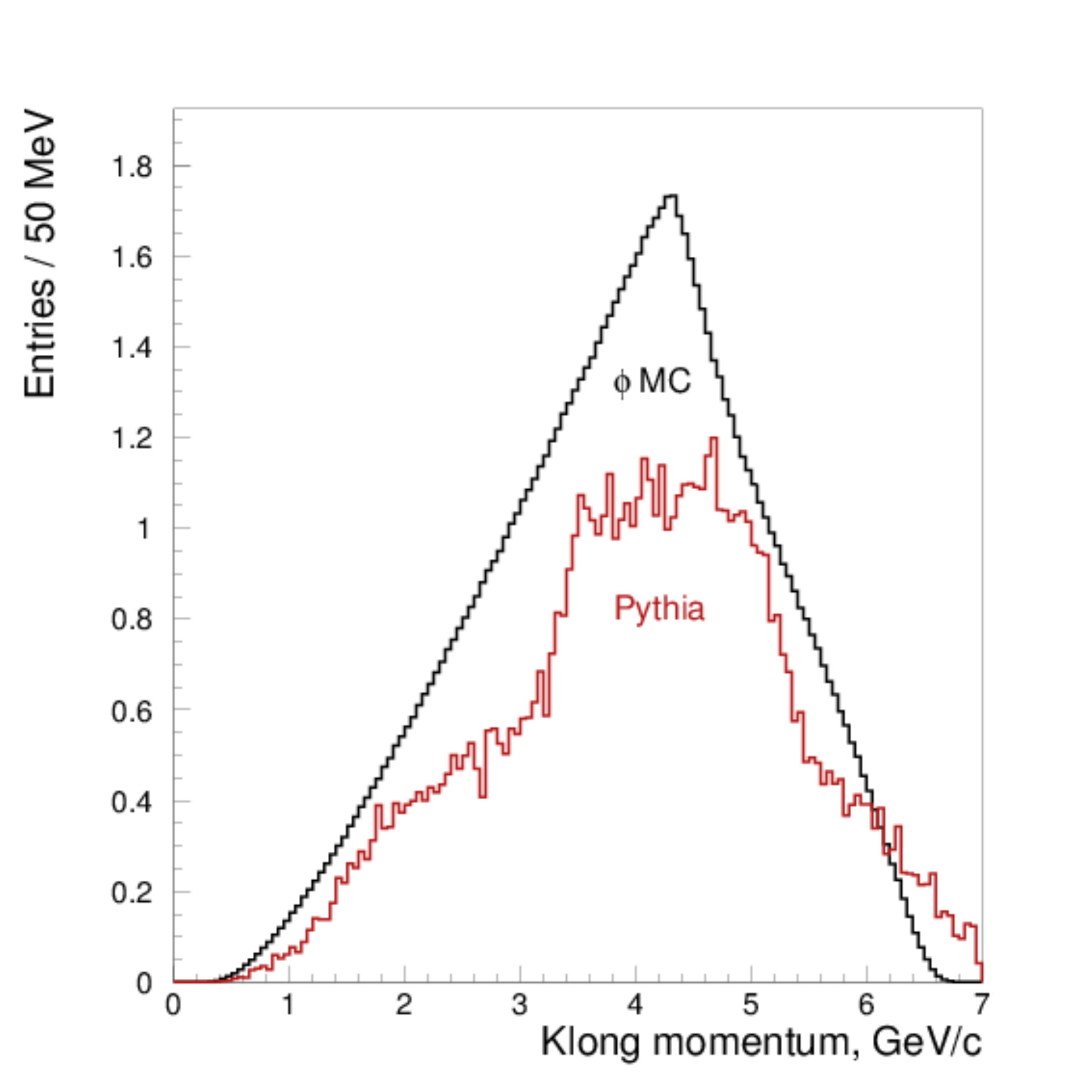} }

    \caption{$K_L$ momentum spectra originating from $\phi$ decays: 
    black histogram - our simulations using 
    DINRREG/Geant3~\protect\cite{geant},
    red histogram - Pythia generator result~\protect\cite{pythia}. } 
    \label{fig:beam}
\end{figure}

\begin{figure}[h!]
\centering
{
    \includegraphics[width=0.6\textwidth,keepaspectratio]{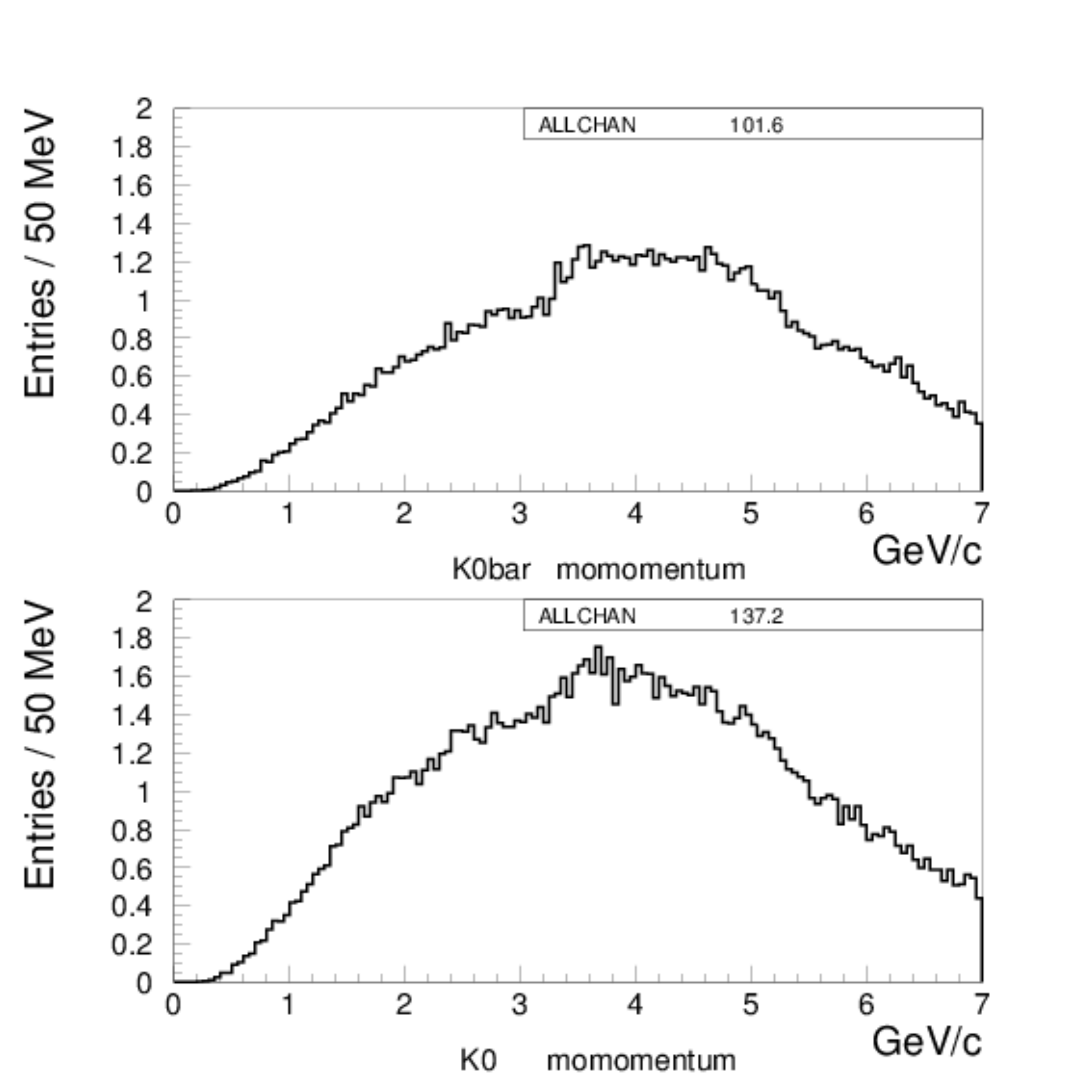} }

    \caption{Momentum spectra from Pythia generator~\protect\cite{pythia}.
    Top plot for $\overline{K^0}$. Bottom plot for $K^0$.} \label{fig:beam1}
\end{figure}

One of the main $K_L$-beam parameters is the momentum distribution (momentum
spectrum as a function of the distance and angle)~\cite{Larin16}. Results
of our simulations for the $K_L$ momentum spectrum for those $K_L$ reaching 
the LH$_2$/LD$_2$ cryotarget is shown in Fig.~\ref{fig:beam}. The spectrum
first increases with $K_L$ momentum up to $\sim4$~GeV/$c$ since the $\phi$ 
decay cone angle decreases at higher $\gamma$-beam and $K_L$ momenta. This 
selects lower $\phi$
production $t$ values, which are more favorable according to the $\phi$
differential cross section. At a certain point, the highest possible
$\gamma$-beam momentum is reached and the $K_L$ momentum spectrum decreases
to the endpoint. For comparison, we selected part of the $K_L$ spectrum
from the Pythia generator that originated only from $\phi$ decays and 
showed it on the same plot (red histogram).
Pythia calculations show that $\phi$ decays yield roughly 30\% of the 
$K_L$ flux.
The number of $K^0$ exceeds the number of $\overline{K^0}$ by 30\%
according to this generator for our conditions. Their momentum spectra
are shown in Fig.~\ref{fig:beam1} separately.


\begin{figure}[h]
\centering
{
    \includegraphics[width=0.45\textwidth,keepaspectratio]{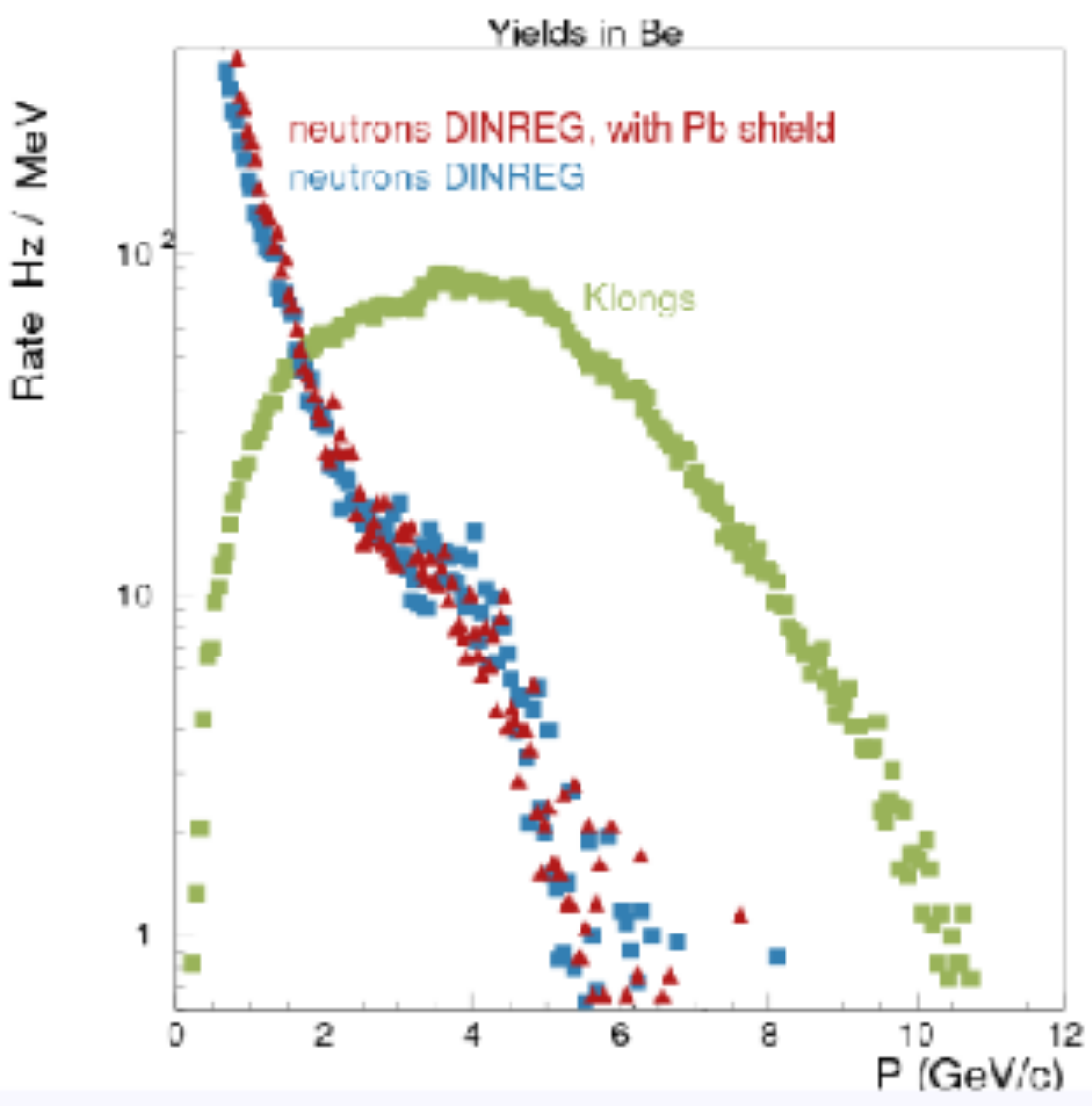} }
{
    \includegraphics[width=0.37\textwidth,keepaspectratio]{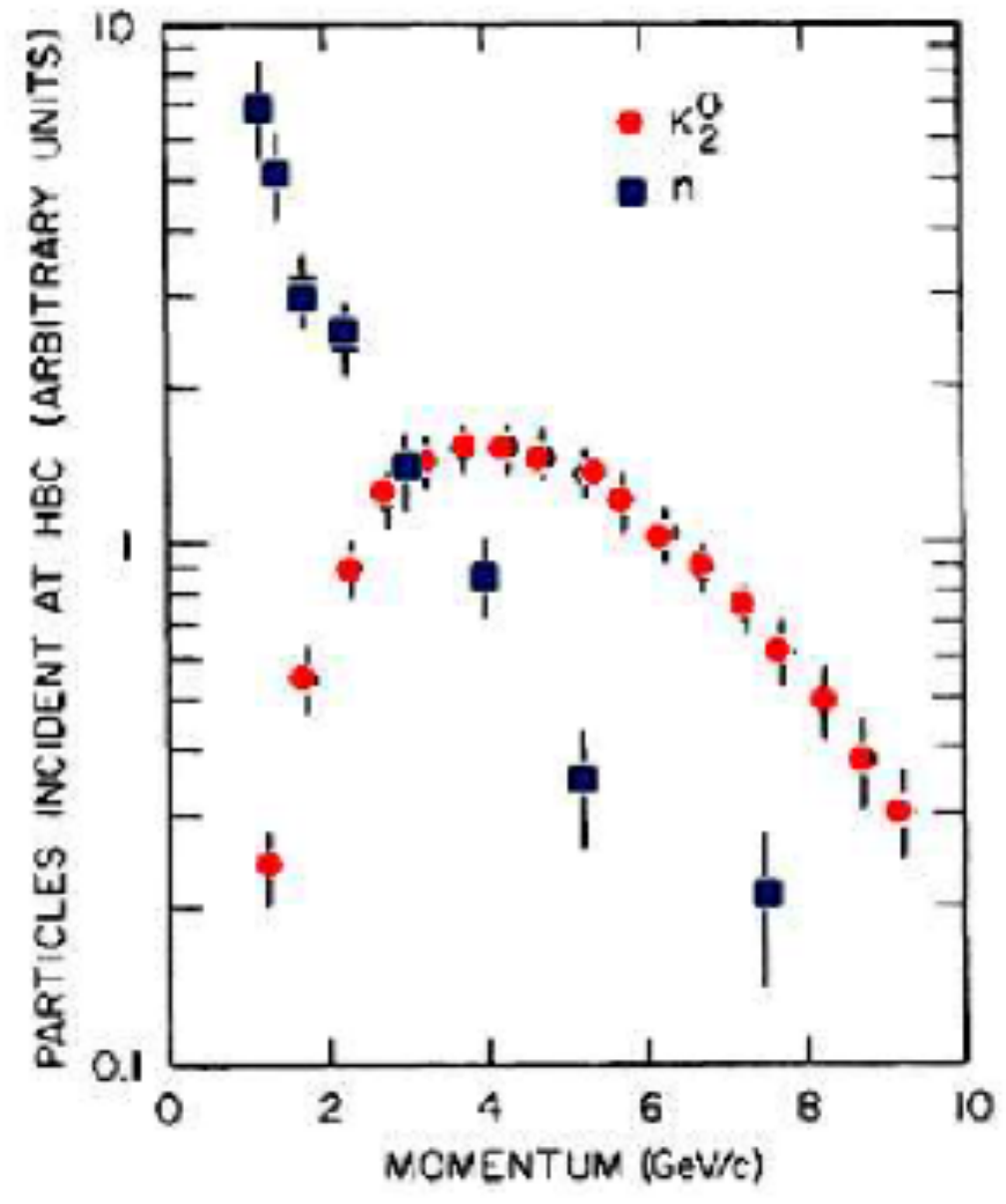} }

  \caption{$K_L$ and neutron momentum spectra.
  Left plot: The rate of $K_L$ (green filled squares) and neutrons (blue 
  and red filled squares and triangles) 
  on LH$_2$/LD$_2$ cryogenic target of Hall~D as a function of their 
  generated momentum, with a total rate of $3\times10^4 K_L$/s.  Neutron 
  calculations were performed using the JLab package 
  DINREG/Geant3~\protect\cite{geant}.
  Right plot: Experimental data from SLAC measurements using a 16~GeV/$c$
  electron beam from Ref.~\protect\cite{Brody}.  The rate of $K_L$
  (red filled circles) and neutrons (black filled squares).}
  \label{fig:neutron}
\end{figure}


To estimate the expected rate of $K_L$s at the LH$_2$/LD$_2$ cryotarget, we 
used the following conditions: 
\begin{itemize}
\item electron beam current 5~$\mu$A,
\item 10\%~R.L. tungsten radiator in CPS,
\item Be-target diameter is 3~cm and length is 40~cm, 
\item and a cryotarget $LH_2/LD_2$ radius of 3~cm
\end{itemize}
which results in a beam flux of about ${\bf 3\times10^{4}~K_L/}$\textbf{s} 
from all production mechanisms at the LH$_2$/LD$_2$ target 
(Fig.~\ref{fig:neutron}).  We simulated the $K_L$ and neutron production 
from $6\times 10^9$ 12-GeV electrons under these conditions for the GlueX 
$K_L$ Facility and the results (Fig.\ref{fig:neutron}(left)) are in 
reasonable agreement with the $K_L$ spectrum measured by SLAC at 16~GeV 
(Fig.~\ref{fig:neutron}(right)).

\clearpage

\subsubsection{$K_L$ Beam Background: Gammas, Muons, and Neutrons}

Background radiation conditions are one of the most important parameters 
of the $K_L$ beam for the JLab GlueX KL Facility~\cite{Larin16}.

\begin{enumerate}
\item \textbf{Gamma Background} 

\begin{figure}[h!]
\centering
{
    \includegraphics[width=0.43\textwidth,keepaspectratio]{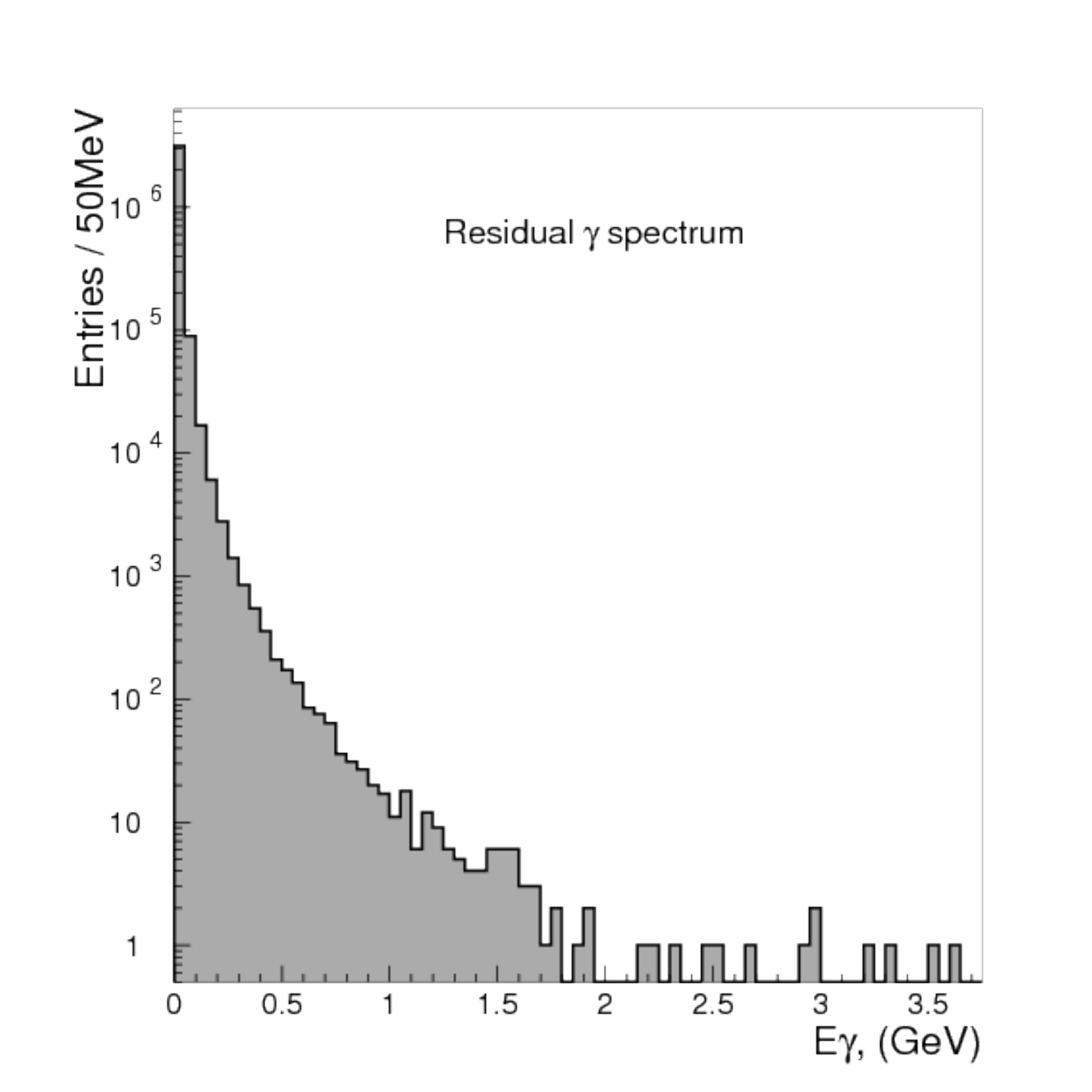} }
{
    \includegraphics[width=0.43\textwidth,keepaspectratio]{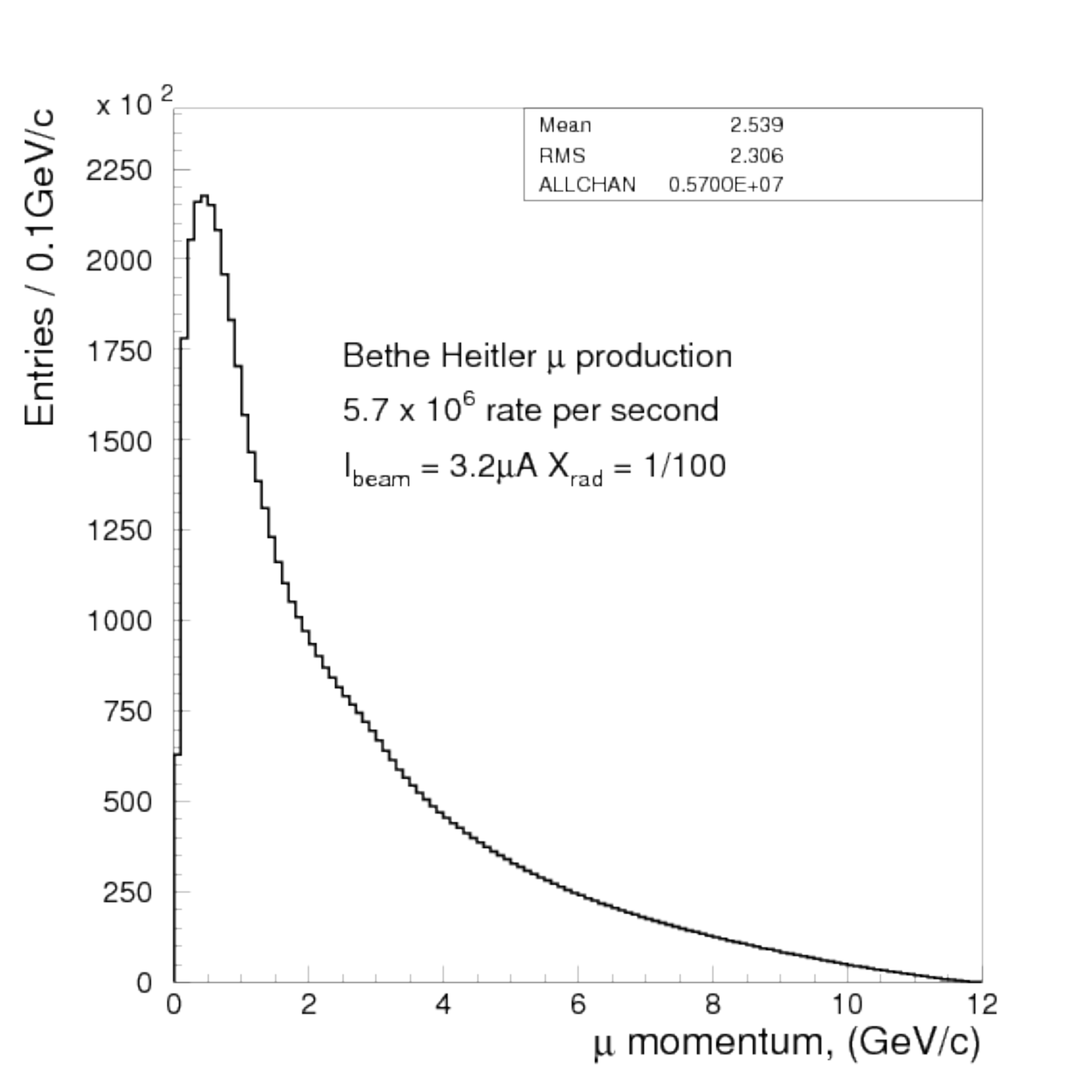} }

    \caption{Left panel: Momentum spectrum of residual $\gamma$s
    just after the concrete shielding.
    Right panel: Muon momentum spectrum for Bethe-Heitler production.} 
    \label{fig:gamma}
\end{figure}

After passing through 30\% R.L. tungsten beam plug and the charged 
background component removed by the sweep magnet, we will have some 
residual $\gamma$ background and neutrons produced by EM showers. The 
momentum spectrum of residual $\gamma$s is shown in 
Fig.~\ref{fig:gamma}(left). It decreases exponentially with increasing 
energy of photons. For the rates, we obtained $\sim 10^5$~s$^{-1}$ for 
$\gamma$s with energy above 50~MeV and $\sim 10^3$~s$^{-1}$ for 
$\gamma$s with energy above 500~MeV.  

\underline{Overall}, the gamma flux for the KLF experiment is 
tolerable.

\item \textbf{Muon Background} 

Following Keller~\cite{Keller}, our Geant4~\cite{Geant4} simulations
included Bethe-Heitler muon background from the Be-production target
and photon dump, both background into the detector and muon dose
rate outside Hall~D. Obviously, most of the muons are produced in the
photon dump. Our calculations show that muons will be swept out of
the $K_L$ beam line; thus, they are not inherently a significant 
background. However, due to their high penetration ability, it might 
be important for purposes of the shielding. We have taken into account 
only the Bethe-Heitler muon production process. Muons from pion decays 
and other production mechanisms will increase the total muon yield 
only slightly. They were not included in our model. The number of 
produced muon in the Be target and lead beam plug is about the same, 
but muons originating in lead have a much softer momentum spectrum. 
The estimated number of produced muons is $\sim 6\times 10^6$~s$^{-1}$.  
Their momentum spectrum is shown in Fig.~\ref{fig:gamma}(right).  The 
above number will increase by factor of 15 with $I_e=5\mu A$ and the 
radiator of 0.1 radiation length.

\underline{To summarize:} Half of muons will have momenta higher than
2~GeV/$c$, $\sim 10\%$ of muons will  have momenta higher than 6~GeV/$c$,
and $\sim 1\%$ of muons with have momenta above 10~GeV/$c$. Overall, the 
muon flux for the KLF experiment is tolerable.

\item \textbf{Neutron Background} 

To estimate the neutron flux in a beam and neuron dose in the
experimental hall from scattered neutrons we used the MCNP6 N-Particle
(MCNP) Transport code~\cite{MCNP}. See Appendix~A4~(Sec.~\ref{sec:A4}) 
for further details. The experimental hall, beam cave,
and photon beam resulted from tungsten radiator were modeled using the
specifications from the layout presented in Fig.\ref{fig:exp}.
Figure~\ref{fig:HallD} shows a graphic model of the experimental setup.
\begin{figure}[h!]
\centering
{
    \includegraphics[width=1\textwidth,keepaspectratio]{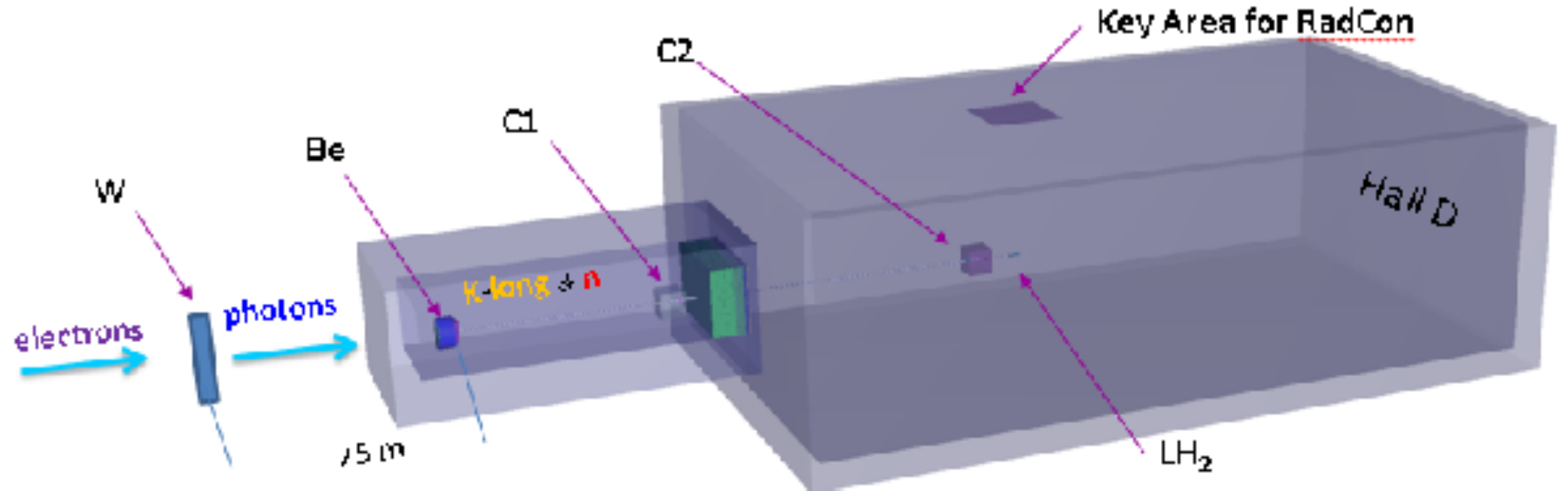} }

    \caption{Schematic view of Hall~D setting for MCNP6 transport
    code~\protect\cite{MCNP} calculations. Beam goes from left to
    right. The model is presented as semi-transparent for
    demonstration purposes. This 3D plot is similar as
    Fig.~\protect\ref{fig:exp} shows.} \label{fig:HallD}
\end{figure}

The physical models implemented in the MCNP6 code take into account
bremsstrahlung photon production, photonuclear reactions, gamma-ray
and neutron multiple scattering processes. We ignored the GlueX
detector setting in these calculations.

The MCNP model simulates a 12~GeV 5$\mu$A electron beam hitting the
tungsten radiator. Electron transport was traced in tungsten radiator, 
vacuum beam pipe for bremsstrahlung photons, and Be target. Neutrons and
gamma rays were traced in all components of the used MCNP model. The 
media outside concrete walls of the beam cave and bremsstrahlung photon 
beam pipe was excluded from consideration to facilitate the calculations.

The tally to estimate neutron fluence at the experimental hall ceiling
just above the LH$_2$/LD$_2$ target, at Key Area for RadCon shown in
Fig.~\ref{fig:HallD}. The neutron dose calculated for the layout
from Fig.~\ref{fig:HallD1} is 14.1$\pm$1.6~mrem/h, 2.7$\pm$0.8~mrem/h
for the layout from Fig.~\ref{fig:HallD4}, and 0.2$\pm$0.07~mrem/h
for the layout from Fig.~\ref{fig:HallD2}. Neutron Fluence-to-Effective
Dose conversion factors from ICRP~116 ~\cite{ICRP} were implemented to
convert neutron fluence to effective dose. The neutron flux at the face
of the $LH_2/LD_2$ target is about $2\times 10^3$ N/(s$\cdot$ cm$^2$)
and is almost independent of the shielding location in the beam cave.

\underline{Overall}, the neutron flux for the KLF experiment is tolerable
and  below the RadCon limit.
\end{enumerate}

\subsubsection{$K_L$ Momentum Determination and Beam Resolution}
\label{sec:Bres}

The mean lifetime of the $K_L$ is 51.16~ns ($c\tau = 15.3$~m)
whereas the mean lifetime of the $K^-$ is 12.38~ns ($c\tau =
3.7$~m)~\cite{PDG2016}.  For this reason, it is much easier to
perform measurements of $K_Lp$ scattering at low beam energies
compared with $K^-p$ scattering.

The momentum of a $K_L$ beam can be measured using time-of-flight
(TOF) - the time
between the accelerator bunch (RF signal from CEBAF) and the reaction in
the LH$_2$/LD$_2$ target as detected by the GlueX spectrometer. 
Thus the TOF resolution is a quadratic sum of  accelerator time and 
GlueX spectrometer time resolutions. Since the accelerator signal
has a very good time resolution ($\sim 150$~ps or better), TOF
resolution will be defined by GlueX detector.  The time resolution of the 
GlueX detectors are discussed in Sec.~\ref{sec:SC}.  
In our calculations, we used time resolutions from 50~ps to 300~ps to show
the dependence of the beam momentum and $W$ resolution.  

Of course, to get TOF information, the electron
beam needs to have a narrow bunch time structure with a bunch
spacing of, at least, 60~ns.  In order to be able to measure the
roughly 20~ns ToF of the elastic protons, the beam for the $G0$
experiment at Hall~C has 32~ns between electron bunches (in
contrast to the usual 2~ns spacing for each experimental hall)
using a 31.1875~MHz pulsed laser to operate the electron
source~\cite{Androic2011}.  One cannot expect a problem with a 60~ns
time structure to delivery an electron beam to any Hall, A,
B, or C~\cite{Spata2016}.
\begin{figure}[h!]
\centering
{
    \includegraphics[width=0.4\textwidth,keepaspectratio]{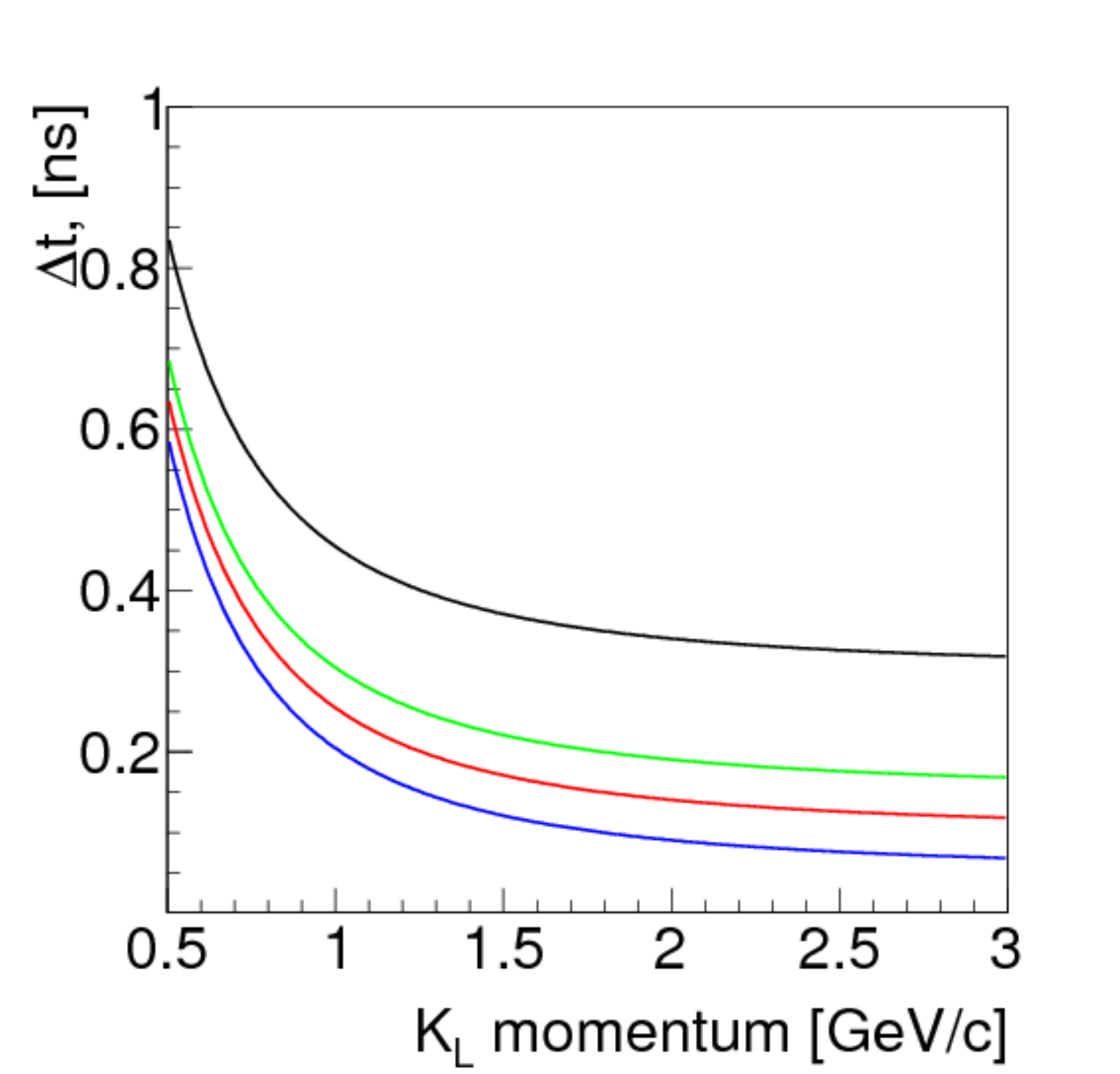} }
{
    \includegraphics[width=0.4\textwidth,keepaspectratio]{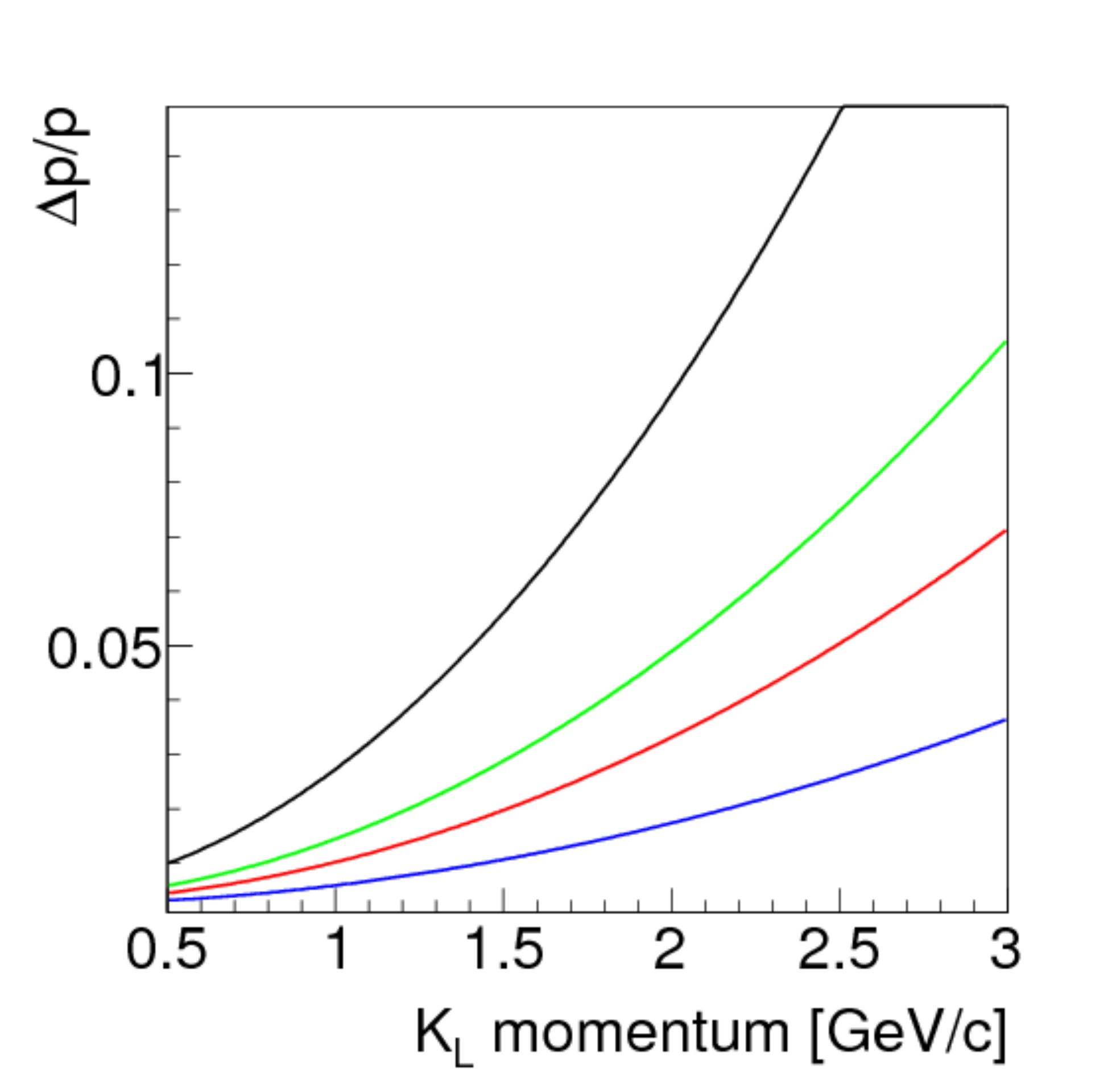} }

  \caption{Left plot: Time resolution, $\Delta t$, for $K_L$ beam 
  as a function of $K_L$ momentum.
  Right plot: Momentum resolution, $\Delta p/p$, as a function of 
  momentum.
  For 300~ps (black), 150~ps (green), 100~ps (red), and 50~ps (blue) 
  time resolutions.} \label{fig:mom}
\end{figure}

\begin{figure}[h!]
\centering
{
    \includegraphics[width=0.45\textwidth,keepaspectratio]{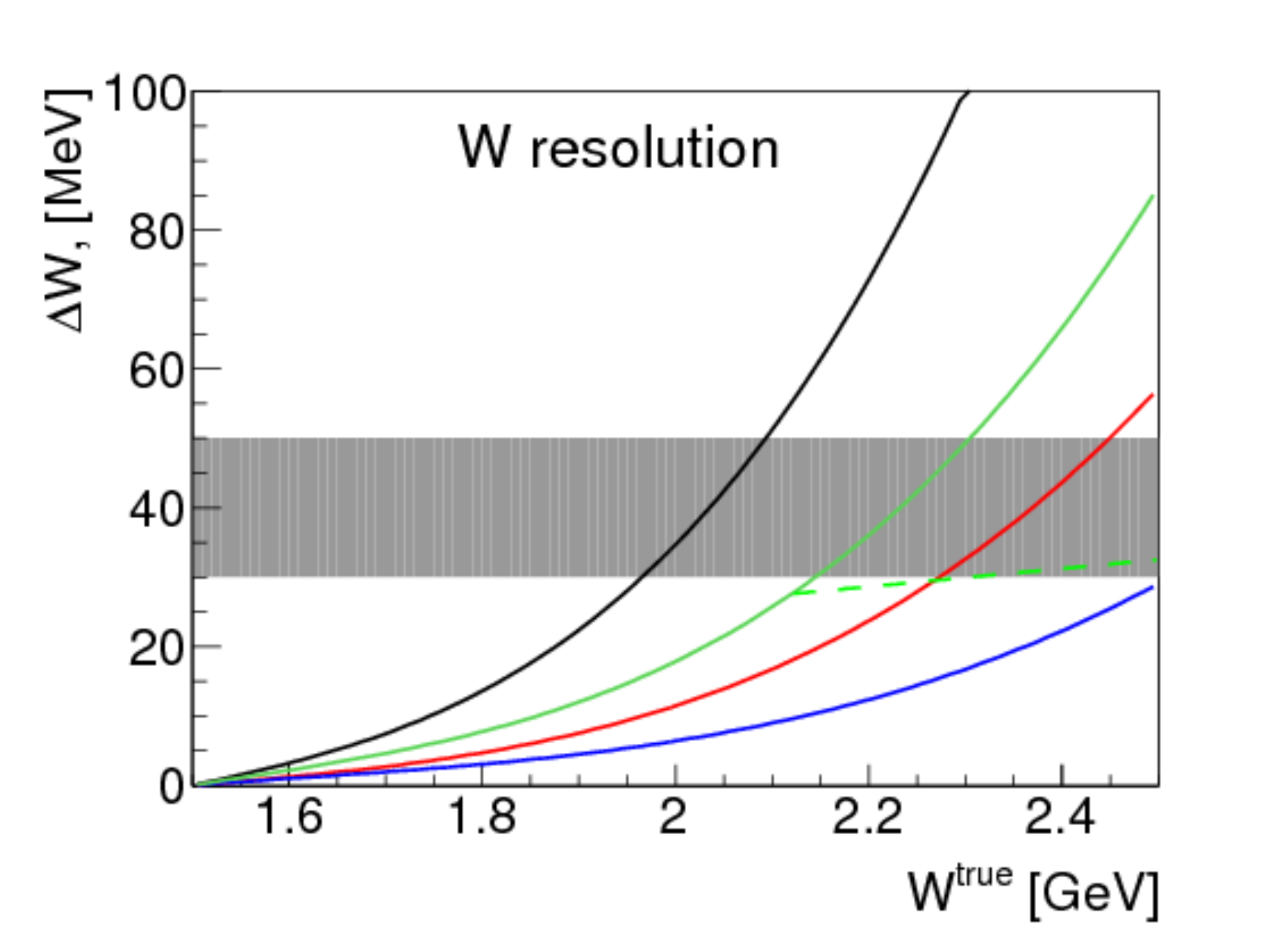} }

  \caption{Energy resolution, $\Delta W/W$, as a function of 
  energy, for 300~ps (black), 150~ps (green), 100~ps (red), and 
  50~ps (blue) time resolution. Green dashed line shows 
  approximate $W$ resolution from reconstruction of final-state 
  particles. Shaded area corresponds to typical hyperon width.} 
  \label{fig:mom1}
\end{figure}

The uncertainty in a neutral kaon production position at lower 
momenta ($p$ < 0.5~GeV/$c$) affects timing resolution caused by the  
TOF difference between the photon and kaon time traversing the
Be~target, however, as $\Delta p/p = \gamma^2\Delta T/T$ momentum 
resolution is below 1\% at lower momenta. Figure~\ref{fig:mom} 
shows TOF, $\Delta t$, (left) and beam momentum resolution, 
$\Delta p/p$ (right) as a function of the $K_L$ beam momentum, 
respectively.  The TOF resolution is flat for momenta higher than 1~GeV/$c$.
The momentum resolution decreases with momentum: for 1~GeV/$c$
it is $\sim$1.4\% and for 2~GeV/$c$ it is $\sim$5\%.  Figure~\ref{fig:mom1} 
shows that for $W < 2.1$~GeV, $\Delta W<$~30~MeV, which is suitable for 
studying low-lying hyperons with widths of $\Gamma = 30$ -- 50~MeV~\cite{PDG2016}.  
For fully reconstructed final states $W$ can be reconstructed directly, which 
provides a better resolution in the region where the TOF method deteriorates, 
$W>2.1$~GeV (see green dashed curve in Fig.~\ref{fig:mom1}).

\subsubsection{GlueX Detector Time Resolution}
\label{sec:SC}

The $K_L$ beam momentum and time resolution is governed by the time
resolution provided by the \gx{} detector from the reconstruction
of charged particles produced in the LH$_2$/LD$_2$ target.  There are
three detector systems that can provide precision timing information
for reconstructed charged particles in \gx{}: the Start Counter (ST),
Barrel Calorimeter (BCAL), and Time of Flight (TOF) detectors.  The
aforementioned detectors, and the charged particle time resolutions
they provide, are discussed in this section.

\begin{figure}[h]
\centering
{
    \includegraphics[width=0.7\textwidth,keepaspectratio]{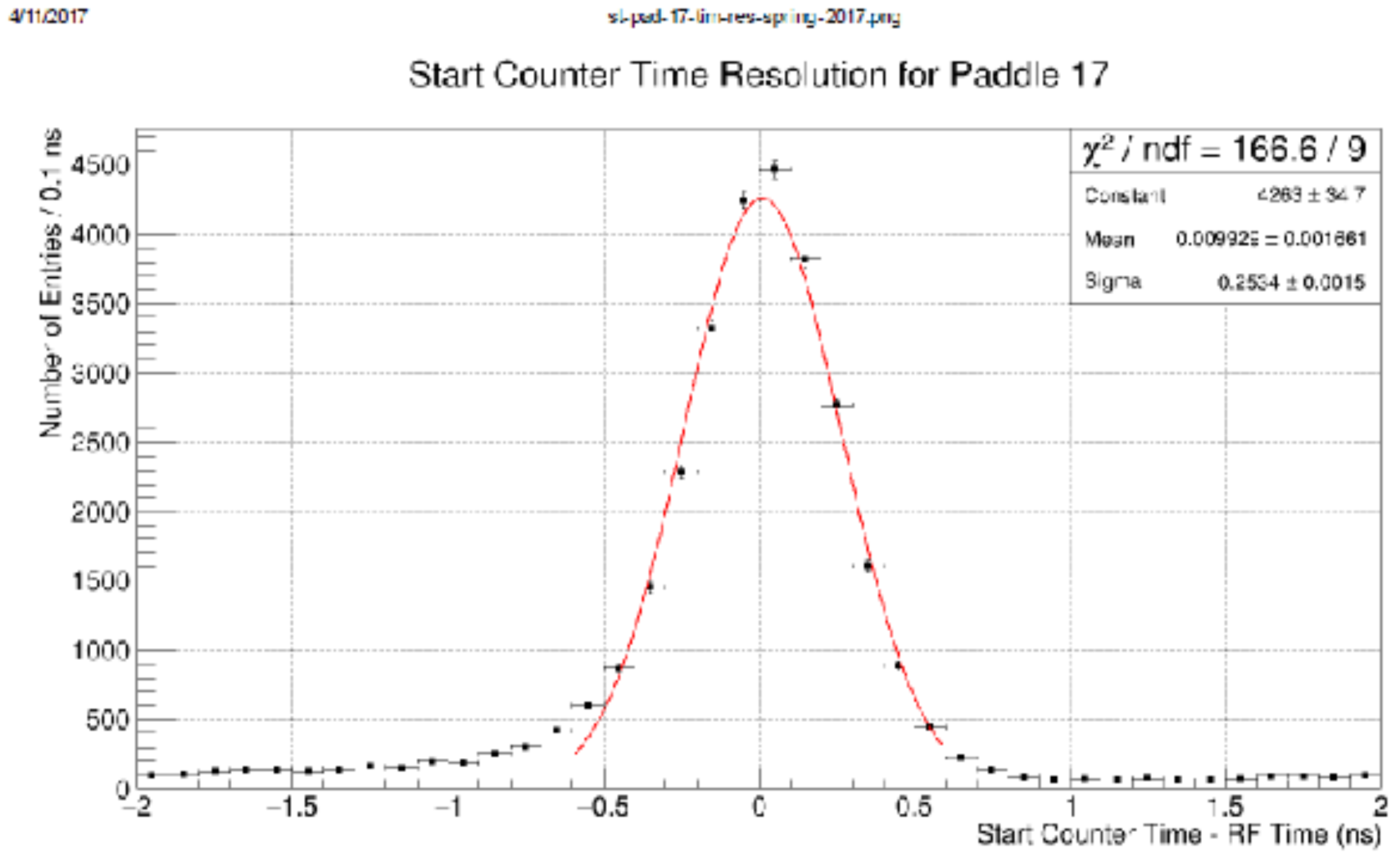} }

	\caption {Time difference between the measured and expected 
	ST time from the Spring 2017 \gx{} run period.  The data were 
	fitted with a Gaussian to determine the current time resolution 
	of $\sim250$~ps.}
\label{fig:st_resolution}
\end{figure}

The \gx{} Start Counter is a cylindrical plastic scintillator detector
surrounding the LH$_2$/LD$_2$ target, with 3~mm thick scintillator bars
and a tapered nose region that bends toward the beamline at the
downstream end.  The scintillation light from each of the 30 scintillator
bars is detected by an array of 4, $3 \times 3~{\mathrm{mm^2}}$
Hamamatsu S10931-050P surface mount silicon photomultipliers
(SiPMs)~\cite{pooser-thesis}.  The time resolution of the ST was
determined to be 250~ps during the 2016 and 2017 \gx{} run periods,
as shown in Fig.~\ref{fig:st_resolution}, and thus provided adequate
separation of the 250~MHz photon beam bunch structure delivered to
Hall~D during that time.  This performance was achieved using the
recommended operating gain and bias voltages supplied by Hamamatsu
to provide both the FADC 250 analog signals and precision F1TDC
discriminator signals used in the \gx{} reconstruction.  For the
$K_L$ program we propose to increase the gain of the ST SiPMs,
thereby increasing the number of detected photoelectrons, as well
as modify the pulse-shape processing electronics.  Similar gain and
readout electronic customization were implemented in the \gx{}
Tagger Microscope, which utilizes an identical SiPM readout system,
and provided timing resolutions of 200~ps.  Implementation of these 
non-invasive modifications to the ST will significantly improve the 
timing resolution.  
we, therefore, assumed a 150~ps resolution as the baseline ST 
performance, which may be achieved with modifications to the current 
device.

Future improvements to the start counter to reduce the time
resolution further will be studied to increase both the light production 
in the scintillators and the light collection efficiency.  A future goal
would be to reach a time resolution of 50-100~ps for the
ST, which would likely require a complete replacement of the current
device.  Increased light production could come
through an increase of the scintillator bar thickness, or a different
choice of scintillator material with a higher light yield and shorter
decay time such as EJ-204.  Improved photodetectors, including
Microchannel Plate PMTs, which also perform well in high magnetic
field environments, could provide higher gain and better efficiency
than the current SiPMs, and will be investigated to assess their
potential impact on the ST performance.

The \gx{} BCAL is a scintillating fiber calorimeter, which provides
timing information for both neutral and charged particles.  The
measured time resolution of the BCAL for charged particles depends
on the reconstructed BCAL energy but had an average value of
$\sim220$~ps during the \gx{} Spring 2017 run period.   For charged
particles with large scattering angles ($11^\circ < \theta <
120^\circ$) this additional measure of the interaction time will
improve the overall $K_L$ time resolution when combined with the
ST measurement.  The \gx{} TOF is composed of two planes of 2.5~cm
thick scintillator bars.  The measured TOF time resolution was 100~ps
from the \gx{} Spring 2017 run period, well below the assumed
performance of the ST. Therefore, for reactions with a charged
particle, which is produced in the forward region $\theta<11^\circ$,
the TOF will be used to provide a better $K_L$ momentum determination
than the ST.

\underline{To summarize}, The simulation studies in this proposal 
(See Sec.~\ref{sec:RC}) 
have assumed a time resolution of 150~ps, which is adequate for the 
proposed physics program.  With the current detector, the overall 
$K_L$ momentum resolution will be determined by utilizing the timing 
information from the ST, BCAL and TOF detectors to ensure that the 150~ps 
specification is achieved.  Finally, we are exploring potential 
upgrades to improve the ST time resolution significantly; however, 
further study is required to understand the impact of such 
improvements on the extracted resonance parameters for the proposed 
hyperon spectroscopy program.

\subsubsection{Measurement of $K_L$ Flux}

The $K_L$ has four dominant decay modes~\cite{PDG2016}:
\begin{enumerate}
	\item $K_L\to \pi^+\pi^-\pi^0, BR=12.54\pm0.05\%$.
	\item $K_L\to \pi^0\pi^0\pi^0, BR=19.52\pm0.12\%$.
	\item $K_L\to \pi^{\pm}e^{\mp} \nu_e, BR=40.55\pm0.11\%$.
	\item $K_L\to \pi^{\pm}\mu^{\mp} \nu_{\mu}, BR=27.04\pm0.07\%$.
\end{enumerate}

In addition, there are several rare decay modes, including the CP-violating
$K_L\to 2\pi$ mode. In three of the four principal decay modes of the
$K_L$, two charged particles are emitted. To measure the flux of the $K_L$
beam at GlueX, we will measure the rate of $K_L$ decays to two
oppositely charged tracks in the Hall~D Pair Spectrometer~\cite{Barbosa2015}
upstream of the GlueX cryotarget. Timing information from the 
pair spectrometer will be used to estimate time of flight elapsed between 
the creation of a $K_L$ in the Be target and its decay to measure momenta 
of decayed kaons. In a long run with high statistics, the $2\pi$ decay mode 
can also be used for a reference to measure independently the flux and 
momenta of decayed kaons and reconstruct the flux of incoming kaons. 
This experiment will employ techniques similar to those used 
in the most precise measurements of $K_L$ flux (see, for example, 
Refs.~\cite{Slayer,Vorsburgh,Brandenburg1973}).  

For the measurement of the $K_L$ flux, we can use regeneration of $K_S$ 
and detecting $\pi^+\pi^-$ pairs in Pair Spectrometer as it was done at 
Daresbury (see Ref.~\cite{Albrow} and references therein).

A measurement of the $K_L$ flux at the $5\%$ level may require additional 
instrumentation, but in general is not essential to the proposed physics program.

\subsection{LH$_2$/LD$_2$ Cryotarget for Neutral Kaon Beam at Hall~D}

\begin{figure}[h!]
\centering
{
    \includegraphics[width=0.7\textwidth,keepaspectratio]{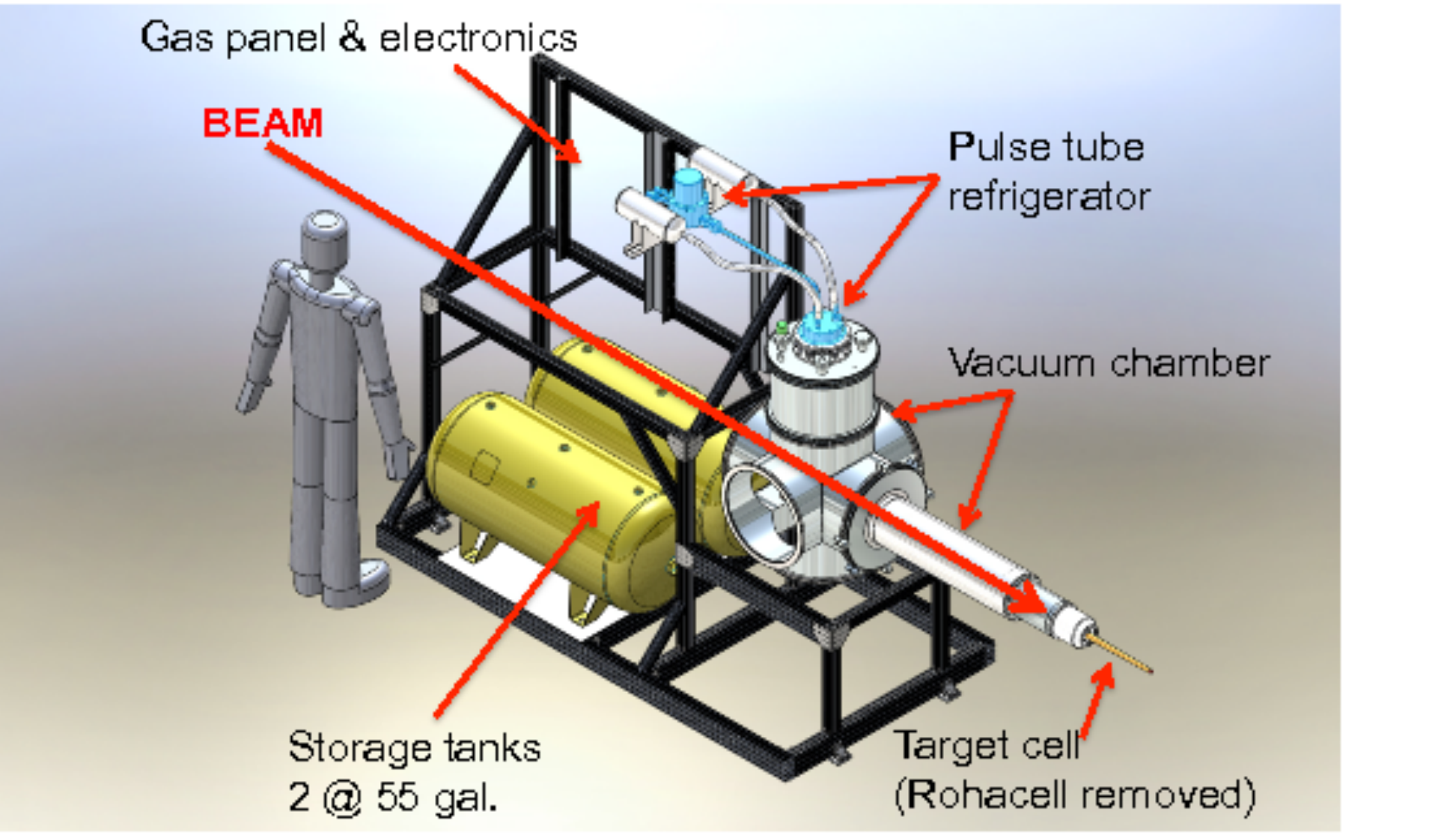} }

  \caption{The GlueX liquid hydrogen target.} \label{fig:lh2a}
\end{figure}

The proposed experiment will utilize the existing GlueX liquid
hydrogen cryotarget (Fig.~\ref{fig:lh2a}) modified to accept a
larger diameter target cell~\cite{Chris2016}. The GlueX target
is comprised of a kapton cell containing liquid hydrogen at a
temperature and pressure of about 20~K and 19~psia, respectively 
The 100~ml cell is filled through a pair of 1.5~m long stainless
steel tubes (fill and return) connected to a small container where
hydrogen gas is condensed from two room-temperature storage
tanks. This condenser is cooled by a pulse tube refrigerator
with a base temperature of 3~K and cooling power of about 20~W
at 20~K. A 100~W temperature controller regulates the condenser
at 18~K.

The entire target assembly is contained within an ``L"-shaped
stainless steel and aluminum vacuum chamber with a Rohacell
extension surrounding the target cell. The Start Counter for the GlueX
experiment fits snugly over this extension. The vacuum chamber,
along with the hydrogen storage tanks, gas handling system, and
control electronics, is mounted on a custom-built beamline cart
for easy insertion into the Hall~D solenoid. A compact I/O
system monitors and controls the performance of the target,
while hardware interlocks on the target temperature and pressure
and on the chamber vacuum ensure the system's safety and
integrity. The target can be cooled from room temperature and
filled with liquid hydrogen in about 5~hours. For empty target
runs, the liquid can be boiled from the cell in about 20
minutes (the cell remains filled with cold hydrogen gas), and
then refilled with liquid in about 40 minutes.
\begin{figure}[h!]
\centering
{
    \includegraphics[width=0.4\textwidth,keepaspectratio]{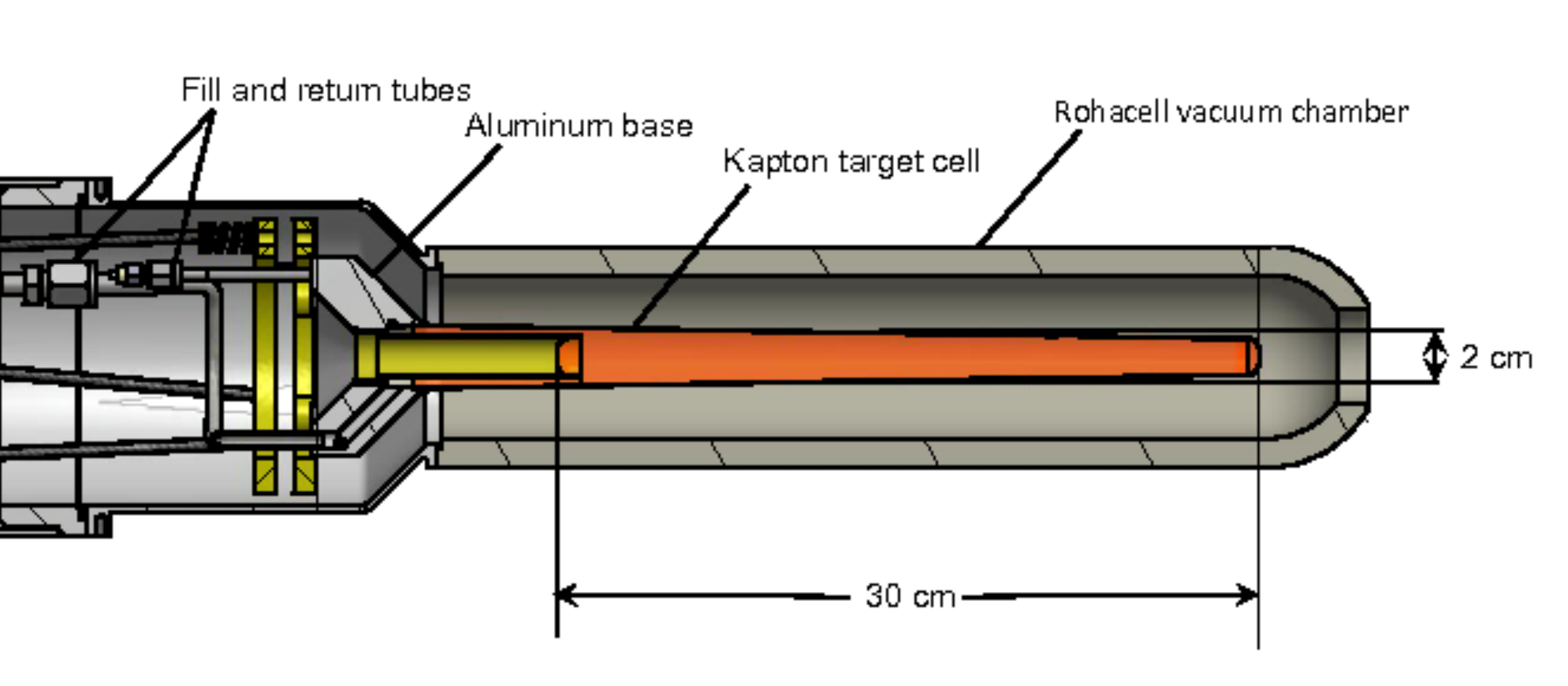} }
{
    \includegraphics[width=0.4\textwidth,keepaspectratio]{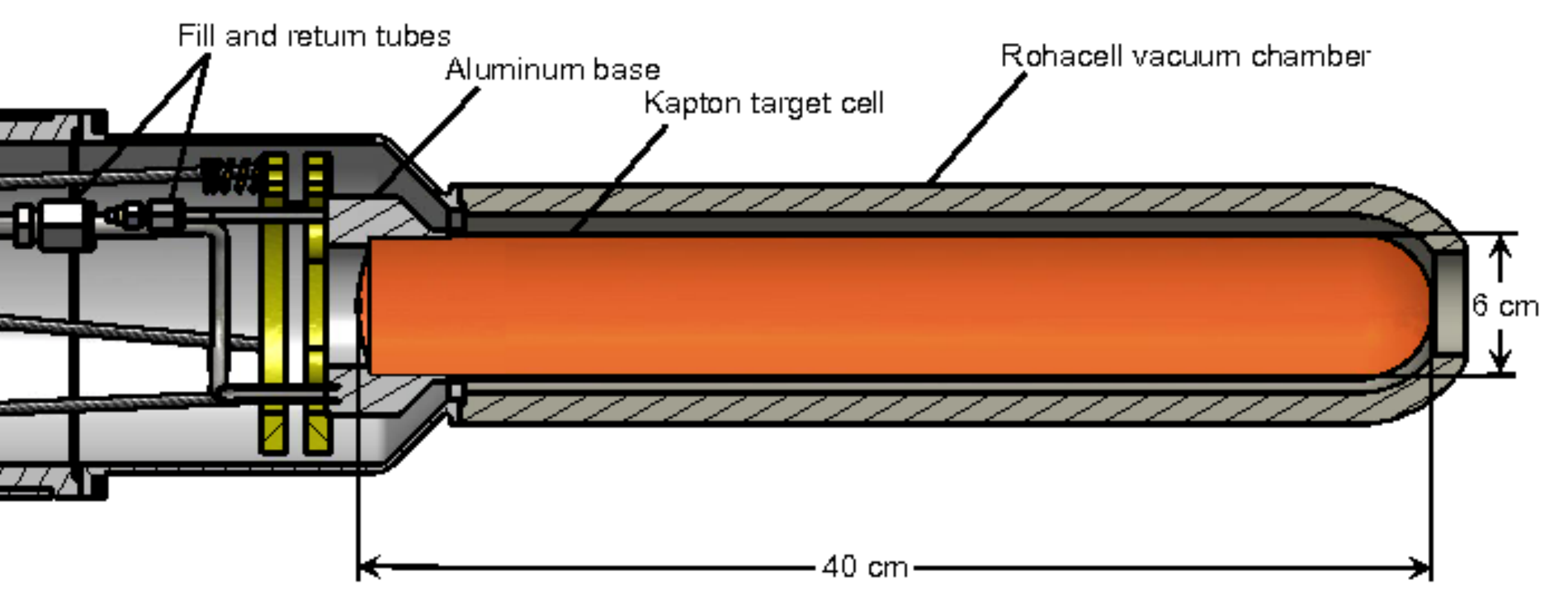} }

  \caption{Left plot: Kapton target cell for the GlueX LH$_2$/LD$_2$
  target. Right plot: Conceptual design for a larger target cell
  for the proposed $K_L$ beam in Hall~D.} \label{fig:lh2b}
  \end{figure}

The GlueX cell (Fig.~\ref{fig:lh2b}) is closely modeled on those
utilized in Hall~B for more than a decade and is a horizontal,
tapered cylinder about 38~cm long with a mean diameter of 2~cm.
The cell walls are 130~$\mu$m kapton glued to an aluminum base.
A 2~cm diameter reentrant beam window defines the length of
LH$_2$/LD$_2$ in the beam to be about 30~cm. Both entrance and exit
windows on the cell are 75~$\mu$m kapton. In normal operation, the
cell, the condenser, and the pipes between them are all filled
with liquid hydrogen. In this manner the liquid can be subcooled
a few degrees below the vapor pressure curve, greatly suppressing
bubble formation in the cell. In total, about 0.4~liter of
LH$_2$ is condensed from the storage tanks, and the system is
engineered to recover this quantity of hydrogen safely back into
the tanks during a sudden loss of insulating vacuum, with a
maximum allowed cell pressure of 49~psia~\cite{Meekins}.

A conceptual design for the neutral kaon beam target is also
shown in Fig.~\ref{fig:lh2b}. The proposed target cell has a
diameter of 6~cm and a 40~cm length from entrance to exit windows,
corresponding to a volume of about 1.1~liter, which will require
filling the existing tanks on the target cart to about 50~psia.
The collaboration will work with the JLab Target Group to
investigate alternative materials and construction
techniques to increase the strength of the cell.  As an example,
the LH$_2$ target cell recently developed for Hall~A is 6.3~cm
in diameter, 18~cm long and has a wall thickness of approximately
0.2~mm.  The cell is machined from a high-strength aluminum alloy,
AL7075-T6, and has a maximum allowed pressure of about 100~psia.
It is expected that minor modifications to the cryotarget's piping
systems will also be required to satisfy the increased volume of
condensed hydrogen.

The proposed system is expected to work equally well with liquid
deuterium, which condenses at a slightly higher temperature than
hydrogen (23.3~K versus 20.3~K at atmospheric pressure). The
expansion ratio of LD$_2$ is 13\% higher, which implies a storage
pressure of about 60~psia. Therefore, the new target cell must be
engineered and constructed to work with either $H_2$ or $D_2$.

\section{Running Condition}
\label{sec:RC}

\subsection{Event Identification, Reconstruction, Acceptances}
\label{sec:MC}

The $K_L$ beam is generated by sampling the momentum distribution of
$K_L$ particles coming from the decays of $\phi$ mesons produced by
interactions of a photon beam with a beryllium target 16~m upstream
of the LH$_2$/LD$_2$ cryotarget. The $K_L$ beam profile was assumed to
be uniform within a 3~cm radius at the LH$_2$/LD$_2$ cryotarget. Due to
the very strong $t$-dependence in the $\phi$ photoproduction
cross section~\cite{Seraydaryan2014} and the $P$-wave origin of the
$\phi\to K_LK_S$ decay, the majority of kaons will be produced at very 
small angles. In the simulation studies discussed in this section, we 
assume a flux of $3\times10^4 K_L$/s on a 40~cm long LH$_2$ target 
for a beamtime of 100 PAC days.

\subsubsection{Simulations and Reconstruction of Various Channels
        Using GlueX Detector} 

\begin{figure}[h!]
\centering
{
    \includegraphics[width=0.75\textwidth,keepaspectratio]{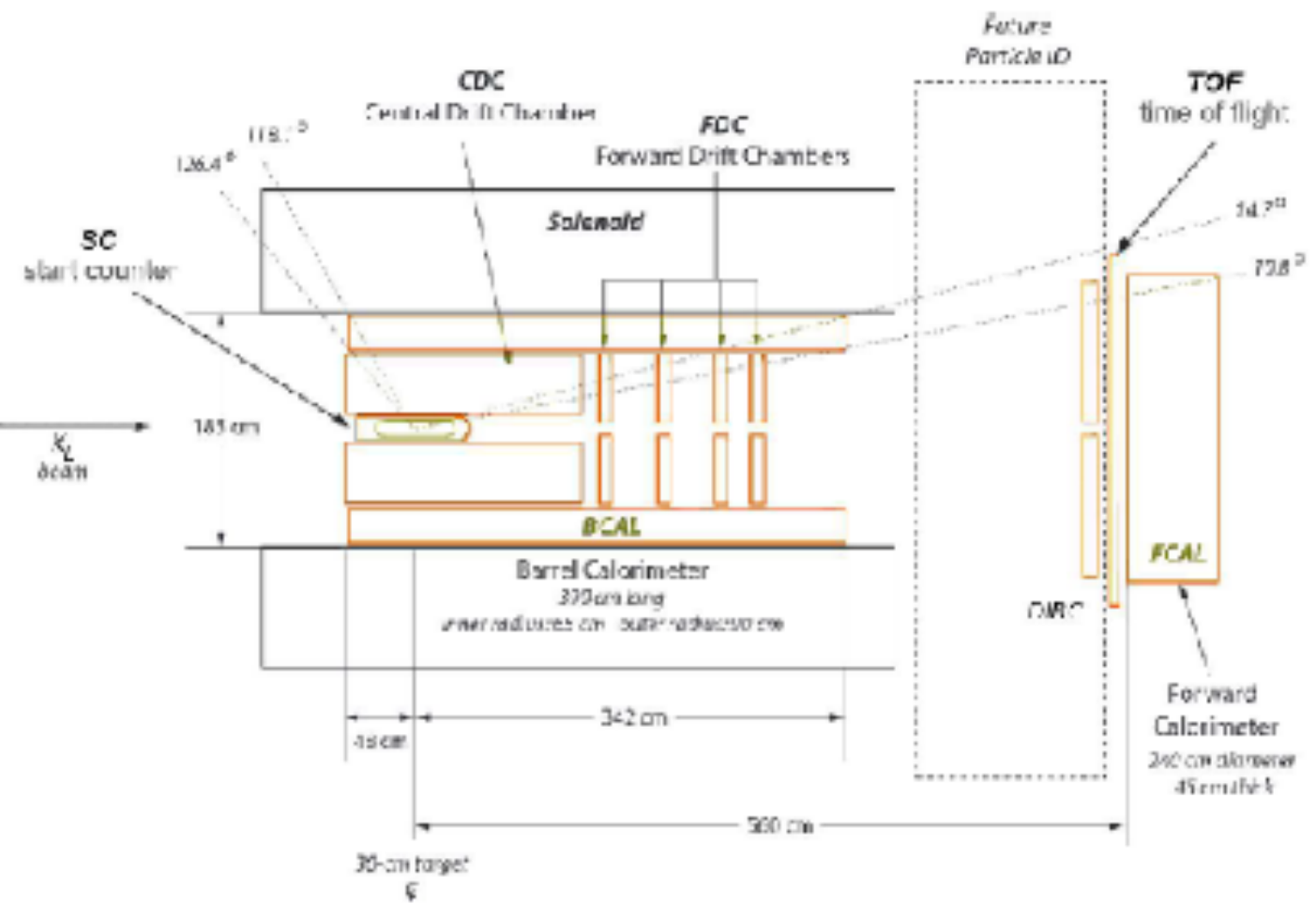} }

    \caption{Schematic view of the GlueX detector.} \label{fig:KLgluex}
\end{figure}

The GlueX detector is a large acceptance detector based on a solenoid
design with good coverage for both neutral and charged particles. The
detector consists of a solenoid magnet enclosing devices for tracking
charged particles and detecting neutral particles, and a forward region
consisting of two layers of scintillators (TOF) and a lead-glass EM
calorimeter (FCAL). A schematic view of the GlueX detector is shown in
Fig.~\ref{fig:KLgluex}.  The magnetic field at the center of the
bore of the magnet for standard running conditions is about 2~T. The
trajectories of charged particles produced by interactions of the beam
with the 40-cm LH$_2$/LD$_2$ cryotarget at the center of the bore of the
magnet are measured using the Central Drift Chamber (CDC) for angles
greater than $\approx 20\textdegree$ with respect to the beam line
Forward-going tracks are reconstructed using the Forward Drift
Chambers (FDC). The timing of the interaction of the kaon beam with
the LH$_2$ cryotarget is determined using signals from the ST, an array
of 30~mm thin (3~mm thick) scintillators enclosing the target region.
Photons are registered in the central region by the Barrel Calorimeter
(BCAL). Detector performance and reconstructions techniques were
evaluated during the main GlueX program. Details can be found
elsewhere~\cite{GlueX15}.

This section describes some simulations of events generated by $K_L$
beam particles interacting with a LH$_2$/LD$_2$ cryotarget at the center
of the solenoid~\cite{Taylor16}. The GlueX detector is used to detect
one or all of the final-state particles. We will be focusing on a few
of the simple two-body reactions, namely
$K_Lp\to\ K_Sp$,
$K_Lp\to\pi^+\Lambda$,
$K_Lp\to K^+\Xi^0$, and
$K_Lp\to K^+n$.

For each topology, one particle (the proton for the $K_S p$ channel,
the $\pi^+$ for the $\pi^+\Lambda$ channel and the $K^+$ for the
$K^+\Xi^0$ channel) provides a rough determination for the position
of the primary vertex along the beam line that is used in conjunction
with the ST to determine the flight time of the $K_L$ from
the beryllium target to the hydrogen target. Protons, pions, and
kaons are distinguished using a combination of $dE/dx$ in the
chambers and time-of-flight to the outer detectors (the Barrel
Calorimeter (BCAL) and two layers of scintillators (TOF)). See
Appendix~A4~(Sec.~\ref{sec:A4}) for further details.

\subsubsection{$K_Lp\to K_Sp$ Reaction} 

\begin{figure}
\centering
{
  \includegraphics[width=0.8\textwidth,keepaspectratio]{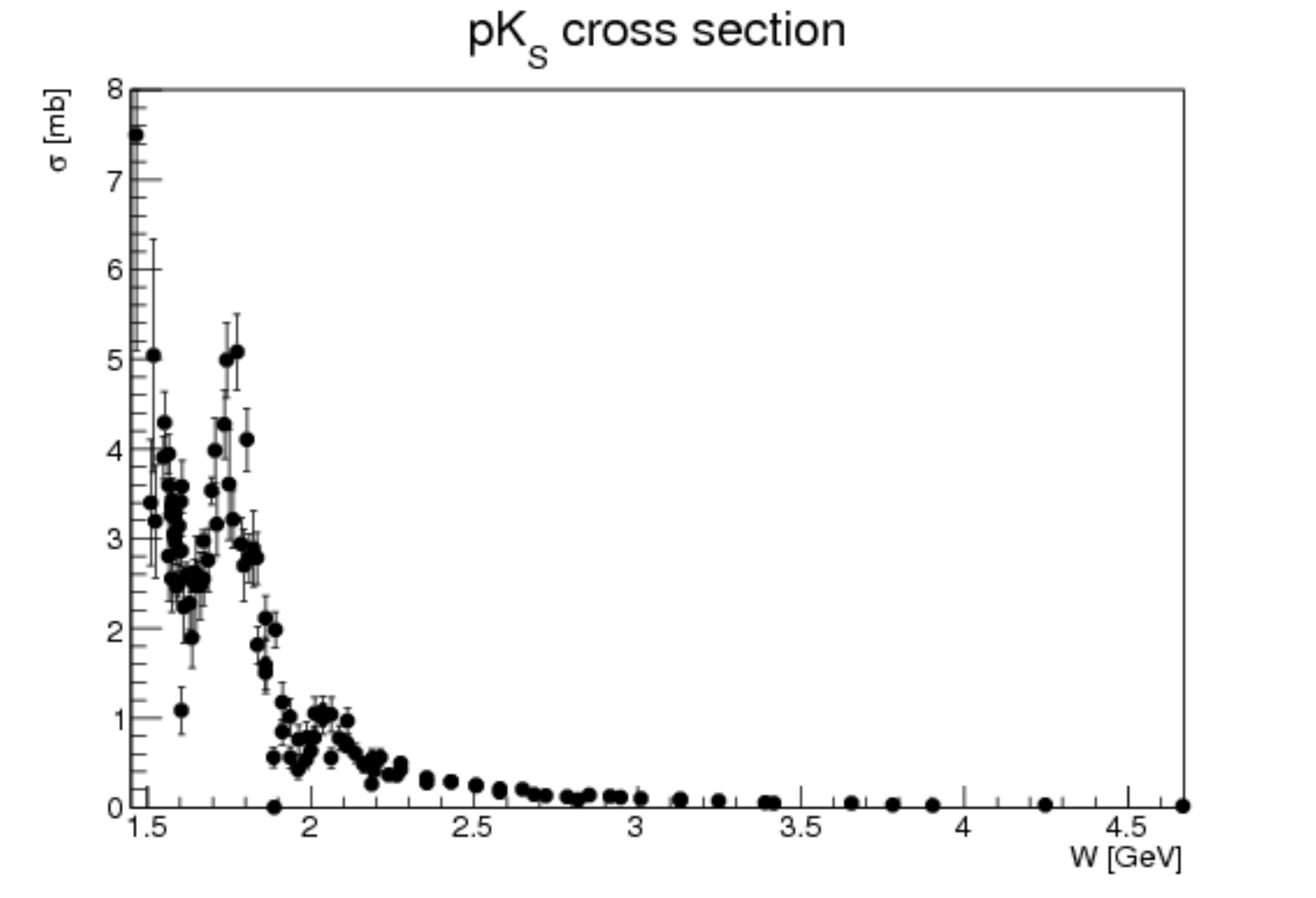}
  }

  \caption{Total cross section for $K_Lp\to K_S p$ as a function 
  of $W$. The measured points are from~\protect\cite{Capiluppi:1982fj} 
  and references therein.}
  \label{fig:pkstotalxsec}
\end{figure}
\begin{figure}
\centering
{
  \includegraphics[width=1\textwidth,keepaspectratio]{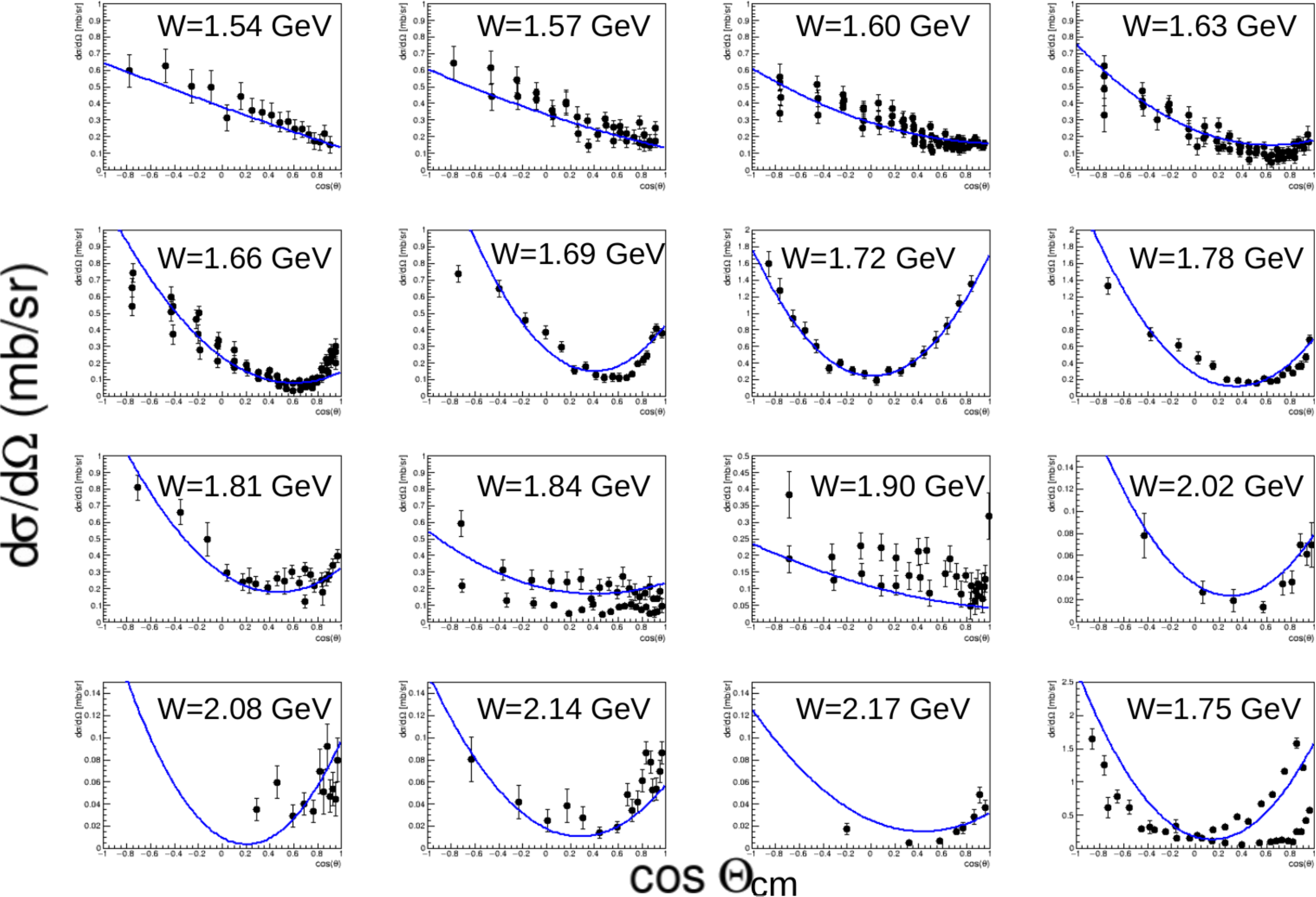}
}

\caption{Differential cross-section plots for $K_Lp \to K_S p$ 
	as a function of $W$. The blue curves are the result of a 
	parametrization of the cross section in terms of Legendre 
	polynomials.  The measured points are 
	from~\protect\cite{Capiluppi:1982fj}. }
\label{fig:pksxsecs}
\end{figure}

\begin{figure}
\centering
{
  \includegraphics[width=0.7\textwidth,keepaspectratio]{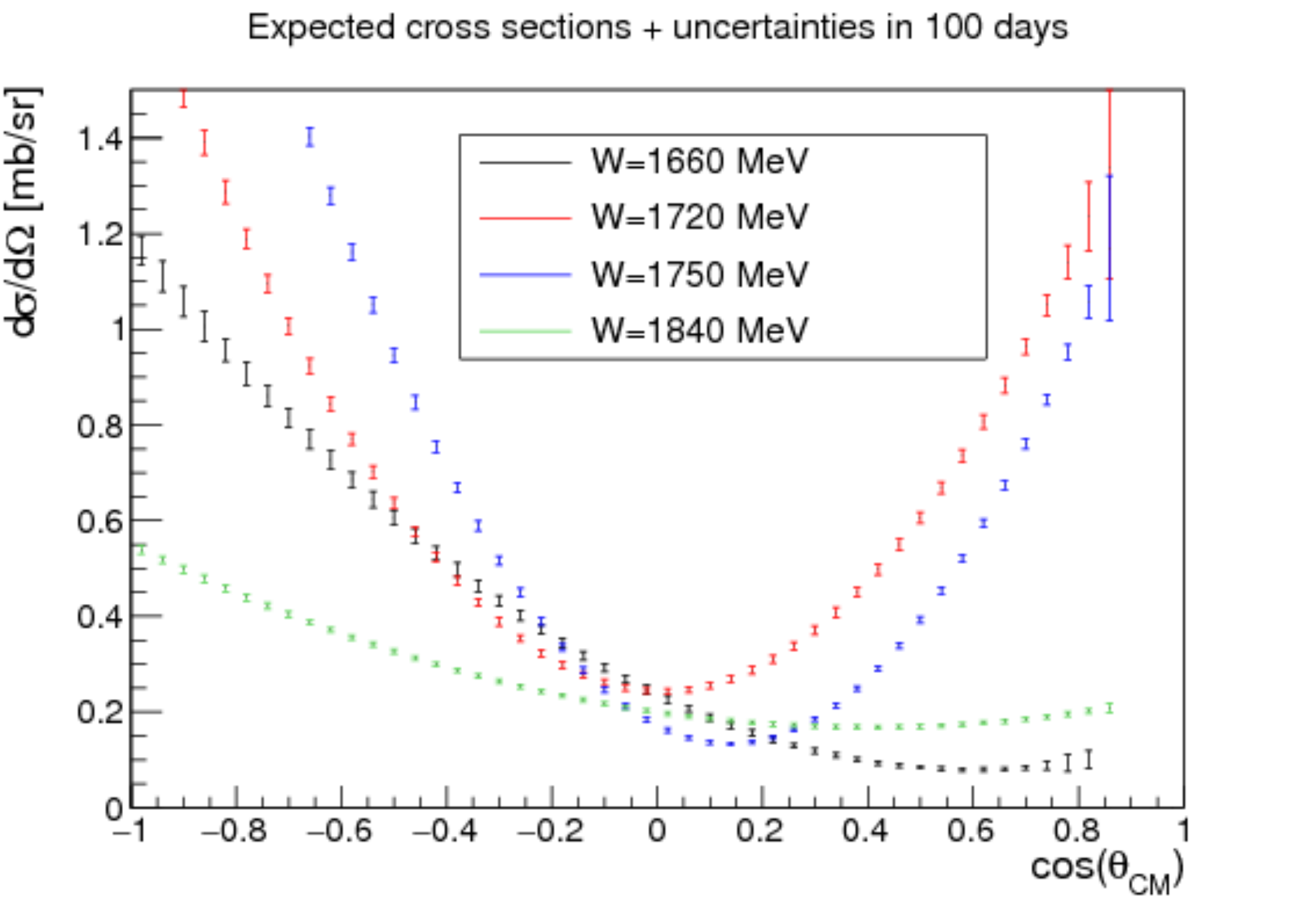}
  }

  \caption{Reconstructed $K_L p \to K_S p$ differential cross sections 
  for various values of $W$ for 100~days of running.}
  \label{fig:pksxsecrecon}
\end{figure}

The total production cross section, shown in Fig.~\ref{fig:pkstotalxsec}, 
is reasonably large; however, for the differential cross section there is 
a fair amount of tension in the existing data sets between different 
measurements, and the angular coverage in some bins is sparse.  
Figure~\ref{fig:pksxsecs} shows the existing differential cross-section 
data for several bins in $W$. The cross section as a function of 
$\cos{\theta_{CM}}$ was parametrized using a set of Legendre polynomials 
(blue curves in Fig.~\ref{fig:pksxsecs}); the weights of each polynomial 
in the set depended on $W$.  This parametrization was used to generate 
$K_L p \to K_S p$ events that were passed through a full 
Geant3-based Monte Carlo of the GlueX detector.  The final-state 
particles were constructed using the standard GlueX reconstruction code. 
We reconstructed the $K_S$ in its $\pi^+\pi^-$ channel.  More details 
about the reconstruction of this channel can be found in
Appendix~A5~(Sec.~\ref{sec:App_PID_KS}). Estimates for statistical errors in 
the measured cross section for 100 days of running at $3\times 
10^4~K_L/s$ as a function of $\cos\theta_{CM}$ for several values of $W$ are 
shown in Fig.~\ref{fig:pksxsecrecon}.  We estimate that for $W<3$~GeV with 
an incident $K_L$ rate of $3\times10^4/s$ on a 40-cm long LH2 target, we 
will detect on the order of 8M $K_S p$ events in the $\pi^+\pi^-$ channel.

\subsubsection{$K_Lp\to\pi^+\Lambda$ Reaction} 

\begin{figure}[!htb]
    \begin{center}
    \includegraphics[angle=0,width=0.45\textwidth,height=6.2cm]{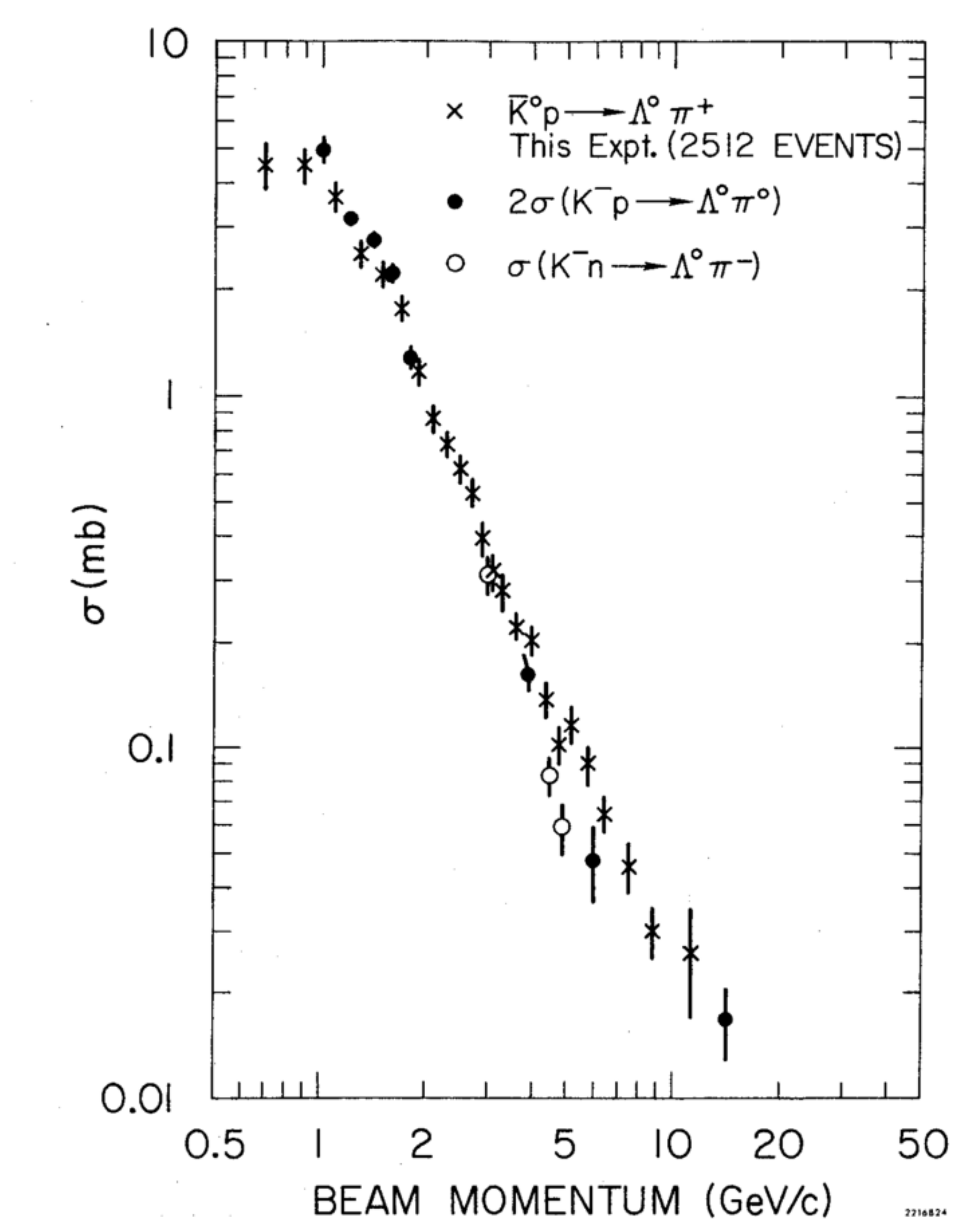}
    \includegraphics[angle=0,width=0.45\textwidth,height=6.2cm]{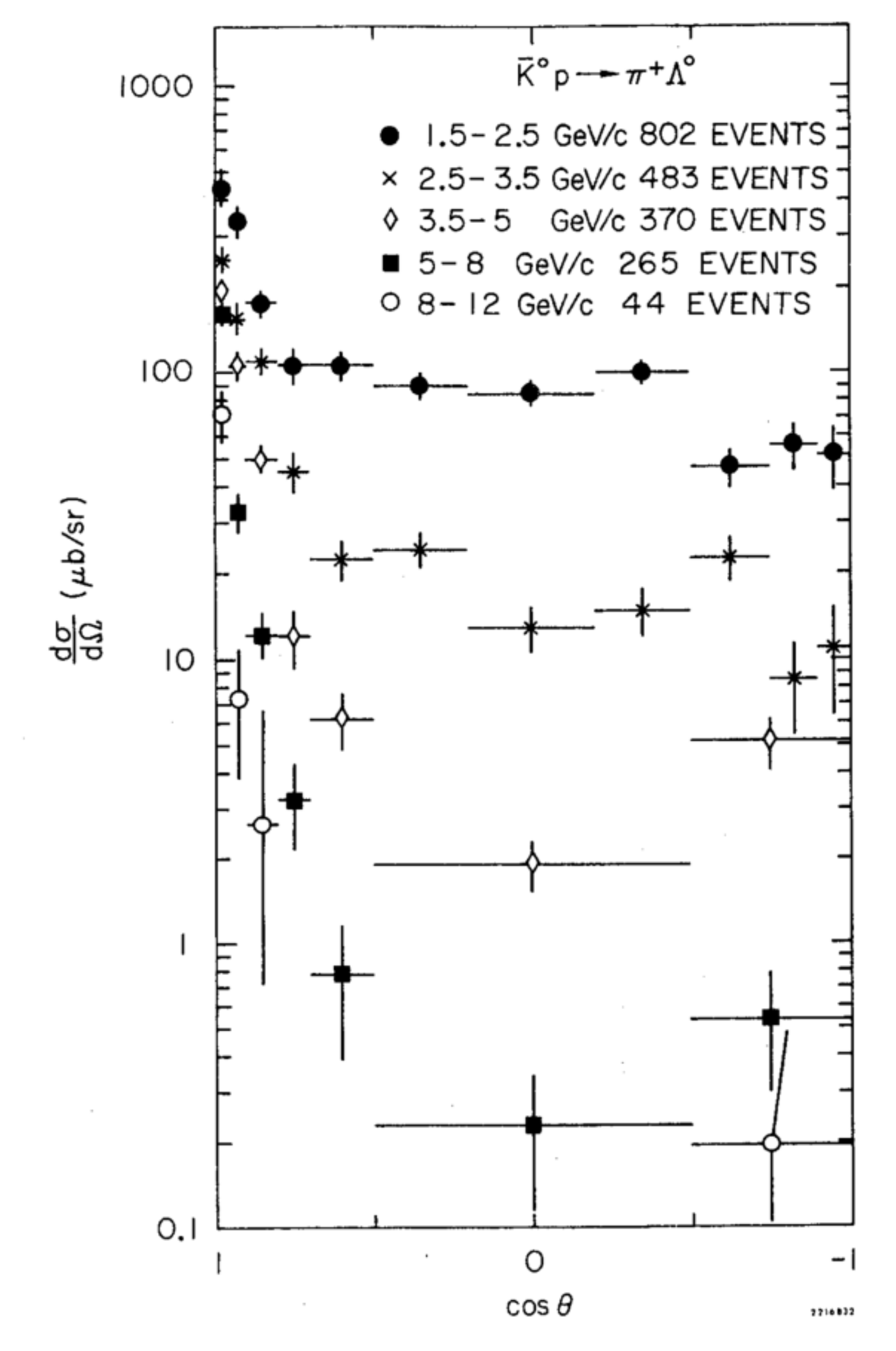}

    \caption[Kinematic coverage]{
	The total cross section for $K_Lp\to\pi^+\Lambda$ reaction as a 
	function of beam momentum~\protect\cite{SLAC177} (left) and the 
	differential cross sections for various beam momentum ranges.
}\label{fig:Lampip1}
    \end{center}
\end{figure}

The $K_Lp\to\pi^+\Lambda$ and $K_Lp\to\pi^+\Sigma^0$ reactions are key to 
studying hyperon resonances - an analog of $N\pi$ reactions for the $N^\ast$ 
spectra. They are also the key reaction to disentangling the weak exchange 
degeneracy of the $K^\ast(892)$ 
and $K^\ast(1420)$ trajectories.  (A general discussion is given in 
Sections~\ref{sec:A1} and \ref{sec:Moskov}). The first measurement of this 
reaction was performed at SLAC 
in 1974~\cite{SLAC177} for $K_0$ beam momentum range between 1~GeV/$c$ to 
12~GeV/$c$, which is shown in Fig.~\ref{fig:Lampip1}.  The total 
number of $\pi^+\Lambda$ events was about 2500~events, which statistically 
limits the measurement.

\begin{figure}[!htb]
\centering{
    \includegraphics[width=0.6\textwidth,keepaspectratio]{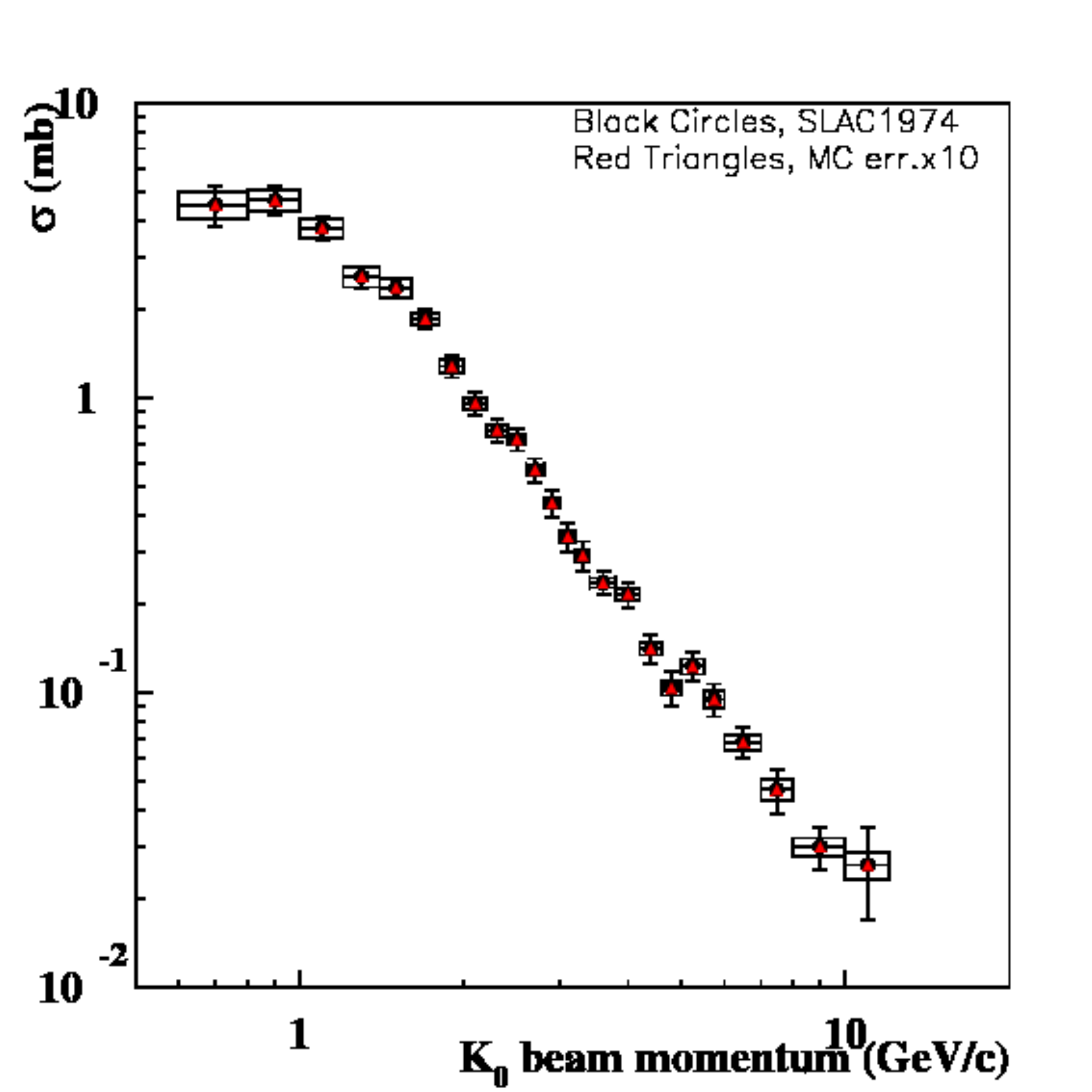} }

    \caption[Estimation of Error]{The total cross-section uncertainty 
    estimate (statistical error only) for $K_Lp\to\pi^+\Lambda$ reaction
    as a function of $K_L$ beam momentum in comparison with SLAC 
    data~\protect\cite{SLAC177}. The experimental uncertainties
    have tick marks at the end of the error bars. The box-shaped error bars 
    in the MC points were increased by a factor of 10.
}\label{fig:comparisonLampPip}
\end{figure}

Figure~\ref{fig:comparisonLampPip} shows our estimate of the 
statistical uncertainty of the $\pi^+\Lambda$ total cross section as 
a function $K_L$ beam momentum. We kept the same momentum bin size 
as the one from the SLAC data. The box-shaped error bars in the MC points
(red triangles) were increased by a factor of 10 for comparison with 
the SLAC data.  The proposed measurements will provide  
unprecedented statistical accuracy to determine the cross section for a 
wide range of $K_L$ momentum.

\subsubsection{$K_Lp\to K^+\Xi^0$ Reaction} 

The study of cascade data will allow us to place stringent constraints on
dynamical coupled-channel models. It was recently found in $N^\ast$ 
spectroscopy 
that many $N^\ast$ resonances do not couple strongly to a $N\pi$ channel, 
but are nicely 
seen in $K\Lambda$ and $K\Sigma$ channels.  The corresponding situation in 
hyperon spectroscopy lead to many $\Lambda^\ast$ and $\Sigma^\ast$ 
resonances 
decaying preferably to a $K\Xi$ channel, see Appendix~A1~\ref{sec:A1} for 
details. 
In addition, cascade data will provide
us with long-sought information on missing excited $\Xi$ states and the
possibility to measure the quantum numbers  of the already established
$\Xi(1690)$ and $\Xi(1820)$ from a double-moments analysis. The expected 
large data sample will allow us to determine the induced polarization
transfer of the cascade with unprecedented precision, which will place 
stringent constraints on the underlying dynamics of the reaction. 
Polarization measurements of hyperons shed light on the contribution from 
individual
quarks to the overall polarization of these states. The polarization of
the ground-state cascade can be measured from its weak decay in a 
straightforward way. With a $K_L$ beam, the study of the reaction  $K_Lp\to
K^+\Xi^0$ is quite simple and an unprecedented statistical sample can be
easily obtained. The statistical uncertainty obtained for 2-fold differential 
polarization observables with 100 days of beam time ($\sim 3\times 10^5$ 
reconstructed events)  is of the order of 0.05-0.1, which is more than an 
order of magnitude smaller from existing measurements, allowing precision 
tests on the underlying dynamics to be performed. 

\begin{figure}[h!]
\centering{
    \includegraphics[width=0.45\textwidth,keepaspectratio]{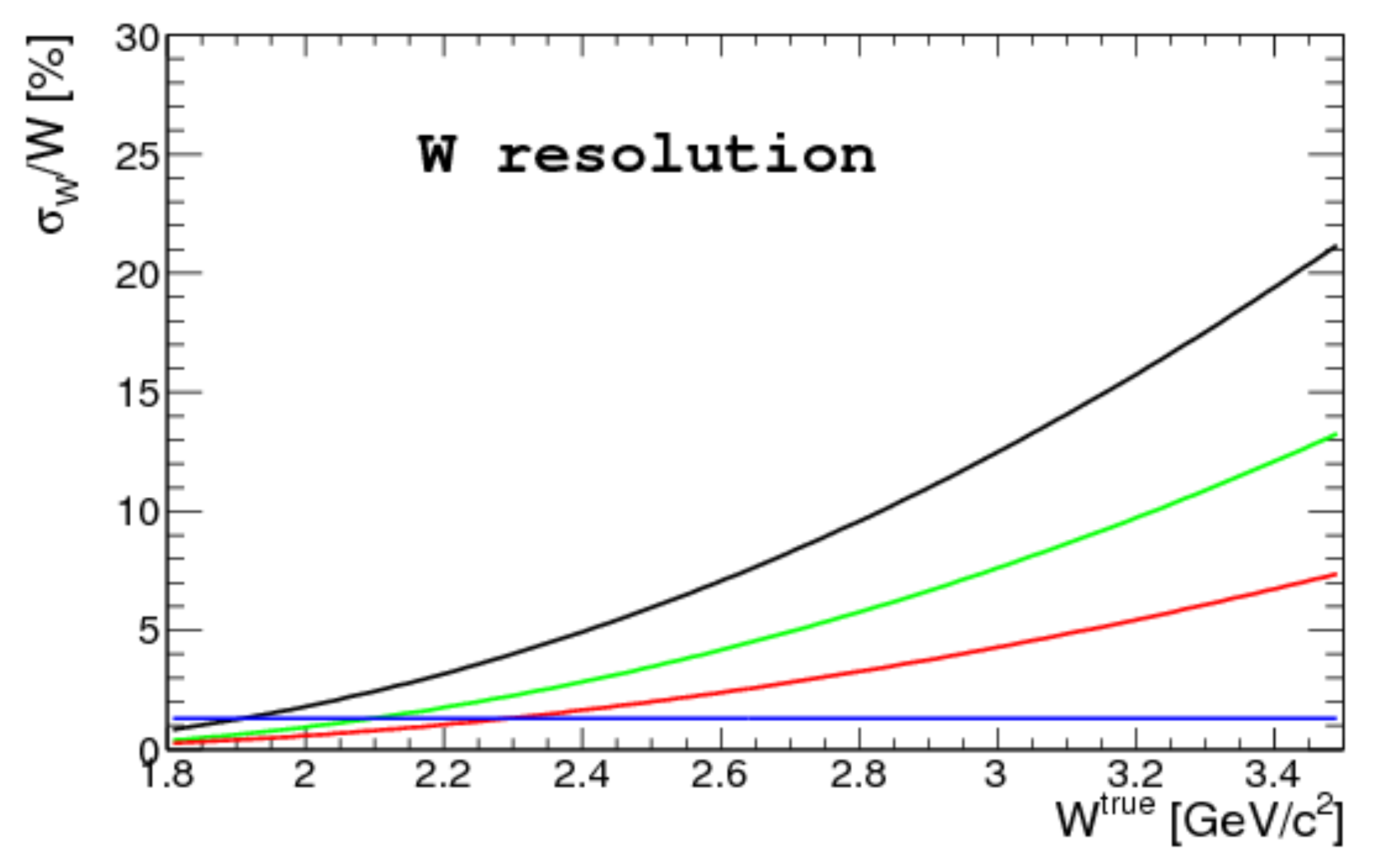}

    \caption{$W$ resolution of $\sigma W/W$, depending on the accuracy 
    	of the time-of-flight for 300~ps (black), 150~ps (green), 100~ps 
        (red), and when $W$ is determined from all detected final-state 
        particles (blue).} \label{fig:KlKpXi2}
}
\end{figure}

Several topologies can be used to reconstruct $K_Lp\to K^+\Xi^0$ events,
thereby enhancing the available statistics. The biggest contribution results 
from requiring the reconstruction of only the $K^+$ in the final state and
reconstructing the reaction using the missing-mass technique. The $\Xi^0$
decays almost 100\% of the time to $\pi^0\Lambda$. By utilizing the large
branching ratios for $\Lambda\to \pi^- p$ and $\pi^0\to\gamma\gamma$ decays, 
we can also fully reconstruct the $\Xi^0$s in the final state using the 
four-momenta
of the detected final-state particles.  Figure~\ref{fig:KlKpXi2} shows the 
expected $W$ resolution for this reaction, depending on the 
accuracy of the time-of-flight for 300~ps (black), 150~ps (green), 100~ps 
(red), and when $W$ is determined from all detected final-state particles 
(blue).

\begin{figure}[h!]
\centering{
    \includegraphics[width=0.9\textwidth,keepaspectratio]{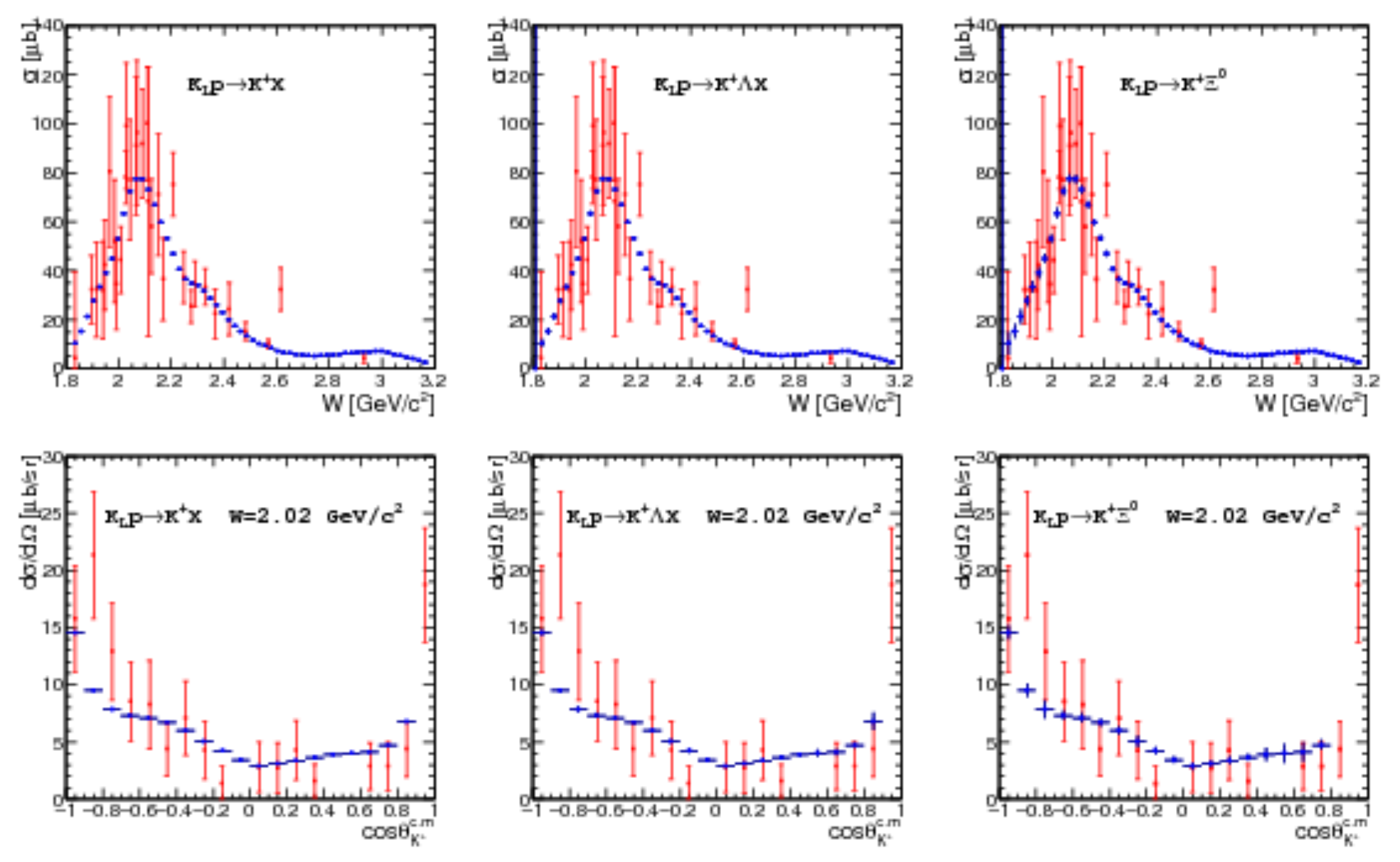}

    \caption{Total and differential cross section statistical uncertainty
        estimates (blue points) for the three topologies (column 1: only 
        $K^+$ reconstructed, column 2: $K^+\Lambda$ reconstructed, and 
        column 3: $K^+\Xi^0$ reconstructed) in comparison with data taken 
        from Ref.~\protect\cite{Ref.DataCasc} (red points). } 
        \label{fig:KlKpXi3}
}
\end{figure}

In 100 days of beamtime with $3\times10^4 K_L/s$ on the target, we expect
$9\times 10^6$ $K_Lp\to K^+\Xi^0$ events. From this, the available
reconstructed events expected is $4\times 10^6$ for Topology~1 $K_Lp\to
K^+X$, $3\times 10^5$ for Topology~2 $K_Lp\to K^+\Lambda X$, and
$4\times 10^4$ for Topology~3 $K_Lp\to K^+\Xi^0$. 
Figure~\ref{fig:KlKpXi3} compares the statistical uncertainties of the 
total and differential cross sections for the reaction $K_Lp\to K^+\Xi^0$ 
with existing data taken from~\cite{Ref.DataCasc} for the three different 
topologies (column~1: only $K^+$ reconstructed, column~2: $K^+\Lambda$ 
reconstructed, and column~3: $K^+\Xi^0$ reconstructed).

\begin{figure}[h!]
\centering{
    \includegraphics[width=0.9\textwidth,keepaspectratio]{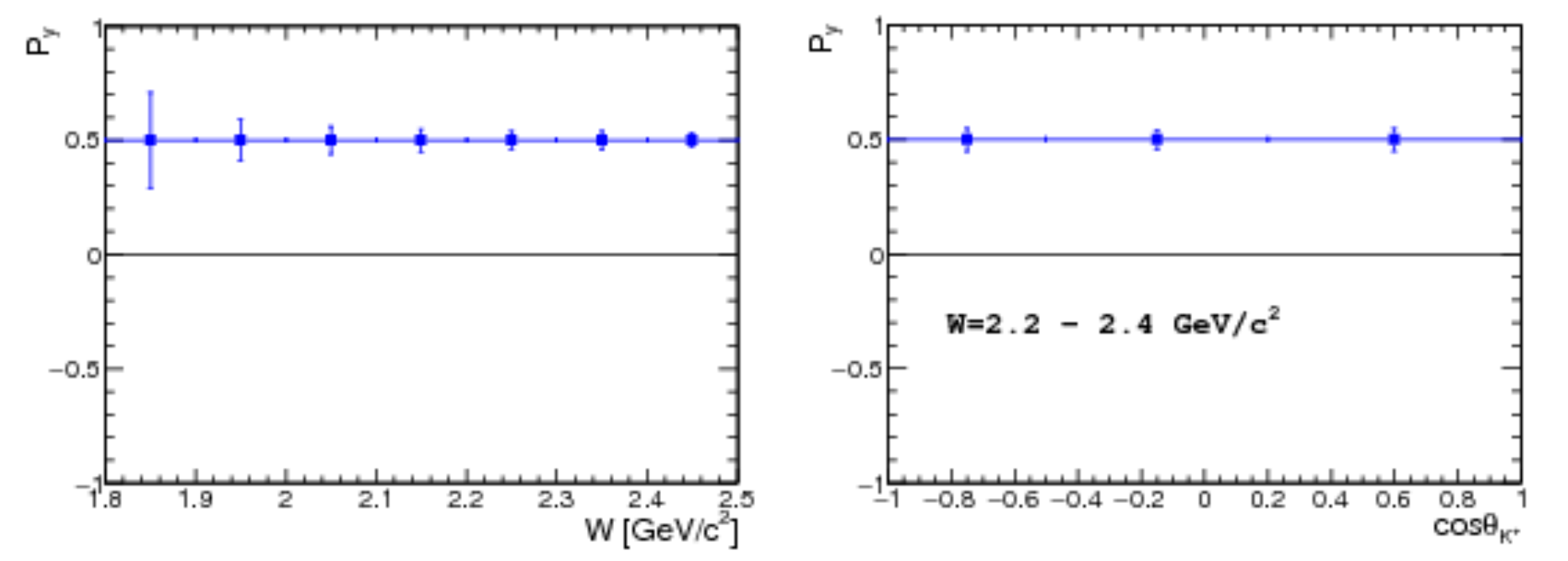}

    \caption{Estimates of the statistical uncertainties of the induced
    polarization of the cascade as a function of $W$ (one-fold differential)
    and $\cos\theta_{K^+}$ (two-fold differential).  
    } \label{fig:KlKpXi4}
}
\end{figure}

These statistics also allow us to determine the cascade
induced polarization by utilizing the fact that the cascade is
self-analyzing with an analyzing power of $-0.406$~\cite{PDG2016}.
Figure~\ref{fig:KlKpXi4} shows the statistical uncertainty estimates
of the induced polarization of the cascade by simple fits to the
acceptance-corrected yields of the pion angular distribution in the
$\Xi^0$ rest frame.

The main background for this reaction would come from the reactions
$K_Lp\to K^+n$ and $K_Lp\to\pi^+\Lambda$, where the $\pi^+$ is
misidentified as a kaon. The former reaction has an order-of-magnitude
higher cross section than $K_Lp\to K^+\Xi^0$; however, the $W$ 
resolution below 2.5~GeV/$c^2$ allows a clean separation of these two 
reactions.  Detection and reconstruction of the $\Lambda$ places 
additional constraints that reduce any background contributions 
significantly. Neutron-induced reactions are not expected to contribute 
significantly to background and, with missing-mass, invariant-mass, and 
time-of-flight cuts, such background contributions can be eliminated.

The KL Facility can be utilized to study excited cascade states 
$K_Lp\to K^+\Xi^\ast$ with $\Xi^\ast\to\pi\Xi$ and $\Xi^\ast\to\gamma\Xi$. 
These excited states should be easily identified and isolated 
using the missing-mass and invariant-mass techniques. A double-moment 
analysis can be employed by reconstructing the entire decay chain and 
establish the spin and parity of these excited 
states~\cite{Ref.DoubleMoments}.

\subsubsection{$K_Lp\to K^+n$ Reaction} 
\label{sec:KPlN}

The $K^0_Lp\to K^+n$ reaction is a very special case in kaon-nucleon
scattering. Due to strangeness conservation, formation of
intermediate resonances is
forbidden for this reaction. The main contribution comes from various
non-resonant processes, which can be studied in a clean and controlled 
way. Similar non-resonant processes can be seen in other
reactions where they can interfere with hyperon production amplitudes, causing
distortion of the hyperon signals. That is why knowledge of the non-resonant
physical background is important not only for the kaon-induced reactions
but for all reactions with strangeness.  The non-resonant nature of the
reaction does not guarantee the absence of bumps in the total
cross section: kaons and/or nucleons can be excited in the intermediate
stage, producing bumps in the total cross section.

\begin{figure}[h!]
\centering
{
    \includegraphics[width=0.45\textwidth,keepaspectratio]{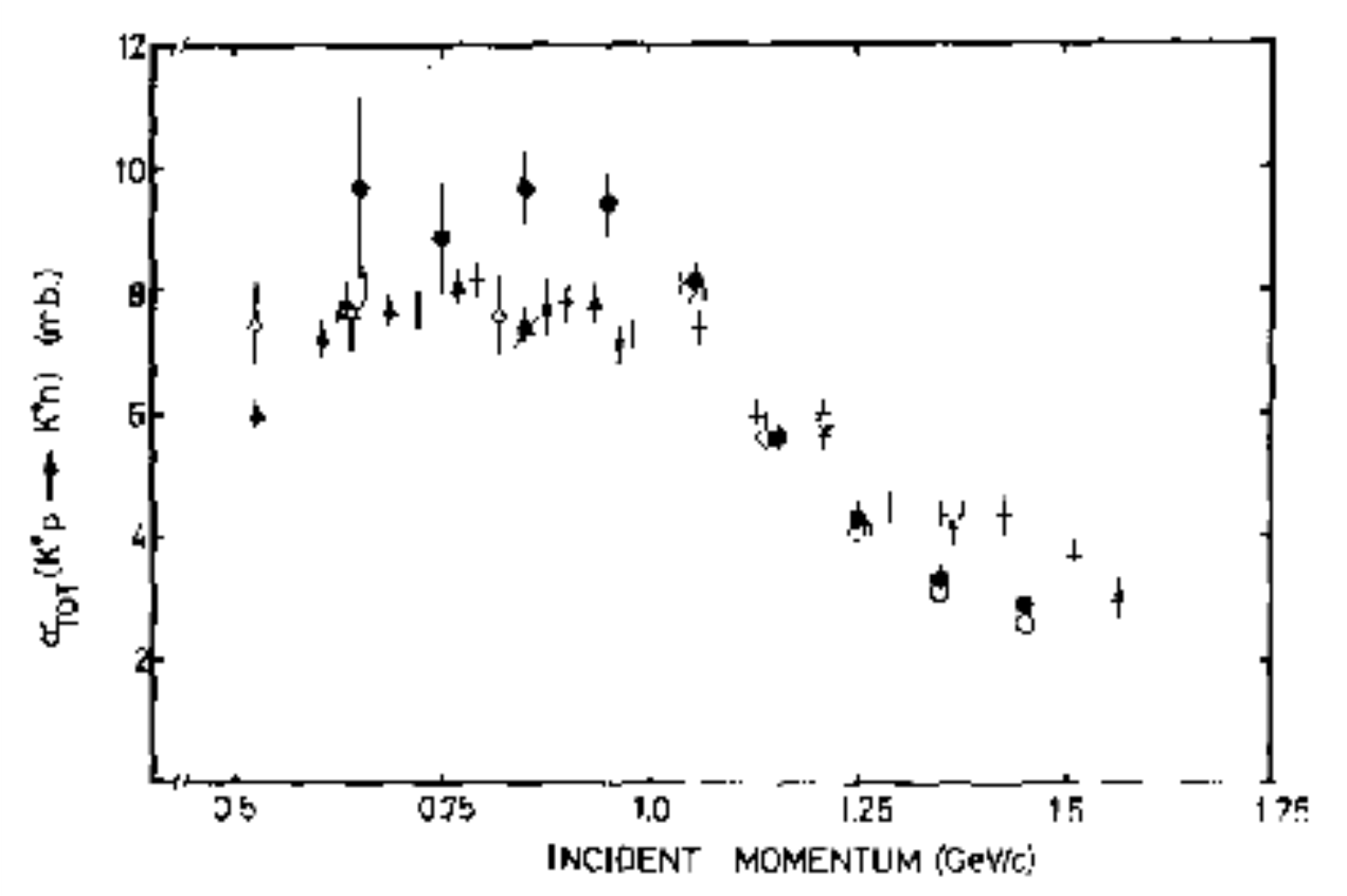} }
{
    \includegraphics[width=0.45\textwidth,keepaspectratio]{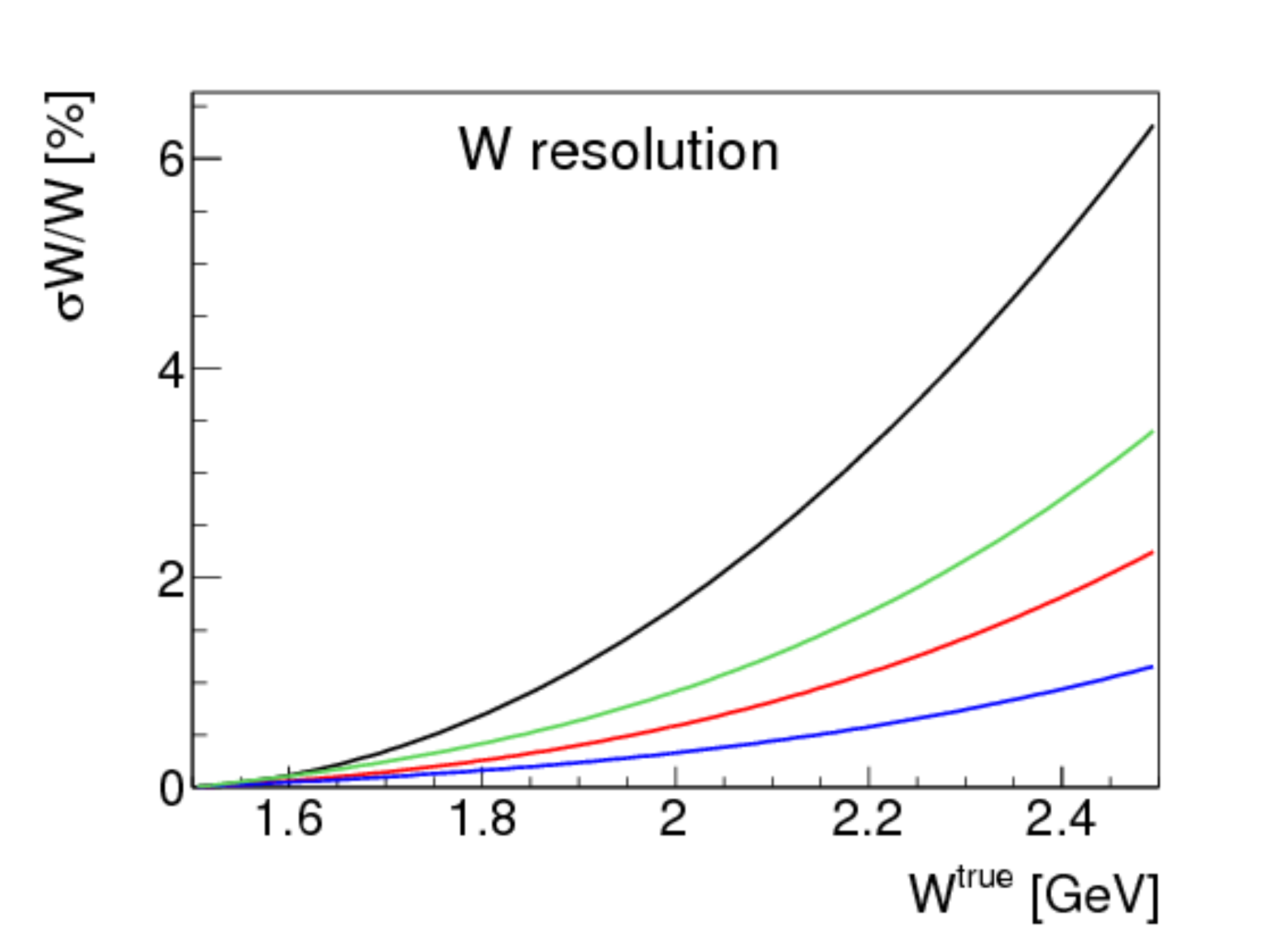} }

    \caption{The total cross section for $K_Lp\to K^+n$ reaction as 
    a function of $K_L$ momentum from Ref.~\protect\cite{ARMITAG1977} 
    (left) and expected $W$ resolution, $\Delta W/W$, depending on 
    time-of-flight accuracy (right) for 300~ps (black), 150~ps (green), 
    100~ps (red), and 50~ps (blue), respectively.} \label{fig:KPln1}
\end{figure}

The reaction $K^0_Lp\to K^+n$ is simple and it has a very high
production cross section, see Fig.~\ref{fig:KPln1}(left); nevertheless, 
the data on this reaction are scarce. It is a bit simpler to perform a 
positive kaon beam scattering for the inverse reaction, but the
necessity of a neutron target with unavoidable many-body and FSI effects
complicates the data analysis. That is why the inverse reaction is also 
not so well known. A fair amount of differential cross-section 
data are available in the range $0.5 < p_{K_L} < 1.5$~GeV/$c$, predominantly 
from bubble 
chambers, see Ref.~\cite{ARMITAG1977}, and there are a few measurements at 
high momenta: $p_K = 5.5$~GeV/$c$~\cite{CLINE1970}, $p_K = 
10$~GeV/$c$~\cite{BAJLLON1978}. In the energy range $2 < W < 3.5$~GeV, 
which can be covered by the KLF experiment with very high statistics, 
there are no data on this reaction at all.

To reconstruct the reaction fully via missing mass tecnique it is enough 
to detect charged kaon.  However at higher $W$, depending on timing 
resolution the $\Delta W$ will blow up, therefore, better timing resolution 
would be preferable.  The Fig.~\ref{fig:KPln1}(right) shows how the timing 
resolution affects $W$ resolution. Future improvements of start counter 
timing resolution will allow to lower W resolution further, such that at 
150~ps we may get $\Delta W$<20~MeV below $W<2.2$~GeV, currently this is 
under the study.  The beam energy is determined by TOF technique utilizing 
the 16-m flight path between the kaon production Be target and the 
reaction hydrogen target. The beam resolution in this case is driven by 
the ST time resolution (Sec.~\ref{sec:Bres}). The present ST time 
resolution leads to a 300~ps vertex time resolution.

In addition to a kaon, one could also detect a neutron; however, due to 
the poor neutron detection efficiency and the large systematic 
uncertainties associated with neutron detection we do not expect any 
improvement in reaction reconstruction in this case.

\begin{figure}[h!]
\centering
{
    \includegraphics[width=0.45\textwidth,keepaspectratio]{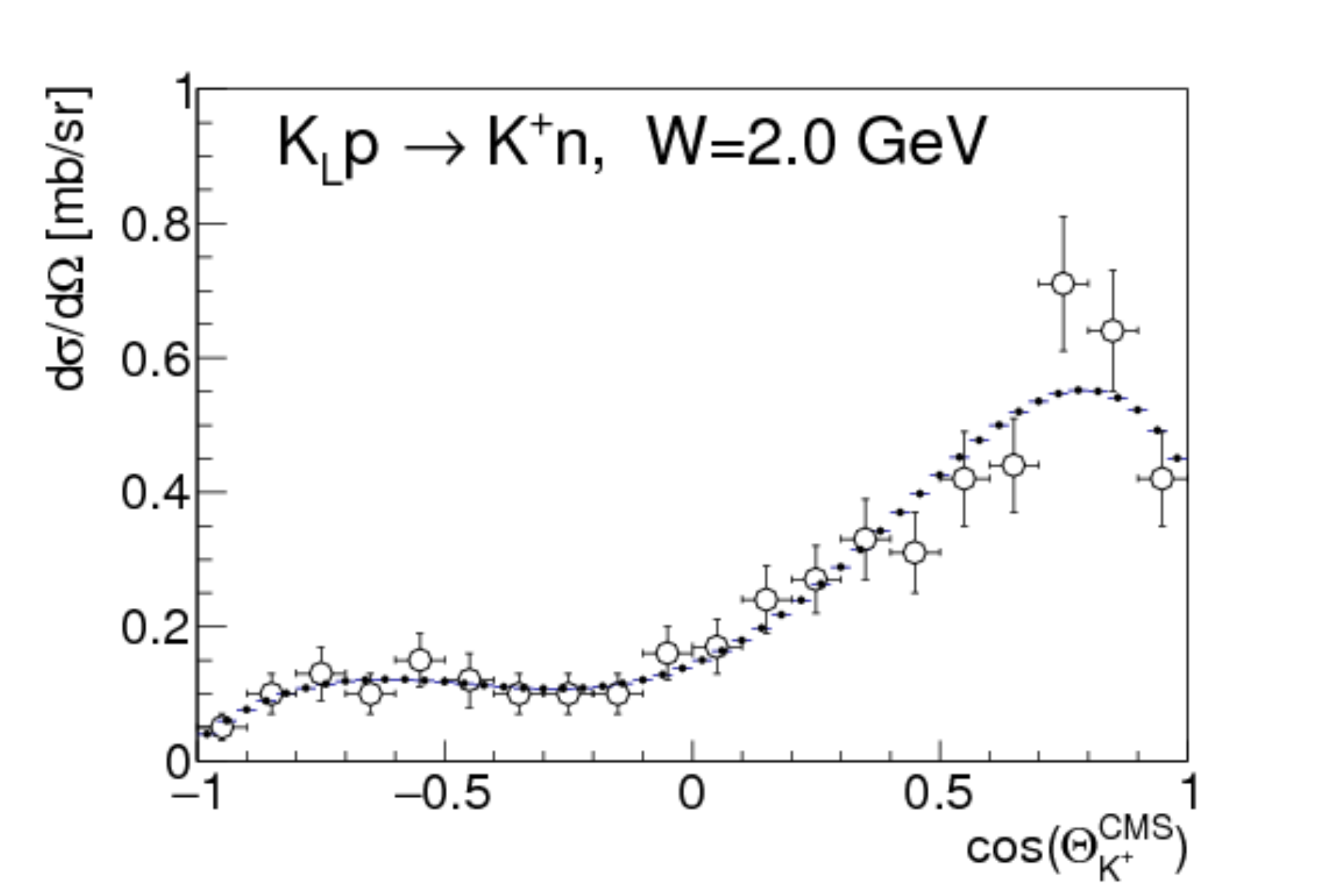} }
{
    \includegraphics[width=0.45\textwidth,keepaspectratio]{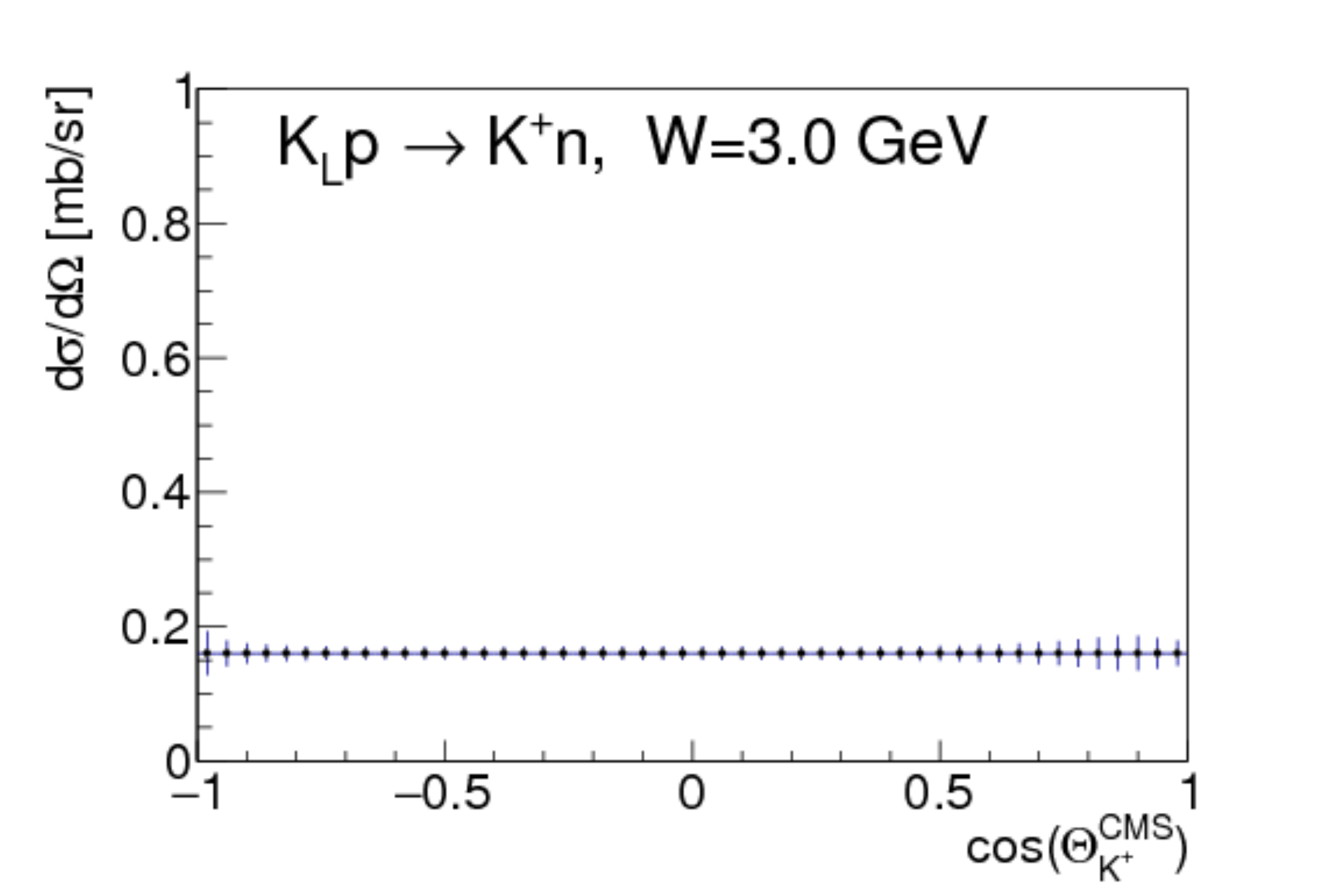} }

  \caption{The cross-section uncertainty estimates (statistical only) 
  for $K_Lp\to K^+n$ reaction for the $W=2$~GeV (left) in comparison 
  with data from Ref.~\protect~\cite{ARMITAG1977} and $W=3$~GeV (right)
  The error bars for the right plot were increased by factor of 10 to 
  make them visible.} \label{fig:KPln2}
\end{figure}

In 100 days of beamtime with $3\times 10^4~K_L$/s on the target, we expect
to detect around 200M $K_Lp\to K^+n$ events. A typical example of the 
expected statistics in comparison to previous data can be seen in
Fig.~\ref{fig:KPln2}(left). The highest flux is expected around $W = 
3$~GeV, where we had to increase statistical errors by a factor of 10 
to make them visible, see Fig.~\ref{fig:KPln2}(right).

There are three major sources of background: $np\to K^+nn$, $np\to\pi^+nn$,
and $K_Lp\to K^+\Xi^0$. Neutron flux drops exponentially with energy (see
Appendix~A4~(Sec.~\ref{sec:A4}) for details) and generally the high-energy 
neutron flux is small, but nonvanishing. If neutrons and $K_L$s have the 
same speed they cannot be separated by time of flight. Neutron-induced
reactions have high cross sections, which is why it is necessary to
consider them as a possible source of background. Fortunately, 
neutron-induced kaon production contributes at the low level of $10^{-3}$, 
which, with missing-mass cuts, can be reduced below $10^{-4}$. Some of 
the pions from $np\to\pi^+nn$ reaction can be misidentified as kaons, 
but with missing mass and time-of-flight cuts we can reduce the 
contribution of this background to a sub-per mill level.  A detailed 
description of various backgrounds can be found in 
Appendix~A5~(Sec.~\ref{sec:A5}). $K_Lp\to K^+\Xi^0$ has a 
cross section 100 times smaller than that for $K_Lp\to K^+n$.  Below 
$W < 2.3$~GeV, $K_Lp\to K^+\Xi^0$ can 
be completely filtered out by a 3$\sigma$ $K^+$ missing-mass cut. At 
high $W$, there is some overlap. One can use conventional background 
subtraction techniques to eliminate it. The $\Xi^0$ often has charged 
particles in its decay chain, which can be used to veto the channel. 
Our studies show that the background from $\Xi^0 \to \pi^0\Lambda\to 
\pi^0\pi^0 n$ can be reduced below $10^{-4}$ level as well.

\subsubsection{Reaction $K_Lp\to K^-\pi^+p$}

\begin{enumerate}
\item \textbf{I=1/2 P-wave Phase Shift Analysis}

The $K\pi$ $S$-wave scattering, below 2~GeV, has two possible
isospin channels, $I = 1/2$ and $I=3/2$. In the same range of
mass, the $P$-wave has one isospin, $I = 1/2$. The $P$-wave is
a narrow elastic wave peaking at $892$ MeV and interpreted as
$K^\ast(892)$. The $I=3/2$ $S$-wave is elastic and repulsive up to
1.7~GeV and contains no known resonances. In the $I=1/2$ $S$-wave,
a peaking broad resonance appears above 1350~MeV. Moreover, some
phenomenological studies~\cite{Anisovich:1997qp,Cawlfield:2006hm,
Delbourgo:1998kg,Scadron:2002mm,Zhou:2006wm} predict the presence
of a resonance with a very large width in the region close to the
$K\pi$ threshold. The hadroproduction of the $K\pi$, using the
$K_L$ facility, provides an adequate environment for studying the
$K^\ast$ states. 

We made a Monte Carlo simulation to study the phase-shift of the
$K\pi$ $P$-wave in the $K_L$ facility. The $K\pi$ production simulation
is based on the model given by Estabrooks et al.~\cite{Estabrooks}.
The model describes the charged kaon reaction $K^-p \to K^-\pi^+ n$.
Therefore in this simulation we assume that the cross-section of $K_L^0p
\to K^-\pi^+p$ is similar to the previous reaction. The model describes
the production mechanism in terms of strongly  degenerate $\pi$-$B$ and
$\rho$-$A2$ Regge exchanges and Regge ``cuts" that have simple structure
in the $t$-channel helicity frame. The $t$-dependent parametrization of
the naturality amplitudes $L_{\lambda}^{\pm}$ for production of a
$K^-\pi^+$ state of invariant mass $m_{K\pi}$, center-of-mass (COM)
momentum $q$, angular momentum $L$, and $t$-channel helicity $\lambda$,
by natural $(+)$ and unatural $(-)$ parity exchange, according
to~\cite{Estabrooks}, is as follows
\begin{eqnarray}
        L_{ 0 } &=& \frac{ \sqrt{-t} }{ m^{2}_{\pi} - t }G_{
        K\pi }^{ L }( m_{K\pi}, t )\;, \\
        L_{ 1 }^{-} &=& \sqrt{\frac{1}{2} L(L+1)}G_{ K\pi }^{ L }( m_{K\pi},
        t )\gamma_{c}( m_{K\pi} )\exp(b_{c}(m_{K\pi})(t - m_{\pi}^2))\;,\\
        L_{ 1 }^{+} &=& \sqrt{\frac{1}{2} L(L+1)} G_{ K\pi }^{ L }( m_{K\pi},
        t ) [ \gamma_{c}( m_{K\pi} ) \exp(b_{c}(m_{K\pi})(t -
        m_{\pi}^2)) \nonumber \\
        &-& 2i\gamma_{a}( m_{K\pi} ) \exp(b_{a}(m_{K\pi})|t^{\prime}| (t
        - m_{\pi}^2))]\;, \\
        L_{\lambda}^{\pm} &=& 0\;, \qquad \lambda \geq 2\;,
\end{eqnarray}
\noindent
where $G_{ K\pi }^{ L }$ is related to the $K\pi$ elastic scattering
amplitude $a_L$ by
\begin{equation}
        G_{ K\pi }^{ L }( m_{K\pi}, t ) = N \frac{ m_{K\pi} }{ \sqrt{q} }
        a_{ L }( m_{K\pi} )\exp(b_{L}(m_{K\pi})(t - m_{\pi}^2) ).
\end{equation}
\noindent
The parameters $\gamma_{a}$, $\gamma_{c}$, $b_a$, $b_c$ and $b_L$ were
determined by fitting the LASS  $K^{-}p \to K^-\pi^+ n$ data.  The $K\pi$
scattering amplitudes are the sum of an isospin-$1/2$ and $3/2$ component,
\begin{equation}
	a_{L} = a^{I=1/2}_L + \frac{1}{2}a_{L}^{I = 3/2}\,,
\end{equation}
\noindent
where $a^{I}_L$ is described by a relativistic Breit-Wigner.

In this study, we include only the $P$-wave component
$K^\ast(892)$ in the simulation. The number of events for $100$ days
of running, at $3 \times 10^4$ $K_L/s$,  in the range of mass $m_{K\pi}
< 1.5~{\rm GeV}$ and negative transfer $4$-momentum $-t < 0.8~{\rm
GeV}^2$, is expected to be around $2$ million events. The $K_{L}$ beam
energy used in this simulation is equal to 7~GeV. The generated Monte
Carlo events, function of the transfer $4$-momentum $-t$ and the
invariant mass $m_{K\pi}$, is illustrated in Fig.~\ref{fig:tmdist}, and the
phase-shift function of $m_{K\pi}$ with mass bin width of 10~MeV and
$|t| < 0.5$ ${\rm GeV}^2$  is illustrated in Fig.~\ref{fig:phase}.
\begin{figure}[h!]
\centering
{
    \includegraphics[width=0.7\textwidth,keepaspectratio]{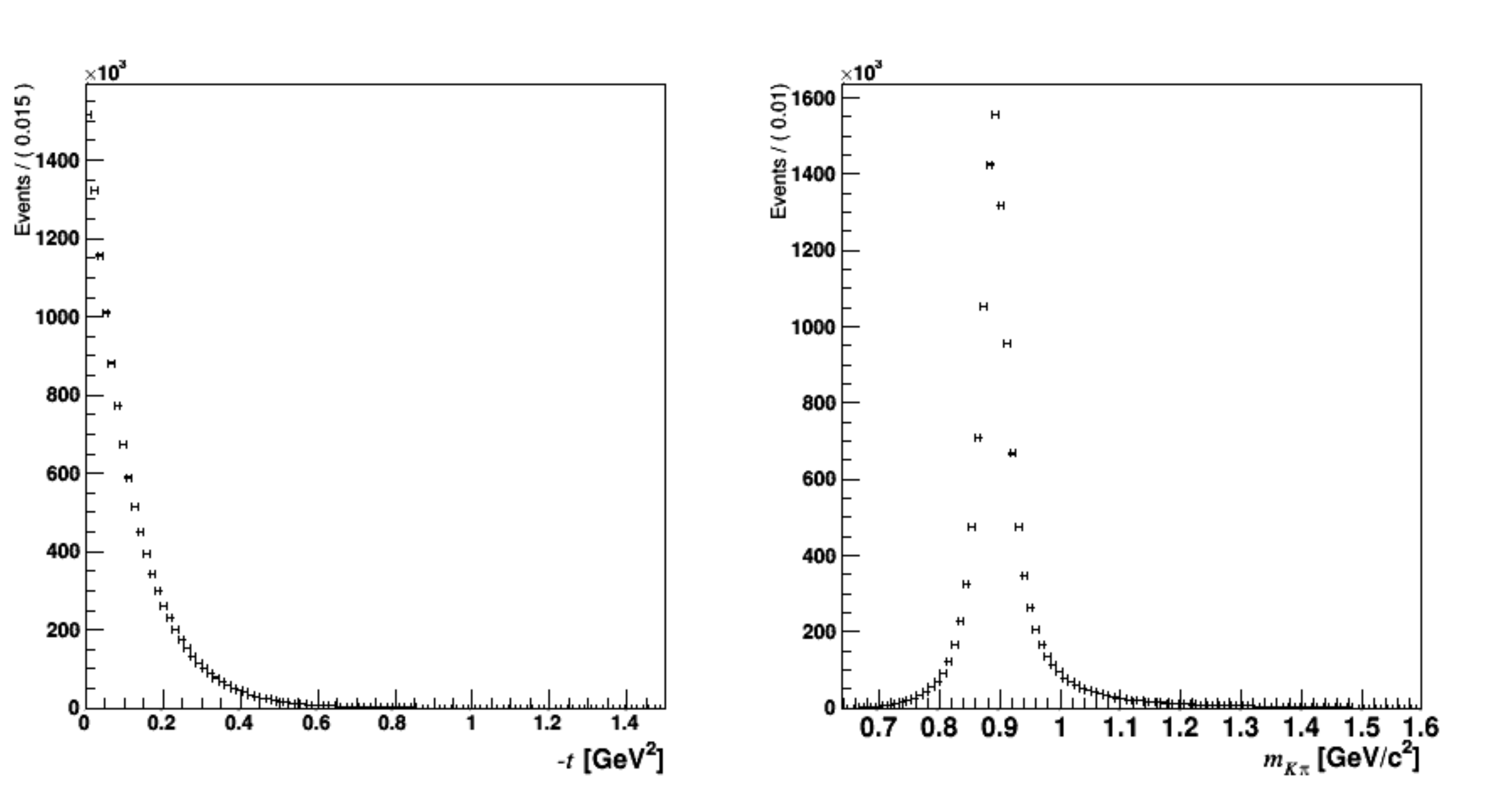} }
{
    \includegraphics[width=0.6\textwidth,keepaspectratio]{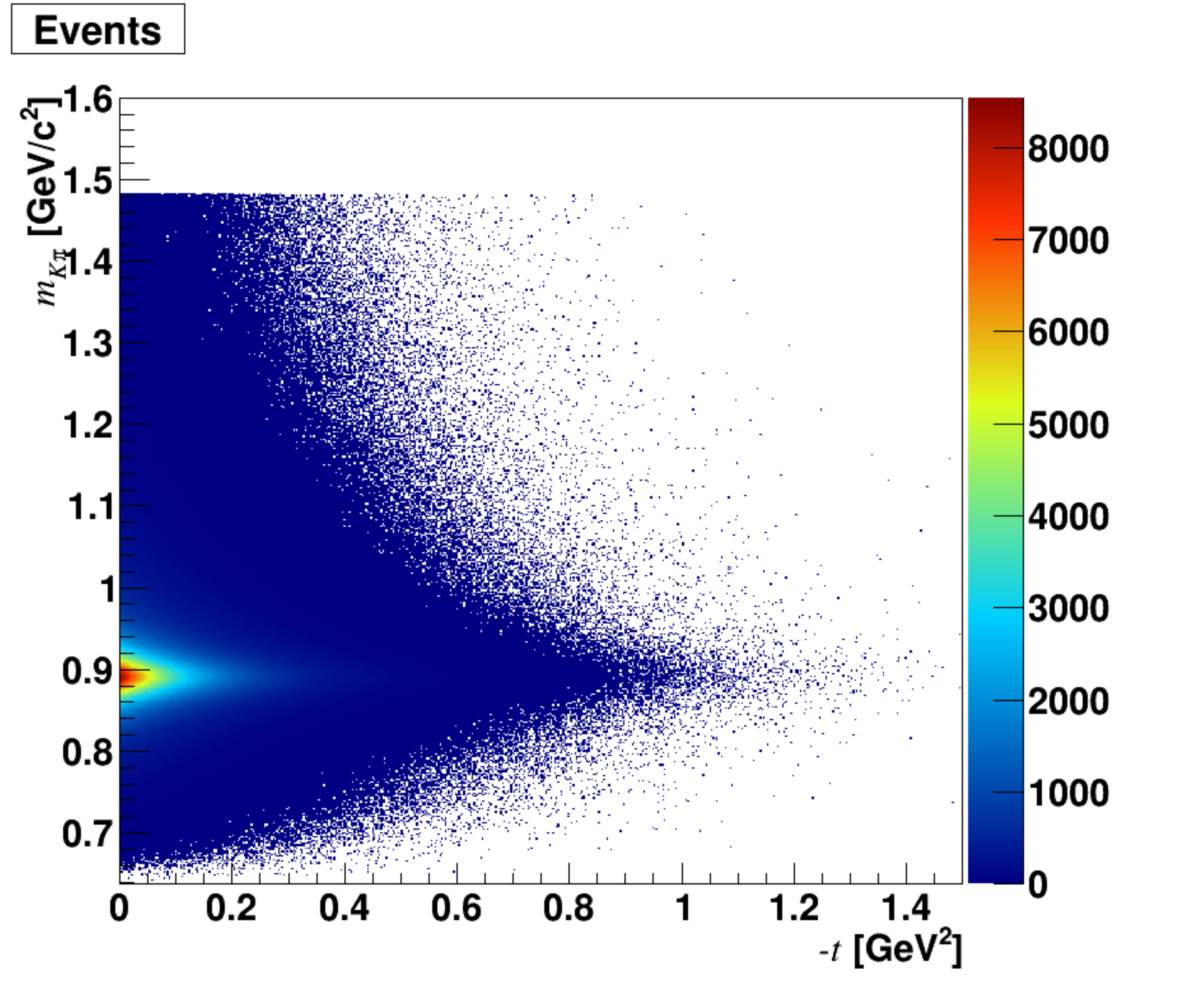} }

	\caption{The generated MC events distribution function of $-t$
	(upper left panel); $m_{K\pi}$ (upper right panel) and
	$m_{K\pi}$ vs $-t$ (lower panel). $m_{K\pi}$ vs $-t$ (lower panel).}
	\label{fig:tmdist}
\end{figure}
\begin{figure}[h!]
\centering
{
    \includegraphics[width=0.40\textwidth,keepaspectratio]{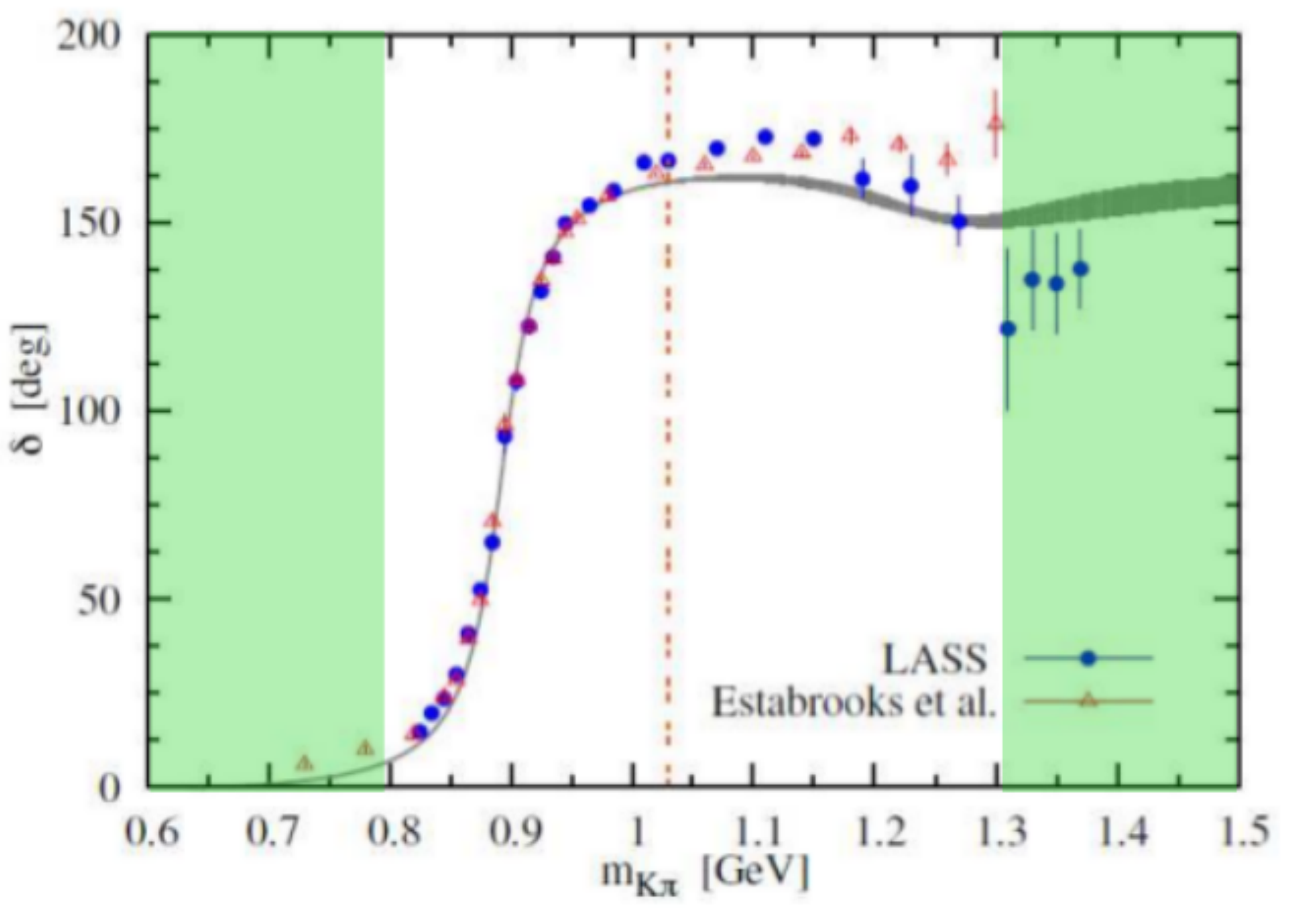} }
{
    \includegraphics[width=0.55\textwidth,keepaspectratio]{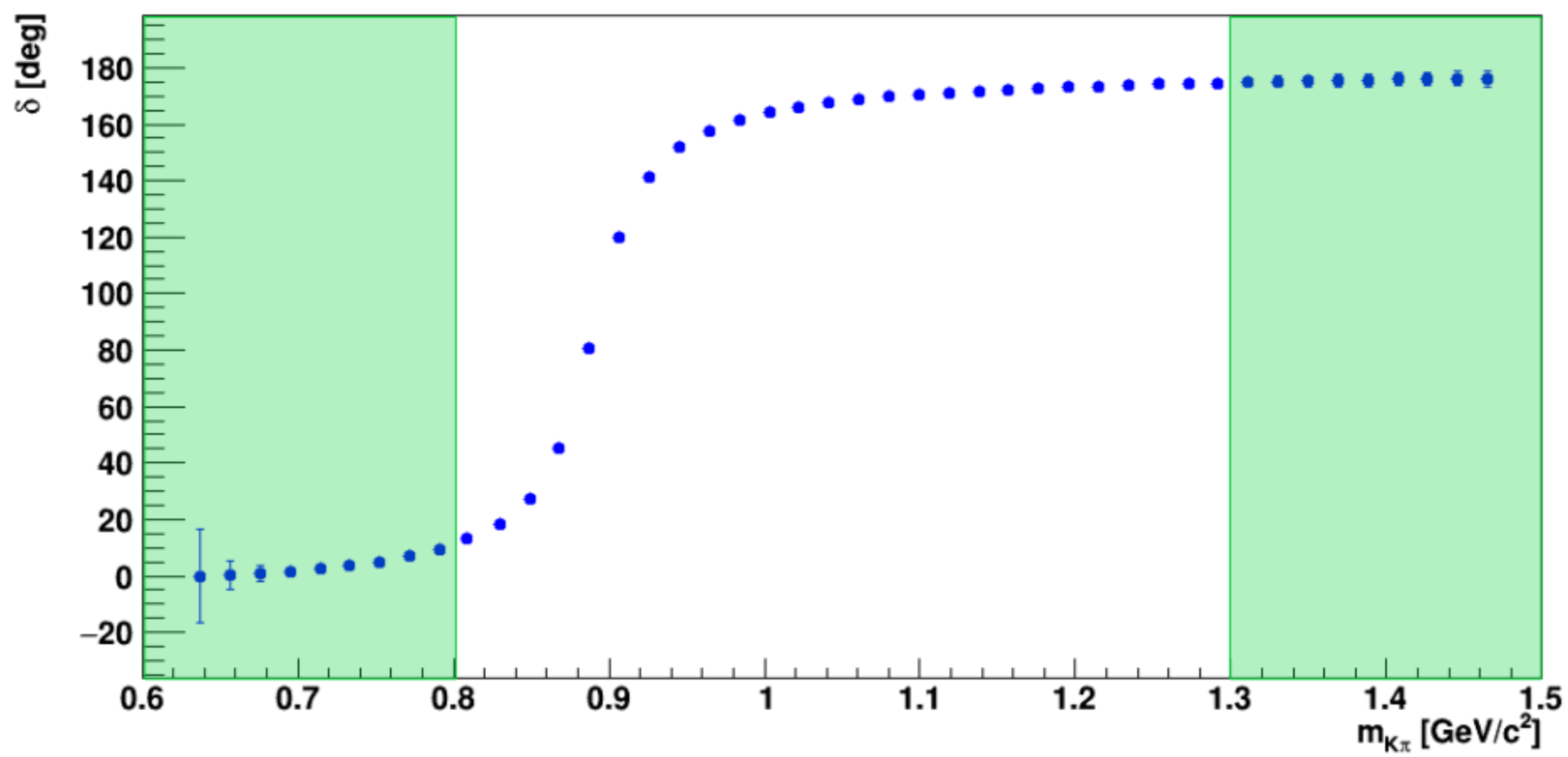} }

        \caption{The $I=1/2$ $K\pi$ scattering $P$-wave phase-shift
        function of $m_{K\pi}$. The left panel shows experimental
        results from LASS~\protect\cite{LASS} and Estabrooks
        \textit{et al.}~\protect\cite{Estabrooks}. The gray
        band represents the fit to the $\tau$ decay data by Boito
        \textit{et al.}~\protect\cite{Boito}. On the right panel,
        we present results of expected measurement for 100 days of
        running. The statistical errors on the right panel are 
        increased by factor of 10 for a better visibility.} 
	\label{fig:phase}
\end{figure}

As one can see  experimental data produced by LASS
experiment~\cite{LASS}  in the mass regions close to the threshold and 
above 1.3~GeV are poorly measured and therefore do not provide constrains 
in these regions.  However, according to this simulation study, the 
proposed $K_L$ facility can provide statistically high precision data,  
which will significanlty constrain the uncertainty of the phase motion in 
the full range of $m_{K\pi}$ from the threshold up to 1.5~GeV.
\end{enumerate}

\subsection{Summary and Beam Time Request}

We propose to perform strange hadron spectroscopy with a secondary $K_L$
beam in the GlueX setup at JLab.  Precise new experimental data (both 
differential cross sections and recoil polarization of hyperons) for 
$K_Lp$ scattering with good kinematic coverage will be obtained. This 
will allow predictions from CQM and LQCD to be checked for all families of 
excited $\Lambda^\ast$, $\Sigma^\ast$, $\Xi^\ast$, and $\Omega^{\ast}$ 
hyperon resonances for the first time.  In addition, it will permit a 
search for the possible 
existence of hybrids in the hyperon sector as predicted by the lattice
calculations~\cite{Dudek:2012ag}.  

A complete understanding of three-quark bound states requires
accurate measurements of the full spectra of hyperons with their 
spin-parity assignments, pole positions, and branching ratios.  
An important 
impact of these strange hyperon spectroscopy measurements is their 
significance for the thermodynamic properties of the early universe 
at freeze-out, which is one of the main physics topics at heavy-ion 
colliders.

Besides hyperon spectroscopy, the experimental data obtained in the 
strange meson sector in the reactions $K_Lp\to K^\pm \pi^\mp p$ and 
$K_Lp\to K_S\pi^\pm n(p)$ will provide precise and statistically 
significant data for experimental studies of the $K\pi$ system. This 
will allow a determination of quantum numbers of strange meson 
resonances in S- (including $\kappa(800)$), P-, D-, and higher-wave 
states.  It will also allow a determination of phase shifts to 
account for final-state $K\pi$ interactions. Measurements of $K\pi$ 
form factors will be important input for Dalitz-plot analyses of 
$D$-meson and charmless $B$ mesons with $K\pi$ in final state. These 
will be important inputs for obtaining accurate an value of the 
CP-violating CKM matrix element $V_{us}$ and testing the unitarity 
relation, in particular through the measurement of the $\tau\to 
K\pi\nu_{\tau}$ decay rate.  

The approval and construction of the proposed facility at JLab will 
be {\it unique in the world}.  The high-intensity secondary beam of 
$K_L$ ($3\times 10^4~K_L$/s) would be produced in electromagnetic 
interactions using the high-intensity and high-duty-factor CEBAF 
electron beam with very low neutron contamination as was done at 
SLAC in the 1970s; but now, with three orders of magnitude higher 
intensity.  The possibility to perform similar studies with charged 
kaon beams is under discussion at J-PARC with intensities similar 
to those proposed for the $K_L$ beam at JLab. If these proposals 
are approved, the experimental data from J-PARC will be complementary 
to those of the proposed $K_L$ measurements.

Below in Table~\ref{tab:sum}, we present the expected statistics for 100 
days of running with a LH$_2$ target in the GlueX setup at JLab.  
The expected statistics for the 5 major reactions are very large. 
There are however, two words of cautions at this stage. These 
numbers correspond to an inclusive reaction reconstruction, which is enough 
to identify the resonance, but might not be enough to uncover its 
nature. The need for exclusive reconstruction to extract polarization 
observables further decrease the expected statistics, e.g., from 4M 
to \textbf{400k} events in the $K\Xi$ case. These statistics, however, 
would allow a precise measurement of the double-differential 
polarization observables with statistical uncertainties on the order 
of 5--10\%.  Secondly, kaon flux has a maximum around $W=3$~GeV, which 
decreases rapidly towards high/low $W$'s.  Thus, the 100 days of beam 
time on the LH$_2$ are essential to maximize the discovery potential of 
the $K_L$ Facility and cover the densely populated hyperon regime at 
low-$W$.
\begin{table}[htb!]

\centering \protect\caption{Expected statistics for differential 
	cross sections of different reactions with LH$_2$ and 
	below $W = 3.5$~GeV for 100~days of beam time.}
\vspace{2mm}
{%
\begin{tabular}{|cc|}
\hline
Reaction             & Statistics \\
                     & (events) \\
\hline
$K_Lp\to\ K_Sp$      &      8M \\
$K_Lp\to\pi^+\Lambda$&     24M \\
$K_Lp\to K^+\Xi^0$   &      4M \\
$K_Lp\to K^+n$       &    200M \\
$K_Lp\to K^-\pi^+p$  &     2M \\
\hline
\end{tabular}} \label{tab:sum}
\end{table}

There are no data on ``neutron" targets and, and for this reason, 
it is hard to make a realistic estimate of the statistics for $K_Ln$ 
reactions. If we assume similar statistics as on a proton target, 
the full program will be completed after running 100~days with 
LH$_2$ and 100~days with LD$_2$ targets.

\section{Cover Letter for KLF Proposal Submission to PAC45}
\label{sec:Cover}

This Proposal follows the Letter of Intent LoI12--15--001,
\textit{Physics Opportunities with Secondary K$_L$ beam at JLab}
presented to PAC43 in 2015.  The Issues and Recommendations 
included in the PAC43 Final Report document read as follow:

\textbf{Issues:} \textit{It is not clear what this experiment can
do that the J-PARC charged kaon program cannot do substantially
better.  An experimental concern is the transverse size of the KLF
beam that must impinge on a 2-3~cm target.  Backgrounds from
neutrons and $K_L$ outside the target acceptance may be important 
in event rates and signal to background rejection.}

\textbf{Recommendation:} \textit{Any proposal would require full
simulations of the beam line and detector to determine the effect 
of backgrounds from neutrons and kaons outside the target 
acceptance.  But it is not clear to the committee if this 
experiment would in any way be competitive with J-PARC or a 
potential Fermilab or CERN program in this energy range. The 
superiority of a neutral beam and/or the GlueX detector for these 
measurements would need to be demonstrated before a future proposal 
would be considered favorably.}

The KLF Collaboration believes that the current proposal addresses 
all the concerns following the recommendations expressed by the 
PAC43:

\begin{enumerate}
\item \textbf{Q1:} \textit{It is not clear what this experiment can
        do that the J-PARC charged kaon program cannot do
        \textit{substantially better}.} \\
        \textbf{A1:} The proposed $K_L$ beam intensity is similar to 
        the proposed charged kaon beam intensity at J-PARC, so there 
        is no reason to expect that J-PARC will do 
        \textit{substantially better}. Using different probes ($K_L$ 
        for JLab and $K^-$ for J-PARC), in principle, we and J-PARC 
        (if charged kaon beam proposal is approved) will be able to 
        collect data for different reactions. To have full experimental 
        information with different final states is important for 
        coupled-channel analyses to determine hyperon parameters. The 
        JLab and J-PARC measurements will be  complementary. \\
        (i) As $c\tau (K^-) = 3.7$~m, while $c\tau(K_L)=15.4$~m, the 
        higher rate of low-momenta kaons with a $K_L$ beam may be an 
        advantage. \\
        (ii) The proposed experiment will have a $K_L$ beam with all 
        momenta simultaneously, while J-PARC has to make many 
        thousand different settings to scan the full range of $W$ 
        distributions in different reactions. \\
        (iii) In the best-case scenario, J-PARC can  start a hyperon
        program in 2024. In Appendix~A6 (Sec.~\ref{sec:A6}), we have 
        presented the ability of other possible facilities as FNAL, 
        J-PARC, Belle, BaBar, $\overline{P}$ANDA, and COMPASS to do 
        hyperon spectroscopy. We do not see a competition factor here 
        for two reasons: 
        a) some of above-mentioned facilities do not yet  
        have secondary kaon beams; 
        b) even if kaon beams are approved and 
        constructed at these facilities, a hyperon spectroscopy program
        will not happen before a decade from now.
\item \textbf{Q2:} \textit{An experimental concern is the transverse 
	size of the KLF beam that must impinge on a 2-3~cm target.  
	Backgrounds from neutrons and KLF outside the target 
	acceptance may be important in event rates and signal to 
	background rejection.}\\
        \textbf{A2:} First of all the collimated beam of  $K_L$ will 
        impinge on the cell of the LH$_2$/LD$_2$ target with $R = 3$~cm 
        radius. All kaons outside of the solid angle defined by the 
        collimators will be absorbed in a 4~m iron shielding in the 
        sweeping magnet and the concrete shielding in front of the 
        GlueX setup. Second, as was shown by our detailed simulations, 
        the rate of neutrons on the GlueX target at momenta $p$ > 
        1~GeV/$c$ is smaller than that of $K_L$. On the other hand, 
        production of strange mesons with neutrons at low momenta 
        kinematically cannot occur due to the threshold, because to 
        conserve strangeness at least two final-state kaons have to 
        be produced. Therefore, the physics background from reactions 
        initiated by neutrons is negligible.

	The rate of neutrons irradiating GlueX setup outside of the 
	target acceptance will be a total on the level of $\sim$100/s 
	with 90$\%$ in the range of energies below 20~MeV, and, 
	therefore, can not cause any background either.        

	From a radiation point of view, our MCNP6 transport-code 
	calculations have shown that the effect of radiation caused 
	by neutrons is below the RadCon limit.
\item \textbf{Q3:} \textit{Any proposal would require full simulations
        of the beam line and detector to determine the effect of
        backgrounds from neutrons and kaons outside the target 
        acceptance.} \\
	\textbf{A3:} See our answers \textbf{A1} and \textbf{A2}.
\item \textbf{Q4:} \textit{But it is not clear to the committee if this
        experiment would in any way be competitive with J-PARC or a 
        potential Fermilab or CERN program in this energy range.} \\
	\textbf{A4:} See our answer \textbf{A1}.
\item \textbf{Q5:} \textit{The superiority of a neutral beam and/or the
        GlueX detector for these measurements would need to be
        demonstrated before a future proposal would be considered
	favorably.} \\
	\textbf{A5:} Our MC simulations have shown that the proposed
	experiment will be able to improve on available world proton 
	target data by three orders of magnitude in statistics.  The 
	proposed experiment will provide the first measurements ever 
	on a neutron using 
	LD$_2$ target. Coupled-channel analyses using both proton and 
	neutron target data promise to find many ``missing" hyperons. 
	We will also significantly improve world data on $K\pi$
	PWA with an impact on other fields of  particle physics.

	The summary of the potential of other facilities is given
	in Appendix~A6~(Sec.~\ref{sec:A6}).
\end{enumerate}

\section{Appendix~A1: Analysis of Three-Body Final States}
\label{sec:A1}

The understanding of baryon properties is hardly possible without
an analysis of reactions with two mesons in the final state. Already 
in the mass region above 1600~MeV, the excited $\Lambda$ hyperons decay
strongly into the $\pi\Sigma(1385)$~\cite{Prevost:1974hf,Cameron:1978en} 
final state while the $\Sigma$-hyperons decay strongly into the 
$\pi\Sigma(1385)$~\cite{Prevost:1974hf} and 
$\pi\Lambda(1405)$~\cite{Timmermans:1976gf} channels. Above 1800~MeV 
almost all known $\Lambda$ and $\Sigma$ hyperons have a dominant 
decay mode defined by production of the vector meson 
$K^\ast(892)$~\cite{Cameron:1978en}. In the $\Sigma$-sector, a number 
of resonances were seen in an analysis of the $\overline{K}\Delta(1230)$ 
final state. It is natural to expect the decay of $J^P=3/2^+$ states 
into the $\pi\Lambda(1520)$~\cite{Cameron:1977jr} channel.

Reactions with two-meson final states provide vital information for 
the analysis of single-meson production reactions.
The singularities that correspond to the opening of the
resonance-meson threshold (branching points) can produce structures
in other channels that can simulate a resonance-like 
signal~\cite{Ceci:2011ae,Burkert:2014wea}. The situation is 
notably more severe in the hyperon sector than in the sector of 
non-strange baryons. Due to the rather small widths of low-mass excited 
hyperons and meson resonances with an $s$-quark, such singularities are 
situated much closer to the physical region and can notably influence  
the data.  Therefore, a combined analysis of the channels with single 
and two mesons in the final state is a must in the search for the
missing resonances.

The combined analysis should help us to understand the structure of 
resonances with masses up to 2.5~GeV and their decay properties. One
of the important tasks is to find nonet partners of the nucleon
states observed in the photo-production reactions in the mass region
around 1900~MeV~\cite{Anisovich:2011fc}. These states have strong
couplings to the $\rho(770)N$ final state and it is natural to
expect that their hyperon partners can be found in an analysis of
the $K^\ast(892)N$ channel.

The analysis of the three-body final state should be done in the
framework of the event-by-event maximum likelihood method, which
allows us to take into account all amplitude correlations in the
multidimensional phase space. It is very important to extract the
polarization observables from the decay of the final hyperons in the
$KN\to\pi\pi\Lambda$ and $KN\to\pi\pi\Sigma$ reactions. One 
possible simplification is connected with an extraction of the
$K^\ast(892)N$ state from the $KN\to K\pi N$ data, where the analysis
can be performed in the framework of the density-matrix-elements
approach. However, the analysis should take into account the
rescattering of the particles in the final state; e.g., triangle 
diagrams that lead to logarithmic singularities in the scattering
amplitude. Due to the small widths of intermediate states, such
singularities can play a more important role than in the case of
nucleon and $\Delta$ excitations. It would be also very important 
to include in the analysis the CLAS photoproduction data with
$K\pi\Lambda$ and $K\pi\Sigma$ final states because there is a 
chance 
that states with a small $KN$ coupling could be observed in these 
reactions.

\section{Appendix~A2: Determination of Pole Positions} 
\label{sec:A2}

In spite of their model dependence, partial-wave Breit-Wigner
parameters have for quite some time been the
preferred connection between experiment and QCD in hadronic spectroscopy. 
More recently, however, pole parameters (e.g., pole positions and residues) 
have justifiably become the preferred connection, and this fact has also
been recognized by the Particle Data Group (PDG) in recent
editions of the {\it Review of Particle Physics}~\cite{PDG2016}. Therefore, 
the extraction of pole parameters from experimental 
data becomes a procedure of utmost importance.

Extraction of pole parameters is usually performed in two ways: (a) in an
energy-dependent way (ED) or (b) in an energy-independent procedure 
through single-energy PWAs (SE). In an ED procedure, one measures as 
many observables as possible to be close to the complete set and then fits
the observables with parameters of a well-founded theoretical model that
describes the reaction in question. Continuity in energy is enforced by 
the features of the theoretical model. In a SE procedure, one again 
measures as many observables as possible but attempts to extract partial 
waves by fitting energy-binned data independently, therefore, 
reducing the theoretical input. A discrete set of partial waves is obtained, 
and the issues of achieving continuity in energy have recently been 
extensively discussed either by introducing the constraints in 
analyticity~\cite{Osmanovic2017} or through angle- and energy-dependent 
phase ambiguity~\cite{Svarc2017}.

In energy-dependent models, pole parameters have been extracted in various
ways. The most natural way is the analytic continuation of theoretical
model solutions into the complex-energy plane. 
Simpler single-channel pole extraction methods have been developed
such as the speed plot~\cite{Hoehler93}, time delay~\cite{Kelkar}, the
N/D method~\cite{ChewMandelstam}, regularization procedures~\cite{Ceci2008},
and Pade approximants~\cite{Padde}, but their success has been limited. In
single-energy analyses the situation is even worse: until recently 
no adequate method has been available for the
extraction of pole parameters. All single-channel methods involve first- or 
higher-order derivatives, so partial-wave data had to be either interpolated 
or fitted with an unknown function, and that introduced additional and, very 
often, uncontrolled model dependencies.

That situation has recently been overcame when a new Laurent+Pietarinen (L+P)
method applicable to both, ED and SE models, has been
introduced~\cite{L+P2013,L+P2014,L+P2014a,L+P2015,L+P2016}. The driving
concept behind the single-channel (and later multichannel) L+P approach
was to replace solving an elaborate theoretical model and analytically
continuing its solution  into the full complex-energy plane, with a local
power-series representation of partial-wave amplitudes having well-defined
analytic properties on the real energy axis, and fitting it to the given
input. In such a way, the global complexity of a model is replaced by a 
much simpler model-independent expansion limited to the regions near the 
real-energy axis, which is sufficient to obtain poles and their residues.  
This procedure gives the simplest function with known analytic
structure that fits the data.  Formally, the introduced L+P method is
based on the Mittag-Leffler expansion\footnote{Mittag-Leffler
expansion~\cite{Mittag-Leffler} is the generalization of a Laurent
expansion to a more-than-one pole situation. For simplicity, we will
simply refer to this as a Laurent expansion.} of partial-wave amplitudes
near the real-energy axis, representing the regular, but unknown,
background term by a conformal-mapping-generated, rapidly converging
power series called a Pietarinen expansion\footnote{A conformal mapping
expansion of this particular type was introduced by Ciulli and
Fisher~\cite{Ciulli,CiulliFisher}, was described in detail and used in
pion-nucleon scattering by Esco Pietarinen~\cite{Pietarinen,Pietarinen1}.
The procedure was denoted as a Pietarinen expansion by H\"{o}hler
in~\cite{Hoehler84}.}. In practice, the regular
background part is usually fitted with three Pietarinen expansion series, 
each representing
the most general function having a branch point at $x_{bp}$, and 
all free parameters are then fitted to the chosen channel input. The
first Pietarinen expansion with branch-point $x_P$ is restricted to an
unphysical energy range and represents all left-hand cut contributions.
The next two Pietarinen expansions describe background in the physical
range with branch points $x_Q$ and $x_R$ defined by the analytic
properties of the analyzed partial wave. A second branch point is
usually fixed to the elastic channel branch point, and the third one
is either fixed to the dominant channel threshold value.
Thus, solely on the basis of general physical assumptions about the analytic
properties of the fitted process (number of poles and number and position
of conformal mapping branch-points) the pole parameters in the complex
energy plane are obtained. In such a way, the simplest analytic function
with a set of poles and branch points that fits the input is
constructed. This method is equally applicable to both
theoretical and experimental input\footnote{Observe that fitting partial-wave 
data coming from experiment is even more favorable.}

The transition amplitude of the multichannel L+P model is parametrized as
\begin{eqnarray} \label{Eq:MCL+P}
        T^a(W)&=&\hspace{-1mm}\sum _{j=1}^{{N}_{pole}} \frac{g^{a}_{j} }
        {W_j-W} \hspace{-0.5mm}
        +\hspace{-0.5mm}\sum_{i=1}^3\sum_{k_i=0}^{K_i^{a}} c^{a}_{k_i}
        \left(\frac{\alpha^a_i\hspace{-1mm}-\hspace{-1mm}\sqrt{x^a_i-W}}
        {\alpha^a_i\hspace{-1mm}+\hspace{-1mm}\sqrt{x^a_i - W }}
        \right)^{k_i}, \nonumber\\[-1ex]
\end{eqnarray}
where $a$ is a channel index, $W_j$ are pole positions in the complex
$W$ (energy) plane, $g^{a}_i$ coupling constants. The $x^{a}_i$ define
the branch points, $c^{a}_{k_i}$, and $\alpha^{a}_i$ are real coefficients.
$K^a_i, \,i=1,2,3$\, are Pietarinen coefficients in channel $a$. The first
part represents the poles and the second term three branch points. The
first branch point is chosen at a negative energy (determined by the fit),
the second is fixed at the dominant production threshold, and the third
branch point is adjusted to the analytic properties of fitted partial wave.

To enable the fitting, a reduced discrepancy function $D_{dp}$ is defined as
\begin{eqnarray} \label{eq:Laurent-Pietarinen}
        D_{dp} &=&\sum _{a}^{all}D_{dp}^a;\qquad  \nonumber \\
        D_{dp}^a &=&  \frac{1}{2 \, N_{W}^a - N_{par}^a} \times
        \sum_{i=1}^{N_{W}^a}
        \left\{ \left[ \frac{{\rm Re} \,T^{a}(W^{(i)})-{\rm Re} \,
        T^{a,exp}(W^{(i)})}{ Err_{i,a}^{\rm Re}}  \right]^2 \right.\nonumber\\
        &&\left.
        +\quad\ \,\left[ \frac{{\rm Im} \, T^{a}(W^{(i)})-{\rm Im} \,
        T^{a,exp}(W^{(i)})}{ Err_{i,a}^{\rm Im}} \right]^2 \right\}
        + {\cal P}^a , \nonumber
\end{eqnarray}
where
\begin{eqnarray}
        {\cal P}^{a} &=& \lambda^a_{k_1} \sum _{k_1=1}^{K^a} (c^a_{k_1})^2 \,
        {k_1}^3 +  \lambda_{k_2}^a \sum _{k_2=1}^{L^a} (c^a_{k_2})^2 \,
        {k_2}^3 + \nonumber
        + \lambda_{k_3}^a \sum _{m=1}^{M^a} (c^a_{k_3})^2 \, {k_3}^3 \nonumber
\end{eqnarray}
is the Pietarinen penalty function, which ensures fast and optimal convergence.
$N_{W}^a$ is the number of energies in channel $a$, $N_{par}^a$ the number
of fit parameters in channel $a$, $\lambda_c^a, \lambda_d^a, \lambda_e^a$
are Pietarinen weighting factors, $Err_{i,a}^{\rm Re, \, Im} \ldots$  errors
of the real and imaginary part, and $c_{k_1}^a, c_{k_2}^a, c_{k_3}^a$ real
coupling constants.
\begin{center}
\begin{figure}[!h]
\centering
{
    \includegraphics[width=1\textwidth,keepaspectratio]{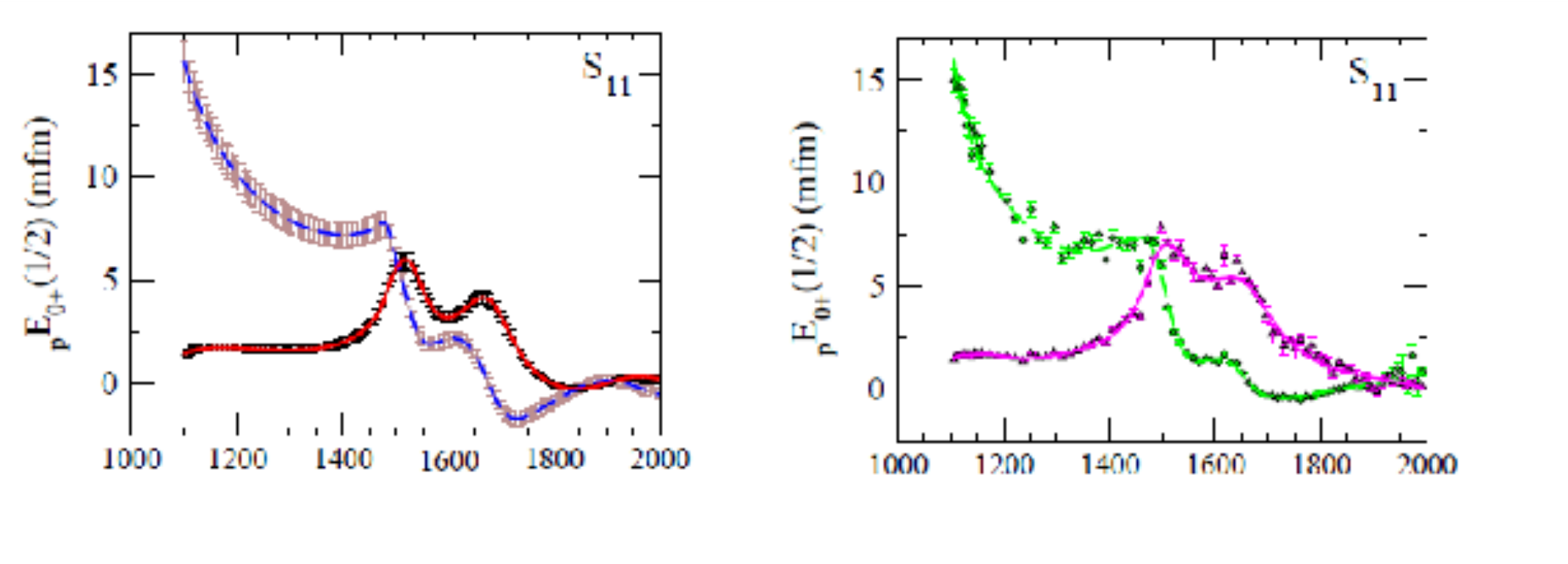} }

    \caption{L+P fit to CM12 GWU/SAID pion photoproduction $_p E_{0+}$
       ED and SE solutions~\protect\cite{Workman:2012jf}.} 
	\label{Fig:L+PCM12}
\end{figure}
\end{center}

In order to obtain reliable answers in the L+P model we have to build 
knowledge
about the analytic structure of the fitted partial wave into the fitting
procedure. Because we are looking for poles, we only have to define which
branch-points to include. Their analytic form will be determined by the
number of Pietarinen coefficients.  As we have only three branch-points at
our disposal we expect that the first branch-point will describe all
subthreshold and left-hand cut processes, the second one is usually fixed to
the dominant channel opening, and the third one is to represent
background contributions of all channel openings in the physical range.
So, in addition to choosing the number of relevant poles, our anticipation
of the analytic structure of the observed partial wave is of great
importance for the stability of the fit.

The L+P model has been successfully applied to both theoretical models and
discreet partial-wave data. As an example, in Fig.~\ref{Fig:L+PCM12}, we
give the achieved quality of the fit for the CM12 GWU/SAID pion
photoproduction amplitudes~\cite{Workman:2012jf}.

\underline{In summary:} Methods of the described L+P model will be used
to extract pole parameters for both ED solutions, obtained by
the method described in Section~\ref{sec:PWA}, and SE solutions
developed independently.

\section{Appendix~A3: Statistics Tools for Spectroscopy of Strange 
	Resonances} 
\label{sec:A3}

Several statistical aspects concerning the analysis of $K_L$ data are
discussed in the following. The proposed experiment will be capable
of producing a large body of consistent data, which is a prerequisite
to carry out statistical analyses. So far, the data in the strangeness
$S=-1$ sector were produced in many different experiments, often from
the 1980s or earlier, with different systematic uncertainties that are,
moreover, unknown in many cases. The problems resemble the situation
in pion-induced inelastic reactions~\cite{Ronchen:2012eg,
Doring:2010ap}. This makes any kind of analysis difficult but
statistical tests, e.g., on the significance of a claimed resonance
signal, are indispensable to carry out meaningful baryon
spectroscopy. Indeed, the search for {\it missing resonances} is not
only a problem of implementing physical principles such as unitarity
in the amplitude but also, to a large extent, a statistical one. This
becomes especially relevant once one searches for states beyond the
most prominent resonances.

\subsection{Minimizing Resonance Content}

Partial-wave analysis, discussed in Section~\ref{sec:PWA}
is needed to extract the physically relevant information from data.
For resonance spectroscopy, one needs the energy dependence of
the amplitude to determine resonance positions and widths. Therefore,
energy-dependent (ED) parametrizations of the partial waves are
fitted either to data or to single-energy (SE) solutions, generated
by conducting partial-wave analysis in narrow energy bins. The
resonance content is usually determined by speed-plot techniques or
analytic continuation of the ED parametrization to complex
scattering energies, where resonances manifest themselves as
poles~\cite{Doring:2009yv}.

Yet, the ED parametrization itself contains, almost always, resonance
plus background terms in one implementation or another. A problem
arises if resonance terms are needed to model missing background
dynamics. Then, false positive resonance signals could be
obtained~\cite{Ceci:2011ae}. Adding resonance terms will always lower
the $\chi^2$ in a given fit, but the question is how significant this
change is.

We plan to address this well-known, yet poorly addressed problem by
applying several statistical analysis tools to the amplitude
parametrization. Some techniques have been used, so far, to address
this problem. For example, in so-called mass scans, the $\chi^2$
dependence on the mass of an additional resonance is
studied~\cite{Dick2003,Anisovich:2011sv}. If $\chi^2$ drops by a 
certain amount at a given energy, potentially in several
reaction channels at once, then a resonance might be responsible.

Beyond mass scans, there exist {\it model selection} techniques
referring to the process of selecting the simplest model with the
most conventional explanation. Here, the conventional/simple
explanation is an (energy-dependent) background and/or threshold
cusps, while the algorithm should penalize unconventional
explanations such as resonances.

Minimizing the resonance content in a systematic way is thus a goal
within partial-wave analysis. For this, the Least Absolute Shrinkage
and Selection Operator (LASSO) technique for model
selection can be applied (which provides a Bayesian posterior-mode
estimate), in combination with cross validation and/or information
theory to control the size of the penalty parameter
$\lambda$~\cite{Tib0,Tib1,Tib2}. The combination of these techniques
effectively suppresses the emergence of resonances except for those
really needed by the data. The numerical implementation is especially
simple because it affects only the calculation of the $\chi^2$.
Trial-and-error techniques, sometimes still applied to check for
resonances in different partial
waves, will become obsolete. Here, one simply starts with an
over-complete resonance set plus flexible backgrounds, and the
algorithm will remove all those resonances not needed by data,
without manual intervention. Apart from cross validation, we will
also consider information theory to regulate $\lambda$ as proposed
in Ref.~\cite{Guegan:2015mea}. In particular, the Akaike and
Bayesian information criteria provide easy-to-use model selection.
Results should be independent of the choice of the criterion.

In 2017, the LASSO technique was, for the first time, used in pion
photoproduction at low energies for the ``blindfolded" selection of
the relevant multipoles and their simplest parametrization to
describe the available data~\cite{Landay:2016cjw}. The analysis of
kaon-induced reactions is closely related. For a recent application
in a different but related context see Ref.~\cite{opti}. Once the
model selection process is finished, uncertainties on resonance
parameters can be obtained by the usual re-sampling techniques.

The existing and proposed partial-wave analysis tools use different
construction principles: resonances are included in the form of bare
states, $K$-matrix poles, or generated from hadron dynamics itself
For the first two classes of approaches, one has at one's disposal
the coupling constants that tune the interaction of a bare singularity
with the meson-baryon continuum. Those are fit parameters that can
be explicitly included in the penalty term. If resonances are
generated from the meson-baryon dynamics itself, the case is a bit
more complicated, because there are no directly accessible tuning
parameters. This parametrization, practiced by the GW/SAID group
for many years (see, e.g., Ref.~\cite{Workman:2012jf}), is, in
principle, the cleanest analysis tool, because resonance generation
does not require manual intervention. Yet, even here the emergence
of resonance terms can be penalized, e.g., through the value of
contour integrals on the second Riemann sheet where resonance poles
are located (a value of zero corresponds then to the absence of
poles).

It should be stressed that the information theory criteria do {\it not}
require a good fit in a frequentist's sense because they merely compare
the relative quality of models. This is especially relevant when it 
comes to the analysis of many different data sets (such as kaon-induced
reactions) in which, e.g., the systematic errors might be underestimated
such that a $\chi^2/{\rm d.o.f.}\approx 1$ is difficult to achieve.

Systematic uncertainties can be treated as in the GW/SAID
approach~\cite{piN} in which the $\chi^2$ is defined as
\begin{equation}
        \chi^2 = \sum_i \left( {{N \Theta_i - \Theta_i^{\rm exp} }\over
        {\epsilon_i}} \right)^2 + \left( {{N-1}\over{\epsilon_N}}
        \right)^2 ,
        \label{chi0}
\end{equation}
where $\Theta_i^{\rm exp}$ is an experimental point in an angular
distribution and $\Theta_i$ is the fit value. Here the overall systematic
error, $\epsilon_N$, is used to weight an additional $\chi^2$ penalty
term due to renormalizaton of the fit by the factor $N$. The statistical
error is given by $\epsilon_i$. Note that the fit function is penalized,
rather than the data, to avoid the bias discussed in
Ref.~\cite{D'Agostini:1993uj}. See also Ref.~\cite{Ball:2009qv} for 
further discussion of this topic.

\subsection{Goodness-of-Fit Tests}

The $\chi^2$ per degree of freedom, $\chi^2_{d.o.f.}$, is usually
considered as a criterion for a good fit, but becomes meaningless if
thousands of data points are fitted (and should be replaced by
Pearson's $\chi^2$ test). Statistical $\chi^2$ tests will
become possible through the new data, putting resonance analysis on a
firmer ground. While $\chi^2$ tests are sensitive to under-fitting, they
are insensitive to over-fitting. Here, the $F$-test~\cite{FTest} is
suitable to test the significance of new fit parameters. That test, can,
thus, be applied to reduce the number of internal parameters in a
partial-wave parametrization, which results in more reliable estimates 
of uncertainties for extracted resonance parameters such as masses, 
widths, and branching ratios.

With increased consistency of data through the KLF experiment,
other goodness-of-fit criteria can also be applied, such as 
Smirnov-Kolmogorov
or Anderson-Darling tests for normality~\cite{AD1952,Stephens1974} or
run tests from non-parametric statistics. For pion photoproduction,
these tests are applied and extensively discussed in
Ref.~\cite{Landay:2016cjw}.

A prerequisite to carry out classical statistical tests is data
consistency. As discussed before, this is unfortunately not always the
case in the $S=-1$ sector. The proposed KLF measurements will produce, 
for the first time, a body of data large enough to enable such tests
reliably.

\subsection{Representation of Results}

As mentioned, ED parametrizations are needed to extract resonance parameters,
but single-energy (SE) fits are useful to search for narrow structures, or 
for other groups to test theoretical models of hadron dynamics. The question 
arises how the partial waves can be presented to allow the theory community 
to carry out their fits. As recently demonstrated~\cite{cova}, SE solutions 
alone carry incomplete statistical information, mainly because they
are correlated quantities. We plan to provide the analysis results in a similar
form as recently done in Ref.~\cite{cova} for elastic $\pi N$ scattering. With
this, the theory community can fit partial waves through so-called {\it
correlated $\chi^2$ fits} obtaining a $\chi^2$ close to the one obtained in a
fit directly to data (see Ref.~\cite{cova} for an extended discussion).  This
format ensures that the maximal information from experiment is transmitted to
theory, allowing to address the {\it missing resonance problem} in the wider
context of questions related to confinement and mass generation, that have been
paramount problems in hadronic physics for decades.

\underline{In summary:} With a large consistent data set from the KLF
experiment, an entire class of statistical tools will become applicable
that is needed to conduct rigorous baryon spectroscopy. With the new data,
the quantitative significance of resonance signals and the quantitative
uncertainties of resonance parameters can be determined.

%
%

\section{Appendix~A4: Neutron Background}
\label{sec:A4}

%
%
%

\begin{figure}[h!]
\centering
{
    \includegraphics[width=0.45\textwidth,keepaspectratio]{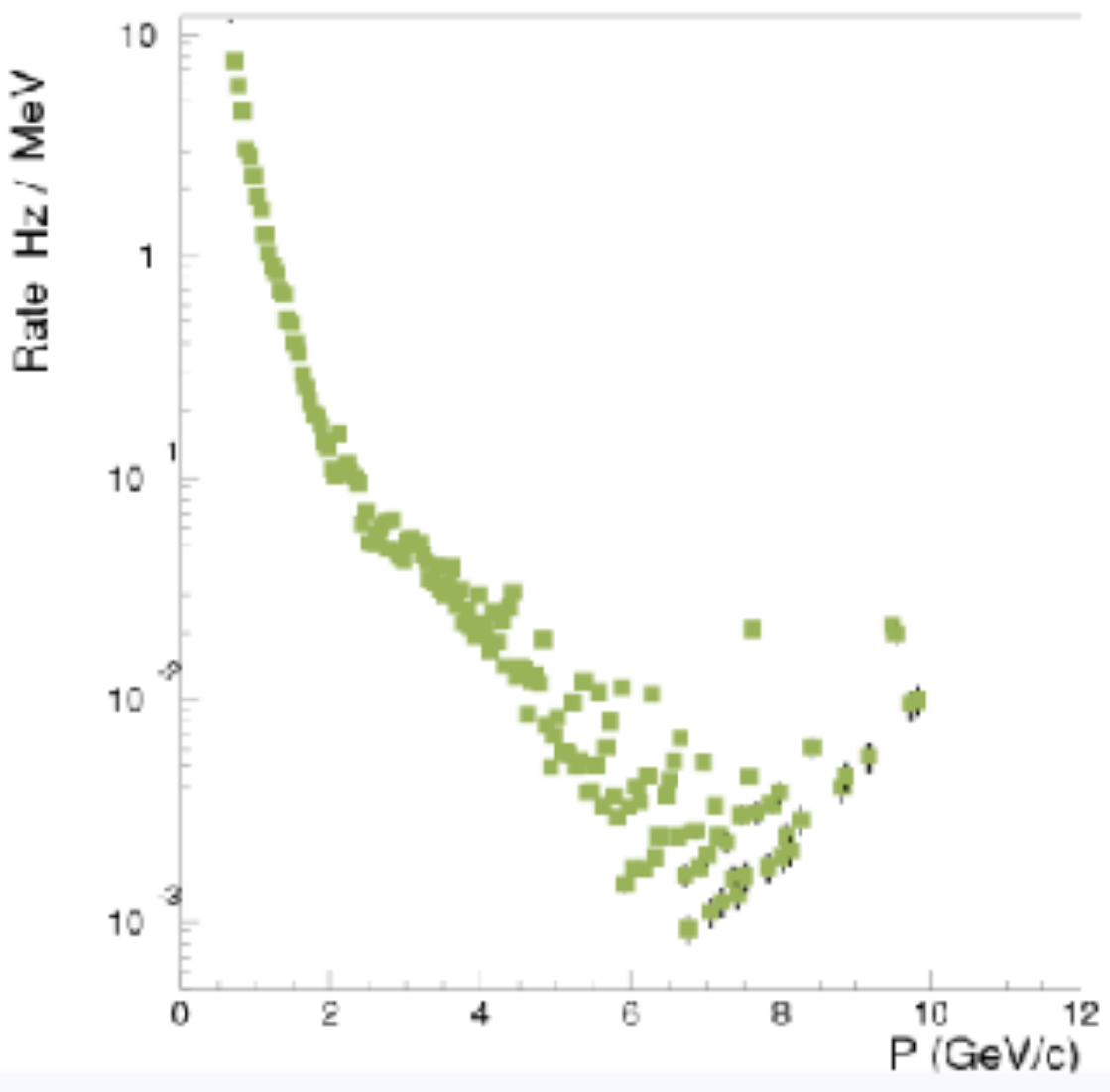} }

          \caption{The $n$ to $K_L$ ratio associated with
          Fig.~\protect\ref{fig:neutron}(left).}
          \label{fig:neutron5}
\end{figure}

\underline{Overall}, our MC simulations for 12~GeV
(Fig.\ref{fig:neutron}(left)) agreed quite well with the neutron yields
measurements that SLAC did for 16~GeV (Fig.~\ref{fig:neutron}(right)).
Note that with a proton beam, the $n/K_L$ ratio is $10^3-10^4$ (see,
for instance, Table~\ref{tab:projectx} in Appendix~A6~(Sec.~\ref{sec:A6}),
while in the JLab case, this ratio is less than 10, as Fig.~\ref{fig:neutron5}
shows.

\begin{figure}[h!]
\centering
{
    \includegraphics[width=0.8\textwidth,keepaspectratio]{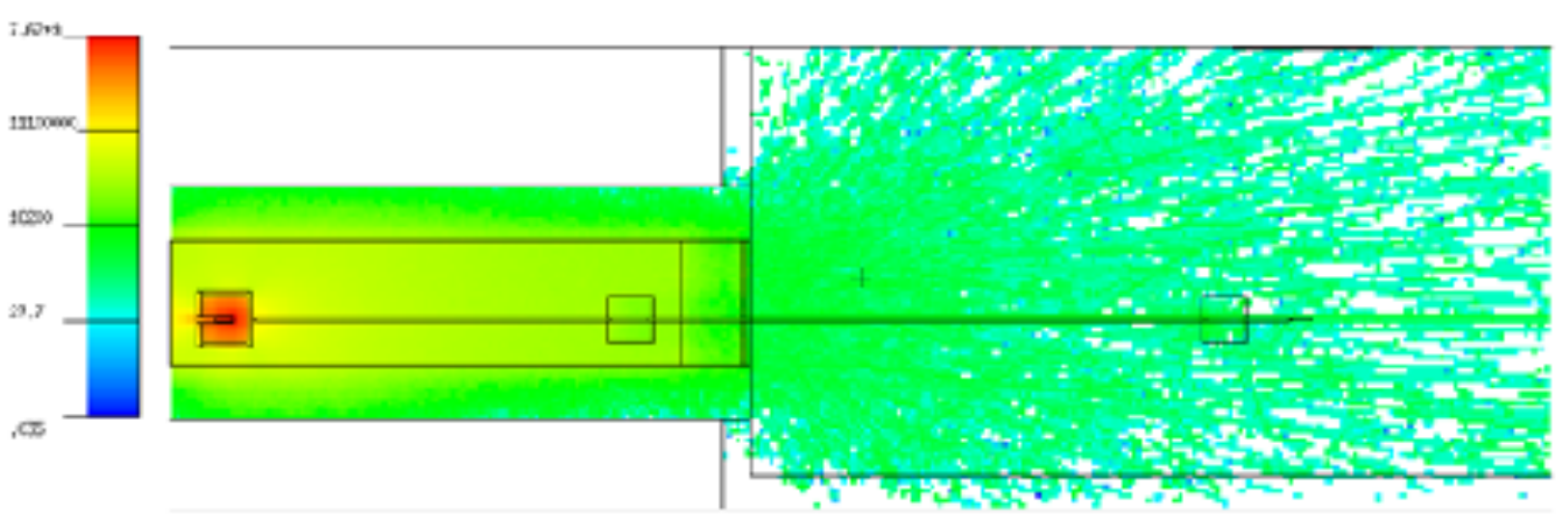} }

    \caption{Vertical cross section of the neutron flux calculated for
    the model where the shielding wall is located at the end of the
    beam cave.} \label{fig:HallD1}
\end{figure}

\begin{figure}[h!]
\centering
{
    \includegraphics[width=0.8\textwidth,keepaspectratio]{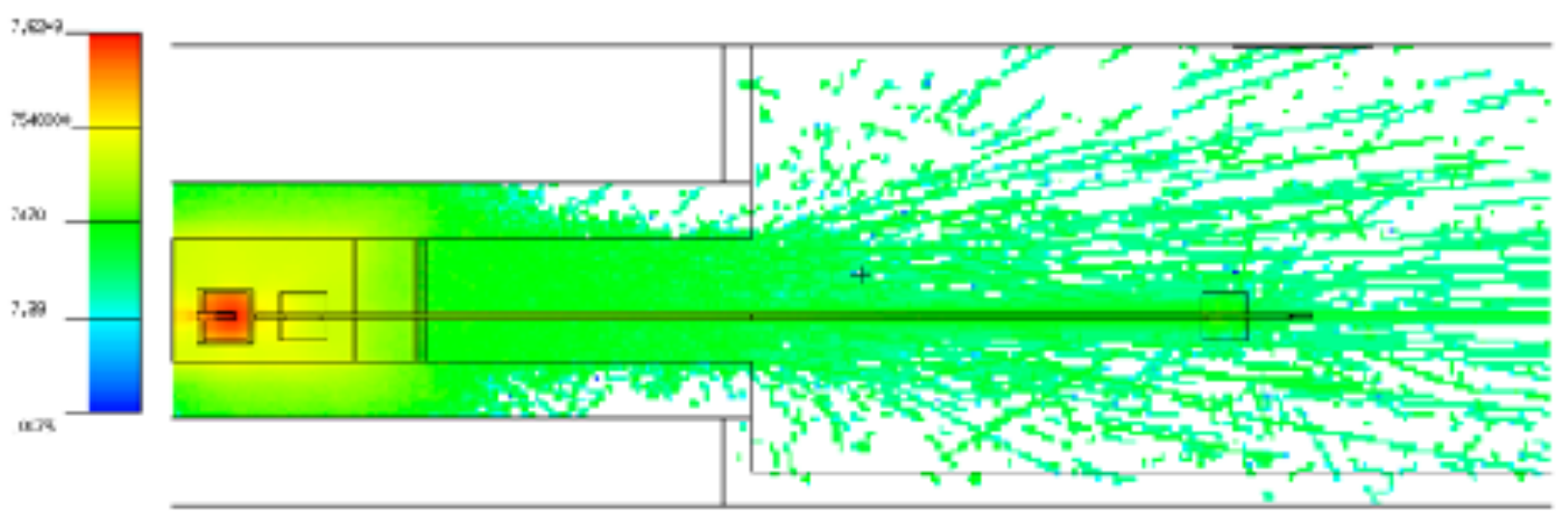} }

     \caption{Vertical cross section of the neutron flux calculated for
     the model where the shielding wall is located as close as possible
     to the first collimator in the beam cave.} \label{fig:HallD4}
\end{figure}

\begin{figure}[h!]
\centering
{
    \includegraphics[width=0.8\textwidth,keepaspectratio]{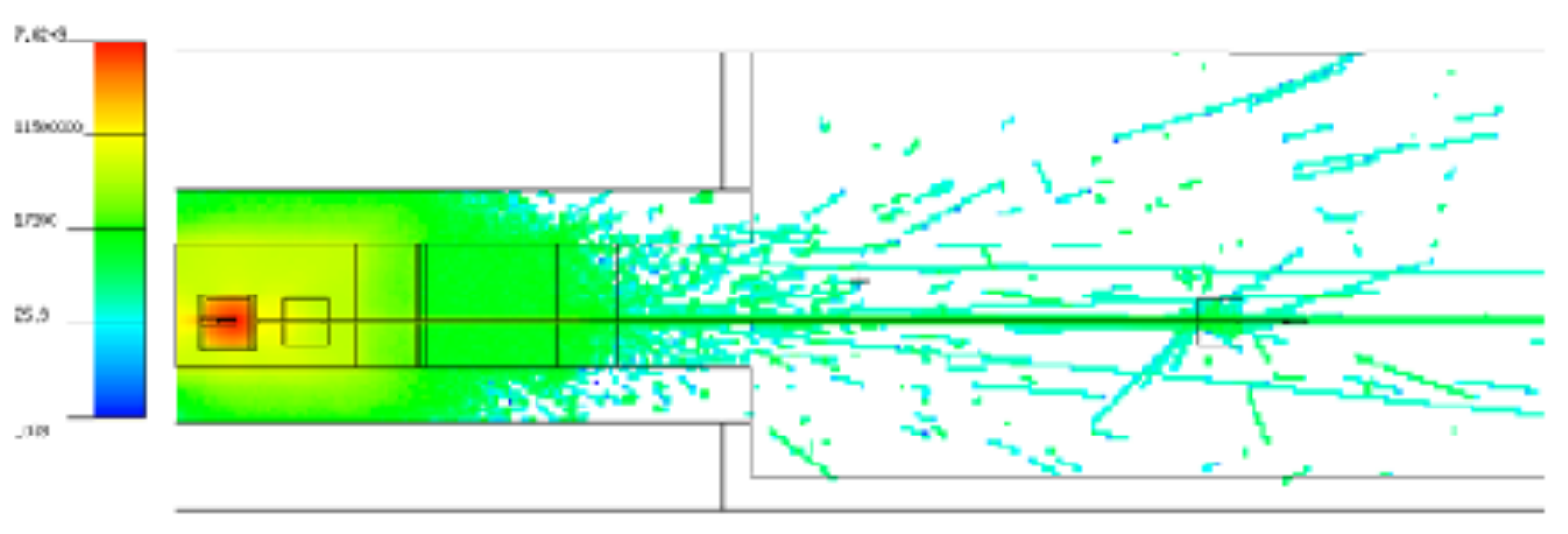} }

    \caption{Vertical cross section of the neutron flux calculated for
    the model where the first shielding wall is located as close as
    possible to the first collimator and Be-target and the second
    shielding wall is located at halfway toward to the end of the
    beam cave.} \label{fig:HallD2}
\end{figure}

For MCNP6 calculations, several neutron tallies were placed along the 
beam and at the experimental hall ceiling for neutron fluence estimation.
Calculations were performed for different shielding modifications
in the beam cave to optimize the neutron dose. Figures~\ref{fig:HallD1},
\ref{fig:HallD4}, and \ref{fig:HallD2} present the vertical
cross section of the neutron flux evolved from beginning to final
configurations considered in the course of this study. Neutron flux
in experimental hall was evaluated for several shielding
configurations in the beam cave. First, the shielding wall was
located at the end of the beam cave; see Fig.\ref{fig:HallD1}. Second,
the shielding wall is located as close as possible to the first
collimator in the beam cave, Fig.~\ref{fig:HallD4}. Also, two labyrinths 
were introduced into the MCNP model, Fig.~\ref{fig:HallD4}, to check 
their influence on anticipated dose rate. No influence was observed 
for proper designed labyrinths with no direct view from experimental 
hall to the source target in the beam cave. Third, the same configuration 
as in Fig.~\ref{fig:HallD4}, but second shielding wall is added at half 
way toward to the end of the beam cave, Fig.\ref{fig:HallD2}
\begin{figure}[h!]
\centering
{
    \includegraphics[width=0.7\textwidth,keepaspectratio]{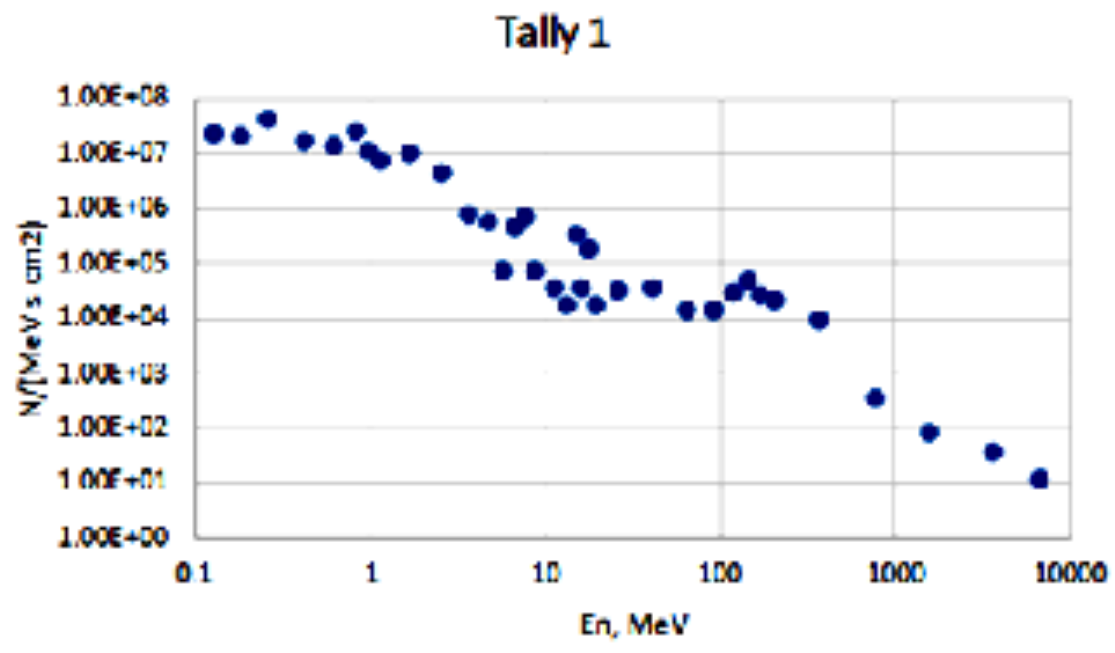} }

    \caption{Energy distribution of neutrons (in N/(MeV$\cdot$s$\cdot$ 
    cm$^2$) units emitted from the Be~target, from calculations with 
    the MCNP6 transport code~\protect\cite{MCNP}.}
    \label{fig:HallD3}
\end{figure}

The energy distribution of neutrons emitted from the Be~target (in 
N/(MeV$\cdot$ s$\cdot$ cm$^2$) units is shown in Fig.~\ref{fig:HallD3}. 

\section{Appendix~A5: Details of Monte Carlo Study}
\label{sec:A5}

\subsection{Particle Identification}
\label{sec:App_PID}

\begin{figure}[h!]
\centering
{
    \includegraphics[width=0.45\textwidth,keepaspectratio]{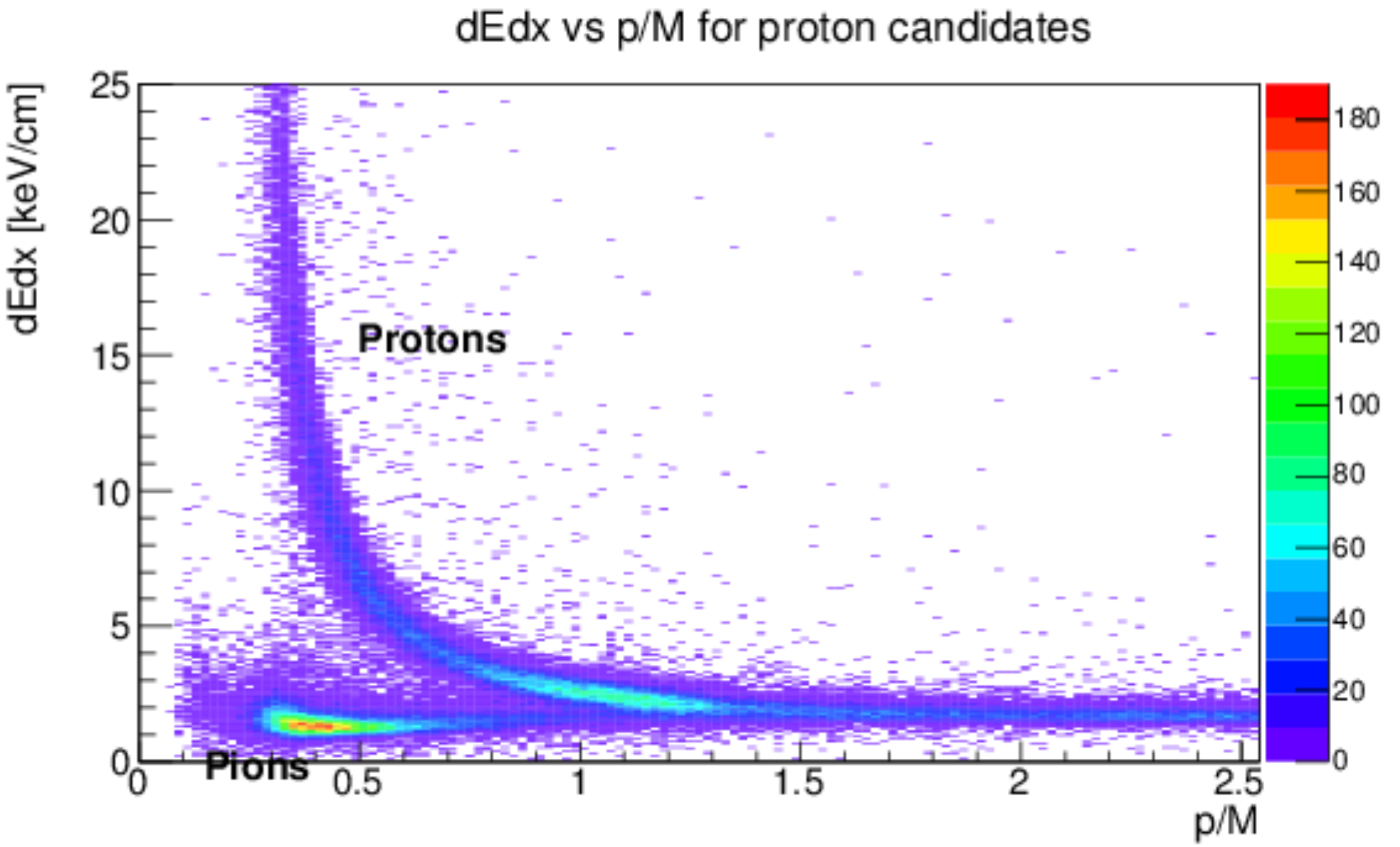} }
{
    \includegraphics[width=0.45\textwidth,keepaspectratio]{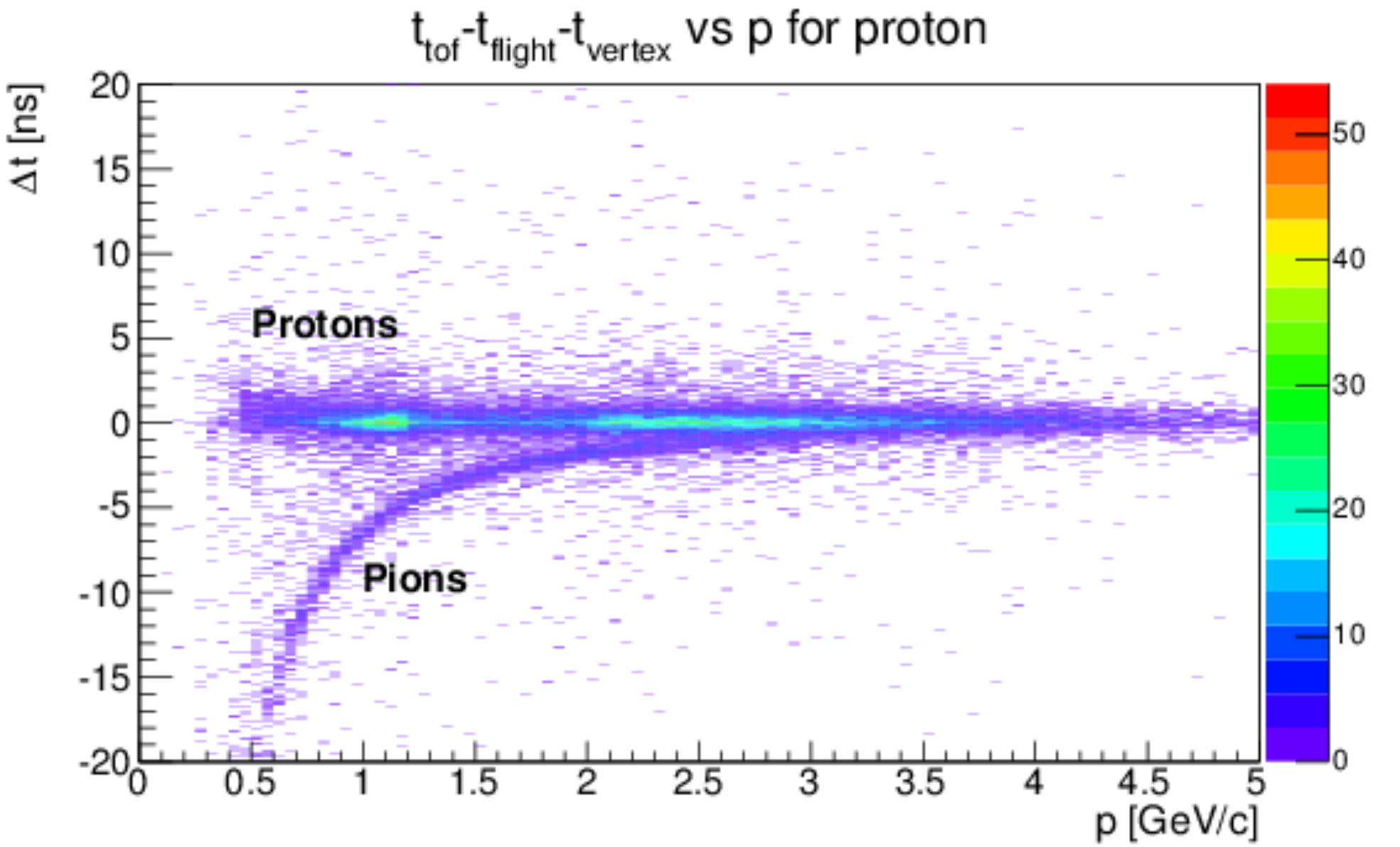} }
{
    \includegraphics[width=0.45\textwidth,keepaspectratio]{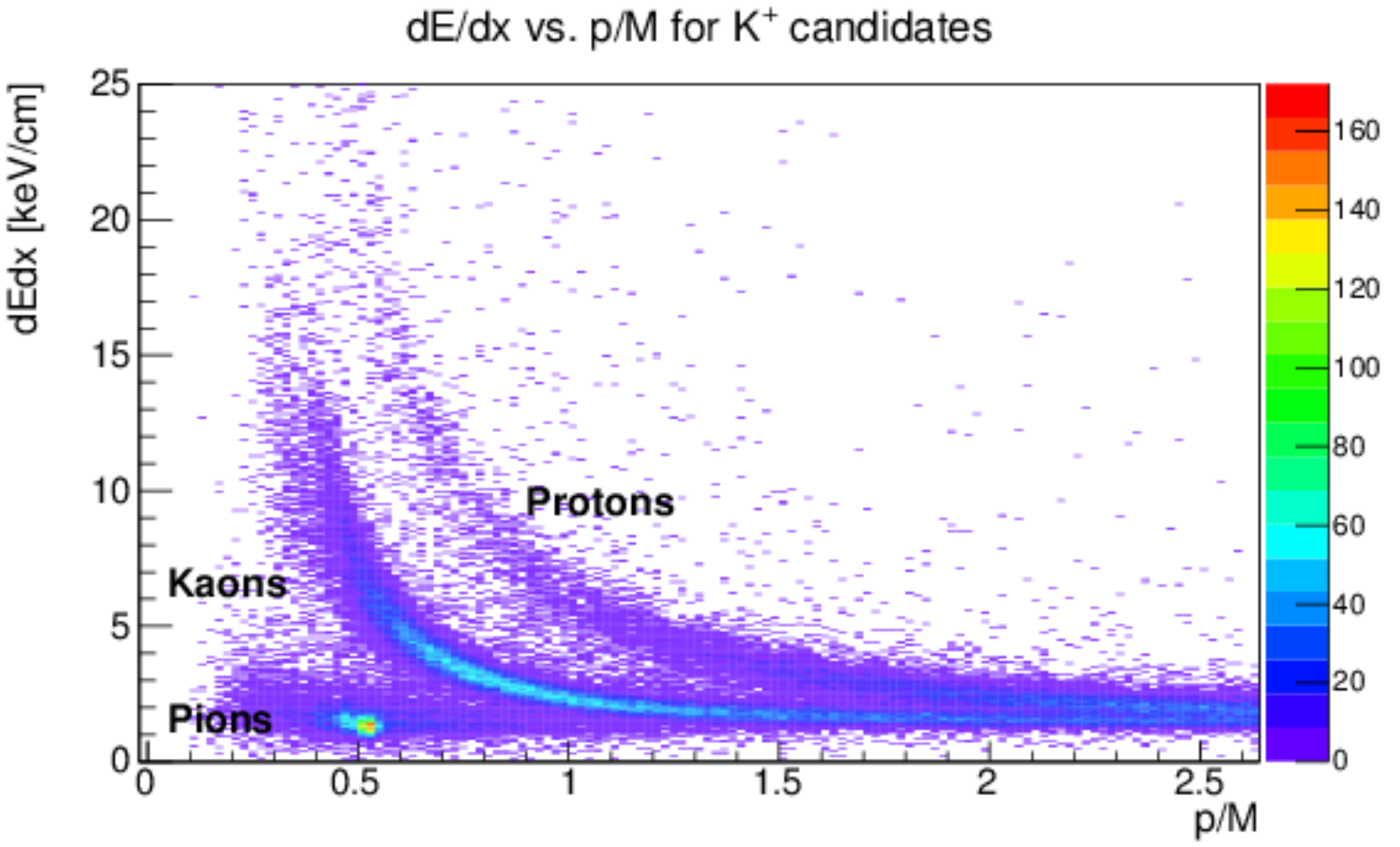} }

    \caption{Particle identification. Top left: $dE/dx$ for the $K_S p$ 
    channel.
    Top right: time difference at the primary ``vertex" for the proton 
    hypothesis for the $K_S p$ channel using the TOF. 
    Bottom plot: $dE/dx$ for the $K^+\Xi$ channel. The proton and pion 
    bands arise from the decay of the $\Lambda$.} \label{fig:A31}
\end{figure}

For each topology, one primary particle (the proton for the $K_S p$ channel,
the $\pi K^+$ for the $\pi^+\Lambda$ channel and the $K^+$ for the $K^+\Xi$
and $K^+n$ channels) provides a rough determination for the position of the
primary vertex along the beam line that is used in conjunction with the 
start counter to determine the flight time and path of the $K_L$ from the 
beryllium target to the hydrogen target. Protons, pions, and kaons are 
distinguished using a combination of $dE/dx$ in the chambers and 
time-of-flight to the outer detectors (BCAL and TOF). The energy loss and 
timing distributions for the $K_S p$ channel are shown in Fig.~\ref{fig:A31}; 
the distributions are similar for the $\pi^+\Lambda$ channel, where a 
proton band arises from the $\Lambda \to \pi^- p$ decay. Also shown 
is the $dE/dx$ distribution for the $K^+\Xi^0$ channel, where a prominent 
kaon band can be seen, along with pion and proton bands arising from 
$\Lambda$ decays.

\begin{figure}[h!]
\centering
{
    \includegraphics[width=0.45\textwidth,keepaspectratio]{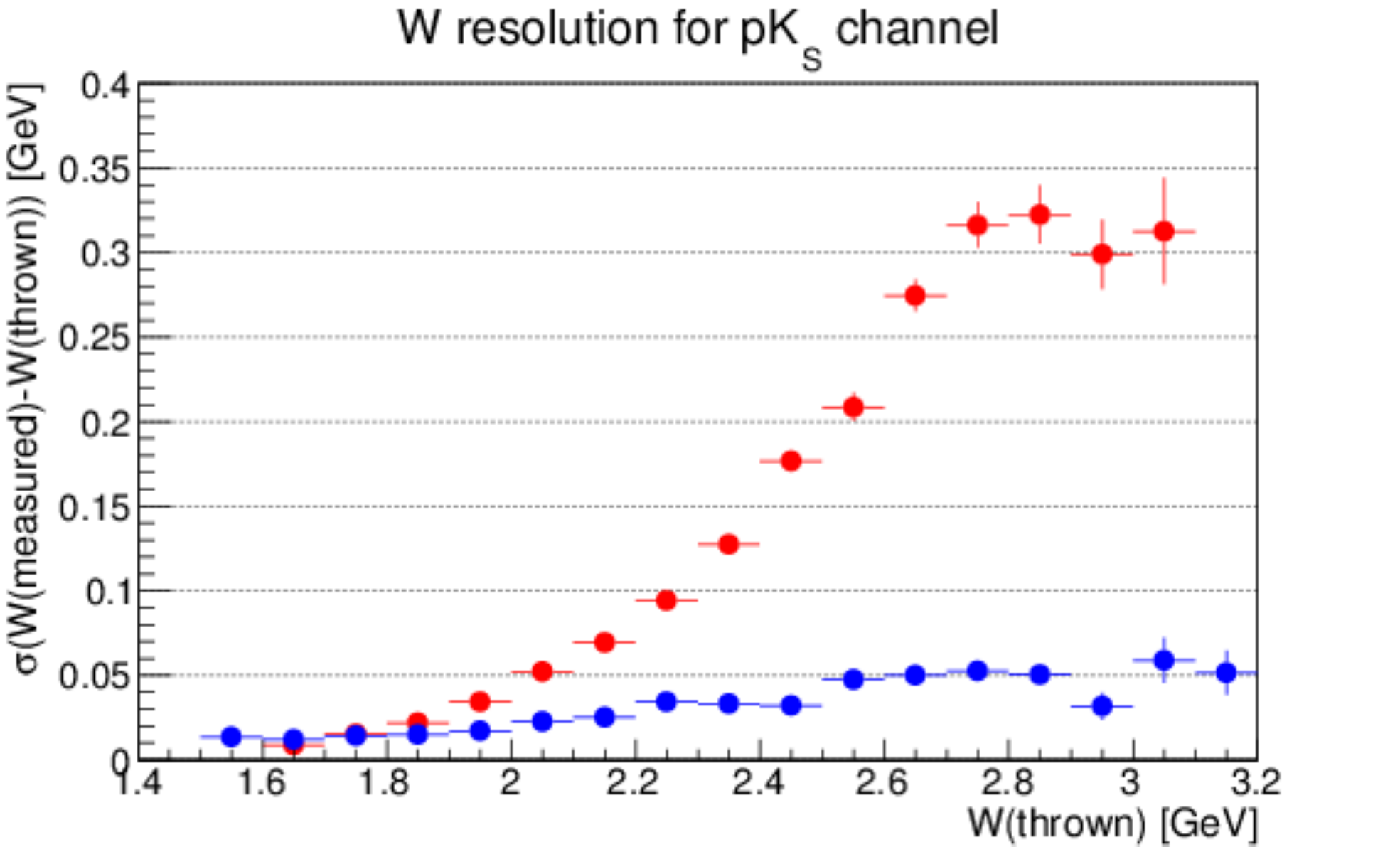} }
    
    \caption{$W$ resolution for the $K_S p$ channel,
    (blue dots) using kinematic fitting after reconstruction 
    of all final state particles; (red dots) using $K_L$ 
    time-of-flight.} \label{fig:A32}
\end{figure}

Since the GlueX detector has full acceptance in $\phi$ for charged
particles and large acceptance in $\theta$ (roughly $1-140^\circ$),
a full reconstruction of events is feasible for the majority of the
channels. That will allow to apply four or more overconstrain 
kinematical fit and improve the resolution considerably.  A typical 
comparison between $W$ reconstruction using the $K_L$ momentum for 
300~ps ST resolution (red dots) and the other using kinematically 
fitted final-state particles for the $K_Sp$ channel (blue dots) is
shown in Fig.~\ref{fig:A32}.
\subsubsection{Details of MC study for $K_Lp\to K_Sp$} 
\label{sec:App_PID_KS}
        
\begin{figure}[h!]
\centering
{
    \includegraphics[width=0.45\textwidth,keepaspectratio]{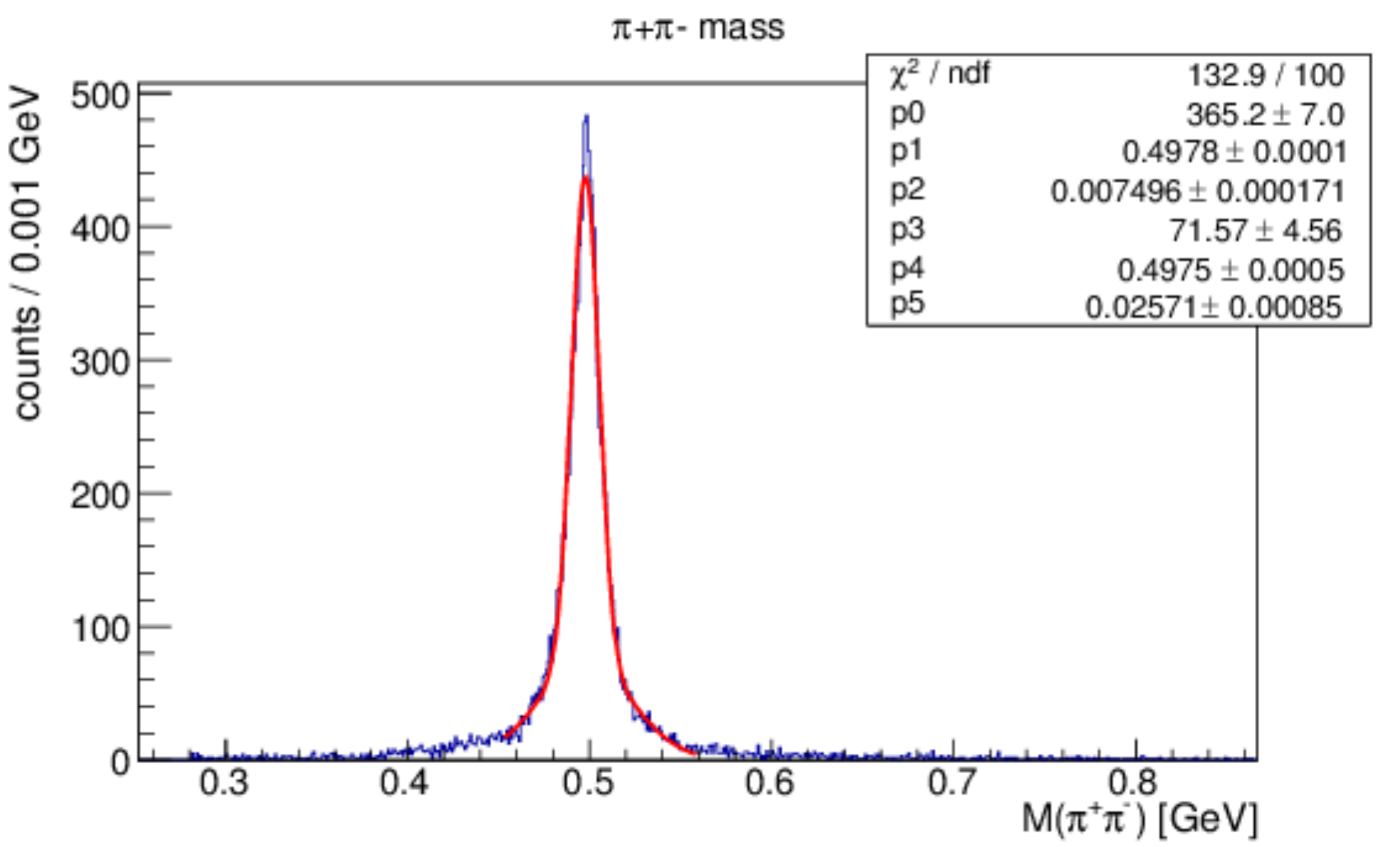} }
{
    \includegraphics[width=0.45\textwidth,keepaspectratio]{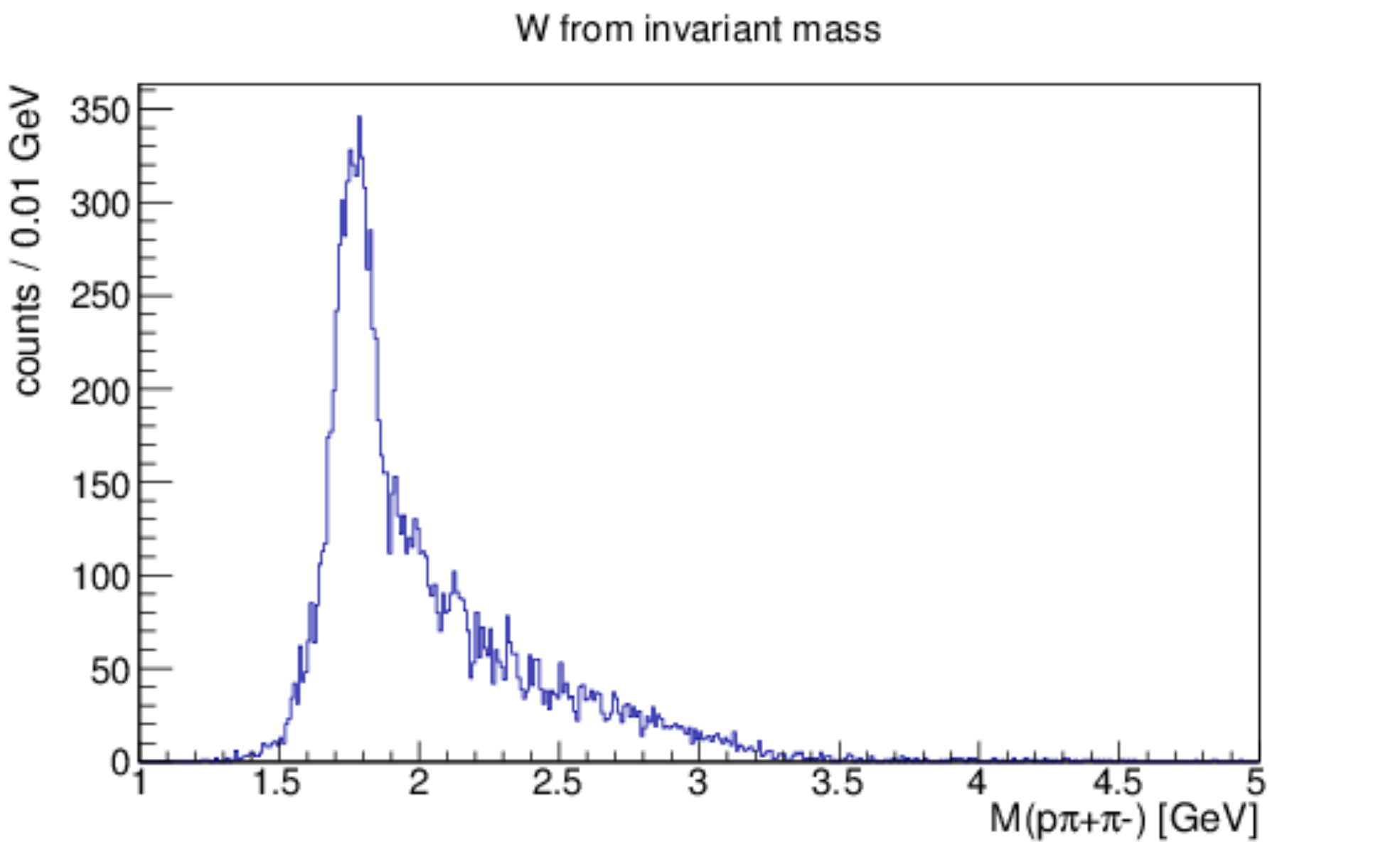} }
{
    \includegraphics[width=0.45\textwidth,keepaspectratio]{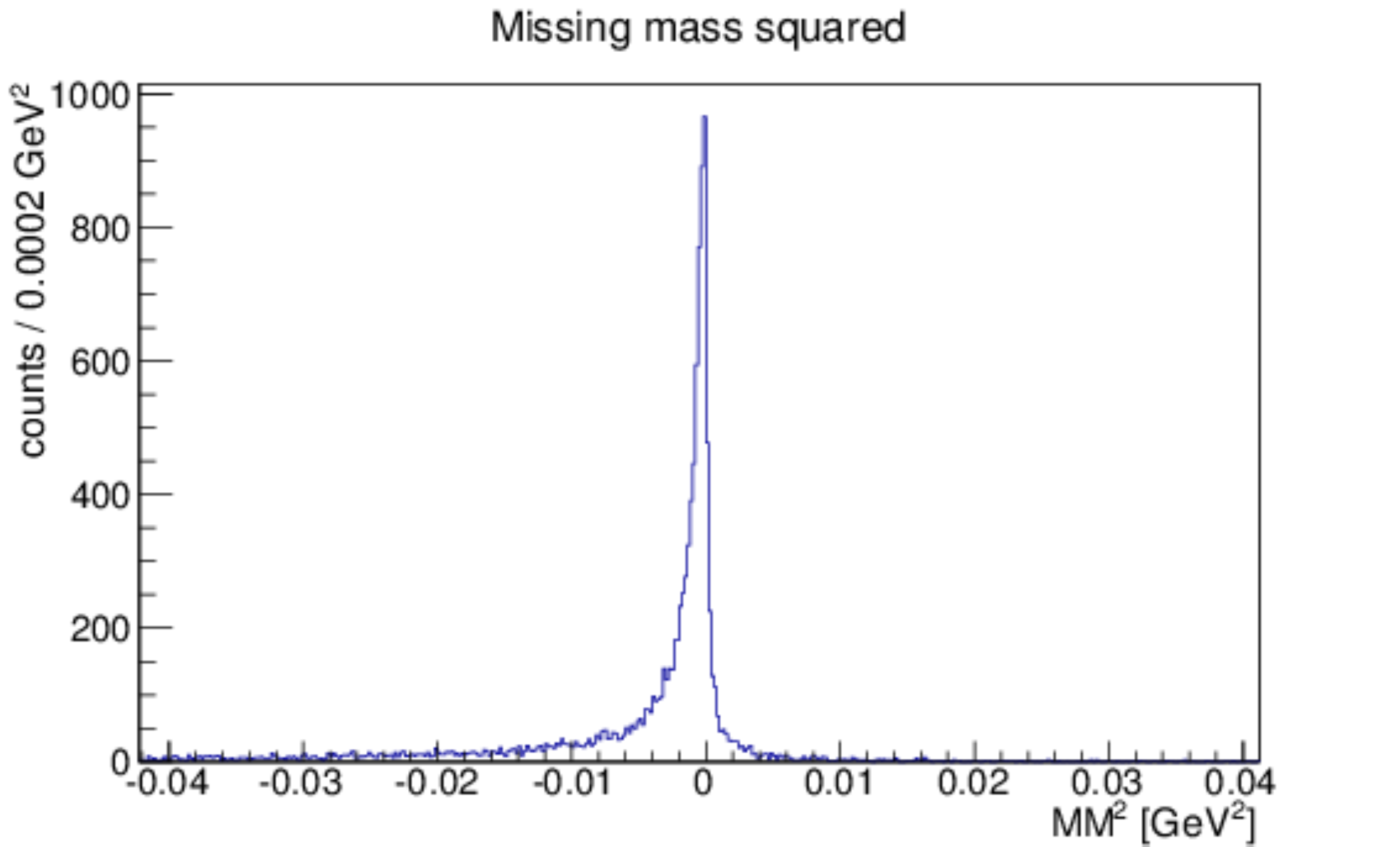} }

    \caption{Full reconstruction for $K_Lp\to K_S p$  and $K_S\to 
    \pi^+\pi^-$.
    Top left: $\pi^+\pi^-$ invariant mass. 
    Top right: $W$ computed from $\pi^+\pi^- p$ invariant mass. 
    Bottom plot: Missing-mass squared for the full reaction.} 
    \label{fig:A33}
\end{figure}

For the $K_S p$ channel, we take advantage of the BR of $69.2\%$ for
$K_S\to\pi^+\pi^-$~\cite{PDG2016}: the invariant mass of the $\pi^+
\pi^-$ pair and $W$ as computed from the four-momenta of the proton
and the two pions is shown in Fig.~\ref{fig:A33}.

\begin{figure}[h!]
\centering
{
    \includegraphics[width=0.45\textwidth,keepaspectratio]{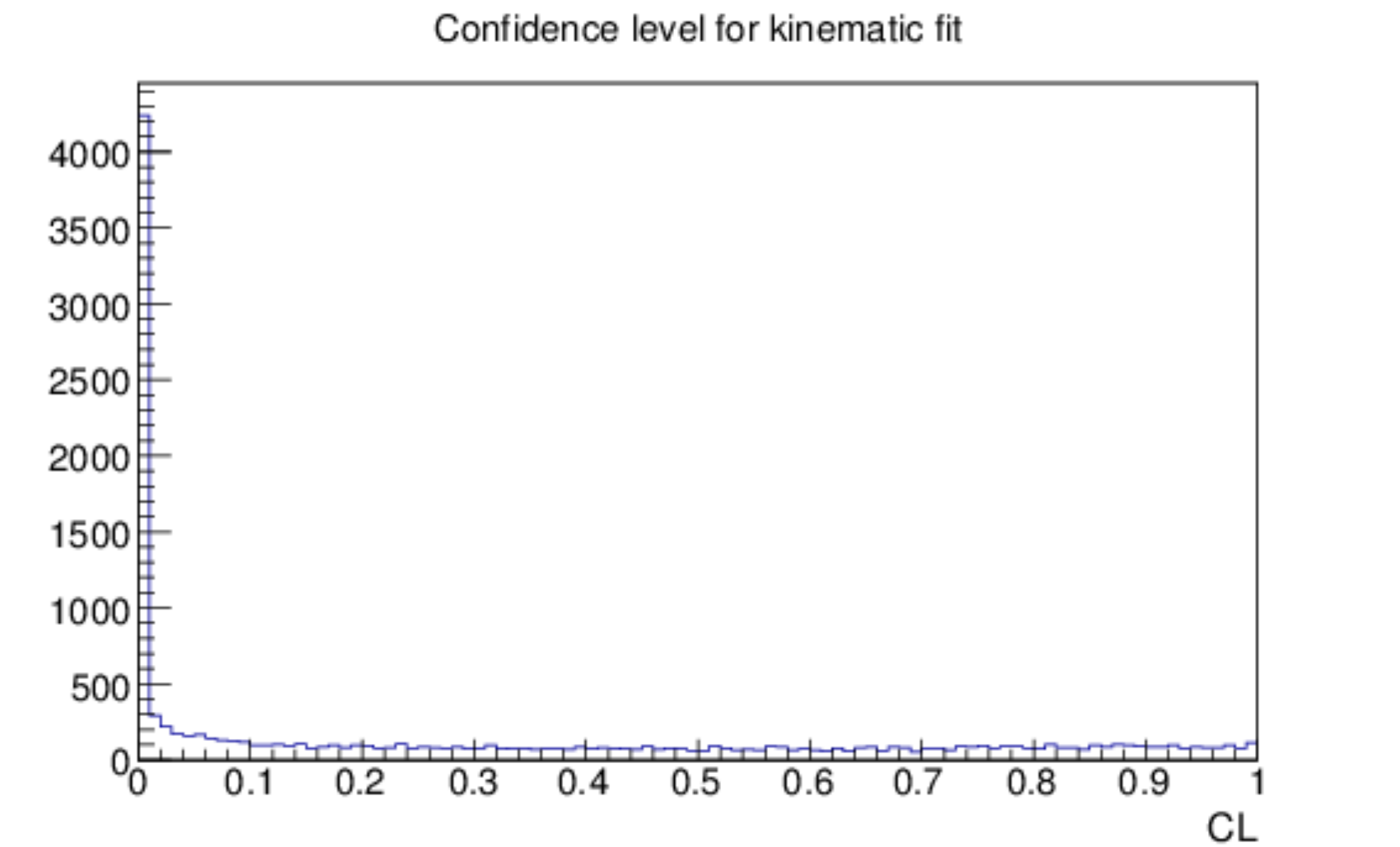} }
{
    \includegraphics[width=0.45\textwidth,keepaspectratio]{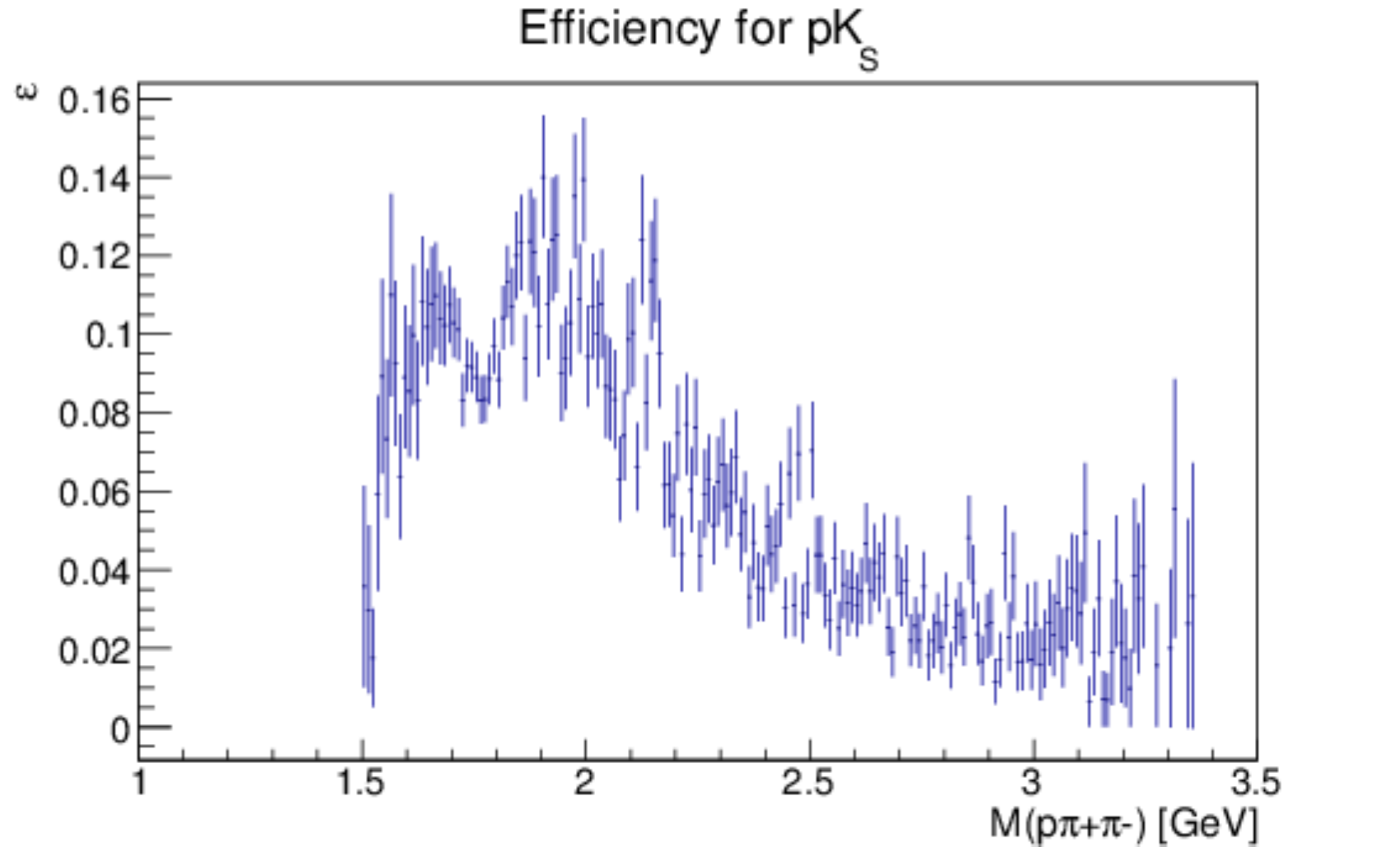} }

      \caption{Left plot: Confidence-level distribution for kinematic
      fit for the $K_S p$ channel. 
      Right plot: Estimate for efficiency for full reconstruction
      of the $K_L p \to K_S p$ and  $K_S \to \pi^+\pi^-$ reaction chain 
      as a function of $W$.} \label{fig:A34}
\end{figure}

After combining the four-momenta of the final-state particles with the
four-momenta of the beam and the target, the missing-mass squared for
the full reaction should be zero, which is also shown in
Fig.~\ref{fig:A33}. Finally, one requires conservation of energy and 
momentum in the reaction by applying a kinematic fit to the data. 
After applying a 0.1 cut on the confidence level of the fit, one 
computed an estimate for the reconstruction efficiency as a function 
of $W$ as shown in Fig.~\ref{fig:A34}. Here the efficiency includes 
the BR for $K_S\to\pi^+\pi^-$. The average reconstruction efficiency 
is about $7\%$.

\subsubsection{Details of MC study for $K_Lp\to\pi^+\Lambda$} 
\label{label:appendixKP}

\begin{figure}[!htb]
   \begin{center}
   \includegraphics[angle=0,width=6cm,height=5.5cm]{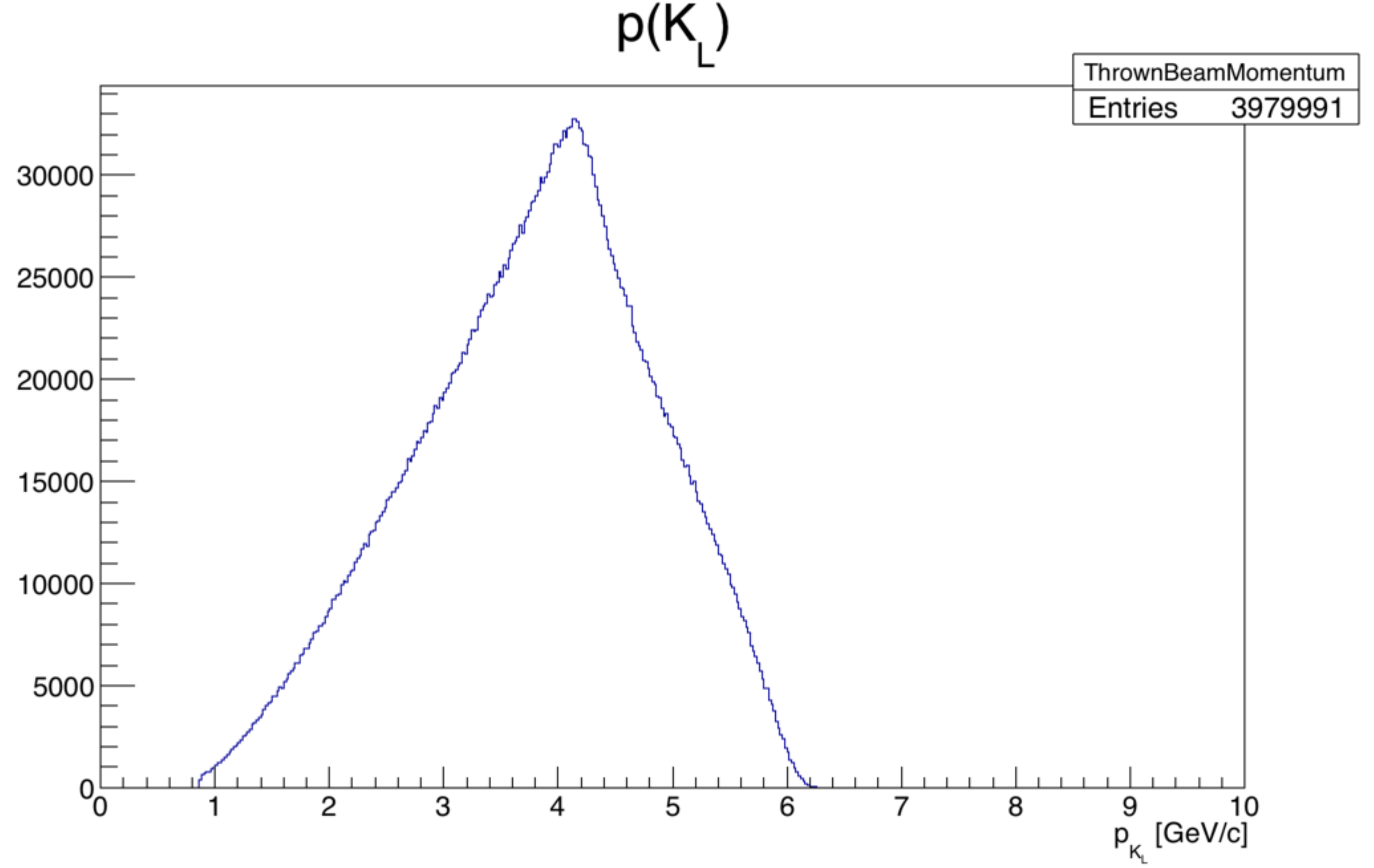}
   \includegraphics[angle=0,width=6cm,height=5.5cm]{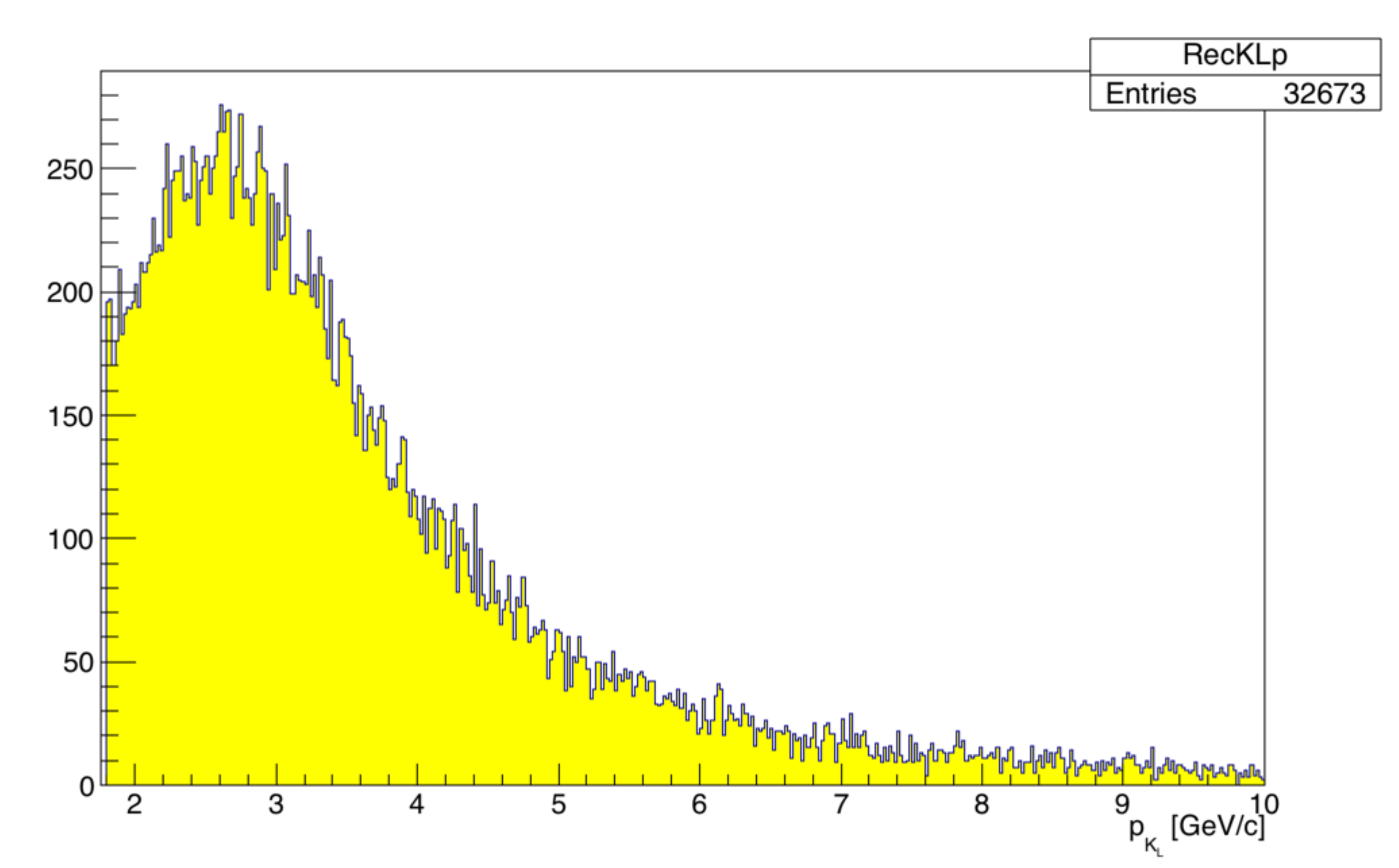}

   \caption{Beam particle ($K_L$) momentum distribution in MC simulation, 
   generated (left) and reconstructed (right).
}\label{fig:KLmom}
   \end{center}
\end{figure}

For our proposed KL Facility in Hall-D, we expect good statistics
of $K_Lp\to\pi^+\Lambda$ for a very wide range of $K_L$ beam momentum.
Figure~\ref{fig:KLmom} shows the $K_L$ beam momentum distributions from the
generated (left) and reconstructed (right) with requiring $\beta_{K_L} > 
0.95$ in time-of-flight.

\begin{figure}[!htb]
    \begin{center}
    \includegraphics[angle=0,width=6cm,height=8cm]{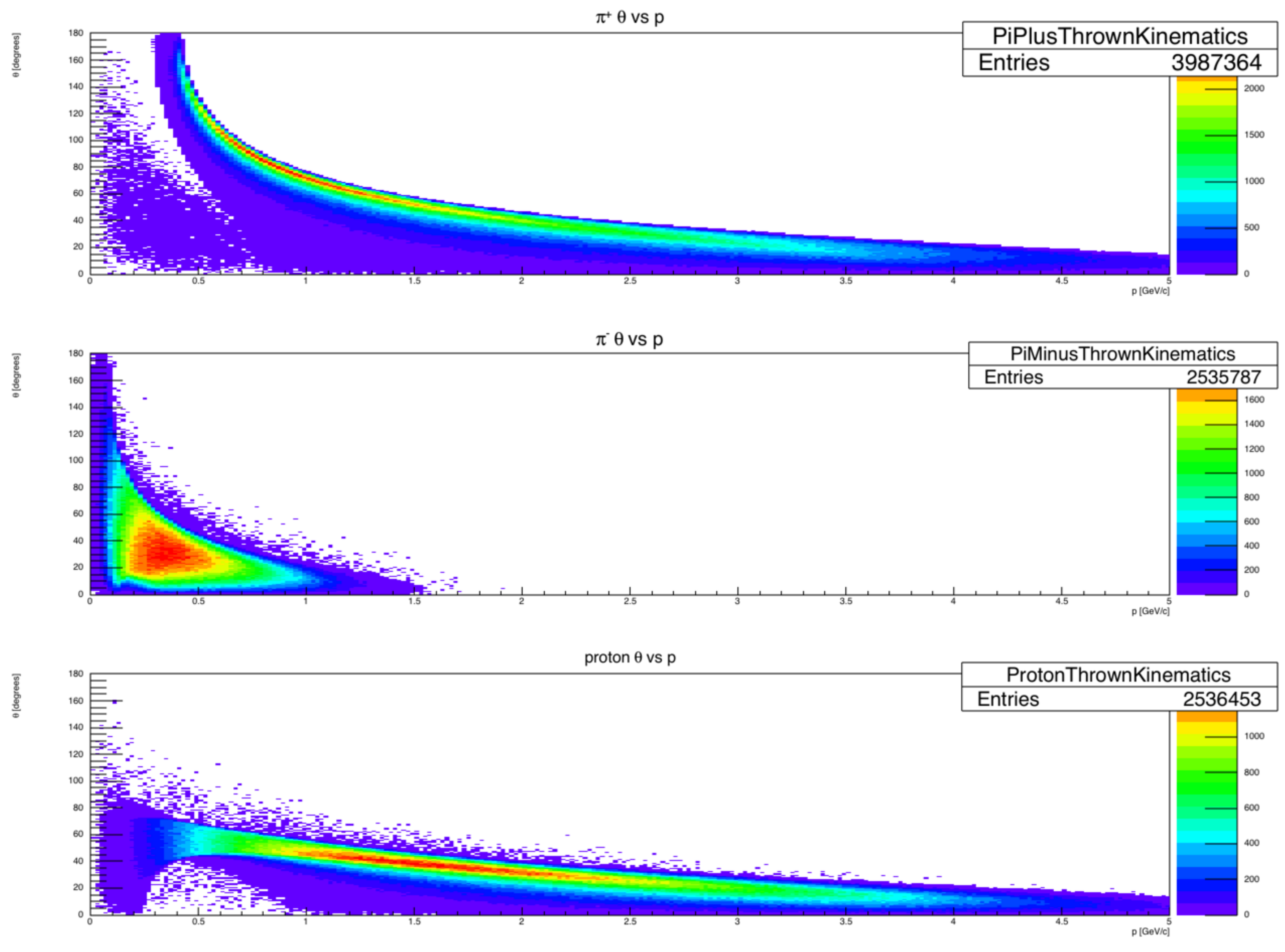}
    \includegraphics[angle=0,width=6cm,height=8cm]{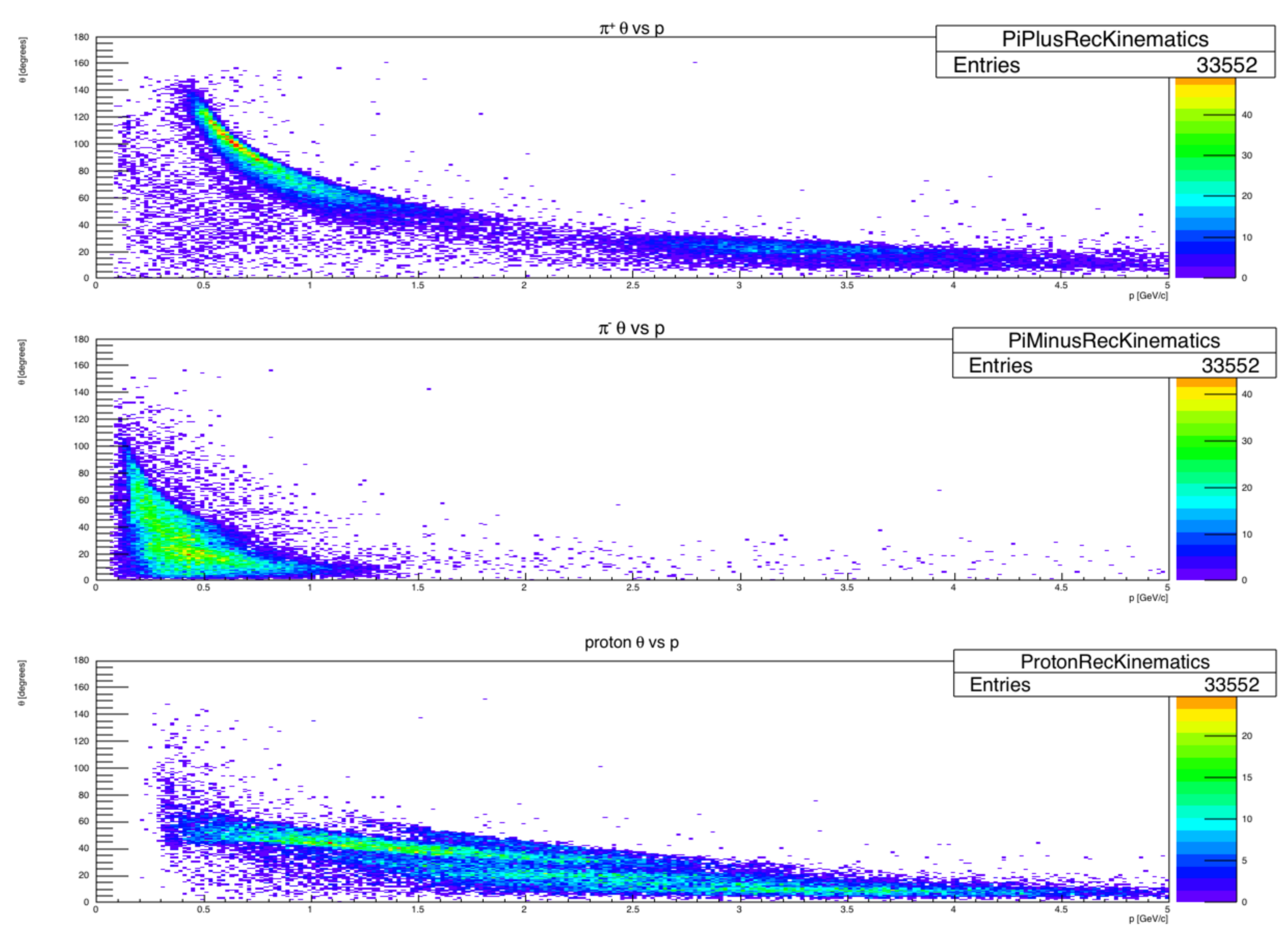}

    \caption[Kinematic coverage]{Momentum and angular distributions of 
    $\pi^+$ (top row), $\pi^-$ (middle row) and proton (bottom row) of 
    the reaction: generated (left column), reconstructed (right column) 
    events.
}\label{fig:Lampip5}
    \end{center}
\end{figure}

\begin{figure}[!htb]
   \begin{center}
   \includegraphics[angle=0,width=6cm,height=5.5cm]{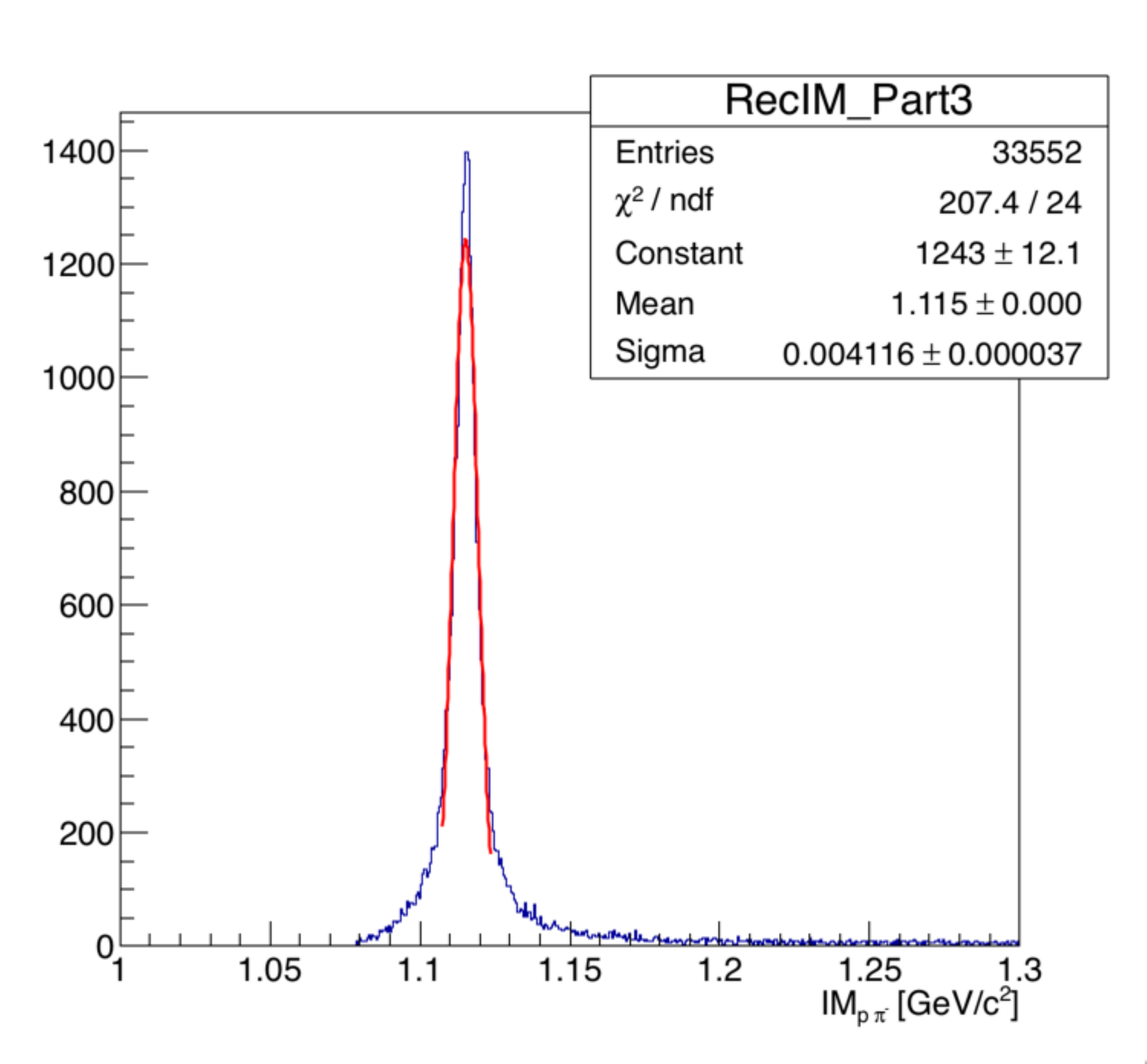}
   \includegraphics[angle=0,width=6cm,height=5.5cm]{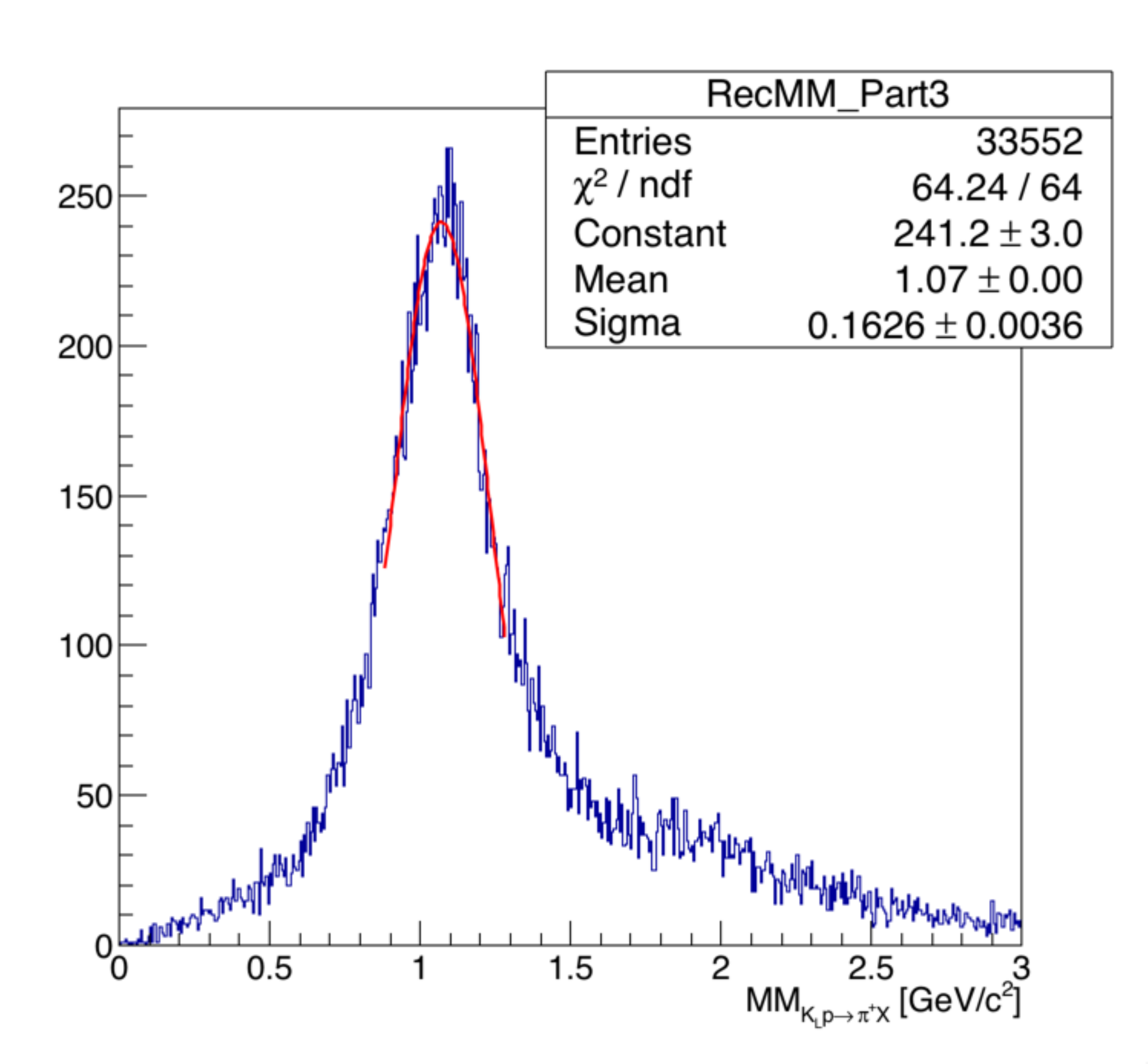}

   \caption[Kinematic coverage]{
        The $\Lambda$ invariant-mass distribution reconstructed from its 
        $\pi^- p$ decay particles (left), and the missing mass of 
        $\pi^+X$ (right).
}\label{fig:Lampip3}
   \end{center}
\end{figure}

We have generated the $K_Lp\to\pi^+\Lambda$ reaction in phase space  
taking into account the realistic $K_L$ beam momentum distribution in the 
event generator.  This momentum spectrum is a function of the distance 
and angle. Then we went through the standard Hall-D full GEANT simulation 
with GlueX detector and momentum smearing. Finally, we utilized the JANA 
for particle reconstruction that we simulated. Figure~\ref{fig:Lampip5} 
shows a sample plot
for polar angle versus momentum distribution of $\pi^+$, $\pi^-$, and
protons from the generated event (left) and reconstructed event (right).  
Figure~\ref{fig:Lampip3} 
shows an example of the reconstructed the $\Lambda$ particle for invariant 
mass (left) and missing mass (right). We 
obtained a 5~MeV invariant-mass resolution and a 150~MeV missing-mass 
resolution. We estimate the expected total number of $\pi^+\Lambda$ 
events as final-state particle within topology of 1$\pi^+$, 1$\pi^-$, and 
1 proton. In 100 days of beam time with $3\times 10^4~K_L$/s on the liquid 
hydrogen target, we expect to detect around 24M $K_Lp\to\pi^+\Lambda$ 
events for $W < 3$~GeV. Such an unprecedented statisitics will improve  
our knowledge of these states through partial-wave analysis.

\begin{figure}[!htb]
    \begin{center}
    \includegraphics[angle=0,width=6cm,height=5.5cm]{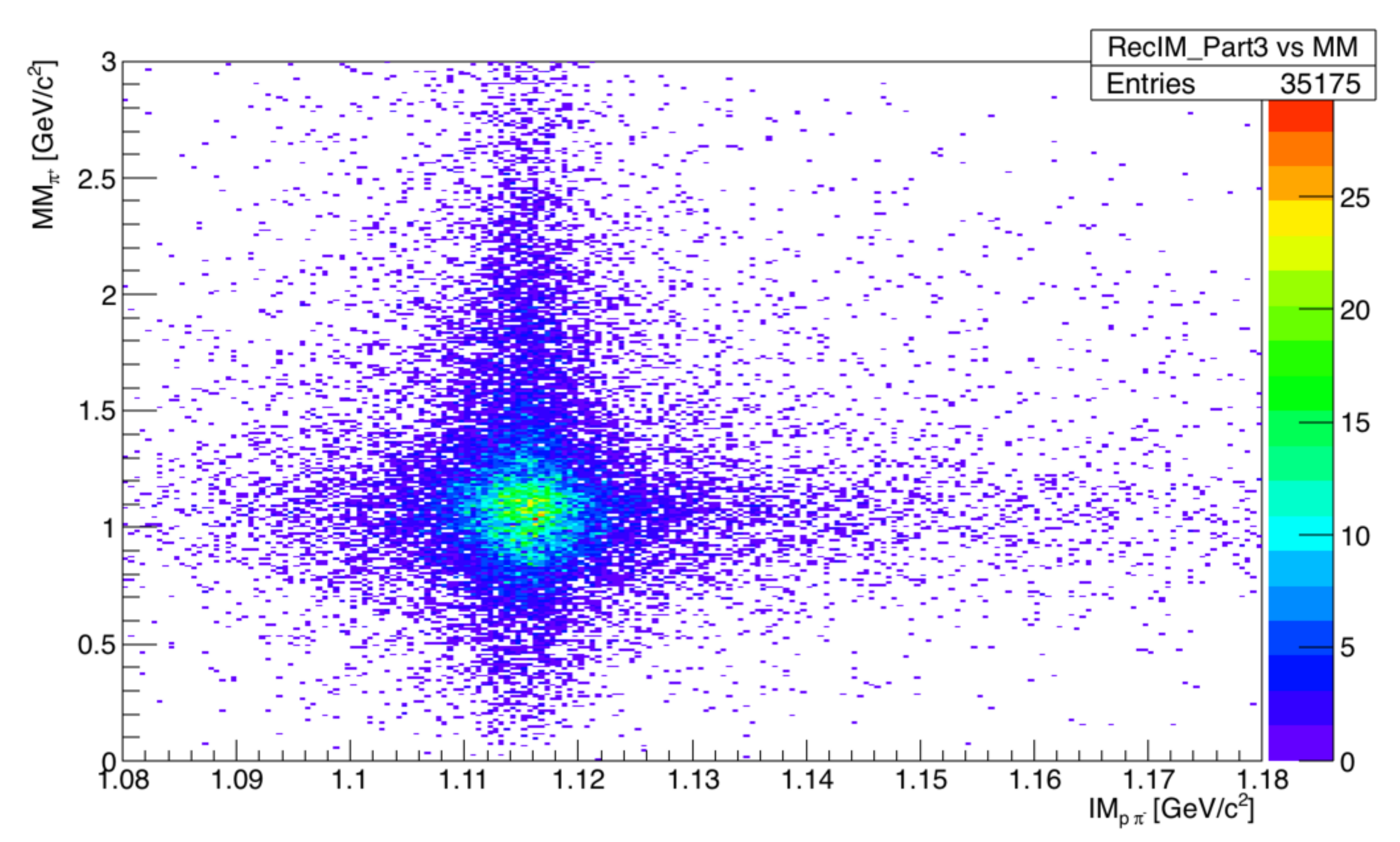}
    \includegraphics[angle=0,width=6cm,height=5.5cm]{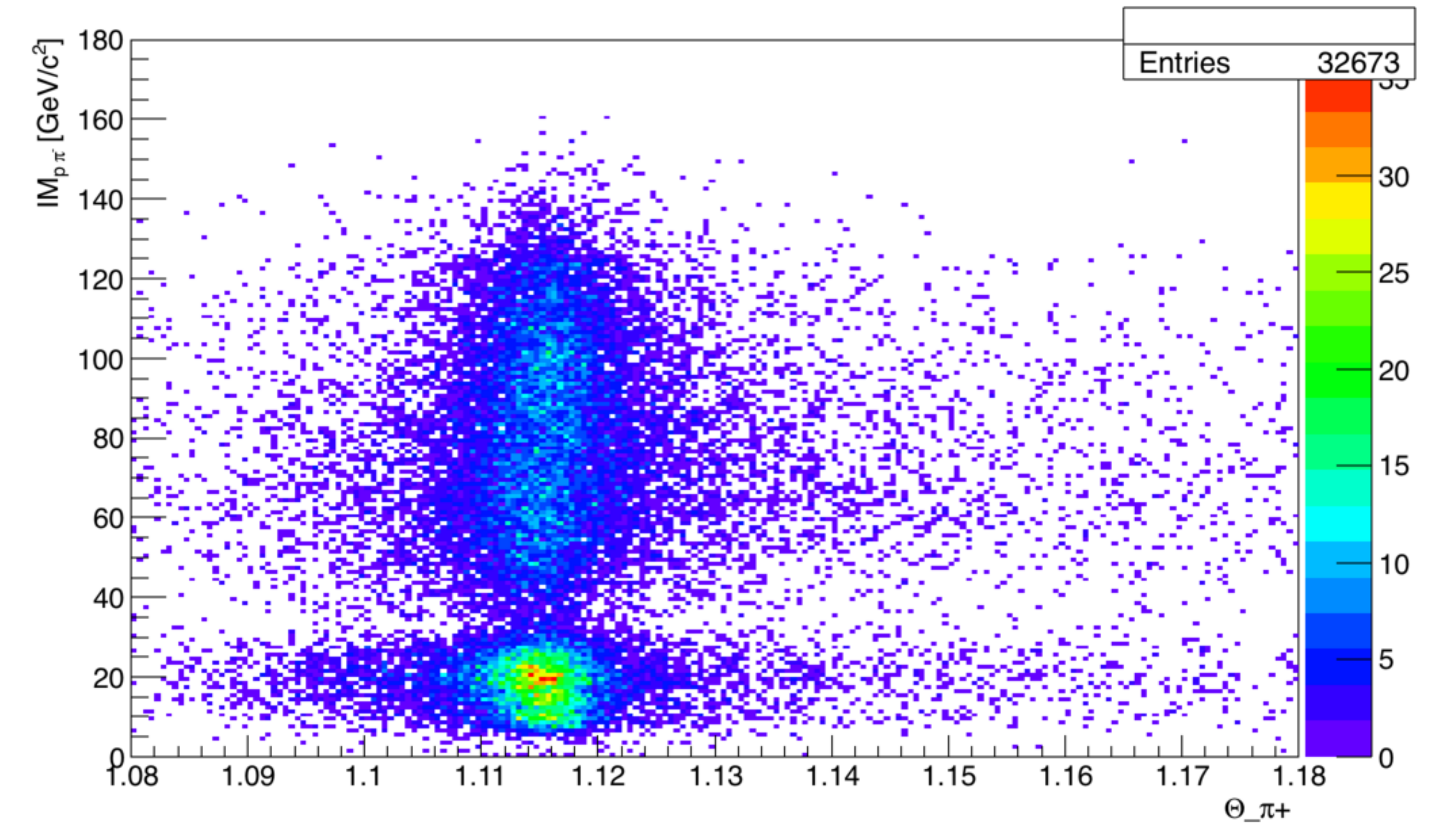}

    \caption[IM_vs_MM]{The $\Lambda$ invariant mass versus missing
    mass of $\pi^+X$ (left) and  the $\theta_{\pi^+}$ angle distribution 
    versus $\Lambda$ invariant mass (right).
}\label{fig:Lampip2}
    \end{center}
\end{figure}

Moreover, Fig.~\ref{fig:Lampip2} (left) shows the correlation between
$\Lambda$ invariant mass from its decay particles ($p$, $\pi^-$) and
missing mass of $\pi^+X$.  The right plot in Fig.~\ref{fig:Lampip2}
shows the $\Lambda$ invariant mass as a function of pion angular
distribution ($\theta_{\pi^+}$). All these plots are based on the 150~ps
time resolution of the ST.

The $K_Lp\to\pi^+\Lambda$ reaction has a relatively high production cross 
section the
order of a few mb in our proposed $K_L$ momentum range (1 -- 6~GeV/$c$).
The beam resolution has been calculated at the time-of-flight vertex
time resolution (150~ps) of the start counter (TOF-ST). The variation of
invariant-mass resolution as a function of $W$ for various TOF-ST
timing resolution (100, 150, 300~ps) is similar to those of other
reactions~\cite{KLmeeting2017}.

The major source of systematic uncertainty for this reaction would be 
mistaken particle identification among $\pi^+$, K$^+$, and proton in the 
final state. However, requiring the reconstructed $\Lambda$ and side-band 
subtraction technique for background will improve this uncertainty 
substantially.
%

\subsubsection{Details of MC study for $K_Lp\to K^+\Xi^0$} 
\label{sec:App_KPlXi}

\begin{figure}[h!]
\centering{
    \includegraphics[width=0.95\textwidth,keepaspectratio]{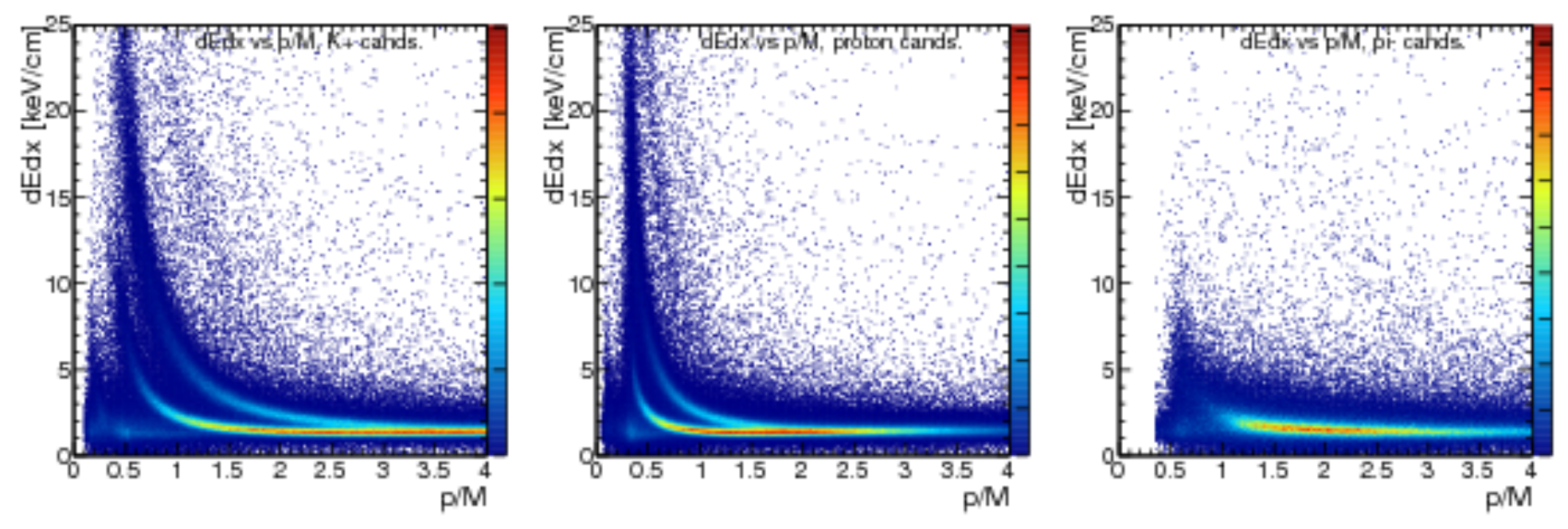}

       \caption{$dE/dX$ distributions used in kaon proton and 
       $\pi^-$ identification for the reconstruction of 
       $K_Lp\to K^+\Xi^0$.} \label{fig:CascPID}
}
\end{figure}

Three topologies can be used to reconstruct this reaction. Topology~1
requires the detection of a $K^+$, topology 2 requires the detection
of a $K^+$ and a $\Lambda$ by utilizing its high branching ration to
a $\pi^- p$ pair (63.9\%), and Topology~3 requires the detection of
the two-photon decay of the $\pi^0$ from $\Xi\to\pi^0\Lambda$.
Particle identification is done via a probabilistic approach involving
$dE/dX$, time-of-flight, and track curvature information as described
in Appendix~A5~(Sec.~\ref{sec:App_PID}). The $dE/dX$ distributions for 
kaon, proton, and $\pi^-$ candidates are shown in 
Fig.~\ref{fig:CascPID}.

At low particle momenta, kaons and protons can be well separated, but 
high-energy particles cannot be unambiguously  differentiated by $dE/dX$
or by ToF information, which leads to particle misidentification. The 
higher the $W$, the
higher ejectile energy we have and the more  misidentification contributions
we have. In this analysis (specifically Topology~2 and 3), these events were
largely removed by making an invariant-mass cut on the $\pi^- p$ pair. 

\begin{figure}[h!]
\centering{
    \includegraphics[width=0.45\textwidth,keepaspectratio]{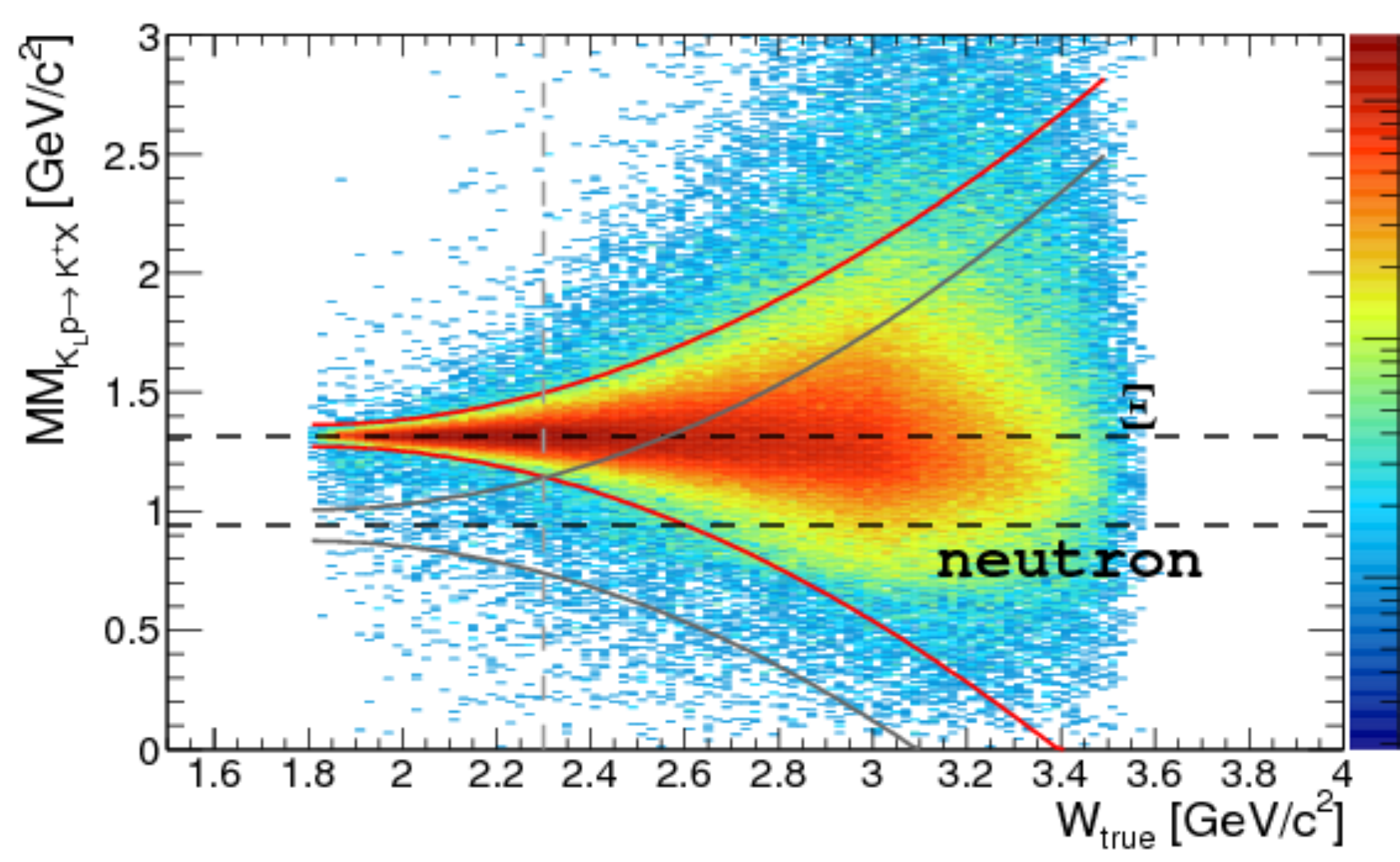}
    \includegraphics[width=0.45\textwidth,keepaspectratio]{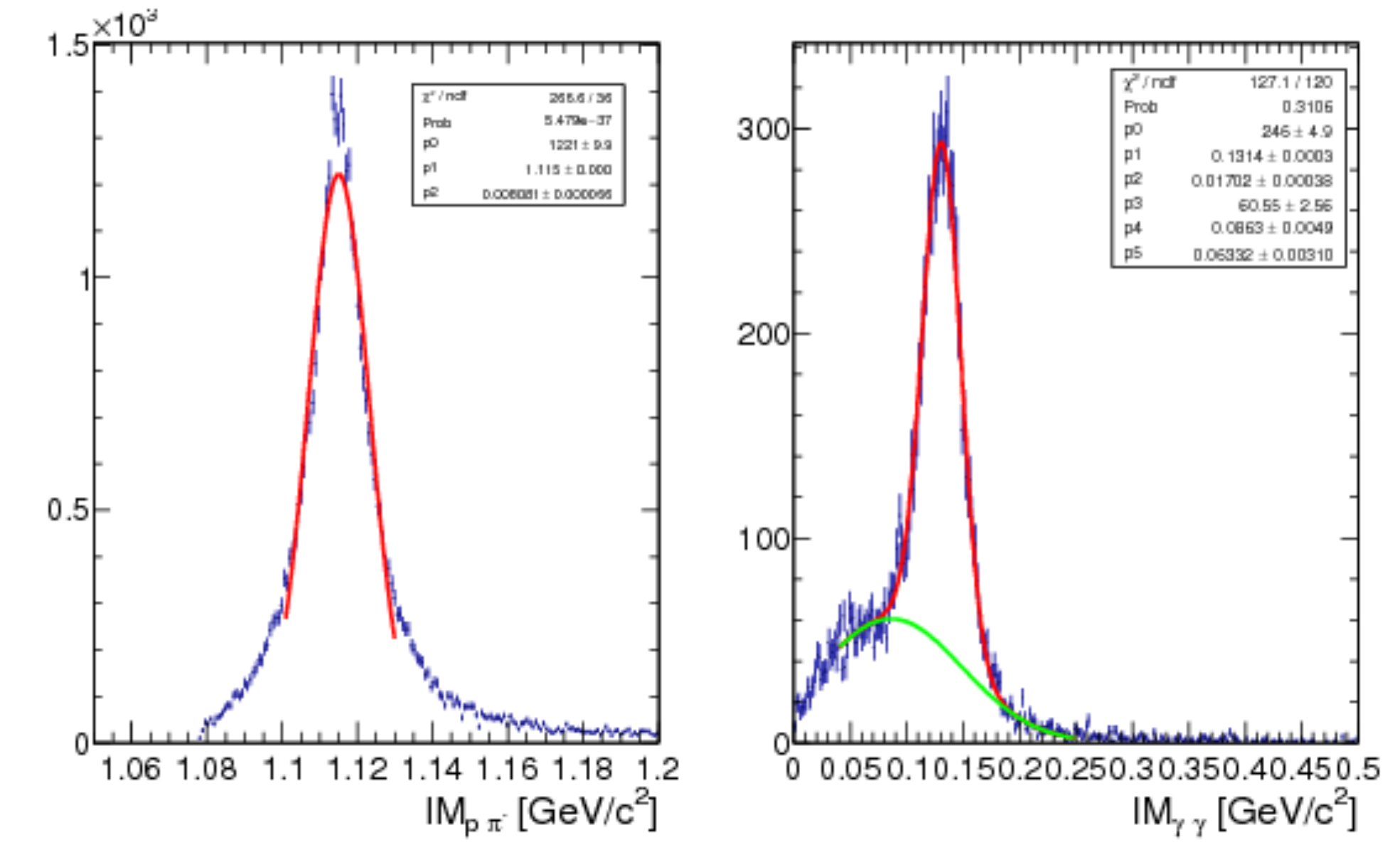}

   \caption{The missing mass of the reaction $K_Lp\to K^+X$ used to
   reconstruct the reaction $K_Lp\to K^+\Xi^0$ (Topology~1), and the 
   invariant mass of $p\pi^-$ pair (Topology~2), and the invariant 
   mass of the two-photon pair (Topology~3). } \label{fig:KlKpXi1}
}
\end{figure}

Figure~\ref{fig:KlKpXi1} shows the missing mass of $K_Lp\to K^+X$ for
simulated data for the reaction  $K_Lp\to K^+\Xi^0$ used in the reconstruction
of all topologies, the invariant-mass distribution of the
$\pi^- p$ pair used to reconstruct Topology~2 ($K_Lp\to K^+\Lambda X$) and 3,
and the invariant-mass of the two-photon pair used to reconstruct Topology~3
($K_Lp\to K^+\Lambda\pi^0$). A $3\sigma$ cut on these distributions allows
us to reconstruct the reaction fully. The left panel of Fig.~\ref{fig:KlKpXi1}
shows the $3\sigma$ $W$-dependent cut applied to select the missing $\Xi^0$
as well as the $W$-dependent $3\sigma$ cut to reconstruct the reaction 
$K_Lp\to K^+n$. (See Appendix~A5~(Sec.~\ref{sec:App_KPlN}) for more details 
on the sources of resolution effects on the missing mass.) The latter is one 
of the major sources of background for our reaction for Topology~1; however, 
the missing-mass resolution (obtained with a vertex-time resolution of 
150~ps) allows a clean separation of these two reactions up to $W = 2.3~$GeV
Above this value, special treatment of the $K_Lp\to K^+n$ background is 
required as discussed in greater detail in 
Appendix~A5~(Sec.~\ref{sec:KPXipol}).

\begin{figure}[h!]
\centering{
    \includegraphics[width=0.9\textwidth,keepaspectratio]{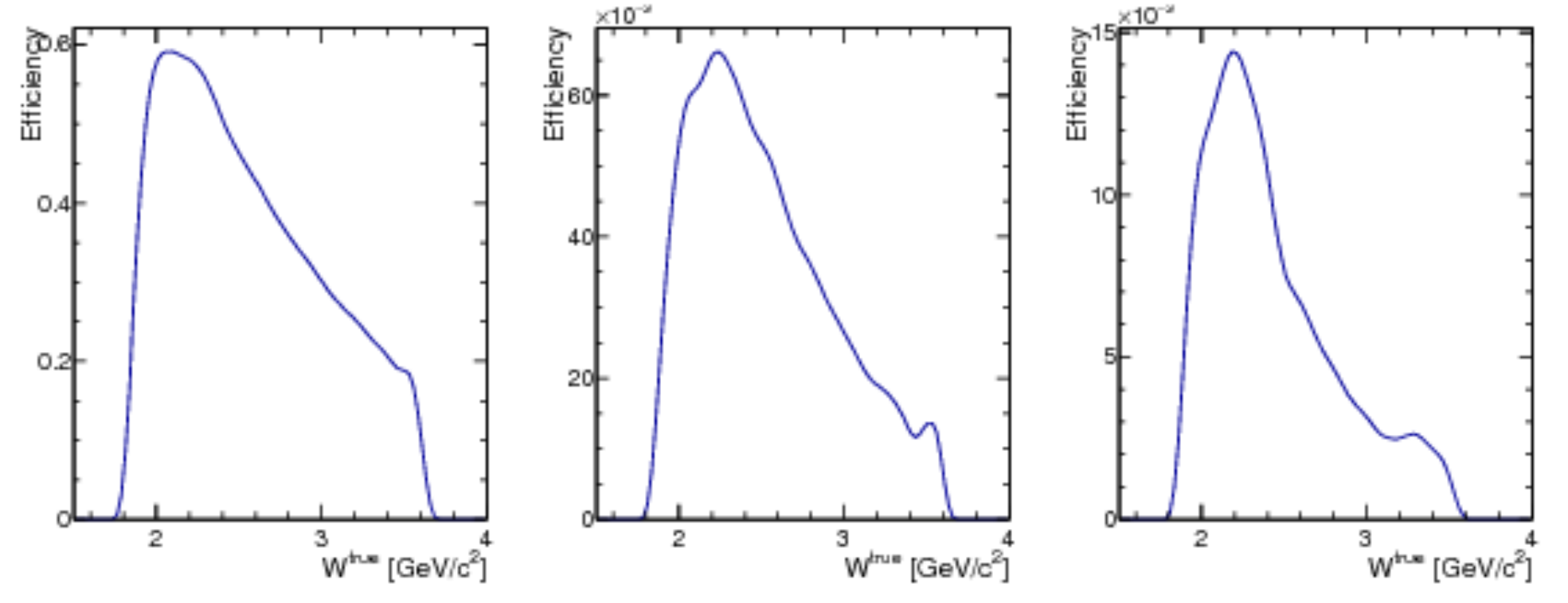}

    \caption{The detection efficiency for the reaction $K_Lp\to K^+\Xi^0$
    for each topology. } \label{fig:KlKpXieff}
}
\end{figure}

The detection efficiency as a function of the true $W$ for each topology 
is shown in Fig.~\ref{fig:KlKpXieff}. As expected, the efficiency is 
highest for Topology~1 reaching a maximum at 60\% for $W = 2.05$~GeV. The 
efficiency for Topology~2 is about an order of magnitude less than 
Topology~1, and Topology~3 detection efficiency is on average 0.8\%.

\paragraph{$K_Lp\to K^+\Xi^0$ background suppression:}
\label{sec:KPXibg}

\begin{figure}[h!]
\centering{
    \includegraphics[width=0.45\textwidth,keepaspectratio]{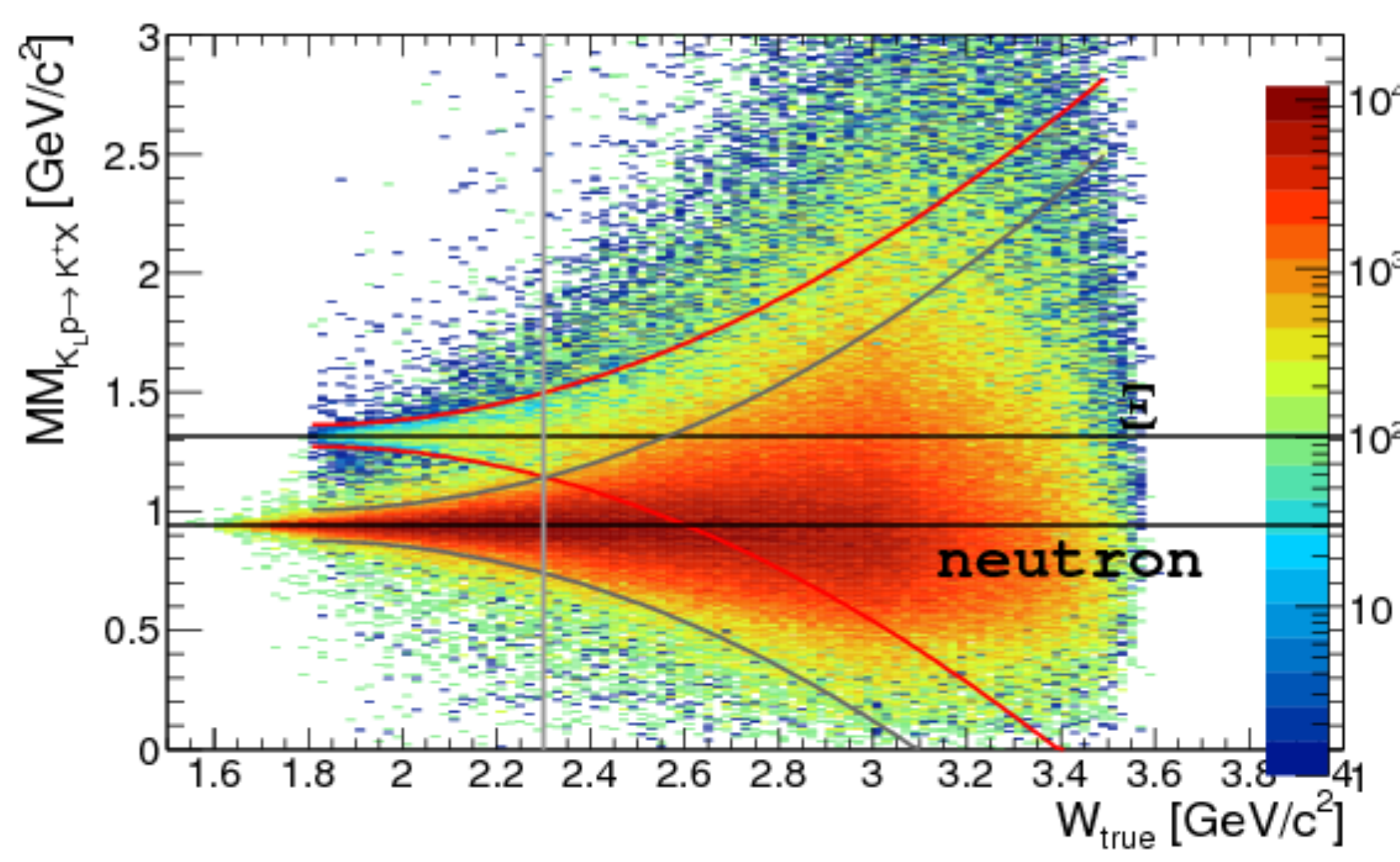}
    \includegraphics[width=0.45\textwidth,keepaspectratio]{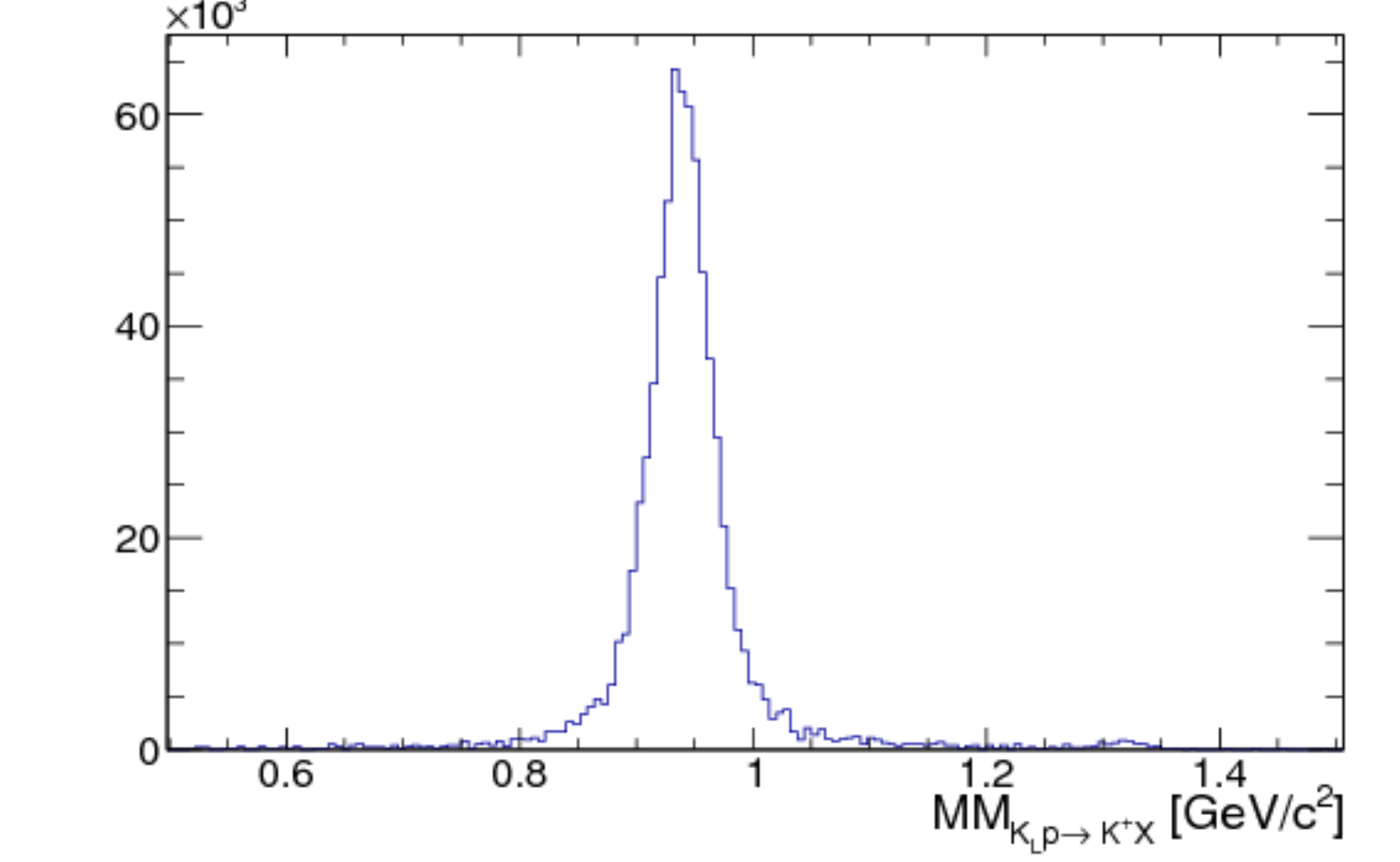}
    
    \caption{The missing mass of the reaction $K_Lp\to K^+X$ used to
    reconstruct the reactions $K_Lp\to K^+\Xi^0$ (Topology~1) and $K_Lp\to
    K^+n$ (which has about 2 orders of magnitude larger cross section). 
    Right panel shows the missing mass at $W=1.9$~GeV. } \label{fig:KlKpXi1a}
}
\end{figure}

\begin{figure}[h!]
\centering{
    \includegraphics[width=0.65\textwidth,keepaspectratio]{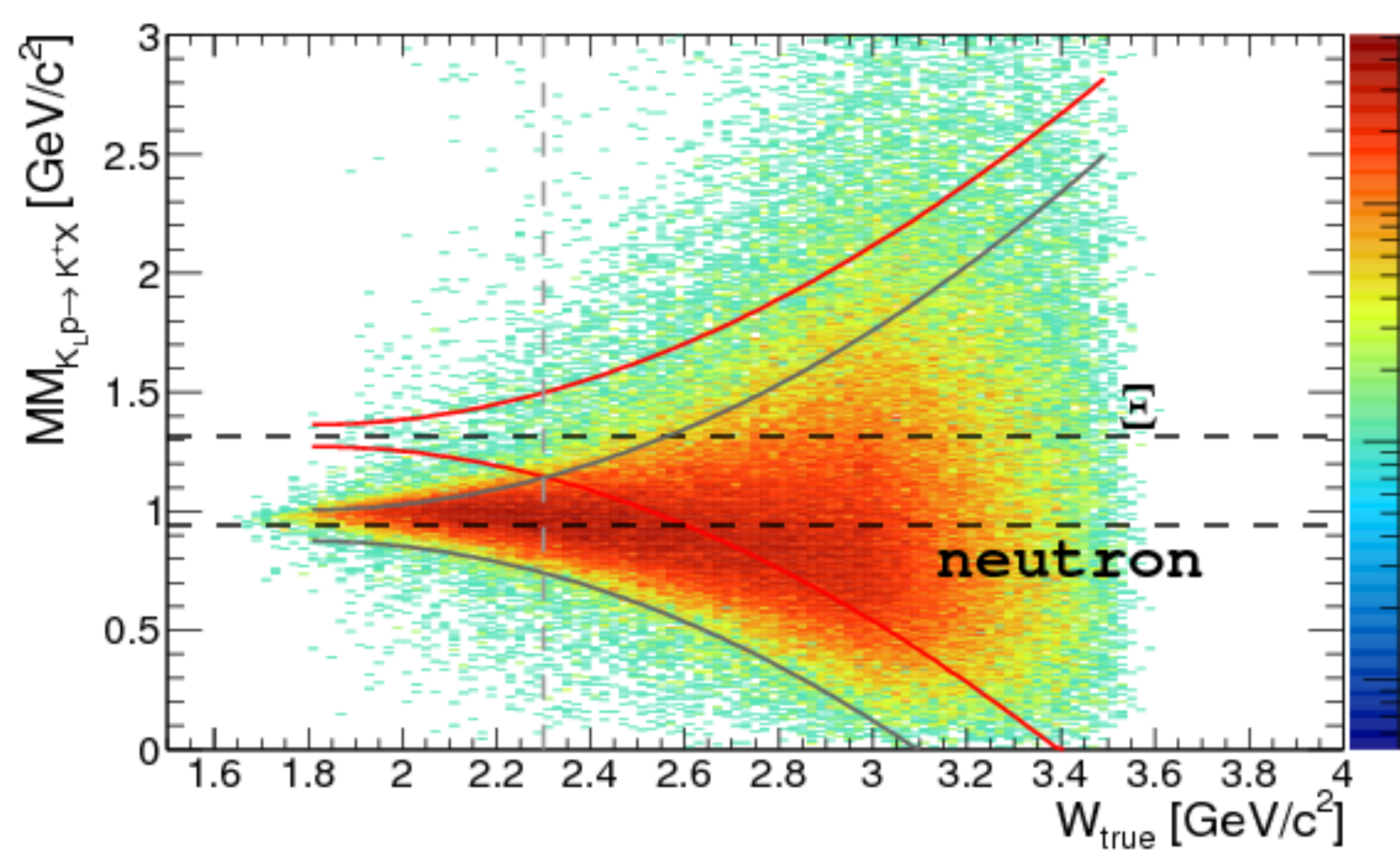}
    
    \caption{The missing mass of the reaction $K_Lp\to K^+X$ for 
    simulated events from the reaction $K_Lp\to \pi^+\Lambda$. The 
    reconstructed events here results from a pion misidentified 
    as a kaon. } \label{fig:KlKpXilambdabg}
}
\end{figure}

Different sources of background will contribute in the three topologies used 
to study this reaction. Disentangling our signal $K_Lp\to K^+\Xi^0$ from 
the reaction $K_Lp\to K^+n$ (for Topology~1), which has two orders of 
magnitude larger cross section is expected to be relatively 
straightforward. As mentioned before, a simple missing-mass cut is 
sufficient to remove any contributions from this reaction for $W < 
2.3$~GeV.  For $W>2.3$~GeV, an s-weight approach (or neuralNets, etc.) 
can be utilized to remove these contribution as the shape of the 
background under any cascade events can be well established from 
simulations.  Figure~\ref{fig:KlKpXi1a} shows the $W$-dependence of the 
missing-mass distribution of $K_Lp\to K^+X$ for the simulated reactions 
$K_Lp\to K^+\Xi^0$ and $K_Lp\to K^+n$ (left panel). The right panel shows  
the missing-mass projection at $W = 1.9$~GeV.
In addition to $K_Lp\to K^+n$, the reaction $K_Lp\to \pi^+\Lambda$ is 
also a source of background events for Topology~1  ($K_Lp\to K^+ X$) and 
2 ($K_Lp\to K^+\Lambda X$). This channel contributes when the final-state
$\pi^+$ is misidentified as a $K^+$. This shifts the missing mass of
$K_Lp\to \pi^+ X$ to values lower than the ones expected, which leads to 
a good separation of this source of background below $W<2.2$~GeV.
Figure~\ref{fig:KlKpXilambdabg} shows the missing-mass distribution of
these misidentified events. Contributions from these events for 
Topology~3 is completely removed by the requirement of two photons in 
the final state that reconstruct the mass of $\pi^0$. For Topology~2, 
coplanarity cuts between the reconstructed (misidentified) $K^+$ and 
$\Lambda$ can reduce contributions, where as a background subtraction 
approach using the missing-mass information can be used to remove any 
contribution at $W>2.2$~GeV.

\paragraph{$\Xi^0$ induced polarization:}
\label{sec:KPXipol}

\begin{figure}[h!]
\centering{
    \includegraphics[width=0.65\textwidth,keepaspectratio]{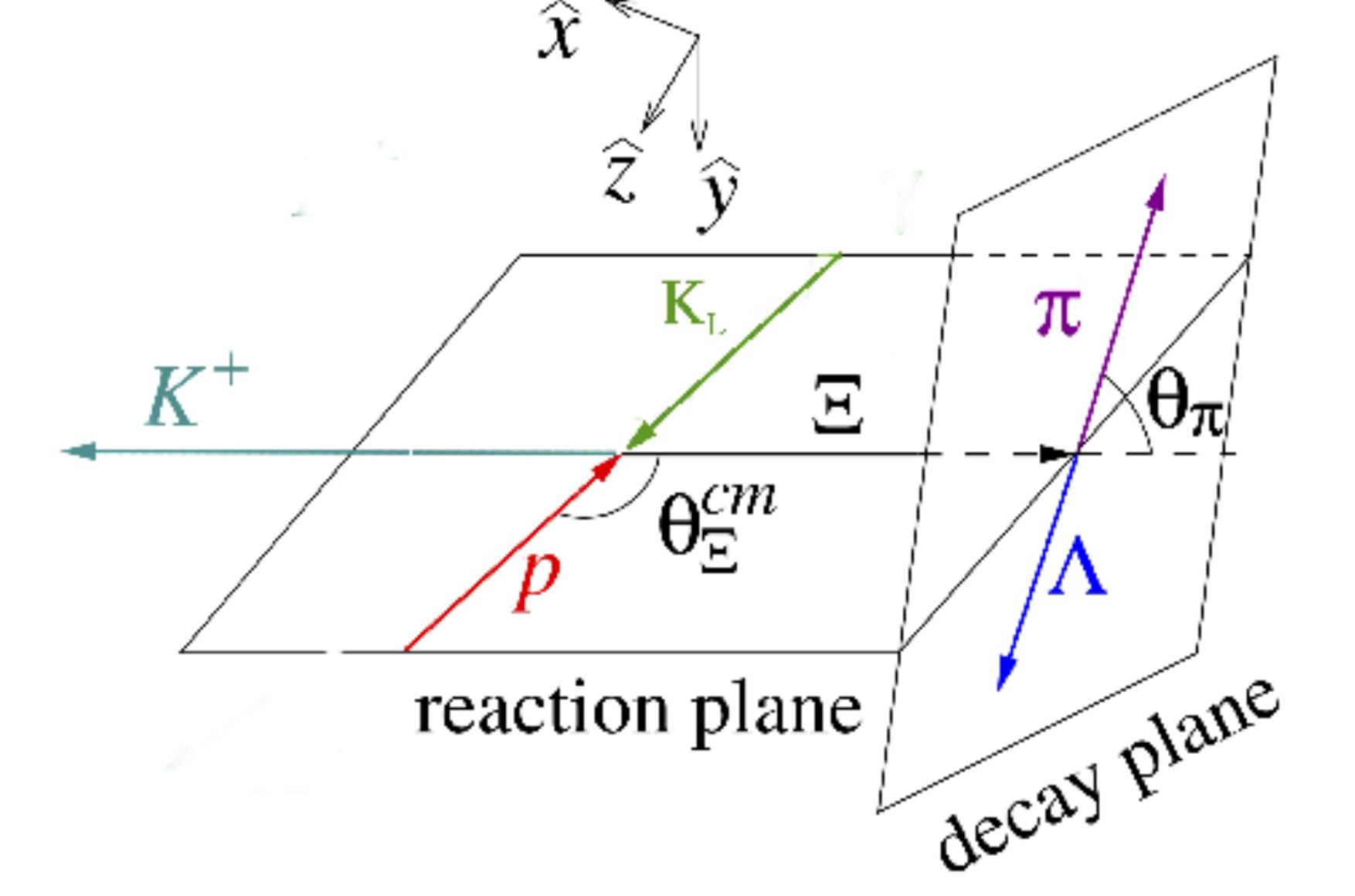}
    
    \caption{ The production plane  form $K_Lp\to K^+\Xi^0$ defined in
    the center-of-momentum system containing the incoming $K_L$ and 
    proton target. The decay plane is defined in the rest-frame of the 
    cascade and contains its decay products. The induced polarization 
    $P^y_{\Xi}$ is defined to lie perpendicular to the reaction plane.} 
    \label{fig:KlKpXiReactPlane}
}
\end{figure}

The parity-violating nature of the cascade's weak decay ($\Xi^0\to
\pi^0\Lambda$) yiels a pion angular distribution given by
\begin{equation}
	n(\theta^y_{\pi})=\frac{N}{2}(1-P^y_{\Xi}\alpha\cos\theta^y_\pi),
\end{equation}
where $P^y_{\Xi}$ is the induced polarization of the cascade, and is
the analyzing power $\alpha=0.406\pm0.013$~\cite{PDG2016}.
Figure~\ref{fig:KlKpXiReactPlane} shows the production plane
defined in the center-of-momentum system containing the incoming $K_L$
and proton target. The decay plane is defined in the rest-frame of the
cascade and contains its decay products.

In terms of four-vectors, conservation of energy and momentum for this reaction 
is written as follows:
\begin{equation}
	\mathcal{P}_{K_L}+\mathcal{P}_{p}=\mathcal{P}_{K^+}
	+\mathcal{P}_{\Xi^0}.
\end{equation}
The production plane is then defined by
\begin{equation}
	\hat{y}=\frac{\vec{P}_{\Xi}\times \vec{P}_{K_L}}{|\vec{P}_{\Xi}\times 
	\vec{P}_{K_L}|}.
\end{equation}
The $\hat{z}$ axis lies along the beam direction
\begin{equation}
	\hat{z}=\frac{ \vec{P}_{K_L}}{|\vec{P}_{K_L}|},
\end{equation}
and thus the $\hat{x}$ axis is defined to give a right-handed coordinate 
system:
\begin{equation}
	\hat{x}=\hat{y}\times\hat{z}.
\end{equation}

The determination of $P^y_{\Xi}$ can be established by linear fits to the
acceptance-corrected pion angular ($\cos\theta^y_{\pi}$) yields. Fitting
these distributions with a first-degree polynomial,
\begin{equation}
	y=a_0(1+a_1\cos\theta^y_{\pi}),
\end{equation}
allows the determination of $a_1$, which gives us the the induced 
polarization
\begin{equation}
	a_1=P^y_{\Xi}\alpha.
\end{equation}
Alternatively, one can determine the induced polarization transfer from
determining the forward-backward asymmetry, $A^y$, of the pion angular
distribution. This asymmetry is defined as
\begin{equation}
	A^y=\frac{N^y_+-N^y_-}{N^y_++N^y_-},
\end{equation}
where $N^y_+$ and $N^y_-$ are the acceptance-corrected yields with 
$\cos\theta^y_\pi$ positive and negative, respectively. The asymmetry is 
related to the induced polarization by
\begin{equation}
	P^y_{\Xi}=\frac{-2A^y}{\alpha}.
\end{equation}
The statistical uncertainty in the asymmetry measurement of $P^y_{\Xi}$ 
is related to the Poisson uncertainty in $N^y_+$ and $N^y_-$. Propagating 
this uncertainty to the uncertainty of $A^y$ gives
\begin{equation}
	\sigma_{A^y}=\frac{2}{(N^y_+ +N^y_-)^2}\sqrt{N^y_+N^y_-(N^y_+ 
	+ N^y_-)}.
\end{equation}
The uncertainty in $P^y_{\Xi}$ is then found by propagating $\sigma_{A^y}$
and $\sigma_{ \alpha}$:
\begin{equation}
	\frac{\sigma_{P^y_{\Xi}}}{P^y_{\Xi}}=\sqrt{(\frac{\sigma_{A^y}}{A^y})^2
	+(\frac{\sigma_{\alpha}}{\alpha})^2}.
\end{equation}

\subsubsection{Details of MC study for $K_Lp\to K^+n$} 
\label{sec:App_KPlN}

\begin{figure}[h!]
\centering{
    \includegraphics[width=0.47\textwidth,keepaspectratio]{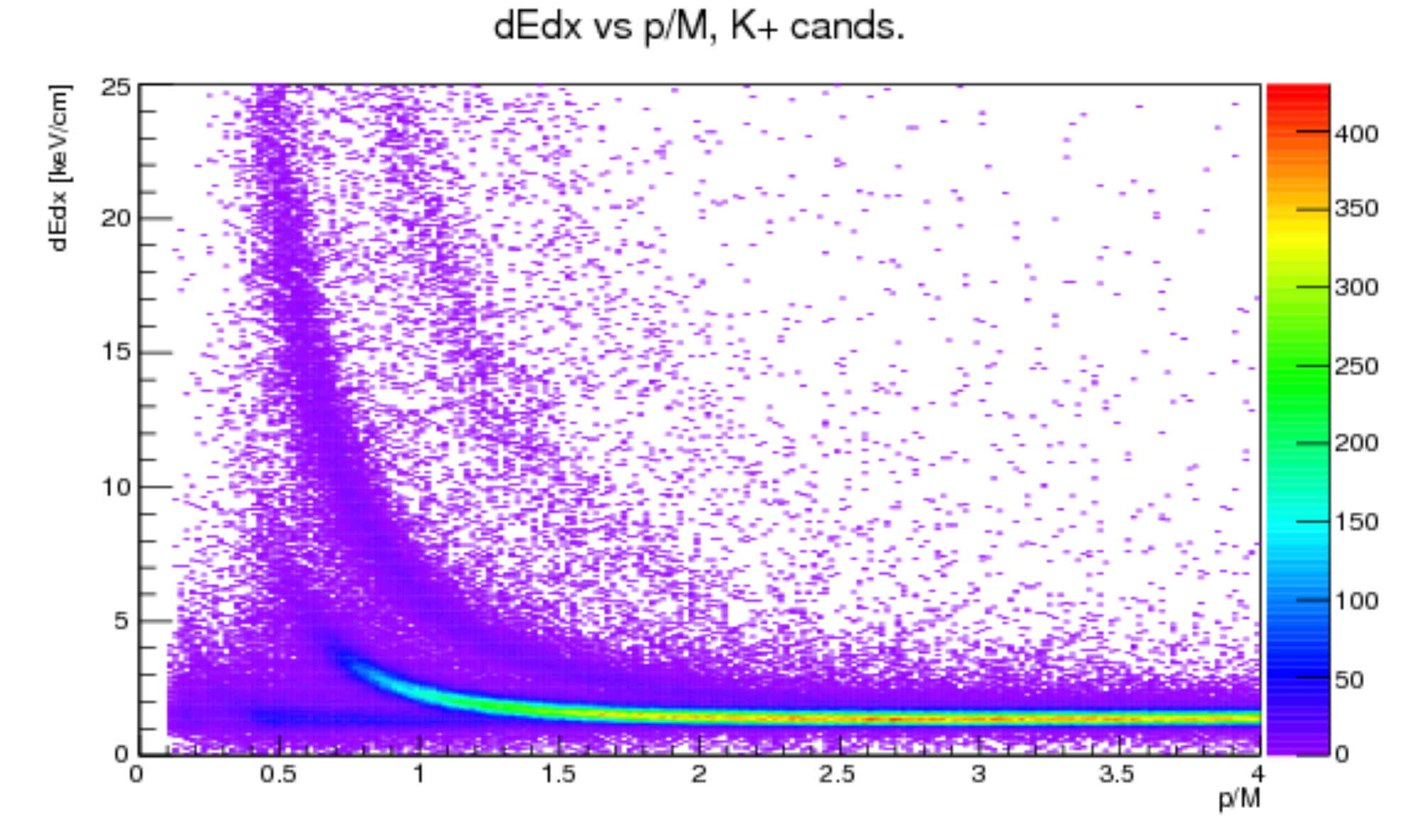}
    \includegraphics[width=0.45\textwidth,keepaspectratio]{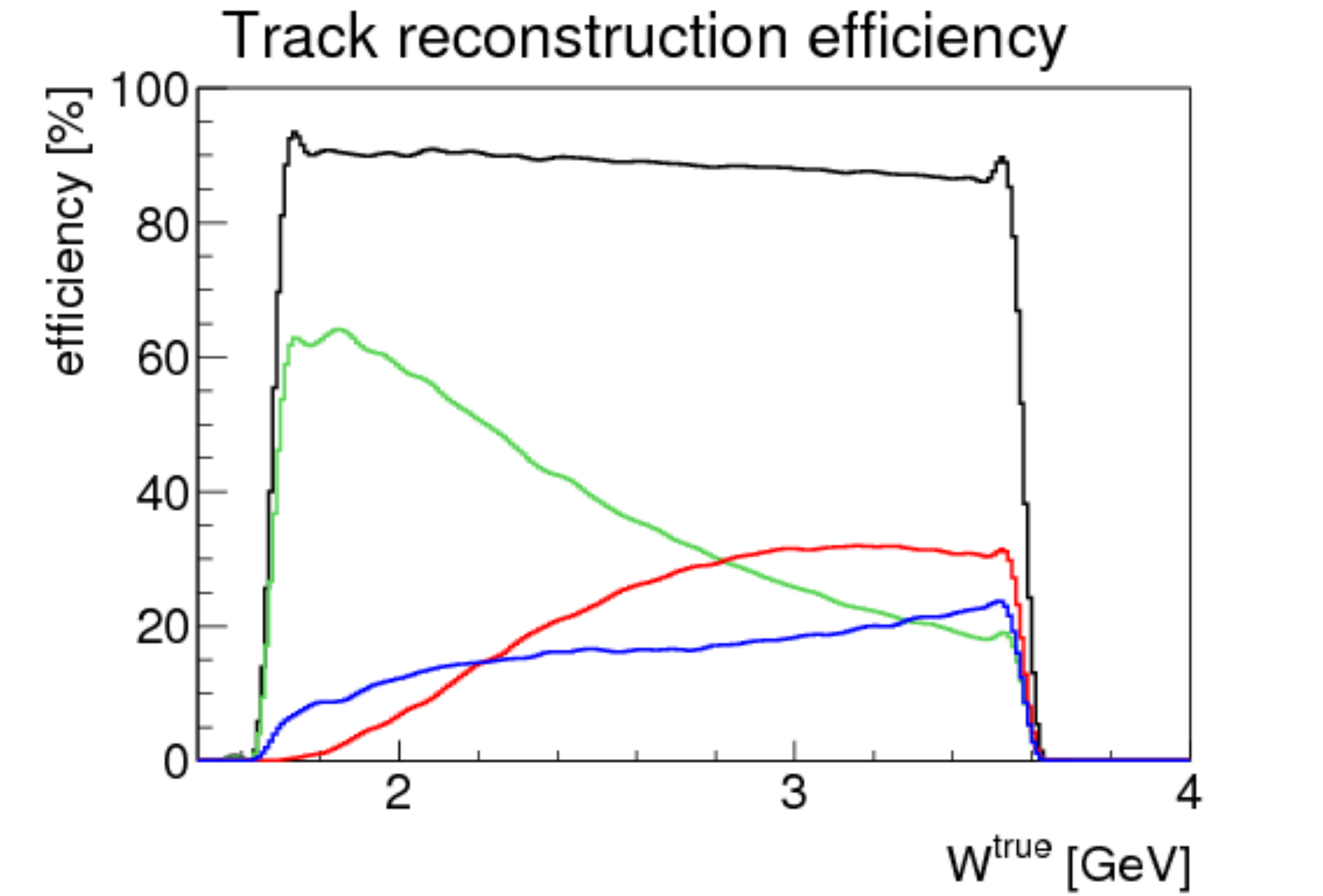}
        
    \caption{Left plot: $dE/dx$ for the $K_Lp\to K^+n$ channel
	Right plot: single charged-particle track detection efficiency
        as a function of $W$ for the $K_Lp\to K^+n$ channel. Any
        charged particle (black), kaon (green), proton (red),
        pion (blue). } \label{fig:KPlNeff}
}
\end{figure}

As described in Section~\ref{sec:KPlN} we used only $K^{+}$ detection
to reconstruct this reaction. Kaon identification is done with a
probabilistic approach involving $dE/dX$, time-of-flight, and track
curvature information; see Appendix~A5~(Sec.~\ref{sec:App_PID}) for 
further details.
Even in pure $K_Lp\to K^+n$ MC case one can have more than one charged particle
track reconstructed due to various reactions in the detector volume.
That is why in addition to the pronounced $K^+$ banana in 
Fig.~\ref{fig:KPlNeff}(left) we see some traces of pion and proton 
bands. At low $K^+$ momenta, kaons can be well separated from pions and 
protons, but high-energy particles cannot be differentiated by 
$dE/dX$ or by ToF information, which leading to particle misidentification. 
The higher 
the $W$, the higher the ejectile energy we have and the more kaons we lose 
due to misidentification; see Fig.~\ref{fig:KPlNeff}(right, green).  In 
our analysis, we restricted ourselves to one and only one reconstructed 
charged-particle track. This condition helps to suppress the background, but 
does not reduce the reconstruction efficiency; see 
Fig.~\ref{fig:KPlNeff}(right, black).

\begin{figure}[h!]
\centering{
    \includegraphics[width=0.45\textwidth,keepaspectratio]{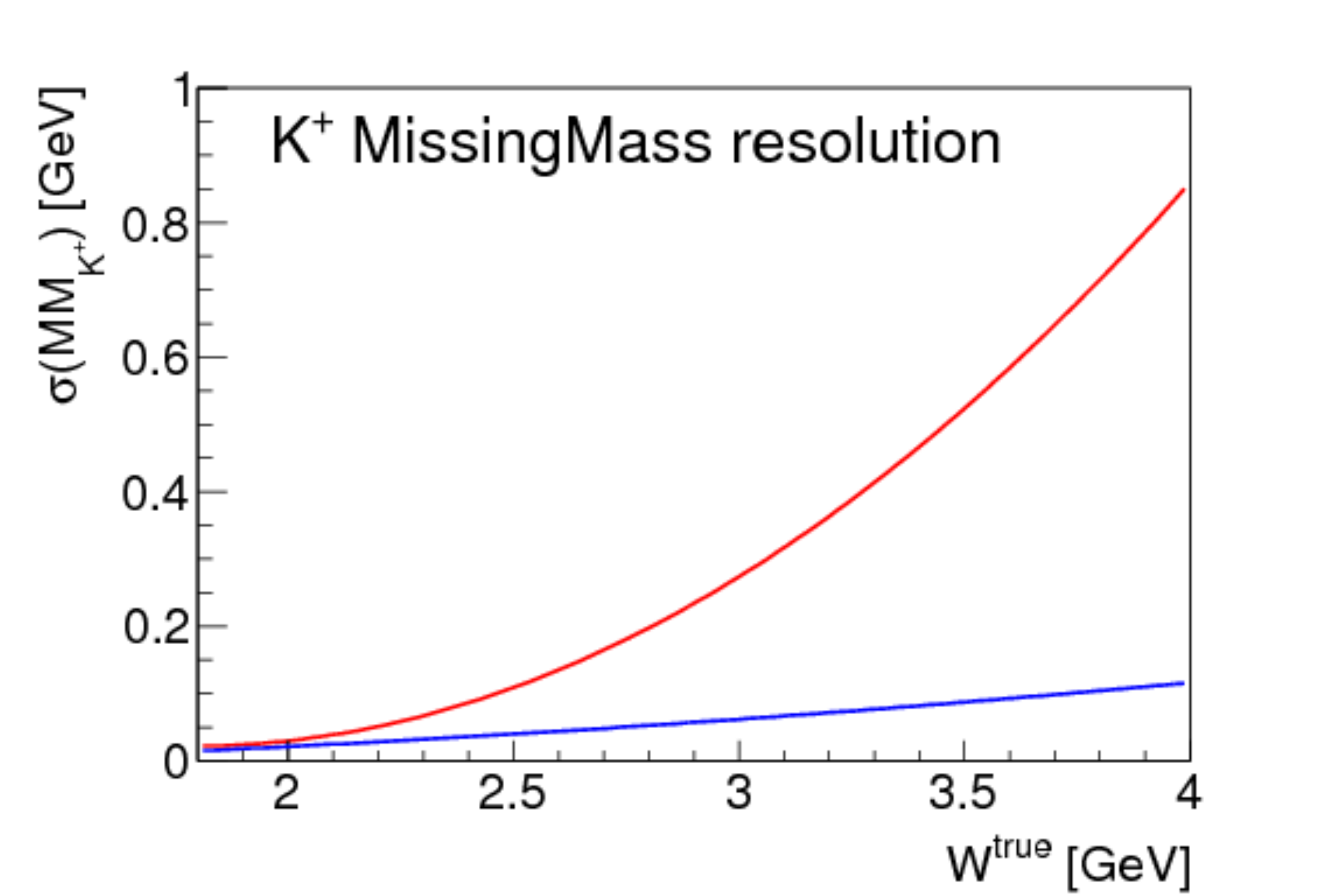}
    \includegraphics[width=0.45\textwidth,keepaspectratio]{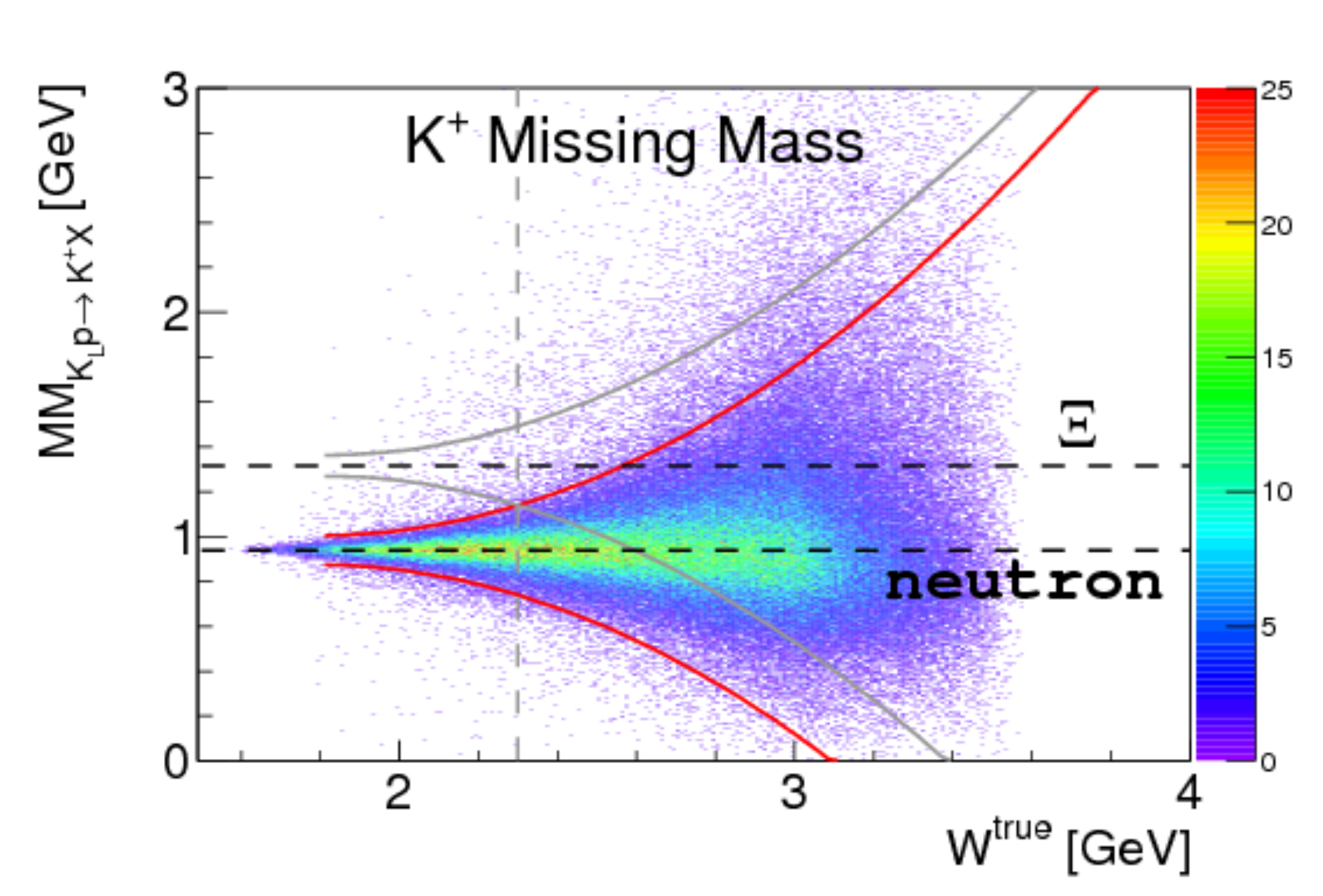}
        
    \caption{Left plot: Full (red) and detector related (blue) $K^+$ 
    missing-mass resolution in terms of $\sigma$. In second case, the true 
    $K_L$ momentum was used to calculate the missing mass. Right plot: 
    $K^+$ missing-mass resolution as a function of $W$. $3\sigma$ 
    missing-mass cuts for the $K_Lp\to K^+n$ (red) and $K_Lp\to 
    K^+\Xi$ (gray) reactions are indicated by solid lines. 
    Horizontal dashed lines show nominal masses of the neutron and $\Xi$ 
    baryon. The vertical gray dashed line indicates the range of pure 
    missing-mass separation between these two reactions.}
    \label{fig:KPlNmiss}
}
\end{figure}

Charged-particle track detection efficiency stays flat over the full range of 
$W$, but kaon reconstruction efficiency drops from about 60\% at low 
$W$ to 20\% at highest available energy. Since the GlueX acceptance is 
large and essentially hole-less, kaon reconstruction efficiency does 
not depend on yet unknown angular distributions.  For the final 
selection of the $K_Lp\to K^+n$ reaction, we used a $3\sigma$ 
missing-mass cut around the neutron's mass; see Fig.~\ref{fig:KPlNmiss}.

Figure~\ref{fig:KPlNmiss} was plotted under the assumption of a 150~ps
vertex time resolution. Both $W$ (Fig.~\ref{fig:KPln1}) and 
missing-mass resolutions are driven by the $K_L$ momentum resolution. That
is why a start counter upgrade is essential. Any further time
resolution improvement below 150~ps would significantly simplify 
reaction analysis and background suppression for all reactions of
interest.

Below $W = 2.3$~GeV, the $K_Lp\to K^+n$ and $K_Lp\to K^+\Xi$ reactions
can be disentangled by $K^+$ missing mass alone. Above this value,
special treatment of the $K_Lp\to K^+\Xi$ background is required.
One may notice that a $3\sigma$ cut for the $K_Lp\to K^+n$ reaction
rises faster than for $K_Lp\to K^+\Xi^0$. This effect has a purely
kinematical explanation - due to the higher mass of the $\Xi^0$ baryon, 
the $K^+$
produced in $K_Lp\to K^+\Xi$ reaction has a lower energy for the
same value of $W$. The lower the $K^+$ energy we have, the better 
missing-mass resolution we get, and the more narrow the missing-mass 
cut one needs to apply.

\paragraph{$K_Lp\to K^+n$ background suppression:}
\label{sec:KPlNbg}

Due to its very high cross section, the $K_Lp\to K^+n$ reaction is 
essentially background free. Due to the extremely high statistics 
expected for this reaction our uncertainties will be dominated by 
systematics. We have identified three major sources of physical 
background: $np\to K^+nn$, $np\to\pi^+nn$, and $K_Lp\to K^+\Xi$ reactions.

Details on $K_Lp\to K^+n$ and $K_Lp\to K^+\Xi$ separation can be found
in Appendix~A5~(Section~\ref{sec:App_KPlXi}). For $W<2.3$~GeV, these two 
reactions can
be separated by a $3\sigma$ $K^+$ missing-mass cut. Above $W=2.3$~GeV, 
one can use standard background suppression techniques - S-weights, 
Q-weights, NeuralNets, etc. $\ldots$. The main decay branch of $\Xi$ is 
$\Xi^0 \to \pi^0\Lambda \to \pi^0\pi^-p$, which leads to several charged 
particles in the final state besides $K^+$; hence filtered out by a 
``one-charge-track-only" selection criterion. Another decay branch 
$\Xi^0\to\pi^0 \Lambda \to \pi^0 \pi^0n$ cannot be filtered out that 
easily; however, due 
to its smaller branching ratio combined with the small $K_Lp\to K^+\Xi$ 
production cross section, this channel only contributes at the level of 
$10^{-3}$ even without any background suppression techniques. 
Further suppression vetoing multiple neutral tracks and/or Q-weight 
should push this background far below $10^{-4}$.

\begin{figure}[h!]
\centering{
    \includegraphics[width=0.6\textwidth,keepaspectratio]{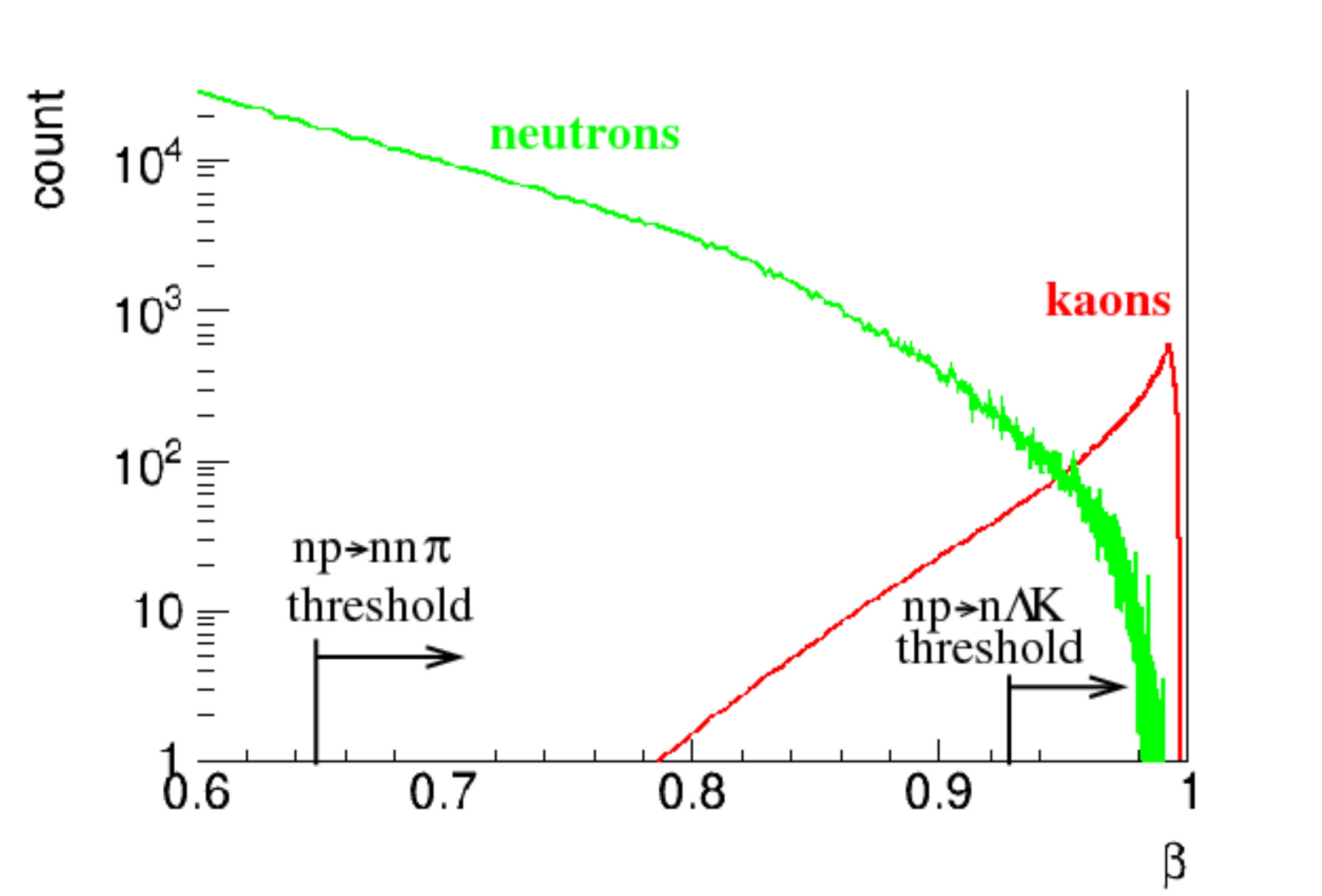}

    \caption{Neutron and $K_L$ fluxes as a function of speed $\beta$.}
    \label{fig:NeutFluxBeta}
}
\end{figure}

Neutron flux drops exponentially with energy (see Appendix~A4~{\ref{sec:A4}} 
for details) and generally the high-energy neutron flux is small, but 
nonvanishing. If neutrons and $K_L$s have the same speed, they cannot be 
separated by time of flight. Neutron-induced reactions have high cross 
sections, which is why one needs to consider them as a possible source of 
background. In Fig.~\ref{fig:NeutFluxBeta}, one can see a comparison of
kaon and neutron fluxes for the worse-case scenario when no neutron
suppression is employed, similar to Fig.~\ref{fig:neutron}(right) in
terms of $\beta$. Particles with the same $\beta$ cannot be separated
by time of flight. At $\beta=0.95$ neutron and kaon fluxes become equal.
This speed corresponds to a neutron momentum of $p_n = 2.9$~GeV/$c$ 
and kaon momentum of $p_K =1.5$~GeV/$c$.

\begin{figure}[h!]
\centering{
    \includegraphics[width=0.45\textwidth,keepaspectratio]{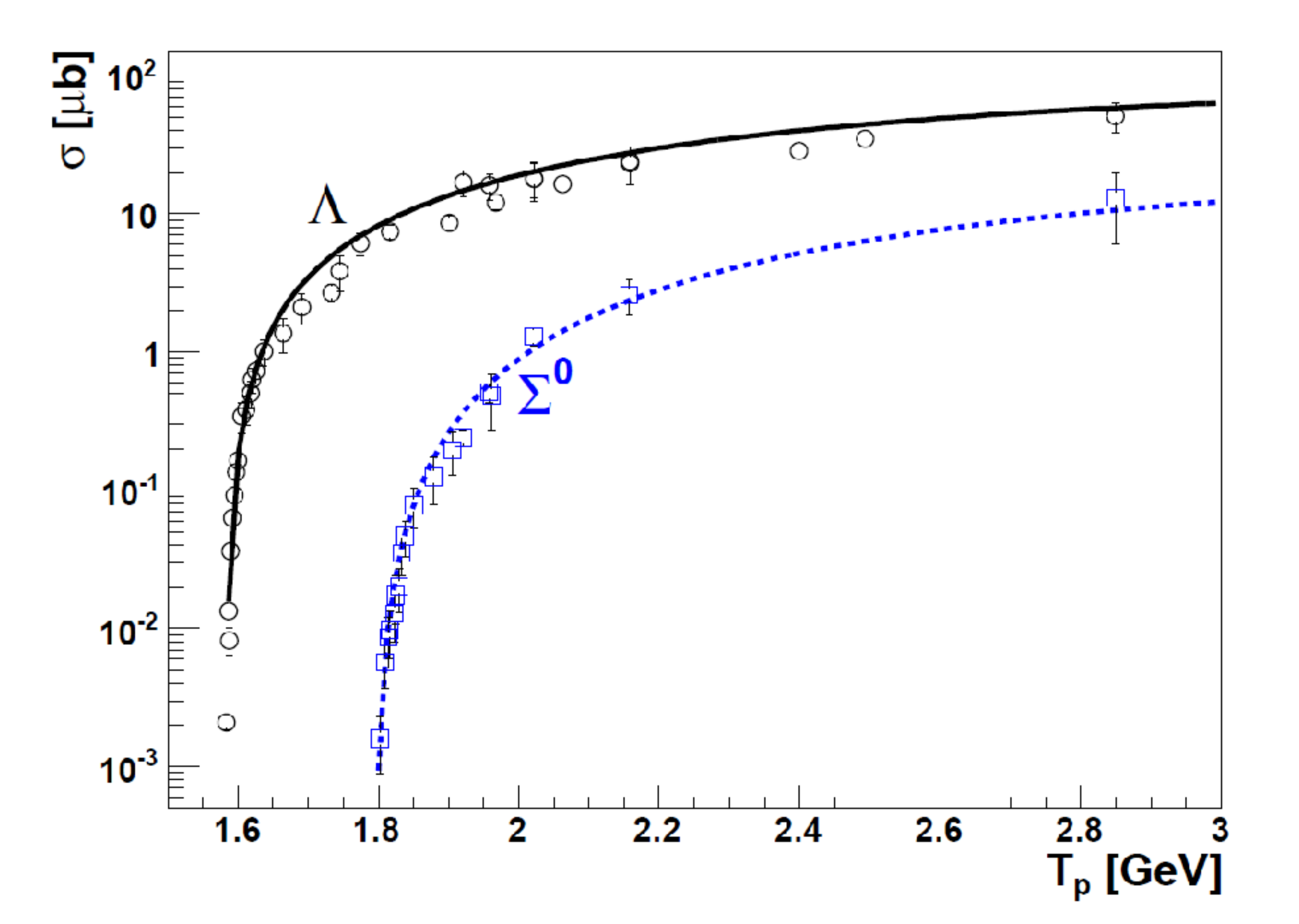}
    \includegraphics[width=0.47\textwidth,keepaspectratio]{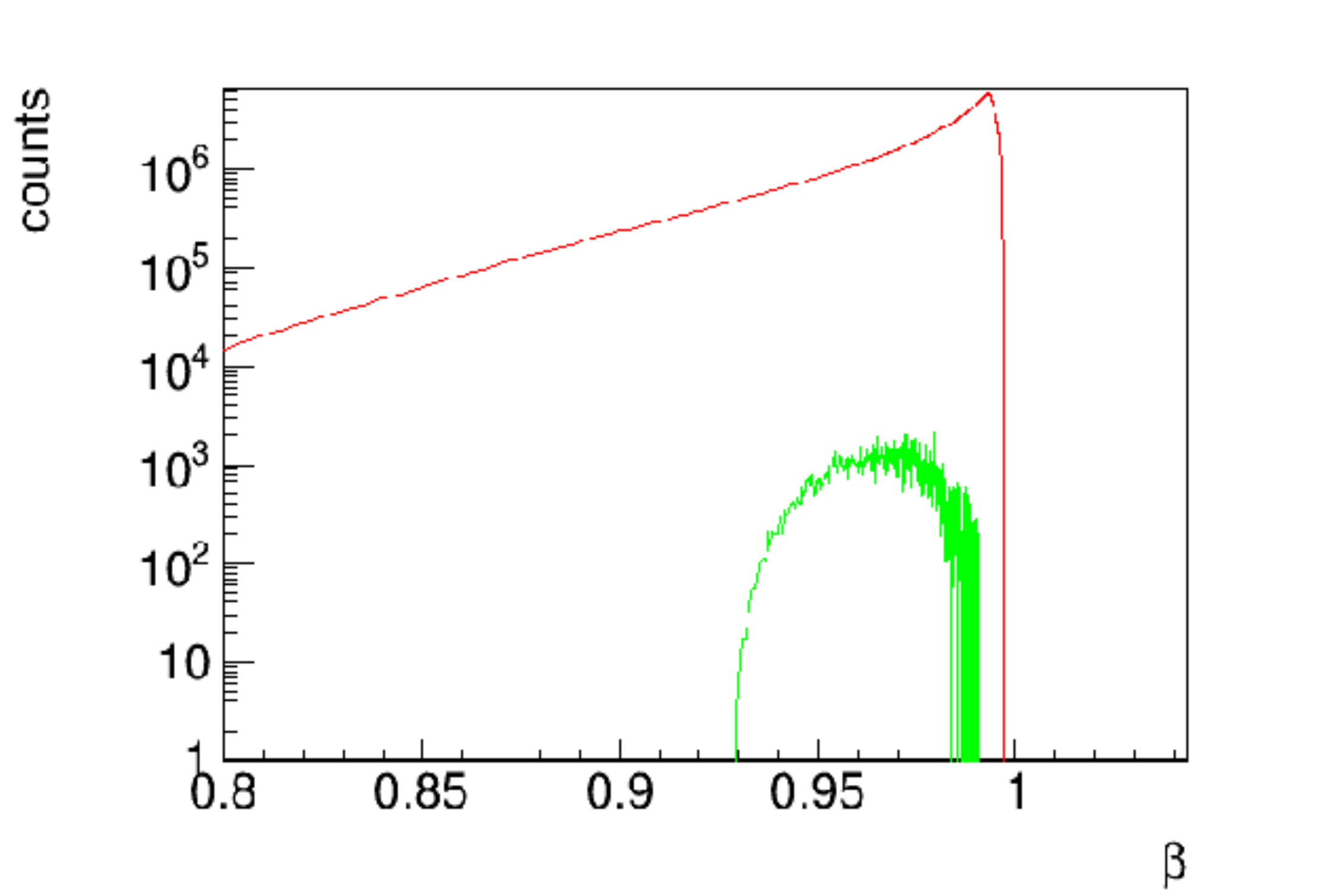}

    \caption{Left plot: $pp\to K^+ \Lambda p$ total cross section from
    Ref.~\protect\cite{Valdau2011}. Right plot: $K^+$ flux as a function 
    of projectile speed $\beta$ for neutron-induced (green) and 
    kaon-induced (red) reactions.}
    \label{fig:KPLambda}
}
\end{figure}

To evaluate the amount of background, we need to fold this flux with
production cross section and reconstruction efficiency.
Let's first consider the $np\to K^+\Lambda n$ background.
Unfortunately, this reaction is not very well measured, so we would
use the $pp\to K^+\Lambda p$ cross-section parametrization together
with the knowledge of $\frac{\sigma(pp\to K^+ \Lambda p)}{\sigma(np\to
K^+ \Lambda n)} = 2$ from Ref.~\cite{Valdau2011}.
In Fig. ~\ref{fig:KPLambda}, one can see the flux of $K^+$s from 
kaon-induced $K_Lp\to K^+n$ reaction in comparison to a neutron-induced
$np\to K^+\Lambda n$ as a function of projectile speeds.

As one can see in Fig.~\ref{fig:KPLambda}, neutron-induced $K^+$
production contributes only in a very narrow range of energies. The 
contribution
is also very small. One can further suppress this type of background by
vetoing charged particles from $\Lambda$ decay and performing a $K^+$
missing-mass cut. Altogether one can suppress this type of background
below $10^{-4}$.

\begin{figure}[h!]
\centering{
    \includegraphics[width=0.6\textwidth,keepaspectratio]{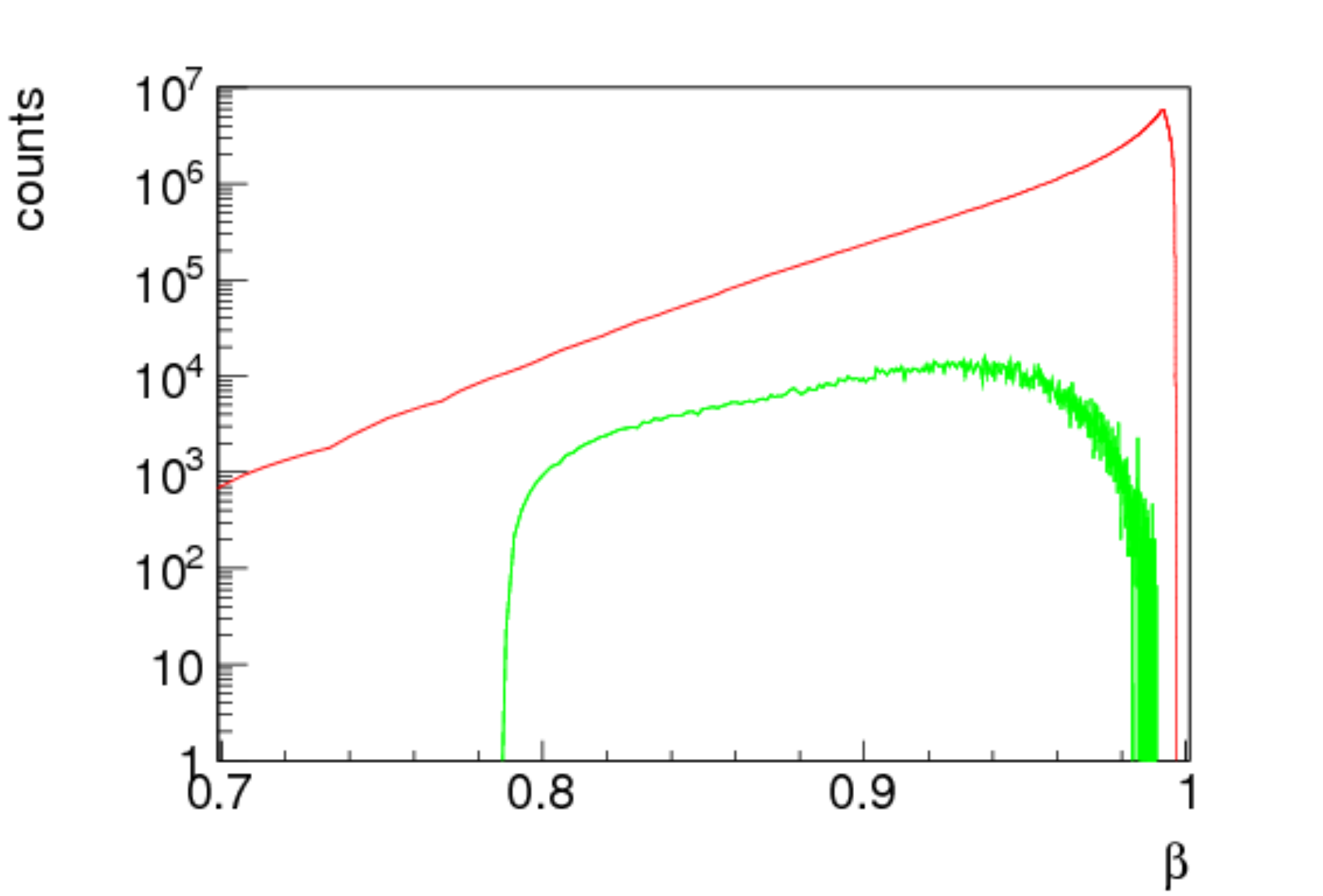}

     \caption{$K^+$ flux as a function of projectile speed for the
     $np\to\pi^+nn$ (green) and $K_Lp\to K^+n$ (red) reactions. Pion
     misidentification efficiency for the neutron-induced reaction is
     extracted from the full MC Geant simulation.}
     \label{fig:NNpi_beta}
}
\end{figure}

The most dangerous type of neutron-induced background originates from
the $np\to\pi^+nn$ reaction with fast $\pi^+$ misidentification as $K^+$.
There are no measurements of $np\to\pi^+nn$ reaction but due to isospin
symmetry one can relate this reaction to an isospin symmetric case
$np\to\pi^-pp$. The later reaction is known, see Ref.~\cite{wasa2017}.
The total cross section for this reaction is about 2~mb. The
$np\to\pi^+nn$ reaction has a much lower threshold compared to $np\to
K^+\Lambda n$, so it can utilize an enormous flux of low-energy neutrons.
However, low-energy neutrons predominately produce low-energy pions,
which can be separated from kaons. The background needs to be
considered only for $\beta>0.8$; see Fig.~\ref{fig:NNpi_beta}. The
background level looks much higher compared to Fig.~\ref{fig:KPLambda},
but it can be severely suppressed with the ``$K^+$" missing-mass cut
since pion kinematics of the three-body $np\to\pi^+nn$ reaction are very
different from $K_Lp\to K^+n$.

\underline{In summary:} Kaon particle identification together with a simple 
$3\sigma$ missing-mass cut and assumption of $K_L$ beam can efficiently 
suppress all physical backgrounds of the $K_Lp\to K^+n$ reaction.

\section{Appendix~A6: Current Hadronic Projects} 
\label{sec:A6}

Past measurements involving kaon scattering measurements were made
at a variety of laboratories, mainly in the 1960s and 1980s when
experimental techniques were far inferior to the standards of today
(short summary is given in Sec.~\ref{sec:data}).  It is important to 
recognize that current projects are largely complementary to the 
proposed Jlab KL hadron beam facility. We \underline{summarize} the 
status of the FNAL, J-PARC, Belle, BaBar, $\overline{P}$ANDA, and 
COMPASS efforts here.

\subsection{Project~X, USA}

The status of Project~X at FNAL~\cite{ProjectX,ProjectXa} is as follows:
First stage of Project~X aims for neutrinos.  Proposed $K_L$ beam can
be used to study rare decays and CP-violation~\cite{Quigg2015}. It
may be impossible to use the FNAL $K_L$ beam for hyperon spectroscopy
because of momentum range and $n/K_L$ ratio (columns 4 and 6 at
Table~\ref{tab:projectx}). In particular, the 8-yr old FNAL LoI
addressed to the CP-violation study proposed to have a neutral kaon
beam rate of $10^{10}$/hr for high energies and very broad energy
binning~\cite{Winstein88}.
\begin{table}[htb!]

\centering \protect\caption{Comparison of the $K_L$ production yield. 
       The BNL AGS kaon and neutron yields are taken from RSVP reviews 
       in 2004 and 2005. The Project~X yields are for a thick target, 
       fully simulated with LAQGSM/MARS15 into the KOPIO beam solid 
       angle and momentum acceptance from Ref.\protect\cite{ProjectXa}.}
\vspace{2mm}
{%
\begin{tabular}{|c|ccccc|}
\hline
Project   & Beam energy & Target      & p($K_L$)   & $K_L$/s        & n/$K_L$ \\
          &    (GeV)    &($\lambda_I$)&  (MeV/$c$) & (into 0.5~msr) & ($E_n >$10~MeV) \\
\hline
BNL~AGS   &   24        & 1.1~Pt      & 300--1200& $60\times 10^6$  & $\sim 1:1000$ \\
Project~X &    3        & 1.0~C       & 300--1200& $450\times 10^6$ &$\sim 1:2700$ \\
\hline
\end{tabular}} \label{tab:projectx}
\end{table}

\subsection{J-PARC, Japan}

While J-PARC has a whole program of charged strange particle and hypernuclear
reactions, the photon beam at GlueX KLF allows unique access to other channels.
J-PARC provides separated secondary beam lines up to 2~GeV/$c$
(Table~\ref{tab:jparc}). The operation of the Hadron Experimental Facility
resumed in April of 2015 following a two-year suspension to renovate the
facility after the accident that occurred in May 2013~\cite{JPARC_accident}.
The primary beam intensity is currently 25~kW, and can be upgraded to 100~kW. 
This will correspond to $\sim 10^9$~ppp (particles per pulse) for
pion beam intensity and to $\sim 10^6$~ppp for kaon beam flux. The $K/\pi$
ratio is expected to be close to 10, which is realized with double-stage
electrostatic separators. One of the main problems in the $K/\pi$ separation
is a high duty-factor of the J-PARC Complex.
\begin{table}[htb!]
\centering \protect\caption{J-PARC Beam line specifications from 
        Ref.~\protect\cite{Ohnishi16}
        }
\vspace{2mm}
{%
\begin{tabular}{|c|c|c|c|}
\hline
Beamline& Paricle                                      &Momentum Range& Typical Beam Intensity\\
        &                                              &              & (40~kW MR operation)\\
\hline
K1.8BR  & $\pi^\pm$, $K^\pm$, and p,$\overline{p}$ (separated)  & $<$1.1~GeV/$c$ &$1.5\times 10^5$ K$^-$/spill at 1~GeV/$c$\\
K1.8    & $\pi^\pm$, $K^\pm$, and p,$\overline{p}$ (separated)  & $<$2.0~GeV/$c$ &$5\times 10^5$ K$^-$/spill at 2~GeV/$c$\\ 
K1.1    & $\pi^\pm$, $K^\pm$, and p,$\overline{p}$ (separated)  & $<$1.1~GeV/$c$ &$1.5\times 10^5$ K$^-$/spill at 1~GeV/$c$\\
High-p  & $\pi^\pm$, $K^\pm$, and p,$\overline{p}$ (unseparated)&up to 20~GeV/$c$&$>\sim 10^7$ $\pi^-$/spill at 20~GeV/$c$\\
        &                                                       &        &$>\sim 10^6$ K$^-$/spill at 7~GeV/$c$\\
        & Primary Proton                                        & 30~GeV &$\sim 10^{11}$ proton/spill\\
\hline
\end{tabular}} \label{tab:jparc}
\end{table}

With $K^-$ beams, currently there is no proposal specific for $S=-1$
hyperons, but the cascades will be studied in the early stage of
E50~\cite{E50}, hopefully in this year, 2018. The $\Delta p/p$ is a
few percent, which is not good to look for narrow hyperons. One can
think that the systematic study for $S=-1$ hyperons even with charged
kaons is desirable and J-PARC folks think that such a study is
definitely needed but currently there is no room to accept a new
proposal to require a long beam line. J-PARC is focusing on
hypernuclei physics~\cite{Naruki2015}.  

                           
There is no $K_L$ beam line for hyperon physics at J-PARC. It is 100\%
dedicated to the study of CP-violation. The momentum is spread out
from 1 to 4~GeV/$c$, there is no concept of $\Delta p/p$ since the
beam cannot be focused with EM devices.

\subsection{Belle, Japan}

The Belle Collaboration at KEK has plenty of $e^+e^-$ data, and people
in Belle [Belle Nuclear Physics Consortium (Belle NPC)] are now
extracting various charm-baryon decay processes, which can be used
for cascade resonance spectroscopy, from those ``raw" $e^+e^-$
data~\cite{Belle}.

\subsection{BaBar, USA}

The BaBar Collaboration at SLAC studied, for instance, properties of 
the $\Xi(1530)^0$ in the decay of $\Lambda^+_C\to (\pi^+\Xi^-)K^+$ and
$\Xi(1690)^0$ in the decay of $\Lambda^+_C\to (\overline{K^0}\Lambda)
K^+$~\cite{BaBar} (see, for instance, a recent overview by
Ziegler~\cite{Ziegler16})

\subsection{$\overline{P}$ANDA, Germany}

The $\overline{P}$ANDA experiment~\cite{PANDA} will measure 
annihilation reactions of antiprotons with nucleons and nuclei in 
order to provide complementary and in part uniquely decisive 
information on a wide range of QCD aspects. The scientific scope 
of $\overline{P}$ANDA is ordered into several pillars: hadron 
spectroscopy, properties of hadrons in matter, nucleon structure 
and hypernuclei. Antiprotons are produced with a primary proton 
beam, collected and phase-space cooled in the CR (Collector Ring), 
and then transferred to the HESR (High Energy Storage Ring) where 
they are stacked, further phase-space cooled, and then directed 
onto an internal target located at the center of the 
$\overline{P}$ANDA detector. The facility will start with a 
luminosity of $10^{31}$~cm$^2$/s and a momentum resolution of 
$\Delta p / p = 10^{-4}$, and later improve to $2\times 10^{32}$ 
and $4\times 10^{-5}$, respectively.  The large cross section 
into baryon-antibaryon final states (e.g., $\sim 1~\mu b$ for 
$\Xi\overline{\Xi}$ or $0.1~\mu b$ for $\Omega\overline{\Omega}$) 
make spectroscopic studies of excited multi-strange hyperons a 
very compelling part of the initial program of $\overline{P}$ANDA, 
which is expected to commence by 2025~\cite{Ritman16}.

\subsection{COMPASS, CERN}

COMPASS is thinking of the physics using an RF-separated beam of
charged kaons. It is still in the discussion stage. The rates, which
were presented as a very first guess by the CERN beamline group
were very interesting for a strangeness physics program via
diffractive production of strange resonances~\cite{Paul16}. 
The cost of a RF-separated beam is high; however, something like this
had been built in the past.

Charged kaons could be used to extend the $\chi$PT investigations into
the strangeness sector (e.g., Primakoff) and the spectroscopy
program. At present, COMPASS filters out kaons in the COMPASS
charged pion beam via Cherenkov detectors but they make up only about 
2.6\% of all beam particles.

The energy of the kaon beam would probably be below 100~GeV but above
40 -- 50~GeV. The latter number is defined by the stability of the
power supplies for the beam line, which after all is about 1~km
long... and of course the decay losses.

\section{Appendix~A7: Additional Physics Potential with a $K_L$ Beam}
\label{sec:A7}

As stated in the summary of Mini-Proceedings of the Workshop on 
Excited Hyperons in QCD Thermodynamics at Freeze-Out 
(YSTAR2016)~\cite{sum}: a very interesting further opportunity for 
the KL Facility is to investigate KL reactions on complex nuclei. 
By selecting events with the appropriate beam momentum together 
with a fast forward-going pion, events can be identified, in 
which a hyperon is produced at low relative momentum to the target 
nucleus or even into a bound state. Baryons with strangeness 
embedded in the nuclear environment, hypernuclei or hyperatoms, 
are the only available tool to approach the many-body aspect of 
the three-flavor strong interaction. Furthermore, appropriate 
events with a forward-going $K^+$ could deposit a double-strange 
hyperon into the remaining nucleus, potentially enabling searches 
for and studies of double-$\Lambda$ hypernuclei.

Similarly, the scattering of kaons from nuclear targets could be a 
favorable method to measure the matter form factor (and, therefore, 
neutron skin) of heavy nuclei, with different and potentially 
smaller systematics than other probes. The character of the neutron 
skin, therefore, has a wide impact and the potential to give 
important new information on neutron star structure and cooling 
mechanisms~\cite{Steiner,Horowitz,Zu,Steiner2,Rutel}, searches for 
physics beyond the standard model~\cite{nstarsand,parity}, the 
nature of 3-body forces in nuclei~\cite{Tsang2,Schwenk}, collective 
nuclear excitations~\cite{Centelles,Carbone,Chen,Tamii} and flows 
in heavy-ion collisions~\cite{Li,Tsang}. Theoretical developments 
and investigations will be required to underpin such a program, but 
the science impact of such measurements is high. 

Further potential exists to search for -- or exclude -- possible 
exotic baryonic states that cannot easily be described by the 
usual three-valence-quark structure. Recent results from LHCb provide 
tantalizing hints for the existence of so-called pentaquarks that 
include a charm valence quark; however, the interpretation of those 
results is under discussion. In contrast, elastic scattering of $K_L$ 
with a hydrogen target gives unambiguous information on the potential 
existence of such states. With the given flux of $K_L$ at the 
proposed facility, a clear proof of existence or proof of absence 
will be obtained within the integrated luminosity required for the 
excited hyperon spectroscopy program that forms the basis of this 
proposal. 

There are two particles in the reaction $K_Lp\to\pi Y$ and $KY$ that 
can carry polarization: the target and recoil baryons. Hence, there 
are two possible double-polarization experiments: target/recoil. The 
total number of observables is three. The formalism and definitions of 
observables commonly used to describe the reaction $K_Lp\to KY$ is 
given in Sec.~\ref{sec:PWA}. Although one cannot easily measure 
recoil polarization with GlueX, the self-analyzing decay of hyperons 
makes this possible. Double-polarization experiments, using, e.g., a 
polarized target like FROST~\cite{Chris2016}, will however be left
for future proposal(s).

The physics potential connected with studies of CP-violating decays 
of the $K_L$ is very appealing; however, that topic is not currently 
the focus of this proposal, since a detailed comparison with the 
competition from existing and upcoming experiments is needed in order 
to identify the most attractive measurements that could be done at 
the proposed KL Facility at JLab.

\section{Appendix~A8: List of New Equipment and of Changes in Existing 
	Setup Required}
\label{sec:A8}

The following major changes to existing are summarized below.
\begin{itemize}
\item Compact photon source (a rough cost estimate is about \$500k).
\item Start Counter upgrade.
\item The upgrade of the $LH_2/LD_2$ cryotarget (a rough cost estimate
        is about \$30k).
\item Modifications of the beam line from the beginning of the 
	collimator cave to the cryogenic target, which includes 
	the Be target, the shielding, new vacuum chambers etc:
\begin{itemize}
	\item Engineering and design: 0.5FTE*Y ME and 1FTE*Y MD\footnote{
	ME,MD,MT - mechanical engineer, designer, technician}.
	\item Equipment: \$300k.
	\item Changeover from the photon to KL beam line: 4-6 months 
	depending on the type of shielding, at a level of 3.5FTE MT.
	\item Changeover from the KL beamline to the photon beam line 
	(after the radiological cooldown of the KL beamline): about 
	4-5 months.
\end{itemize}
\end{itemize}

%



\newpage
\section{References}
\begin{thebibliography}{99}
\bibitem{LRP} \textit{The 2015 Long Range Plan for Nuclear Science},
        http://science.energy.gov/np/nsac/ .
\bibitem{PDG2016} C.~Patrignani \textit{et al.} (Particle Data Group),
        Chin.\ Phys.\ C\ \textbf{40}, 100001 (2016).
\bibitem{Nefkens97} B.M.K.~Nefkens, $\pi N$ Newsletter, \textbf{14}, 150 
	(1997).
\bibitem{KI80} R.~Koniuk and N.~Isgur, Phys.\ Rev.\ Lett.\ \textbf{44}, 
	845 (1980).
\bibitem{Ghoul16} H.~Al~Ghoul \textit{et al.} (GlueX Collaboration), AIP\ 
	Conf.\ Proc.\ \textbf{1735}, 020001 (2016), \textit{Proceedings 
	of the 16th International Conference on Hadron Spectroscopy} 
	(Hadron2015), Newport News, VA, Sept. 2015, edited by M. Pennington.
\bibitem{AlekSejevs13} A.~AlekSejevs \textit{et al.} (GlueX Collaboration), 
	arXiv:1305.1523 [nucl--ex].
\bibitem{Qiang2010} Y.~Qiang, Ya.I.~Azimov, I.I.~Strakovsky, 
        W.J.~Briscoe, H.~Gao, D.W.~Higinbotham, and V.V.~Nelyubin,
        Phys.\ Lett.\ B\ \textbf{694}, 123 (2010).
\bibitem{Briscoe2015} W.J.~Briscoe, M.~D\"oring, H.~Haberzettl, 
        D.M.~Manley, M.~Naruki, I.I.~Strakovsky, and E.~Swanson, 
        Eur.\ Phys.\ J.\ A\ \textbf{51}, 129 (2015)
\bibitem{KL2016} Web page of the Workshop on \textit{Physics with Neutral 
	Kaon Beam at JLab} (KL2016), JLab, Newport News, VA, USA, Feb. 2016: 
	https://www.jlab.org/conferences/kl2016/ contains presentations.
\bibitem{YSTAR2016} Web page of the Workshop on \textit{Excited Hyperons in 
	QCD Thermodynamics at Freeze-Out} (YSTAR2016), JLab, Newport News, VA,
        USA, Nov. 2016: https://www.jlab.org/conferences/YSTAR2016/ contains
        presentations.
\bibitem{HIPS2017} Web page of the Workshop on \textit{New Opportunities with
        High-Intensity Photon Sources} (HISP2017), CUA, Washington, DC,
        USA, Feb. 2017: https://www.jlab.org/conferences/HIPS2017/
        contains presentations.
\bibitem{ProcKL}M.~Albrow \textit{et al.}, Mini-Proceedings of the 
	Workshop on \textit{Physics with Neutral Kaon Beam at JLab} (KL2016), 
	JLab,  Newport News, VA, USA, Feb. 2016, edited by M.~Amaryan, 
	E.~Chudakov, C.~Meyer, M.~Pennington, J.~Ritman, and I.~Strakovsky, 
	arXiv:1604.02141 [hep--ph].
\bibitem{LoI} \textit{Physics opportunities with secondary $K_L$ beam
        at JLab}, Spokesperson: M.~Amaryan (GlueX Collaborations), JLab
        LoI12--15--001, Newport News, VA, USA, 2015.
\bibitem{VSP} \textit{Photoproduction of the very strangest baryons on the 
	proton target in CLAS12}, Spokespersons: L.~Guo, M.~Dugger, J.~Goetz, 
	E.~Pasyuk, I.I.~Strakovsky, D.P.~Watts, N.~Zachariou, and V.~Ziegler 
	(Very Strange Collaboration for CLAS Collaboration), JLab Proposal 
	E12--11--005A, Newport News, VA, USA, 2013.
\bibitem{KY} \textit{Nucleon resonance structure studies via exclusive KY 
	electroproduction at 6.6~GeV and 8.8~GeV}, Spokespersons: D.S.~Carman, 
	R.~Gothe, and V.~Mokeev (CLAS Collaboration), JLab E12--16--010A, 
	Newport News, VA, USA, 2016.
\bibitem{ProcYS} P.~Alba \textit{et al.}, Mini-Proceedings of the Workshop 
	on \textit{Excited Hyperons in QCD Thermodynamics at Freeze-Out} 
	(YSTAR2016), JLab,  Newport News, VA, USA, Nov. 2016, edited by 
	M.~Amaryan, E.~Chudakov, K.~Rajagopal, C.~Ratti, J.~Ritman, and 
	I.~Strakovsky, arXiv:1701.07346 [hep--ph].
\bibitem{ProcHI} S.~Ali \textit{et al.}, Mini-Proceedings of the Workshop
	on \textit{New Opportunities with High-Intensity Photon Sources},
	CUA, Washington, DC, USA, Nov. 2016, edited by T.~Horn, C.~Keppel, 
	C.~Munoz-Camacho, and I.~Strakovsky, arXiv:1704.00816 [nucl--ex].
\bibitem{kappa} C.~Amsler, S.~Eidelman, T.~Gutsche, C.~Hanhart, S.~Spanier, 
	and N.A.~T\"ornqvist, in: Ref.~\protect\cite{PDG2016};
	S.~Descotes-Genon, and B.~Moussallam, Eur.\ Phys.\ J.\ C\
	\textbf{48}, 553 (2006).
\bibitem{Julia07} J.~Julia-Diaz \textit{et al.}, Phys.\ Rev.\ C\ \textbf{75},
         015205 (2007).
\bibitem{Sato09} T.~Sato and T.S.-H.~Lee, J.\ Phys.\ G\ \textbf{36},
         073001 (2009).
\bibitem{Liu17} Z.-W.~Liu \textit{et al.}, Phys.\ Rev.\ D\ \textbf{95},
         014506 (2017).
\bibitem{Oller2001} J.~A.~Oller and U.-G.~Mei{\ss}ner, Phys.\ Lett.\ B\ 
	\textbf{500}, 263 (2001).
\bibitem{Maxim16} A.~Ciepl\'y \textit{et al.}, Nucl.\ Phys.\ A\ \textbf{954},
        17 (2016).
\bibitem{Moriya:2013eb} K.~Moriya {\it et al.} (CLAS Collaboration),
       Phys.\ Rev.\ C\ {\bf 87}, 035206 (2013).
\bibitem{Mai:2014xna} M.~Mai and U.~G.~Mei\ss ner, Eur.\ Phys.\ J.\ A\
        {\bf 51}, 30 (2015).
\bibitem{Kamleh16} W.~Kamleh, PoS \textbf{CD15}, 037 (2016).
\bibitem{Scoccola88} N.N.~Scoccola \textit{et al.}, Phys.\ Lett.\ B\
        \textbf{201}, 425 (1988); B\ \textbf{220}, 658 (1989).
\bibitem{Callan88} C.~Callan \textit{et al.}, Phys.\ Lett.\ B\ \textbf{202},
        269 (1988).
\bibitem{Glozman96} L.Ya.~Glozman and D.O.~Riska, Phys.\ Rept.\
        \textbf{268}, 263 (1996).
\bibitem{Liu14} K.-F.~Liu \textit{et al.}, PoS LATTICE2013, 507 (2014).
\bibitem{Coester98} F.~Coester \textit{et al.}, Nucl.\ Phys.\ A\
        \textbf{364}, 335 (1998).
\bibitem{Luscher:1990ux} M.~Luscher, Nucl.\ Phys.\ B\ \textbf{354}, 531 (1991).
\bibitem{Rummukainen:1995vs} K.~Rummukainen and S.A.~Gottlieb, Nucl.\ Phys.\ B\ 
	\textbf{450}, 397 (1995).
\bibitem{Aoki:2007rd} S.~Aoki \textit{et al.}, Phys.\ Rev.\ D\ \textbf{76},
	094506 (2007).
\bibitem{Feng:2010es} Xu~Feng, K.~Jansen, and D.B.~Renner, Phys.\ Rev.\ D\ 
	\textbf{83}, 094505 (2011). 
\bibitem{Dudek:2012xn} J.J.~Dudek, R.G.~Edwards, and C.E.~Thomas, Phys.\ Rev.\ 
	D\ \textbf{87}, 034505 (2013).
\bibitem{Guo:2016zos} D.~Guo, A.~Alexandru, R.~Molina, and M.~D\"oring,
	Phys.\ Rev.\ D\ {\bf 94}, 034501 (2016).
\bibitem{Alexandrou:2017mpi} C.~Alexandrou {\it et al.},
	arXiv:1704.05439 [hep--lat].
\bibitem{Bulava:2016mks} J.~Bulava, B.~Fahy, B.~H\"orz, K.~J.~Juge, 
	C.~Morningstar, and C.~H.~Wong, Nucl.\ Phys.\ B\ {\bf 910}, 842 (2016).
\bibitem{Lang:2011mn} C.B.~Lang, D.~Mohler, S.~Prelovsek, and M.~Vidmar,
	Phys.\ Rev.\ D\ {\bf 84}, 054503 (2011);
	Erratum: [Phys.\ Rev.\ D\ {\bf 89}, 059903 (2014)]
\bibitem{Guo:2012hv} P.~Guo, J.J.~Dudek, R.G.~Edwards, and A.P.~Szczepaniak, 
	Phys.\ Rev.\ D\ \textbf{88}, 014501 (2013).
\bibitem{Briceno:2012yi} R.A.~Briceno, and Z.~Davoudi, Phys.\ Rev.\ D\ 
	\textbf{88}, 094507 (2013).
\bibitem{Meissner:2014dea} U.-G.~Mei{\ss}ner, G.~Rios, and A.~Rusetsky,
	Phys.\ Rev.\ Lett.\ \textbf{114}, 091602 (2015).
\bibitem{Liu:2005kr} C.~Liu, X.~Feng, and S.~He, Int.\ J.\ Mod.\ Phys.\ A\ 
	{\bf 21}, 847 (2006).
\bibitem{Lage:2009zv} M.~Lage, U.~G.~Mei\ss ner, and A.~Rusetsky,
	Phys.\ Lett.\ B\ {\bf 681}, 439 (2009).
\bibitem{Wilson:2015dqa} D.J.~Wilson, R.A.~Briceno, J.J~Dudek, R.G.~Edwards, 
	and C.E.~Thomas, Phys.\ Rev.\ D\ \textbf{92}, 094502 (2015).
\bibitem{Wilson:2014cna} D.J.~Wilson,  J.J~Dudek, R.G.~Edwards, and 
	C.E.~Thomas, Phys.\ Rev.\ D\ \textbf{91}, 054008 (2015).
\bibitem{Dudek:2014qha} J.J~Dudek, R.G.~Edwards, and C.E.~Thomas,
	Phys.\ Rev.\ Lett.\ \textbf{113}, 182001 (2014).
\bibitem{Edwards:2011jj} R.G.~Edwards, J.J~Dudek, D.G.~Richards, and
	S.J.~Wallace, Phys.\ Rev.\ D\ \textbf{84}, 074508 (2011).
\bibitem{Engel:2013ig} G.~P.~Engel {\it et al.} (BGR Collaboration),
	Phys.\ Rev.\ D\ {\bf 87}, 074504 (2013).
\bibitem{Dudek:2012ag} J.J.~Dudek and R.G.~Edwards, Phys.\ Rev.\ D\ 
        \textbf{85}, 054016 (2012).
\bibitem{Edwards2013} R.G.~Edwards, N.~Mathur, D.G.~Richards, and
        S.J.~Wallace (Hadrons Spectrum Collaboration), Phys.
        Rev.\ D\ \textbf{87}, 054506 (2013).
\bibitem{Bellwied:2013cta} R.~Bellwied, S.~Borsanyi, Z.~Fodor, S.D.~Katz, 
	and C.~Ratti, Phys.\ Rev.\ Lett.\  {\bf 111}, 202302 (2013).
\bibitem{Borsanyi:2010bp} S.~Borsanyi {\it et al.} (Wuppertal-Budapest 
	Collaboration), JHEP\ {\bf 1009}, 073 (2010).
\bibitem{Bazavov:2014pvz} A.~Bazavov {\it et al.} (HotQCD Collaboration),
	Phys.\ Rev.\ D {\bf 90}, 094503 (2014).
\bibitem{Floris:2014pta} M.~Floris, Nucl.\ Phys.\ A\ {\bf 931}, 103 (2014).
\bibitem{Adamczyk:2013dal} L.~Adamczyk {\it et al.} (STAR Collaboration),
	Phys.\ Rev.\ Lett.\  {\bf 112}, 032302 (2014).
\bibitem{Adamczyk:2014fia} L.~Adamczyk {\it et al.} (STAR Collaboration),
	Phys.\ Rev.\ Lett.\  {\bf 113}, 092301 (2014).
\bibitem{Alba:2014eba} P.~Alba, W.~Alberico, R.~Bellwied, M.~Bluhm, 
	V.~Mantovani Sarti, M.~Nahrgang, and C.~Ratti, Phys.\ Lett.\ B\ 
	{\bf 738}, 305 (2014).
\bibitem{Adam:2016emw} J.~Adam {\it et al.} (ALICE Collaboration),
	arXiv:1606.07424 [nucl--ex].
\bibitem{Dashen:1969ep} R.~Dashen, S.~K.~Ma, and H.~J.~Bernstein, Phys.\ 
	Rev.\ {\bf 187}, 345 (1969).
\bibitem{Venugopalan:1992hy}  R.~Venugopalan and M.~Prakash, Nucl.\ 
	Phys.\ A {\bf 546}, 718 (1992).
\bibitem{Karsch:2003zq} F.~Karsch, K.~Redlich, and A.~Tawfik, Phys.\ 
	Lett.\ B {\bf 571}, 67 (2003).
\bibitem{Tawfik:2004sw} A.~Tawfik, Phys.\ Rev.\ D {\bf 71}, 054502 (2005).
\bibitem{Hagedorn:1976ef} R.~Hagedorn, Prog.\ Sci.\ Culture {\bf 1}, 395 
	(1976).
\bibitem{Capstick:1986bm} S.~Capstick and N.~Isgur, Phys.\ Rev.\ D\ {\bf 34}, 
	2809 (1986).
\bibitem{Ebert:2009ub} D.~Ebert, R.N.~Faustov, and V.O.~Galkin,
	Phys.\ Rev.\ D\ {\bf 79}, 114029 (2009).
\bibitem{Majumder:2010ik} A.~Majumder and B.~Muller, Phys.\ Rev.\ Lett.\  
	{\bf 105}, 252002 (2010).
\bibitem{Bazavov:2014xya} A.~Bazavov {\it et al.}, Phys.\ Rev.\ Lett.\  
	{\bf 113}, 072001 (2014).
\bibitem{Alba:2017mqu} P.~Alba {\ it et al.}, arXiv:1702.01113 [hep--lat].
\bibitem{Giannini:2015zia} M.~M.~Giannini and E.~Santopinto, Chin.\ J.\
        Phys.\  {\bf 53}, 020301 (2015).
\bibitem{Amsler:2008zzb} C.~Amsler {\it et al.} (Particle Data Group),
	Phys.\ Lett.\ B\ {\bf 667}, 1 (2008).
\bibitem{Santopinto:2014opa} E.~Santopinto and J.~Ferretti, Phys.\ Rev.\
        C\ {\bf 92}, 025202 (2015).
\bibitem{Durham} The Durham HEP Reaction Data Databases (UK) (Durham
        HepData): http://durpdg.dur.ac.uk/hepdata/reac.html .
\bibitem{NINA} M.G.~Albrow \textit{et al.}, Nucl.\ Phys.\ B\
        \textbf{23}, 509 (1970).
\bibitem{Albrow} M.G.~Albrow, in: \textit{Workshop on Physics with
        Neutral Kaon Beam at JLab: mini-Proceedings}, arXiv:1604.02141
        [hep--ph] (February, 2016), p.~5.
\bibitem{Brody} A.D.~Brody \textit{et al.}, Phys.\ Rev.\ Lett.\
	\textbf{22}, 966 (1969).
\bibitem{DJ} S.D.~Drell and M.~Jacob, Phys.\ Rev.\ \textbf{138}, B1312
        (1965).
\bibitem{SAID-website} The George Washington University INS Data 
        Analysis Center (SAID); http://\-gwdac.\-phys.\-gwu.edu .
\bibitem{Hoehler84} G.~H\"ohler, \textit{Pion-Nucleon Scattering},
        Landoldt-B\"ornstein, Vol. I/9b2, edited by H.~Schopper
        (Springer-Verlab, Berlin, 1983).
\bibitem{piN} R.A.~Arndt, W.J.~Briscoe, I.I.~Strakovsky, and 
        R.L.~Workman, Phys.\ Rev.\ C\ \textbf{74}, 045205 (2006)
\bibitem{Mark2016} D.M.~Manley, in: \textit{Workshop on Physics with
        Neutral Kaon Beam at JLab: mini-Proceedings}, arXiv:1604.02141
        [hep--ph] (February, 2016), p.~42.
\bibitem{Zhang2013a} H.~Zhang, J.~Tulpan, M.~Shrestha, and 
        D.M.~Manley, Phys.\ Rev.\ C\ \textbf{88}, 035204 (2013).
\bibitem{Zhang2013b} H.~Zhang, J.~Tulpan, M.~Shrestha, and 
        D.M.~Manley, Phys.\ Rev.\ C\ \textbf{88}, 035205  (2013).
\bibitem{Hassall1981} J.K.~Hassall \textit{et al.}, Nucl.\ Phys.\
        B\ \textbf{189}, 397 (1981).
\bibitem{Kamano2014} H.~Kamano, S.X.~Nakamura, T.-S.H.~Lee, and
        T.~Sato, Phys.\ Rev.\ C\ \textbf{90}, 065204 (2014).
\bibitem{Kamano2015} H.~Kamano, S.X.~Nakamura, T.-S.H.~Lee, and 
         T.~Sato, Phys.\ Rev.\ C\ \textbf{92}, 025205 (2015)
\bibitem{Jackson:2015dva} B.~C.~Jackson, Y.~Oh, H.~Haberzettl, and 
	K.~Nakayama, Phys.\ Rev.\ C {\bf 91}, 065208 (2015).
\bibitem{Wilson2015} D.J.~Wilson, J.J.~Dudek, R.G.~Edwards, and 
        C.E.~Thomas, Phys.\ Rev.\ D\ \textbf{91}, 054008 (2015).
\bibitem{Dudek2014} J.J.~Dudek \textit{et al.} (Hadron Spectrum 
        Collaboration), Phys.\ Rev.\ Lett.\ \textbf{113}, 182001 
        (2014).
\bibitem{Ikeda:2012au} Y.~Ikeda, T.~Hyodo, and W.~Weise, Nucl.\ Phys.\ 
	A\ {\bf 881}, 98 (2012).
\bibitem{Guo:2012vv} Z.~H.~Guo and J.~A.~Oller, Phys.\ Rev.\ C\ {\bf 87}, 
	035202 (2013).
\bibitem{Cieply:2011nq} A.~Ciepl\'{y} and J.~Smejkal, Nucl.\ Phys.\ A\ 
	{\bf 881}, 115 (2012).
\bibitem{Mai2012} M.~Mai and U.-G.~Mei{\ss}ner, Nucl.\ Phys.\ A\ {\bf 900},
        51 (2013).
\bibitem{Bruns:2010sv} P.C.~Bruns, M.~Mai, and U.-G.~Mei\ss ner,
	Phys.\ Lett.\ B\ {\bf 697}, 254 (2011).
\bibitem{Estabrooks} P.~Estabrooks \textit{et al.}, Nucl.\ Phys.\ B\
	\textbf{133}, 490 (1978).
\bibitem{LASS} D.~Aston \textit{et al.}, Nucl.\ Phys.\ B\ \textbf{296},
	493 (1988).
\bibitem{LASS1} D.~Aston \textit{et al.}, Nucl.\ Phys.\ B\ \textbf{201},
        169 (1988).
\bibitem{LASS2} D.~Aston \textit{et al.}, Nucl.\ Phys.\ B\ \textbf{292},
        693 (1988).
\bibitem{a46} J.~Gasser and U.-G.~Mei{\ss}ner, Nucl.\ Phys.\ B\ 
	\textbf{357}, 90 (1991).
\bibitem{a47} U.-G.~Mei{\ss}ner and J.~Oller, Nucl.\ Phys.\ A\
	\textbf{679}, 671 (2001).
\bibitem{a48} J.~Oller, E.~Oset, and J.~Palomar, Phys.\ Rev.\ D\ 
	\textbf{63}, 114009 (2001).
\bibitem{a49} M.~Frink, B.~Kubis, and U.-G.~Mei{\ss}ner, Eur.\ Phys.\ 
	J.\ C\ \textbf{25}, 259 (2002).
\bibitem{a50} J.~Bijnens and P.~Talavera, Nucl.\ Phys.\ B\
	\textbf{669}, 341 (2003).
\bibitem{a51} T.A.~L\"ahde and U.-G.~Mei{\ss}ner, Phys.\ Rev.\ D\
	\textbf{74}, 034021 (2006).
\bibitem{a52} V.~Bernard and E.~Passemar, JHEP \textbf{04}, 001 (2010).
\bibitem{a53} Z.-H.~Guo, J.~Oller, and J.~Ruiz~de~Elvira, Phys.\ Rev.\ D\
	\textbf{86}, 054006 (2012).
\bibitem{a54} J.F.~Donoghue, J.~Gasser, and H.~Leutwyler, Nucl.\ Phys.\ B\
	\textbf{343}, 341 (1990).
\bibitem{a58} V.~Bernard and E.~Passemar, Phys.\ Lett.\ B\ \textbf{661},
	95 (2008).
\bibitem{Descotes} S.~Descotes-Genon and B.~Moussallam, Eur.\ Phys.\
	J.\ C\ \textbf{48}, 553 (2006).
\bibitem{Pelaez} J.R.~Pelaez and A.~Rhodas, arXiv:1703.07661.
\bibitem{Kubis2015} F.~Niecknig and B.~Kubis, JHEP {\bf 1510}, 142 (2015).
\bibitem{Doring:2013wka} M.~D\"oring, U.-G.~Mei{\ss}ner, and W.~Wang,
        JHEP {\bf 1310}, 011 (2013).
\bibitem{Boito} D.R.~Boito, R.~Escribano, and M.~Jamin, JHEP\
        \textbf{1009}, 031 (2010).
\bibitem{Bouttiker} P.~B\"uttiker, S.~Descotes-Genon, and B.~Moussallam,
	Eur.\ Phys.\ J.\ C\ \textbf{33}, 409 (2004).
\bibitem{report1}
	https://wiki.jlab.org/cuawiki/images/3/32/Sergey$_-$Abrahamyan$_-$WACS$_-$NPS$_-$2014$_-$update.pdf 
\bibitem{proposalWACS}
	\textit{Polarization observables in wide-angle Compton scattering 
	at photon energies up to 8~GeV}, Spokespersons: B.~Wojtsekhowski,
	S.~Abrahamyan, and G.~Niculescu (Neutral Particle Spectrometer 
	Collaboration), JLab Proposal PR12--15--003, Newport News, VA, USA, 
	2015.
\bibitem{PDBW} P.~Degtyarenko and B.~Wojtsekhowski, in:
        \textit{Workshop on Physics with Neutral Kaon Beam at JLab:
        mini-Proceedings}, arXiv:1604.02141 [hep--ph] (February, 2016),
        p.~214.
\bibitem{cpsW} T.~Horn, C.~Keppel, C.~Munoz-Camacho, I.~Strakovsky,
	\textit{Workshop on New Opportunities with High-Intensity Photon 
	Sources: mini-Proceedings}, arXiv:1704.00816 [nucl--ex] (February,
	2017), p.~61.
\bibitem{pythia} We used modified version of Pythia package for the
         GlueX Collaboration at JLab Hall~D,
         http://home.thep.lu.se/torbjorn/Pythia.html 
\bibitem{Titov2003} A.~Titov and T.-S.H.~Lee, Phys.\ Rev.\ C\ 
        \textbf{67}, 065205 (2003).
\bibitem{McClellan1971} G.~McClellan \textit{et al.}, Phys.\ Rev.\ 
        Lett.\ \textbf{21}, 1593, (1971).
\bibitem{Titov2007} A.~Titov and B.~Kampfer, Phys.\ Rev.\ C\
        \textbf{76}, 035202 (2007).
\bibitem{Mibe2007} T.~Mibe \textit{et al.} (CLAS Collaboration), 
        Phys.\ Rev.\ C\ \textbf{76}, 052202 (2007).
\bibitem{Brandenburg1973} G.W.~Brandenburg \textit{et al.}, Phys.\ 
        Rev.\ D\ \textbf{7}, 708 (1973).
\bibitem{Larin16} I.~Larin, in: \textit{Workshop on Physics with Neutral 
	Kaon Beam at JLab: mini-Proceedings}, arXiv:1604.02141 [hep--ph]
        (February, 2016), p.~198.
\bibitem{geant} Application Software Group, GEANT - \textit{Detector 
        Description and Simulation Tool}, CERN Program Library Long Writeup 
        W5013, CERN, Geneva, Switzerland (1994).
\bibitem{Keller} L.~Keller, private communication, 2015.
\bibitem{Geant4} J.~Allison \textit{et al.} (Geant4 Collaboration),
        Nucl.\ Instrum.\ and Meth.\ A\ \textbf{835}, 186 (2016).
\bibitem{MCNP} T.~Goorley \textit{et al.}, Nucl.\ Tech.\ \textbf{180}, 298.
        (2012); https://mcnp.lanl.gov/ .
\bibitem{ICRP} ICRP 116 Publication, \textit{Conversion Coefficients for 
	Radiological Protection Quantities for External Radiation 
	Exposures}, Annals of the ICRP, \textbf{40}, No 2-5 (2010)
\bibitem{Androic2011} D.~Androic \textit{et al.}, Nucl.\ Instrum.\
        Meth.\ A\ \textbf{646}, 59 (2011).
\bibitem{Spata2016} M.~Spata, private communication, 2016.
\bibitem{pooser-thesis} E.~Pooser, Ph.D. Thesis, Florida International 
	University (2016).
\bibitem{Barbosa2015} F.~Barbosa, C.~Hutton, A.~Sitnikov, A.~Somov,
        S.~Somov, and I.~Tolstukhin, Nucl.\ Instrum.\ and Meth.\ A\
        \textbf{795}, 376 (2015).
\bibitem{Slayer} G.A.~Slayer\textit{et al.}, Phys.\ Rev.\ \textbf{169},
	1045 (1968).
\bibitem{Vorsburgh} K.G.~Vorsburgh \textit{et al.}, Phys.\ Rev.\ D\
	\textbf{6}, 1834 (1972).
\bibitem{Chris2016} C.~Keith, in: \textit{Workshop on Physics with Neutral
        Kaon Beam at JLab: mini-Proceedings}, arXiv:1604.02141 [hep--ph]
        (February, 2016), p.~223.
\bibitem{Meekins} D.~Meekins, TGT-CALC-401-007: \textit{Hall~D
        Cryogenic Target: General calculations for relief of the LH$_2$
        target}.
\bibitem{Seraydaryan2014} H.~Seraydaryan \textit{et al.}, Phys.\
        Rev.\ C\ \textbf{89}, 055206 (2014).
\bibitem{GlueX15} H.~Al~Ghoul \textit{et al.} (GlueX Collaboration)
        arXiv:1512.03699v4 [nucl--ex].
\bibitem{Taylor16} S.~Tylor, in: \textit{Workshop on Physics with Neutral Kaon
        Beam at JLab: mini-Proceedings}, arXiv:1604.02141 [hep--ph]
        (February, 2016), p.~205.
\bibitem{Capiluppi:1982fj} P.~Capiluppi, G.~Giacomelli, G.~Mandrioli, 
	A.M.~Rossi, P.~Serra-Lugaresi, and L.~Zitelli, IFUB-81-25.
\bibitem{SLAC177} R.J.~Yamartino, Ph.~D Thesis, SLAC Standford University,
        May (1974).
\bibitem{Ref.DataCasc} D.A.~Sharov, V.L.~Korotkikh, and D.E.~Lanskoy, Eur.\
       Phys.\ J.\ A\ {\bf 47}, 109 (2011).
\bibitem{Ref.DoubleMoments} S.F.~Biagi \textit{et al.}, Z\ Phys.\ C\ {\bf 34},
       175 (1987).
\bibitem{ARMITAG1977} J.C.M.~Armitage \textit{et al.}, Nucl.\ Phys.\
         B\ \textbf{123}, 11 (1977).
\bibitem{CLINE1970} D.~Cline, J.~Penn, and D.D.~Reeder, Nucl.\ Phys.\ B\
       \textbf{22}, 247 (1970).
\bibitem{BAJLLON1978} P. Bajllon \textit{et al.}, Nucl.\ Phys.\ B\
       \textbf{134}, 31 (1978).
\bibitem{Anisovich:1997qp} A.V.~Anisovich and A.V.~Sarantsev,
	Phys.\ Lett.\ B\ \textbf{413}, 137 (1997).
\bibitem{Cawlfield:2006hm} C.~Cawlfield \textit{et al.}, Phys.\ Rev.\ D\
	\textbf{74}, 031108 (2006).
\bibitem{Delbourgo:1998kg} R.~Delbourgo and M.D.~Scadron, Int.\ J.\ Mod.\ Phys.\ 
	A\ \textbf{13}, 657 (1998).
\bibitem{Scadron:2002mm} M.D.~Scadron, F.~Kleefeld, G.~Rupp, and E.~van~Beveren,
	Nucl.\ Phys.\ A\ \textbf{724}, 391 (2003).
\bibitem{Zhou:2006wm} Z.Y.~Zhou and H.Q.~Zheng, Nucl.\ Phys.\ A\
	\textbf{775}, 212 (2006).
\bibitem{Prevost:1974hf} J.~Prevost {\it et al.} (CERN-Heidelberg-Saclay 
       Collaboration), Nucl.\ Phys.\ B\ {\bf 69}, 246 (1974).
\bibitem{Cameron:1978en} W.~Cameron {\it et al.} (Rutherford-London 
       Collaboration), Nucl.\ Phys.\ B\ {\bf 143}, 189 (1978).
\bibitem{Timmermans:1976gf} J.~Timmermans {\it et al.} 
       (Amsterdam-CERN-Nijmegen-Oxford Collaboration), Nucl.\ Phys.\ B\ 
       {\bf 112}, 77 (1976).
\bibitem{Cameron:1977jr} W.~Cameron {\it et al.} (Rutherford-London 
       Collaboration), Nucl.\ Phys.\ B\ {\bf 131}, 399 (1977).
\bibitem{Ceci:2011ae} S.~Ceci, M.~D\"oring, C.~Hanhart, S.~Krewald, 
	U.-G.~Meissner, and A.~Svarc, Phys.\ Rev.\ C {\bf 84}, 015205 
	(2011).
\bibitem{Burkert:2014wea} V.D.~Burkert {\it et al.}, arXiv:1412.0241 
       [nucl--ex].
\bibitem{Anisovich:2011fc} A.V.~Anisovich, R.~Beck, E.~Klempt, V.A.~Nikonov,
       A.V.~Sarantsev and U.~Thoma, Eur.\ Phys.\ J.\ A {\bf 48}, 15 (2012).
\bibitem{Osmanovic2017} H.~Osmanovi\'{c}, Talk at The International Workshop 
       on Partial Wave Analyses and Advanced Tools for Hadron Spectroscopy 
       (PWA9/ATHOS4, Bad Honnef near Bonn (Germany), March, 2017.
\bibitem{Svarc2017} A.~\v{S}varc, Talk at The International Workshop on 
       Partial Wave Analyses and Advanced Tools for Hadron Spectroscopy 
       (PWA9/ATHOS4, Bad Honnef near Bonn (Germany), March, 2017.
\bibitem{Hoehler93} G.~H\"{o}hler, $\pi$N Newsletter \textbf{9}, 1 (1993).
\bibitem{Kelkar} N.G.~Kelkar and M.~Nowakowski,  Phys.\ Rev.\ A\ \textbf{78},
       012709 (2008), and references therein.
\bibitem{ChewMandelstam} G.F.~Chew and S.~Mandelstam, Phys.\ Rev.\
       \textbf{119}, 467 (1960).
\bibitem{Ceci2008} S.~Ceci, J.~Stahov, A.~\v{S}varc, S.~Watson, and
       B.~Zauner, Phys.\ Rev.\ D\ \textbf{77}, 116007 (2008).
\bibitem{Padde} P.~Masjuan, J.~Ruiz de Elvira, and J.~Jos\'{e}
       Sanz-Cillero, Phys.\ Rev.\ D\ \textbf{90}, 097901 (2014).
\bibitem{L+P2013} A.~\v{Svarc}, M.~Had\v{z}imehmedovi\'{c},
       H.~Osmanovi\'{c}, J.~Stahov, L.~Tiator, and R.L.~Workman, Phys.\ 
       Rev.\ C\ \textbf{88}, 035206 (2013).
\bibitem{L+P2014} A.~\v{Svarc}, M.~Had\v{z}imehmedovi\'{c},
       R.~Omerovi\'{c}, H.~Osmanovi\'{c}, and J.~Stahov,  Phys.\ Rev.\
       C\ \textbf{89}, 45205 (2014).
\bibitem{L+P2014a} A.~\v{Svarc}, M.~Had\v{z}imehmedovi\'{c},
       H.~Osmanovi\'{c}, J.~Stahov, L.~Tiator, and R.L.~Workman,
       Phys.\ Rev.\ C\ \textbf{89}, 065208 (2014).
\bibitem{L+P2015} A.~\v{Svarc}, M.~Had\v{z}imehmedovi\'{c},
       H.~Osmanovi\'{c}, J.~Stahov, and R.L.~Workman, Phys.\ Rev.\
       C\ \textbf{91}, 015207 (2015).
\bibitem{L+P2016} A.~\v{Svarc}, M.~Had\v{z}imehmedovi\'{c},
       H.~Osmanovi\'{c}, J.~Stahov, L.~Tiator, and R.L.~Workman,
       Phys.\ Lett.\ B\ \textbf{755}, 452 (2016).
\bibitem{Mittag-Leffler}Michiel Hazewinkel: \emph{Encyclopaedia
       of Mathematics}, Vol.\textbf{6},  Springer, 31. 8. 1990, pg.~251.
\bibitem{Ciulli}S.~Ciulli and J.~Fischer, Nucl.\ Phys.\ \textbf{24},
       465 (1961).
\bibitem{CiulliFisher}I.~Ciulli, S.~Ciulli, and J.~Fisher, Nuovo\
       Cimento\ \textbf{23}, 1129 (1962).
\bibitem{Pietarinen} E.~Pietarinen, Nuovo\ Cimento\ Soc.\ Ital.\ Fis.\
       \textbf{12A}, 522 (1972).
\bibitem{Pietarinen1} E.~Pietarinen, Nucl.\ Phys.\ B\ \textbf{107}, 21
       (1976).
\bibitem{Workman:2012jf} R.L.~Workman, M.W.~Paris, W.J.~Briscoe, and 
       I.I.~Strakovsky, Phys.\ Rev.\ C\ \textbf{86}, 015202 (2012).
\bibitem{Ronchen:2012eg} D.~R\"onchen, M.~D\"oring, F.~Huang, 
         H.~Haberzettl, J.~Haidenbauer, C.~Hanhart, S.~Krewald, 
         U.-G.~Mei{\ss}ner, and K Nakayama,
         Eur.\ Phys.\ J.\ A\ {\bf 49}, 44 (2013).
\bibitem{Doring:2010ap} M.~D\"oring, C.~Hanhart, F.~Huang, S.~Krewald, 
         U.-G.~Mei{\ss}ner, and D.~R\"onchen, Nucl.\ Phys.\ A\ {\bf 
         851}, 58 (2011).
\bibitem{Doring:2009yv} M.~D\"oring, C.~Hanhart, F.~Huang, S.~Krewald, 
         and U.-G.~Mei{\ss}ner, Nucl.\ Phys.\ A\ {\bf 829}, 170 (2009).
\bibitem{Dick2003} R.A.~Arndt, I.I.~Strakovsky, and R.L.~Workman, Phys.\ 
       Rev.\ C\ \textbf{68}, 042201(R) (2003),
       Ya.I.~Azimov, R.A.~Arndt, I.I.~Strakovsky, and R.L.~Workman, Phys.\ 
       Rev.\ C\ \textbf{68}, 045204 (2003).
\bibitem{Anisovich:2011sv} A.V.~Anisovich, E.~Klempt, V.A.~Nikonov, 
         A.V.~Sarantsev, H.~Schmieden, and U.~Thoma,
         Phys.\ Lett.\ B\ {\bf 711}, 162 (2012).
\bibitem{Tib0} R.~Tibshirani, J.R.~Statist. Soc.\ B\ {\bf 58}, 267 (1996).
\bibitem{Tib1}
    {\it The Elements of Statistical Learning: Data Mining, Inference,
    and Prediction}, T.~Hasti, R.~Tibshirani, J.~Friedman, Springer 2009.
    second ed.; E-book available at
    http://statweb.stanford.edu/~tibs/ElemStatLearn/index.html .
\bibitem{Tib2}
    {\it An Introduction to Statistical Learning},  Gareth James,
    Daniela Witten, Trevor Hastie and Robert Tibshirani, Springer
    2015; 6th printing; E-book available at 
    http://www-bcf.usc.edu/~gareth/ISL/ .
\bibitem{Guegan:2015mea} B.~Guegan, J.~Hardin, J.~Stevens, and M.~Williams,
       JINST {\bf 10}, P09002 (2015).
\bibitem{Landay:2016cjw} J.~Landay, M.~D\"oring, C.~Fern\'andez-Ram\'irez, 
       B.~Hu, and R.~Molina, Phys.\ Rev.\ C\ {\bf 95}, 015203 (2017).
\bibitem{opti} D.~Agadjanov, M.~D\"oring, M.~Mai, U.-G.~Mei{\ss}ner, and
        A.~Rusetsky, arXiv:1603.07205 [hep--lat].
\bibitem{D'Agostini:1993uj} G.~D'Agostini, Nucl.\ Instrum.\ Meth.\ A\ {\bf
        346}, 306 (1994).
\bibitem{Ball:2009qv} R.D.~Ball {\it et al.} (NNPDF Collaboration), JHEP
        {\bf 1005}, 075 (2010).
\bibitem{FTest} L.~Wilkinson and G.E.~Dallal, Technometrics {\bf 23},
        377 (1981).
\bibitem{AD1952} T.~W.~Anderson, and D.~A.~Darling, Annals of
        Mathematical Statistics {\bf 23}, 193 (1952).
\bibitem{Stephens1974} M.~A.~Stephens, Journal of the American Statistical
        Association {\bf 69}, 730 (1974).
\bibitem{cova} M.~D\"oring, arXiv:1603.07265 [nucl--th].
\bibitem{KLmeeting2017} KLong bi-weekly group meeting,
        https://wiki.jlab.org/klproject/index.php/March-1st,-2017.
\bibitem{Valdau2011} Yu.~Valdau  \textit{et al.}, Phys.\ Rev.\ C\ {\bf 84},
        055207 (2011).
\bibitem{wasa2017} P.~Adlarson \textit{et al.}, arXiv:1702.07212.
\bibitem{ProjectX} \textit{Project~X Physics Study},
        https://indico.fnal.gov/event/projectxps12 .
\bibitem{ProjectXa} \textit{Project X: Physics Opportunities}, 
       Part 1 edited by S.D.~Holmes, arXiv:1306.5022 [physics.acc--ph];
       Part 2 edited by A.S.~Kronfeld and R.S.~Tschirhart, arXiv:1306.5009 
       [physics.acc--ph];
        and Part 3 edited D.M.~Asner, P.C.~Bhat, S.~Henderson, R.~Plunkett,
        arXiv:1306.5024 [physics.acc--ph].
\bibitem{Quigg2015} C.~Quigg, private communication, 2015.
\bibitem{Winstein88} B.~Winstein \textit{et al.}, \textit{High precision, high
        intensitivity $K^0$ physics at the main injector}, FNAL LoI 0804, 1988.
\bibitem{JPARC_accident} \textit{Summary of the Report from the 
        Working Group for The External Expert Panel on the 
        Radioactive Material Leak Accident at the Hadron 
        Experimental Facility of J-PARC};
        http://j-parc.jp/\-en/\-topics/\-HDAccident20130827$\underline{~~}$02.p
\bibitem{Ohnishi16} H.~Ohnishi, in: \textit{Workshop on Physics with
        Neutral Kaon Beam at JLab: mini-Proceedings}, arXiv:1604.02141
        [hep--ph] (February, 2016), p.~22.
\bibitem{E50} H.~Noumi \textit{et al.}, \textit{Charmed baryon spectroscopy 
       experiment at J-PARC}, J-PARC Proposal E50, 2012.
\bibitem{Naruki2015} M.~Naruki, private communication, 2015.
\bibitem{Belle} K.~Abe \textit{et al.} (Belle Collaboration), Phys.\ Lett.\
        B\ \textbf{524}, 33 (2002); T.~Lesiak \textit{et al.} (Belle
        Collaboration), Phys.\ Lett.\ B\ \textbf{605}, 237 (2004).
\bibitem{BaBar} B.~Aubert \textit{et al.} (BaBar Collaboration), Phys.\
        Rev.\ D\ \textbf{78}, 034008 (2008); B.~Aubert \textit{et al.}
        (BaBar Collaboration), Phys.\ Rev.\ Lett.\ \textbf{97}, 112001
        (2006); B.~Aubert \textit{et al}. (BaBar Collaboration),
        Phys.\ Rev.\ Lett.\ \textbf{95}, 142003 (2005).
\bibitem{Ziegler16} V.~Ziegler, in: \textit{Workshop on Physics with
        Neutral Kaon Beam at JLab: mini-Proceedings}, arXiv:1604.02141
        [hep--ph] (February, 2016), p.~113.
\bibitem{PANDA} W.~Erni \textit{et al.} ($\overline{P}$ANDA
        Collaboration), arXiv:0903.3905 [hep--ex].
\bibitem{Ritman16} J.~Ritman, invited talk at \textit{Excited Hyperons
        in QCD Thermodynamics at Freeze-Out} Workshop, see 
        Ref.~\protect\cite{YSTAR2016}.
\bibitem{Paul16} S.~Paul, private communication, 2016.
\bibitem{sum}  J.~Goity, P.~Huovinen, J.~Ritman, and A,~Tang, in:
        Workshop on \textit{Excited Hyperons in QCD Thermodynamics at
        Freeze-Out: mini-Proceedings}, arXiv:1701.07346 [hep--ph],
        (November, 2016) p.~158.
\bibitem{Steiner}A.W.~Steiner, M.~Prakash, J.M.~Lattimer, and P.J.~Ellis, 
	Phys.\ Rep.\ \textbf{411}, 325 (2005).
\bibitem{Horowitz}C.J.~Horowitz and J.~Piekarewicz, Phys.\ Rev.\ Lett.\ 
	\textbf{86}, 5647 (2001).
\bibitem{Zu}J.~Xu \textit{et. al.}, Astrophys.\ J.\ \textbf{697}, 1549 
	(2009).
\bibitem{Steiner2}A.W.~Steiner, J.M.~Lattimer, and E.F.~Brown, Astrophys.\ 
	J.\ \textbf{722}, 33 (2010).
\bibitem{Rutel}B.G.~Todd-Rutel and J.~Piekarewicz, Phys.\ Rev.\ Lett.\ 
	\textbf{95}, 122501 (2005).
\bibitem{nstarsand}De-Hua~Wen, Bao-An~Li, and Lie-Wen~Chen, Phys.\ Rev.\ 
	Lett.\ \textbf{103}, 211102 (2009).
\bibitem{parity}S.J.~Pollock and M.C.~Welliver, Phys.\ Lett.\ B\ 
	\textbf{464}, 177 (1999).
\bibitem{Tsang2}M.B.~Tsang \textit{et. al.}, Phys.\ Rev.\ C\ \textbf{86}, 
	015803 (2012).
\bibitem{Schwenk}K.~Hebeler, J.M.~Lattimer, C.J.~Pethick, and A.~Schwenk, 
	Phys.\ Rev.\  Lett.\ \textbf{105}, 161102 (2010).
\bibitem{Centelles}M.~Centelles, X.~Roca-Maza, X.~Vi\~nas, and M.~Warda, 
	Phys.\ Rev.\ Lett.\ \textbf{102}, 122502 (2009).
\bibitem{Carbone}A.~Carbone \textit{et. al.}, Phys.\ Rev.\ C\ \textbf{81}, 
	041301(R) (2010).
\bibitem{Chen}L.W.~Chen \textit{et. al.}, Phys.\ Rev.\ C\ \textbf{82}, 
	024321 (2010).
\bibitem{Tamii}A.~Tamii \textit{et. al.}, Phys.\ Rev.\ Lett.\ \textbf{107}, 
	062502 (2011).
\bibitem{Li}B.A.~Li, L.W.~Chen, and C.M.~Ko, Phys.\ Rep.\ \textbf{464}, 
	113 (2008).
\bibitem{Tsang}M.B.~Tsang \textit{et. al.}, Phys.\ Rev.\ Lett.\ 
	\textbf{102}, 122701 (2009).
\end{thebibliography}
\end{document}